\documentclass[a4paper, amsfonts, amssymb, amsmath, reprint, showkeys, nofootinbib, twoside, floatfix]{revtex4-2}

\usepackage[english]{babel}
\usepackage[utf8]{inputenc}
\usepackage[caption=false]{subfig}
\usepackage{graphicx}
\usepackage{amsmath}
\usepackage{amsfonts}
\usepackage{amssymb}
\usepackage{color}
\usepackage{bm}
\usepackage{mathrsfs}
\usepackage{epstopdf}
\usepackage{url}
\usepackage{footnote}
\usepackage{textcomp}
\usepackage{dsfont}
\usepackage{ulem}
\usepackage{hyperref}
\usepackage{enumerate}  
\usepackage{aas_macros}

\usepackage[usenames]{xcolor}

\newcommand{\nv}[1]{\textcolor{black}{#1}}

\begin{document}
\title[Stellar flybys shape planet-forming discs]{Close encounters: How stellar flybys shape planet-forming discs}

\author{Nicol\'as Cuello}
    \email[Corresponding author: ]{nicolas.cuello@univ-grenoble-alpes.fr}
    \affiliation{Univ. Grenoble Alpes, CNRS, IPAG / UMR 5274, F-38000 Grenoble, France}
    
\author{Fran\c cois M\'enard}
    %\email[Correspondence email address: ]{francois.menar@univ-grenoble-alpes.fr}
    \affiliation{Univ. Grenoble Alpes, CNRS, IPAG / UMR 5274, F-38000 Grenoble, France}

\author{Daniel J. Price}
    %\email[Correspondence email address: ]{daniel.price@monash.edu}
    \affiliation{School of Physics and Astronomy, Monash University, Clayton Vic 3800, Australia}
    
%\date{} % Leave empty to omit a date

\begin{abstract}
%Please provide an abstract of 150 to 250 words.
We review the role of stellar flybys and encounters in shaping planet-forming discs around young stars, based on the published literature on this topic in the last 30 years. Since most stars $\leq~2$~Myr old harbour protoplanetary discs, tidal perturbations affect planet formation. First, we examine the probability of experiencing flybys or encounters: More than 50\% of stars with planet-forming discs in a typical star forming environment should experience a close stellar encounter or flyby within 1000~au. Second, we detail the dynamical effects of flybys on planet-forming discs. Prograde, parabolic, disc-penetrating flybys are the most destructive. Grazing and penetrating flybys in particular lead to the capture of disc material by the secondary to form a highly misaligned circumsecondary disc with respect to the disc around the primary. One or both discs may undergo extreme accretion and outburst events, similar to the ones observed in FU~Orionis-type stars. Warps and broken discs are distinct signatures of retrograde flybys. Third, we review some recently observed stellar systems with discs where a stellar flyby or an encounter is suspected --- including UX~Tau, RW~Aur, AS~205, Z~CMa, and FU~Ori. Finally, we discuss the implications of stellar flybys for planet formation and exoplanet demographics, including possible imprints of a flyby in the Solar System in the orbits of trans-Neptunian objects and the Sun's obliquity.
\end{abstract}

%\keywords{protoplanetary discs; planet formation; star formation; dynamics; hydrodynamics; celestial mechanics; observations; exoplanets; Solar System}

\maketitle

%%%%%%%%%%%%%%%%%%%%%%%%%%%%%%%%%%%%%%%%%%%%%%%%%%%%%%%%%%
\section{Introduction}
\label{sec:introduction}

Most, if not all, stars are born in clusters \citep{Lada&Lada2003}. In active star forming regions of the Milky Way, we observe circumstellar discs with a large spread in disc orientation, mass, and size \citep{Tobin+2016, Segura-Cox+2018, Tychoniec+2018, Tobin+2020, Tobin+2022, Manara+2022}. In such environments, dynamical interactions between discs and stars --- either on bound or unbound orbits --- are common and lead to efficient disc truncation and stripping. Episodic, misaligned gas accretion can modify the discs' angular momentum budget \citep{Bate2018, Kuffmeier+2020} and deeply affect their subsequent evolution. The fate of stars and discs are intimately connected. 

The fraction of stars in multiple stellar systems --- defined as \textit{multiplicity fraction} --- is initially very high ($\gtrsim 50\%$) and decreases with time \citep{Raghavan2010, Chen+2013, Duchene&Kraus2013, Offner+2022, Gaia+2022-multi}, though still remaining above 50\% for F, G and K-type stars on the main sequence in the solar neighbourhood \citep{DuquennoyMayor1991, Offner+2022}. \citet{Larson1972} first formulated the hypothesis that all stars could be born in multiple stellar systems. This could explain the growing number of observed multiple stellar systems in surveys, possible thanks to new generations of high resolution instruments on large telescopes \nv{and state-of-the-art data analysis techniques}.

Young multiple systems form in non-hierarchical configurations, which decay towards more stable orbital arrangements \citep{Reipurth+2014, Tokovinin2021}. If one were to form triple stellar systems by randomly placing the stars within a volume, more than 98\% of these would be in unstable configurations \citep{Anosova1986} that would eventually decay (typically in about 100 crossing times). The lowest-mass member is the most likely to be ejected in the decay of a triple system \citep{Reipurth+2010}. However, sometimes the third body has insufficient energy to escape and returns. In this case, multiple encounters occur before the third star is finally ejected. The process of hierarchical decay implies that \nv{a fraction of} single stars could have escaped from unstable multiples. This naturally produces a population of unbound perturbers \citep[e.g.][]{Schoettler+2020} roaming in the galaxy.

In the following, we define a \textit{flyby} to occur when a star on a parabolic or hyperbolic orbit (with eccentricity $e\geq1$) perturbs another star. By definition, this event happens only once for a given pair of stars. By contrast, we define a given star experiencing repeated encounters with another star as a stellar \textit{encounter}. In this case, $e~<~1$. If discs are present around the stars, they experience periodic tidal perturbations at a frequency equal to the binary or triple orbital frequency.

When stars are young ($\leq 2$--$3$ Myr old), there is a $\gtrsim 40\%$ chance that either star participating in an encounter or a flyby harbours a disc \citep{Mamajek2009,Pfalzner2013}. Since discs are much more radially extended compared to stars, they experience strong tidal perturbations. Stellar multiplicity thus translates (theoretically) into smaller disc sizes and masses (compared to single stars), along with a large spread of mutual orbital inclinations \citep{Bate2018, Wurster+2019, Bate2019, Lebreuilly+2021}. This is observed: Surveys in mm-continuum emission have consistently found that discs in multiple systems are more radially truncated in mm-continuum emission compared to those around single stars \citep{Manara+2019, Akeson+2019, Czekala+2019, Zurlo+2020, Zurlo+2021}.

\citet{Clarke&Pringle1993} first investigated the dynamical effects of tidal encounters on discs considering different orbital inclinations (prograde, orthogonal, and retrograde) and periastron distances. Their simulations revealed that flybys lead to i) tidal stripping, ii) spiral arms, iii) disc truncation, iv) capture of material, v) warping, and vi) ejection of unbound material.

Our aim in this review is to outline the role of stellar flybys and encounters in shaping planet forming discs. We ultimately wish to understand the effects of environment and multiplicity on planet formation. Our review is structured as follows: In Section~\ref{sec:flybys} we examine when and where stellar flybys and encounters occur in order to estimate the likelihood of these events. In Section~\ref{sec:discdynamics}, we review the dynamical effects of flybys on planet-forming discs with special emphasis on disc morphology, evolution, and accretion as a function of orbital parameters. Then, in Section~\ref{sec:observations}, we report some recent observations of discs where a stellar flyby or an encounter is suspected. Finally, in Section~\ref{sec:implications}, we discuss the main implications of flybys and encounters for planet formation. We conclude in Section~\ref{sec:conclusions}.

%%%%%%%%%%%%%%%%%%%%%%%%%%%%%%%%%%%%%%%%%%%%%%%%%%%%%%%%%%
\section{When and where do stellar flybys and encounters occur?}
\label{sec:flybys}

Stellar flybys and encounters occur because the majority of stars form in a clustered environment, and stars spend their early life in regions of enhanced stellar density \cite[and references therein]{Lada&Lada2003, Porras+2003, Winter+2018a, Galli+2019}. The rate at which stars in a given cluster experience stellar flybys depends on the average stellar density $\langle \rho_* \rangle$. For massive young stellar clusters in the Milky Way, $\langle \rho_* \rangle$ varies from $0.01$ to several $10^5$ $M_\odot\,{\rm pc}^{-3}$ and the typical cluster radii range between 0.1 pc and several tens of pc \citep{Pfalzner2013}.

In the following, we assume that one of the stars (denoted as the primary, with mass $M_1$) harbours a protoplanetary disc. We define $r_{\rm peri}$ as the periastron distance of the perturber (denoted as the secondary star, with mass $M_2$). The mass ratio $q=M_2/M_1$ determines the intensity of the gravitational perturbation experienced by the disc \citep{Pfalzner2003, Cuello+2019b}. Typically, the circumprimary disc has a radial extension (\nv{denoted as} $R_{\rm disc}$) ranging between a few tens to hundreds of astronomical units (au) at most. It is therefore reasonable to introduce a critical minimum distance below which one considers that a flyby has occurred. Following previous works on stellar clusters \citep[e.g.][]{Pfalzner2013, Winter+2018b}, we consider that a stellar flyby or encounter occurs whenever the distance between two stars is less than $1000$ au. Using this definition and for typical cluster values, one may compute the average time (denoted $\tau$) until a star undergoes a flyby with a given periastron distance $r_{\rm peri}$ \citep{Davies2011}:
\begin{equation} \label{eq:tau}
    \tau \approx 33\,{\rm Myr} \left( \frac{100\,{\rm pc}^{-3}}{n} \right)  \bigg(\frac{v_\infty}{1\,{\rm km/s}} \bigg) \bigg(\frac{10^3\,{\rm au}}{r_{\rm peri}}\bigg) \bigg(\frac{M_\odot}{m} \bigg) ,
\end{equation}
where $n$ is the stellar number density in pc$^{-3}$, $v_\infty$ is the mean relative velocity at infinity of the stars in the cluster, and $m$ is the total mass of the stars involved in the flyby. As expected, $\tau$ decreases for increasing values of~$n$: Flybys are more frequent in dense environments. The dependence on stellar mass implies that more massive stars experience flybys more frequently \nv{--- which is partially due to gravitational focussing. Remarkably}, massive stars also have a higher multiplicity fraction compared to less massive stars \citep{Duchene&Kraus2013, Moe+2017}. Other more refined estimates have been proposed to compute either the probability for one star to suffer \textit{at least} one flyby occurring in a given interval of time \citep{Munoz+2015}, or the probability that a star experiences a flyby closer than a given separation \citep{Winter+2018b}. For the latter, in a 3-Myr old cluster of uniform density composed of stars of 1~$M_\odot$ and with a velocity dispersion of 4~${\rm km\,s}^{-1}$, the probability of a flyby at less than 1000 au is above 70\% (for stellar densities above 500~${\rm pc}^{-3}$) \citep[see their figure 7]{Winter+2018b}.

\begin{figure}[ht!]
    \centering
    \includegraphics[width=8.6cm]{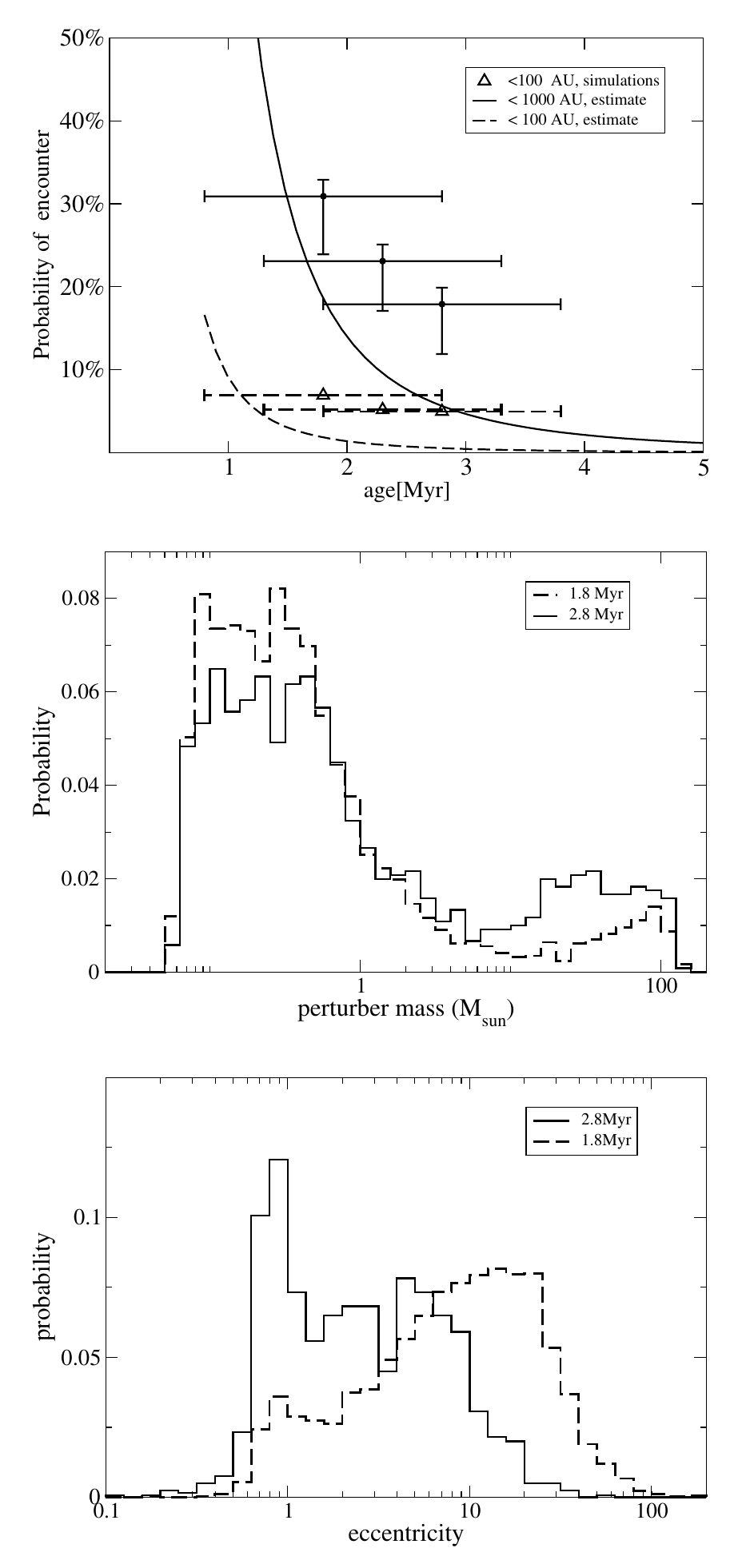}
    \caption{Figures extracted from the work by Pfalzner \citep{Pfalzner2013}. Results obtained for Solar-type stars ($0.8<M<1.2\,M_\odot$) in leaky clusters, located within 0.5 pc from the cluster centre. \textit{Top:} Percentage of stars experiencing a flyby as a function of cluster age. The dots and triangles show $N$-body simulation results for different minimum separations. The solid and dashed lines show the analytical estimates obtained with Eq~\ref{eq:tau}. \textit{Middle:} Distribution of the perturber mass during flybys at 1.8 (dashed line) and 2.8 Myr (solid line). \textit{Bottom}: Eccentricity distribution of flybys at 1.8 (dashed line) and 2.8 Myr (solid line). \nv{Credit: Pfalzner 2013, A\&A, 549, A82, reproduced with permission $\copyright$ ESO.}}
    \label{fig:plots-Pfalzner2013}
\end{figure}

A related question is: \textit{What fraction of stars with discs experience at least one stellar flyby or encounter in a given environment?} Answering it is non-trivial since it depends on the detailed stellar distribution within the cluster, which is not accounted for in Equation~\ref{eq:tau}. Yet, using $N$-body simulations it is possible to estimate the probability of encounter as a function of the cluster age; the mass distribution; and the eccentricity distribution for the stellar perturbers. Fig.~\ref{fig:plots-Pfalzner2013} shows the results obtained by Pfalzner \citep{Pfalzner2013} when modelling a typical leaky cluster \nv{($\langle \rho_* \rangle< 1$--$10^3 M_\odot\, {\rm pc}^{-3}$)} or OB association --- similar to the birth cluster of the solar system. The top panel shows the percentage of solar-type stars ($0.8\,M_\odot<M_1<1.2\,M_\odot$), located within 0.5 pc from the cluster centre, that experience a stellar flyby. The data points are extracted from simulations; while the solid and dashed lines show the probabilities for flybys at distances below 1000~au and 100~au (respectively), as a function of time using Equation~\ref{eq:tau}. Notice that the analytical estimate underestimates the percentage of stars perturbed by flybys. Extrapolating to young stars ($\lesssim$1~Myr), this percentage is close to 50\%. It then decreases with cluster age, although the fraction is still around 20\% at 3~Myr. Therefore, \textit{more than half of the stars \nv{undergo encounters within 1000~au which could potentially affect the planet-forming disc}}. At face value, this statement implies that --- even if stellar clusters eventually disperse --- planet-forming discs perturbed by young nearby stars constitute the rule rather than the exception (within the entire disc population). So, potentially \textit{all} planetary systems are shaped by flybys passing within 1000 au!

The middle panel of Fig.~\ref{fig:plots-Pfalzner2013} shows the perturber mass distribution for stellar flybys around solar-type stars from \citep{Pfalzner2013}, in leaky clusters at two different ages (1.8 and 2.8~Myr). This calculation assumes that the stellar masses in the cluster follow an initial mass function (IMF) with an average stellar mass of 0.5~$M_\odot$. The mass distribution obtained is bimodal, with a clear prevalence of less massive perturbers ($\leq1\,$M$_\odot$). The second peak is less significant and it is located between 10 and 100~$M_\odot$, according to the cluster age. This mass distribution directly informs us about the distribution of $q=M_2/M_1$ for flybys around solar-type stars. The most likely values of $q$ range between 0.1 and 1, with values larger than a few being at least three times less probable. Last, the bottom panel of Fig.~\ref{fig:plots-Pfalzner2013} gives the distribution \nv{of the eccentricities of the encounters} for the same encounters as in the middle panel. Remarkably, for 2.8-Myr old clusters, the most likely eccentricity is around 1 --- as if each star of a given pair \nv{had been} initially approaching from infinity. \nv{This occurs because the mean eccentricity of an encounter is directly related to the cluster density. Thus, because the cluster density declines with age, the density is lower at 2.8 Myr and only the most strongly gravitationally focused encounters (i.e. with $e \approx 1$) have pericentres at $< 1000$ au.} The situation is different for younger clusters (1.8 Myr) where hyperbolic flybys ($e \gtrsim 3$) are preferred, with the most likely eccentricity $\approx 5$--$30$. So, as clusters evolve, parabolic flybys ($e=1$) become the most probable for stars with discs.

One may argue that stellar flybys should be rare in young low-mass clusters. \nv{However,} \citet{Pfalzner+2021b} showed that the mean and \nv{perhaps} also the central stellar density of low-mass clusters are approximately the same as in high-mass clusters --- because high mass clusters are also larger. This counter-intuitive fact implies that close stellar flybys are also common in low-mass clusters. \nv{The claim that low mass clusters contain central regions with stellar densities as high as $10^4 M_\odot \,{\rm pc}^{-3}$ is more controversial and depends on the density profile assumed for the cluster.} Nevertheless, dedicated surveys of nearby molecular clouds \citep{Hatchel+2005, Joncour+2017, Joncour+2018, Robitaille+2020} have revealed that star formation not only occurs in clusters but also within dense filamentary structures of gas --- both at large and small scales. For instance, in Perseus, roughly 80\% of the star formation takes place in clusters with stellar densities over 1 $M_\odot\,{\rm pc}^{-3}$. Half of these young stars form in clusters containing less than a few tens of members \citep{Hatchel+2005}. In addition, using state-of-the-art algorithms such as {\sc dbscan}\footnote{Density-Based Spatial Clustering of Applications with Noise.}, it is possible to reliably extract over-dense structures from the sky-projected spatial distribution of stars. In Taurus for instance, half of the entire stellar population is found within the so-called NESTs (Nested Elementary STructures). Individually, one NEST contains between 4 and 23 stars with a surface density of stars as high as 2500 pc$^{-2}$ for a mean value of 340 pc$^{-2}$ \citep{Joncour+2018}. Remarkably, half of these NESTs account for about 75\% of the Class 0 and I objects in that region. Therefore, at least three-quarters of young stellar objects (YSOs) --- harbouring planet-forming discs --- are born in locally dense regions of high stellar density where stellar flybys and encounters are frequent.

Stellar flybys and encounters are also expected for more evolved Class~III and Main Sequence stars. The debris disc around the triple stellar system HD 141569 is a fine example, showing dynamical hints of an ongoing stellar flyby \citep{Ardila+2005, Reche+2009, White+2018}. Both Mamajek \citep{Mamajek+2015} and more recently Pfalzner \& Vincke \citep{Pfalzner+2020} have hypothesised a stellar flyby in the Solar System to account for the $7^{\rm o}$-tilt of the ecliptic plane and the orbital arrangement of the Trans-Neptunian Objects (TNOs) \nv{as first suggested by \cite{Heller1993, Ida+2000, Kenyon+2004, Kobayashi+2005}}. Moreover, with the advent of high-precision astrometric and radial velocity measurements from Gaia \citep{Gaia+2016} and ground-based telescopes, it is now possible to sample the stellar trajectories in the solar neighbourhood \nv{\cite{Bailer-Jones2015, Bailer-Jones2018, Bailer-Jones+2018, Ma+2022, Hansen2022, Shuai+2022}.} We further discuss the potential implications for the Solar System in Sect.~\ref{sec:implications}.

One should also consider that unstable configurations constitute the majority of young multiple stars ($<1$~Myr) at all separations \citep{Reipurth&Mikkola2012}. Although triples are likely born compact (at separations of less than 1000~au), these systems can rapidly develop hierarchical architectures in less than 1~Myr. The most extreme wide systems take tens to hundreds of million years to decay towards more stable arrangements. So, hierarchical decay is the favoured scenario for stellar encounters within multiple stellar systems. Typically, for a triple system, the \nv{most weakly bound} stellar body is dynamically scattered into a distant orbit and proceeds to periodically perturb the remaining inner binary. The latter is expected to become a close binary after losing the energy of ejection to the third star. A fraction of wide binaries in the field could be loosely bound triples with periodic stellar encounters \nv{\citep{Tokovinin2014, Halbwachs2017, Riddle2015, Clarke2020}}. In some cases, the outer star becomes unbound and becomes a runaway star. In a survey of the Orion Nebula using Gaia DR2, \citet{McBride+2019} detected (conservatively) 26 candidate runaway stars likely to have had a stellar encounter within the last 1~Myr. Interestingly, a large fraction of apparent single stars in the field may \nv{have been} part of multiple stellar systems in the past before being ejected. Hence, runaway stars supply an alternative source of potential perturbers in star forming regions. \nv{It is worth noting that runaway stars are characterised by their high velocity; and hyperbolic flybys impart only a small impulse to the disc (as discussed in the Section below). Therefore, although runaway stars may have many encounters, their effect on disc truncation is minor.}

%%%%%%%%%%%%%%%%%%%%%%%%%%%%%%%%%%%%%%%%%%%%%%%%%%%%%%%%%%
\section{Impact of flybys on disc dynamics and evolution}
\label{sec:discdynamics}

\begin{figure*}
    \centering
    \includegraphics[width=15cm]{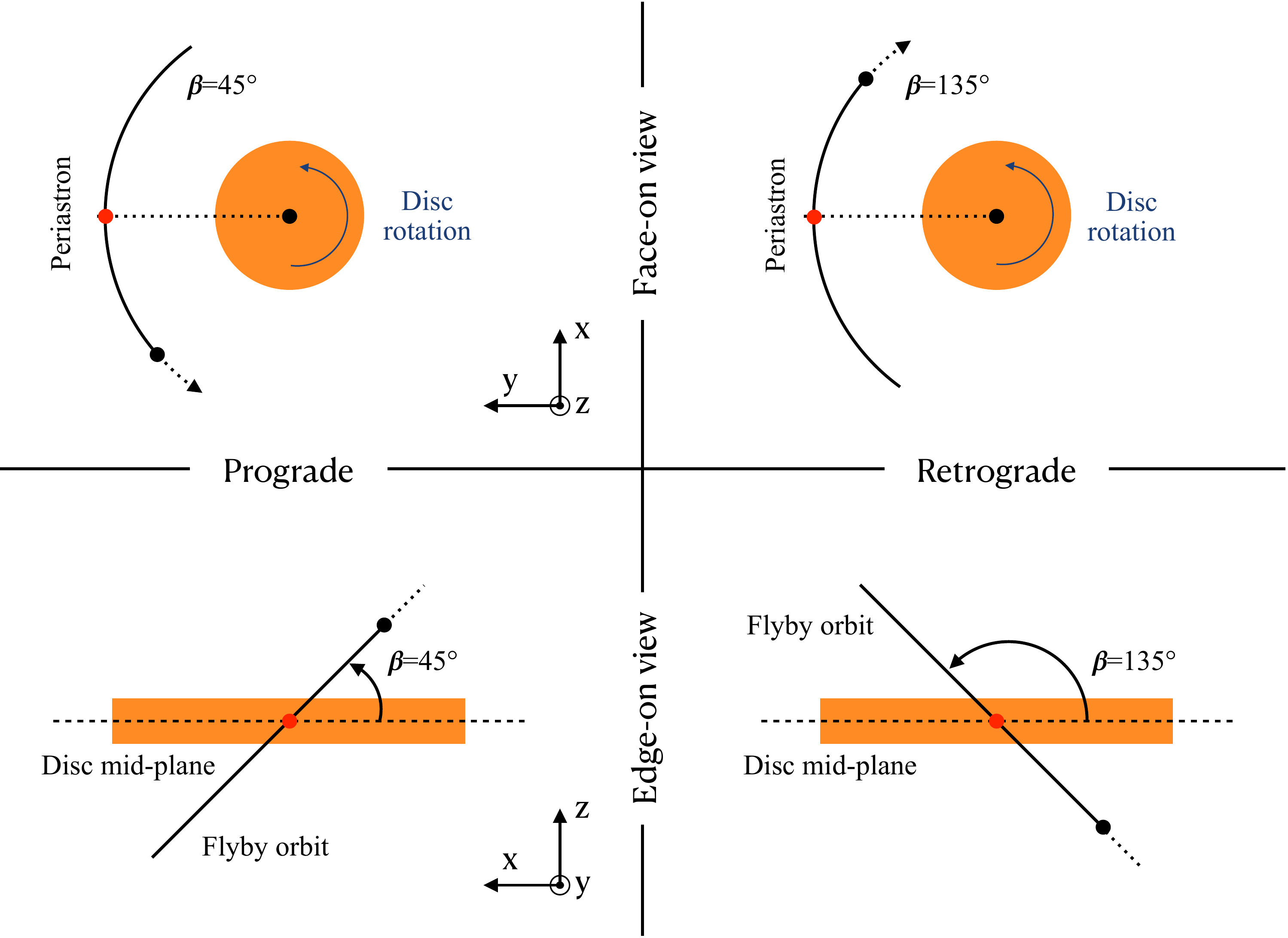}
    \caption{Prograde ($\beta=45^{\rm o}$, left) and retrograde ($\beta=135^{\rm o}$, right) flybys, from face-on (top) and edge-on (bottom) views. $\beta$ is the angle between the disc mid-plane and the flyby orbit. The periastron (red circle) is located along the $y$-axis. The stars are represented as black dots. See Sects.~\ref{sec:pro} and \ref{sec:retro} for details.}
    \label{fig:prograde-retrograde}
\end{figure*}

\begin{figure*}
    \centering
    \includegraphics[width=17cm]{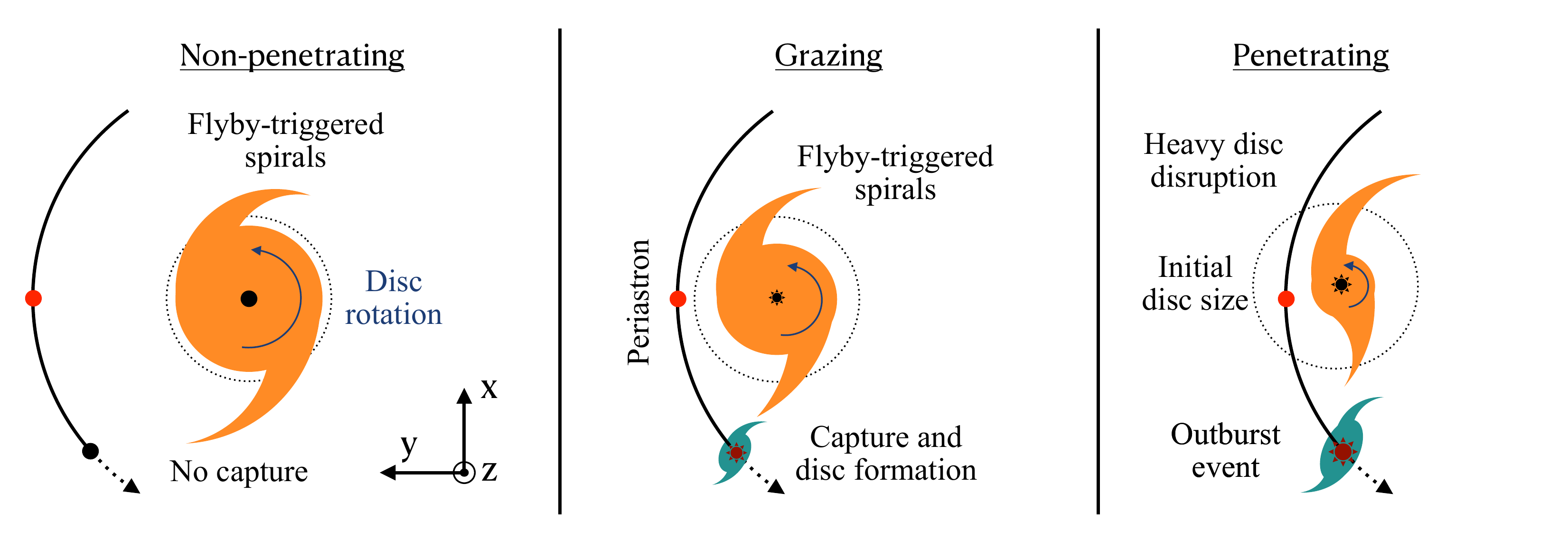}
    \caption{Types of prograde flybys: non-penetrating (left), grazing (middle), disc-penetrating (right). The periastron is marked with a red circle and stars are represented as black dots when quiescent and with star symbols when potentially accreting. During the flyby, the material in between the two stars is often called the \textit{stellar bridge}. See Sect.~\ref{sec:grazingpenetrating} for details.}
    \label{fig:prograde-flybys}
\end{figure*}

A flyby can be classified as prograde or retrograde depending on the motion of the perturber star with respect to the rotation sense of the disc. We define a prograde (retrograde) flyby as one where the normal vector to the flyby orbit and the disc rotation axis lie in the same (opposite) hemisphere. A polar flyby is the special case where these two vectors are nearly orthogonal. In the following, we place ourselves in a three-dimensional orthogonal frame where the disc mid-plane lies in the $xy$-plane with its centre in $(0,0,0)$. For simplicity, periastron is always located along the $y$-axis, which means that we rotate the perturber's orbit about that axis. We define $\beta$ as the angle between the disc mid-plane and the flyby orbital plane. This gives three different types of flybys: prograde ($-90^{\rm o}<\beta<90^{\rm o}$), polar ($\beta=90^{\rm o}$), and retrograde ($90^{\rm o}<\beta<270^{\rm o}$). Fig.~\ref{fig:prograde-retrograde} shows the examples for $\beta=45^{\rm o}$ and $\beta=135^{\rm o}$ flybys, which are discussed in Sect.~\ref{sec:pro} and \ref{sec:retro}.

We introduce a second classification by defining the penetration factor $r_{\rm flyby}=r_{\rm peri}/R_{\rm disc}$. This quantity determines whether the flyby is disc-penetrating ($r_{\rm flyby}<1$), grazing ($\geq 1$) or non-penetrating ($\gg 1$). Section~\ref{sec:grazingpenetrating} discusses these three cases, sketched in Fig.~\ref{fig:prograde-flybys}. Section~\ref{sec:dust} details the specific effects of stellar flybys on dust dynamics. Section~\ref{sec:variability} discusses the deep connection between flybys, variability, accretion and outburst events. For simplicity, we focus on \textit{parabolic} orbits ($e=1$, see motivation in Sect.~\ref{sec:flybys}) for which the exchange of angular momentum during the flyby is maximal. Hyperbolic flybys ($e > 1$) occur at faster speeds in the disc reference frame, which translates into lower amplitude spiral arm formation and less captured material by the perturber (due to the smaller impulse). Finally, in Sect.~\ref{sec:encounters}, we show how discs that experience encounters with stars on highly eccentric orbits ($0.8 \leq e <1$) are equivalent to discs subject to multiple flybys.

\subsection{Prograde flybys}
\label{sec:pro}

\begin{figure*}
    \centering
    \includegraphics[width=13cm]{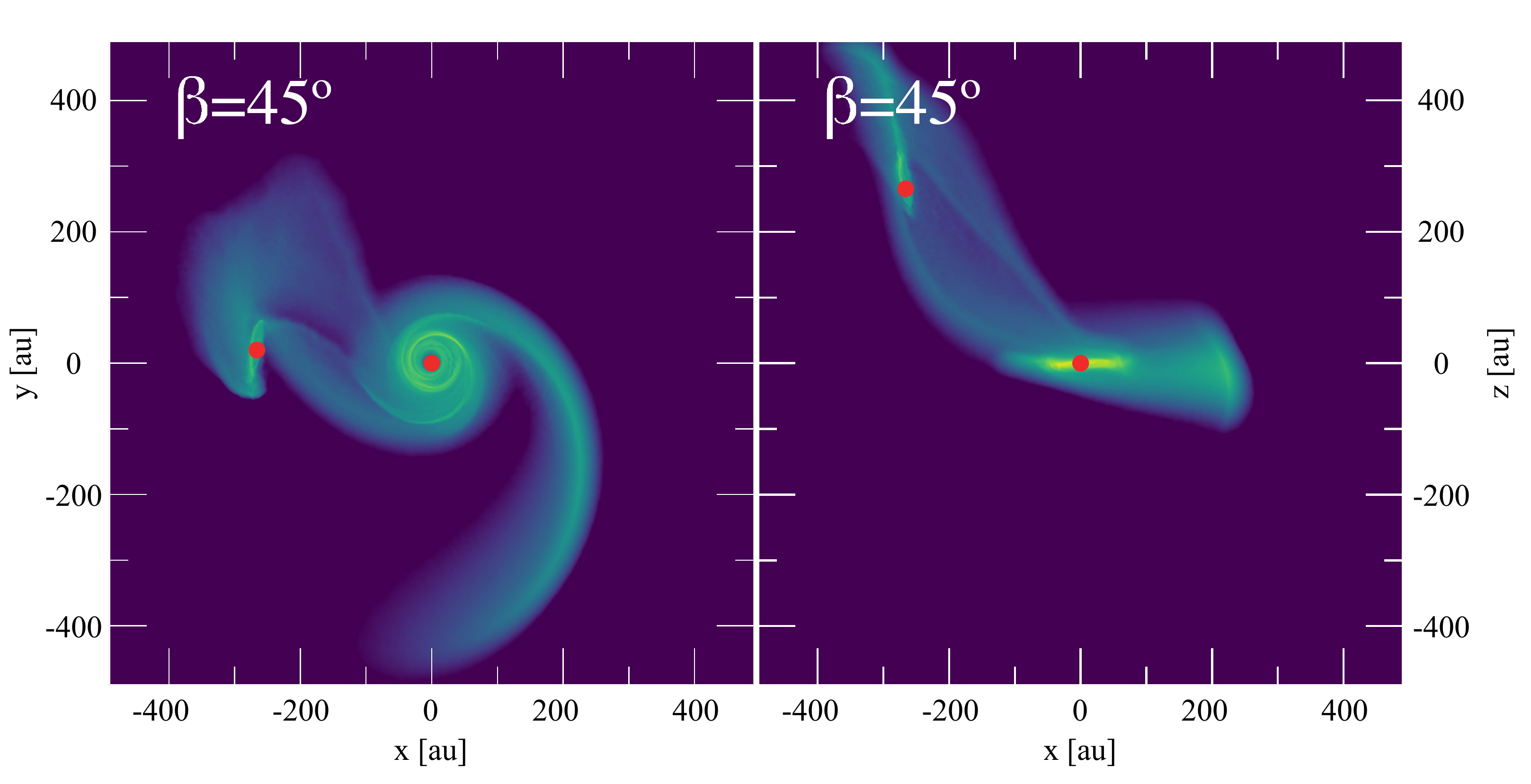}
    \caption{Bridge of material formed between the two stars during an inclined, prograde flyby, showing column density in original disc plane (left) and perpendicular to the disc plane (right). If the orbit is inclined the bridge of material is formed out of the plane. Adapted from \citep{Cuello+2019b}.}
    \label{fig:bridge}
\end{figure*}

\paragraph*{\textbf{Spiral arms.}} For $-90^{\rm o}<\beta<90^{\rm o}$, two prominent spirals form shortly after the passage at periastron \citep{Clarke&Pringle1993, Ostriker1994, Pfalzner2003, Munoz+2015, Cuello+2019b}. The spiral arm between the two stars appears as a bridge of material and is a characteristic signature of an ongoing stellar flyby. \nv{The situation is analogous to close encounters of galaxies \citep[e.g.][]{Toomre1972, DOnghia+2010}.} By contrast, companions on bound orbits ($e<1$) are expected to efficiently truncate the disc after a few passages at periastron, which \nv{prevents} the formation of such extended \textit{bridges}. For inclined prograde flybys, the bridge is formed in a different plane compared to the spiral arm on the other side of the disc (Figure~\ref{fig:bridge}, adapted from Figure 3 in \cite{Cuello+2019b}). This leaves distinct observational signatures at different wavelengths \citep{Cuello+2020}. In scattered light ($\lambda \sim 1 \mu$m), the misaligned bridge (with respect to the disc mid-plane) will appear brighter compared to the spiral arm on the opposite side, which is nearly coplanar with the disc.

Flyby-induced spiral arms are (by definition) triggered once in the disc lifetime and dissipate in a few kyr --- for typical disc sizes and flyby encounter parameters \citep{Cuello+2019b}. This corresponds to a few times the \nv{disc rotation} period at $r_{\rm peri}$ around the primary. The amplitude of these spirals depends on the perturber mass --- more massive perturbers excite more prominent arms as opposed to lower-mass perturbers. The pattern speed is close to the angular velocity of the perturber at periastron. Coplanar and inclined prograde flybys trigger similar spiral arms in the disc. Pitch angles \citep{BinneyTremaine2008} evolve with time as the spirals wind up and eventually disappear, once the perturber has left the disc vicinity.

\paragraph*{\textbf{Disc eccentricity.}} The disc eccentricity \nv{induced by the flyby} (up to values of 0.2) leaves a long-lasting signature \citep[see their Figure~10]{Cuello+2019b}.  By contrast, spirals triggered by a circular, bound perturber are steady in shape and co-orbit with \nv{the} perturber \citep{Dong+2016}. For eccentric perturbers, spirals are more complex and their shape also evolve with time \citep{Zhu+2022}. However, irrespective of the eccentricity of the bound companion, the disc is expected to exhibit spirals potentially during its entire lifetime. Hence, the spirals' morphology and the disc eccentricity can be used to search for potential perturbers around discs where a flyby is suspected (see examples of observed systems in Sect.~\ref{sec:observations}).

\paragraph*{\textbf{Disc truncation.}} Prograde flybys are more destructive than retrograde. This is due to the absence of orbital resonances in the retrograde case \citep{Clarke&Pringle1993, Winter+2018b}. Besides spiral arms, the most characteristic feature of prograde orbits is disc truncation \citep{Breslau+2014, Vincke+2015, Vincke+2016, Bhandare+2016}. These \nv{studies} explored a wide parameter space and estimated the final disc size ($r_{\rm final}$) as a function of $r_{\rm peri}$ and $q$ for different flyby parameters. \nv{Averaging over the different orbital inclinations, Bhandare et al. \citep{Bhandare+2016} obtained the following relation: $r_{\rm final} \approx 1.6\,q^{-0.2}\,r_{\rm peri}^{0.72}$. This empirical formula is based on N-body calculations, which neglect gas. The disc size is defined as the radial location for which the steepest gradient is measured. We note that the scaling with $r_{\rm peri}^{0.72}$ is puzzling as it renders the problem scale-dependent. However, for prograde coplanar flybys, Breslau et al. \citep{Breslau+2014} find the following equation}:
\begin{equation}\label{eq:rdisc}
    r_{\rm final} \approx 0.28\,q^{-0.32}\,r_{\rm peri}\,\,\,.
\end{equation}
\nv{Given the linear dependence on $r_{\rm peri}$, we favour this relation over the one obtained averaging over all the possible inclinations. However, when applying Eq.~\ref{eq:rdisc}, one must keep in mind that there is a dependence on the flyby orbital inclination: prograde orbits are more efficient in truncating the disc than retrograde ones.} Using 3D hydrodynamical simulations and computing the characteristic disc radius as the position at which the enclosed mass reaches 63.2\% of the total disc mass, Cuello et al. \cite{Cuello+2019b} reported up to 30\% larger disc sizes.

\paragraph*{\textbf{Capture of disc material.}} During prograde flybys, if the perturber gets close enough to the disc outer edge (i.e. $r_{\rm flyby} \sim 1$) it can capture disc material. The transfer of mass takes place near periastron \nv{and} the captured material forms a circumsecondary disc around the perturber \citep{Clarke&Pringle1993, Munoz+2015, Cuello+2019b}. This process depends on the flyby orbital parameters. However, thanks to the Virtual Pericentre Positions (VPPs) framework proposed by Breslau et al. \citep{Breslau+2017}, it is possible to compute which particles from the initially unperturbed disc will be captured by the perturber. The material from the outer disc regions, which happens to be in the angular sector close to the pericentre when the flyby occurs, is more likely to end up bound to the perturber as a second generation discs (discussed further in Sect.~\ref{sec:implications}). The distribution of orbital parameters of the captured material (e.g. eccentricity and inclination) gives information about the likely flyby orbit \citep{Jilkova+2016}. In the case where the perturber also has a disc, the resulting structures are more complex \citep{PicognaMarzari2014, Munoz+2015}. To first order, this scenario is equivalent to adding two effects: i) the primary star perturbation to the circumsecondary disc, and ii) the secondary star perturbation to the circumprimary disc. Importantly, the presence of a disc around the perturber prior to the passage at pericentre can force the material to orbit at a different inclination compared to the original disc plane. In other words, the disc angular momentum favours accretion in a preferred orbital plane. This is in agreement with recent numerical experiments \citep{Dullemond+2019, Kuffmeier+2020}, which study misaligned accretion from the environment onto discs. Such processes leave distinctive kinematical features such as warped spirals, twisted velocity channel maps, and tail-like structures in discs \citep[see examples in Sect.~\ref{sec:observations}]{Cuello+2020, Vorobyov+2020, Pinte+2022}.

\subsection{Retrograde flybys}
\label{sec:retro}

\paragraph*{\textbf{Spiral arms.}} For $90^{\rm o}<\beta<270^{\rm o}$, the transfer of angular momentum is less efficient compared to prograde encounters. Hence, the disc dynamical response to the tidal perturbation is less violent. However, if $r_{\rm flyby} \sim 1$, then spirals can be triggered in the disc, though with lower amplitude than those triggered by prograde flybys \citep{Cuello+2019b}. Interestingly, this does not occur for exactly coplanar retrograde flybys ($\beta=180^{\rm o}$) \citep[see their figure 2]{Cuello+2019b}. Also, significant capture of disc material only occurs if the flyby is penetrating ($r_{\rm flyby} \leq 1$).

\paragraph*{\textbf{Warped discs}} are a characteristic signature of retrograde flybys. External perturbers on misaligned orbits warp the disc \citep{Terquem+1993}. \citet{Xiang-Gruess2016} \nv{--- using smoothed particle hydrodynamics simulations ---} showed that retrograde, inclined, parabolic flybys can generate long-lived disc inclinations of up to $60^{\rm o}$. In addition, if the perturbed disc contains an embedded planet --- massive enough to carve a deep gap and separate the disc into two separate regions --- then a retrograde flyby can in principle generate two disconnected misaligned discs. This leads to a disc configuration similar to the one obtained through the mechanism of circumbinary disc breaking \citep{Nixon+2013}. This phenomenon is easily observed since such warps and broken inner discs cast shadows \citep{Marino+2015, Mayama+2018, Nealon+2019, Nealon+2020c, Facchini+2018, MontesinosCuello2018, Cuello+2019a, Bohn+2022}. Also, the emergence of warps in planet-forming discs should leave characteristic chemical signatures \citep{Young+2021, Young+2022}, \nv{changing the molecular abundances.} %which irreversibly change the dust composition.

\paragraph*{\textbf{Disc truncation}} is less dramatic for retrograde flybys \citep{Bhandare+2016, Winter+2018b, Cuello+2019b}. For fixed $r_{\rm flyby}$ and $q$, retrograde orbits result in disc sizes up to twice as large compared to prograde orbits \citep{Bhandare+2016}.

\paragraph*{\textbf{Disc eccentricity.}} Since eccentricity excitation is less efficient for retrograde flybys, discs perturbed by retrograde flybys exhibit low eccentricities (below 0.1) and recircularise faster. Unless the disc is observed as the perturbation occurs, it is difficult to observationally identify a disc that has undergone a retrograde flyby. The best hint of a hypothetical retrograde flyby would be a significantly warped disc around a single star --- where the presence of an inner and outer stellar companion has been ruled out. This is why, to date and to the best of our knowledge, no retrograde stellar flyby has been proposed for a disc-harbouring star (see Table~\ref{tab:flybys}).

\subsection{Non-penetrating, grazing, and disc-penetrating stellar flybys}
\label{sec:grazingpenetrating}

In this subsection, we focus on \textit{prograde} encounters only, as these generate the most characteristic features. Unless specified otherwise, we assume that $\beta=45^{\rm o}$ and that the flyby occurs between two equal-mass stars ($q~=~1$), as shown in Fig.~\ref{fig:prograde-flybys}.

\paragraph*{\textbf{Non-penetrating flybys.}} If $r_{\rm flyby} \gg 1$, the dynamical effect of the flyby on the disc is mild and there is no capture of material by the perturber. \nv{For disc sizes of the order of 100 au, this means that the distance of minimum approach between the two stars ranges between a few hundred and a few thousand au.} At most, the disc develops a couple of spirals that are expected to quickly dissipate \citep{Cuello+2019b}. Since the process of angular momentum exchange is less efficient compared to grazing and disc-penetrating flybys, the stellar orbits remain practically unchanged \citep{Munoz+2015}. It is challenging to identify an ongoing non-penetrating flyby observationally unless the disc exhibits any of the features described in Sect.~\ref{sec:pro}. The only known example of a non-penetrating flyby is Z~CMa, where a relatively low-mass stellar companion has been recently detected \citep{Dong+2022}. Despite the large (projected) distance between the disc and the perturber (almost 5000~au), the latter happens to be in the azimuthal sector suggested by the disc southern spiral. We discuss Z~CMa further in Sect.~\ref{sec:observations}.

\paragraph*{\textbf{Grazing flybys.}} If $r_{\rm flyby}\sim1$, then as the perturber nears periastron it starts capturing material. This process leads to the formation of a second generation disc, fed from the outermost regions of the circumprimary disc \citep{Breslau+2017, Cuello+2019b}. The bridge of disc material that connects both stars shortly after the periastron is strongly misaligned with respect to the disc mid-plane (unless the flyby is perfectly coplanar). The best example of an ongoing grazing flyby is UX~Tau  \citep{Menard+2020, Zapata+2020}. Besides the detection of several unambiguous flyby signatures (bridge, truncation, spirals), there is a strong hint of capture around the perturber (UX~Tau~C): both discs are highly misaligned with respect to each other (see Sect.~\ref{fig:UXTau}). Other young stellar systems exhibit similar disc features (see Table~\ref{tab:flybys}, Sect.~\ref{sec:observations}, and Fig.~\ref{fig:flyby-gallery}).

\paragraph*{\textbf{Disc-penetrating flybys.}} The connection between outburst events --- such as the one observed almost a century ago in FU Orionis --- and binarity has been suggested by several authors \citep[see also Sects.~\ref{sec:variability} and \ref{sec:observations}]{Bonnell+1992, Aspin+2003, Pfalzner2008}. The most extreme flybys occur for $r_{\rm flyby}<1$, where the perturber penetrates the disc. Such events are destructive but less frequent since --- for typical protoplanetary disc sizes --- it requires $r_{\rm peri}<100$~au. Fig.~\ref{fig:plots-Pfalzner2013} shows that, for solar-type stars with ages between 1 and 2 Myr located within 0.5 pc from the cluster centre, the occurrence probability is below 15\% \citep{Pfalzner2013}. Vorobyov et al. \citep{Vorobyov+2020} and Borchert et al. \citep{Borchert+2022} recently studied penetrating flybys in detail using 3D hydrodynamical simulations. Besides the strong perturbation of the disc which leads to spiral formation and truncation, they found that penetrating flybys produce \nv{the rapid onset of} accretion bursts --- similar to the ones observed in FU Orionis-type stars --- which translate into spectacular luminosity increases of up to three orders of magnitude. In this case, no thermal instability \citep{Lin+1985, Clarke+1990} is required to suddenly increase the stellar accretion rate. Importantly, the main accretion burst occurs on the \textit{perturber}. Borchert et al. \citep{Borchert+2022} showed that accretion rates exceeding $10^{-5}$ M$_\odot / {\rm yr}$ onto the perturber can be maintained for more than 100 years, as captured material from the environment falls onto the circumsecondary disc \nv{and cancels angular momentum.} \nv{Remarkably, in disc-disc flybys, 50 to 100\% of the accreted material onto the perturber is \textit{alien} --- originally orbiting the {\it other} star \citep{Borchert+2022b}}. During a penetrating flyby, even if $M_2<M_1$, the secondary star can easily accrete more mass than the primary. This is for instance the case of FU Orionis (the star that gave its name to an entire class of strong accretors' \cite{Hartmann+1996,Audard+2014}). One important aspect to highlight: FU Orionis is not a single star, but a binary stellar system where the stellar orbits are loosely constrained \citep{Wang+2004}.

\subsection{Dust dynamics during stellar flybys}
\label{sec:dust}

\begin{figure*}
    \centering
    \includegraphics[width=18cm]{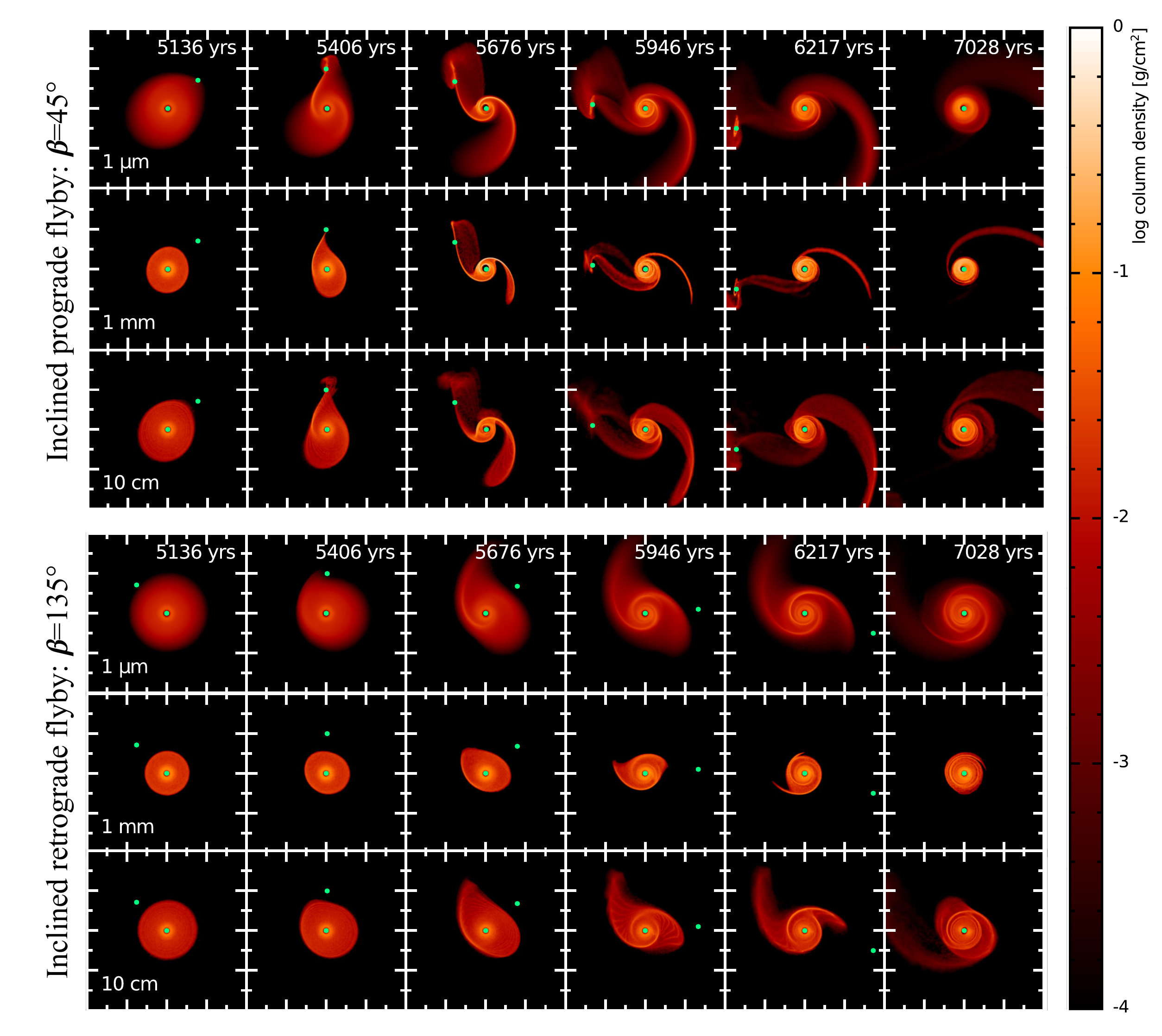}
    \caption{Dust evolution in a grazing flyby, showing column density in 1~$\mu$m (top row), 1~mm (middle row) and 10~cm (bottom row) grains. The upper and bottom panels show the cases of an inclined prograde ($\beta=45^{\rm o}$) and an inclined retrograde flyby ($\beta=135^{\rm o}$), respectively. Time increases from left to right. The perturber goes through periastron at 5406 yrs. Each panel has 400~au in width and in height. The disc rotation is anticlockwise. Sink particles (green dots) represent the stars in the simulation and are large for visualisation purposes only. At periastron, the 1-mm dust disc is more compact than the 1~$\mu$m one due to radial drift. The spirals are sharper for mm-sized particles compared to other grain sizes because of gas drag. Micron-sized particles behave as gas particles given that they are strongly coupled to the gas; while 10 cm particles, poorly coupled to the gas, behave as N-body test particles.}
    \label{fig:dust}
\end{figure*}

Assuming that --- prior to the flyby --- a protoplanetary disc evolved long enough to experience aerodynamical sorting \citep{Pignatale+2017}, then the flyby gravitational perturbation is expected to have different effects on dust population with non-identical Stokes numbers (noted St, ratio between stopping time and orbital period \citep{Weidenschilling1977, Fouchet+2005, Laibe+2012}). For typical protoplanetary discs, there are three main coupling regimes: strongly coupled grains with St$\ll1$ ($0.1\,\mu$m-$10\,\mu$m), marginally coupled particles with St$\sim1$ ($100\,\mu$m-$10$~cm), and decoupled solids with St$\gg1$ ($s>1$~m). In order to properly explore dust dynamics during flybys, given the degree of asymmetry, 3D hydrodynamical simulations with interacting gas and dust are required (see Fig.~\ref{fig:dust}). Such calculations show that \citep{Cuello+2019b}:
\begin{itemize}
    \item $\mu$m-sized grains closely follow the gas density, exhibiting similar features (spirals, warps, capture);
    \item Particles with ${\rm St}\sim1$ (typically mm-sized grains) form a more compact disc due to radial-drift;
    \item  Particles with ${\rm St}\sim1$ are more efficiently concentrated within gas over-densities such as spiral arms;
    \item Large ($\geq10$ cm) particles, being more decoupled, are almost unaffected by radial-drift and behave as test particles in a gas-free environment;
    \item During grazing encounters, mainly $\mu$m-sized grains are captured by the perturber, given that these are preferentially populating the outer disc regions.
\end{itemize}

Since tidal truncation depends on dust-to-gas coupling, the final disc sizes for particles with ${\rm St}\sim1$ are smaller compared to the ones obtained for well-coupled grains ${\rm St}\ll1$. This mechanism critically depends on the amount of time available for radial-drift to occur before the flyby. For an evolutionary time of a few tens of thousand years, the difference in size is expected to be of the order of 10 au \citep{Cuello+2019b}. Longer evolutionary times should translate into larger size discrepancies. At any rate, smaller (or equivalently lighter) dust particles will be subject to stronger flyby perturbations. In particular, if capture occurs, then the circumprimary disc mainly loses gas and $\mu$m-sized particles from the outer regions. This impacts the dust-to-gas ratios ($\epsilon$) within the circumprimary and circumsecondary discs: due to gas removal, we expect high values of $\epsilon$ in the perturbed disc; whereas, by construction, $\epsilon$ should be low for the gas-rich captured disc. In the absence of robust dust-to-gas ratio measurements for both discs, the presence or absence of continuum thermal emission is a good proxy. For example, in a grazing flyby, we do not expect continuum emission around the perturber since the captured disc should be devoid of mm-sized grains (see UX~Tau in Sect.~\ref{sec:observations}).

The excitation of spirals and eccentricity causes particles to mix and collide within the disc, which directly affects the subsequent composition and growth of solids \citep{Pignatale+2017, Pignatale+2019, Cuello+2019b, Silsbee+2021}. In particular, shortly after the encounter, the radial surface density of the gaseous disc is steeper compared to the unperturbed disc \citep[see their figure 11]{Cuello+2019b}. The steeper the radial profile, the faster the radial drift of particles with ${\rm St}\sim1$. Hence, stellar flybys are able to potentially accelerate the motion of solids towards the inner disc regions. This appears to be problematic if one wishes to preserve mm-sized grains. However, this also translates into an increase of the dust-to-gas ratio (by a factor of at least 2, \citep{Cuello+2019a}), which can favour the formation of self-induced dust traps \citep{Gonzalez+2017, Vericel+2020}. In summary, flybys modify the radial, vertical, and azimuthal distributions of solids within the disc. They can also melt dust, see Section below.

\subsection{Variability, outbursts, accretion, and thermal effects}
\label{sec:variability}

Young stars undergoing extreme accretion events where their luminosities increase by factors ranging between 10 and 1000 are called FU-Orionis objects \citep{Vitton&Errico2005}. In 1936, FU Orionis --- the prototype of this class --- brightened by 6 magnitudes in less than a year and remained bright ever since with a slowly decreasing luminosity \citep{Aspin+2003, Reipurth&Aspin2004}. Over the last years, several explanations have been proposed to explain this intriguing phenomenon \citep{Forgan+2010, Audard+2014, Vorobyov+2021}, among which gravitational perturbations by other stars. Bonnell \& Bastien first suggested the link between accretion events and stellar flybys and encounters \citep{Bonnell+1992}. Since then, this scenario has been studied as a potential explanation for FU-Orionis events. Here, we focus on stellar accretion induced by flybys.

For grazing flybys, regardless of the formation of a second generation disc, the tidal perturbation by the secondary can trigger an accretion event onto the primary star \citep{Pfalzner2008, Forgan+2010}. This process depends on the exchange of angular momentum during the flyby. Therefore, prograde orbits lead to more dramatic accretion episodes compared to retrograde ones. In fact, for a given set of parameters, the highest accretion rate of the primary star ($\dot M_1$) is measured for $\beta=45^{\rm o}$. Also, $\dot M_1$ increases with increasing values of $q$ \citep[see their figures~14 and D4]{Cuello+2019b}. The sudden increase of $\dot M_1$ is delayed with respect to the perturber's passage at periastron. The time delay for the accretion depends on the specific orbital parameters, but ranges between a few tens and several hundreds of years \citep{Cuello+2019b}. In this scenario, most of the accreted material onto the primary comes from the \nv{original circumprimary} disc \citep{Shakura+1973}. Provided that the accretion rate is proportional to the disc mass \citep{Clarke&Pringle2006}, then a mass increase by a factor of 100 corresponds to an accretion rate increase by the same factor. Hence, heavier (lighter) discs should lead to higher (reduced) accretion rates. During flybys,  the rise in accretion is caused by a combination of misaligned accretion and eccentricity excitation within the disc. This eventually translates into a burst of luminosity due to accretion roughly equal to:
\begin{equation}
    L_{\rm acc} \approx \frac{1}{2} \frac{G M_1 \dot M_1}{R_1} \,\,\,,
\end{equation}
where $G$ is the gravitational constant and $R_1$ the stellar radius of the primary. Therefore, to trigger a luminosity increase by a factor of 100, the mass accretion rate must increase by the same amount. Such extreme events are possible, but require rise time-scales of at least a few tens of years \citep{Forgan+2010, Cuello+2019b}, which is at odds with FU Orionis ($\sim~1$~yr). Regarding the duration of the accretion events, for prograde flybys, these typically last of the order of 100 years as the disc material progressively circularises, \nv{with the mass accretion rate slowly decreasing with time}.

Instead, for penetrating flybys, a significant amount of material is captured around the perturber close to the periastron \citep{Forgan+2010, Borchert+2022, Vorobyov+2021}. As this happens, the stellar accretion rate of the secondary ($\dot M_2$) rises almost immediately (in less than 1~yr) and remains high for more than a hundred years \citep{Forgan+2010, Borchert+2022} --- in agreement with FU Orionis. The accretion peak for $\dot M_1$ is at least one order of magnitude lower than $\dot M_2$ \citep[see their figure~3]{Forgan+2010, Borchert+2022} and it is delayed (as expected). Therefore, the fact that \textit{both stars accrete material and undergo an outburst episode} constitutes the `smoking gun' of a disc-penetrating flyby. As for grazing flybys, different flyby parameters ($\beta$, $e$, $q$, $r_{\rm flyby}$) are able to produce a broad variety of outbursts --- in terms of duration and intensity \citep{Borchert+2022b}. Such violent events can cause a dramatic increase of the temperature within the disc, with local temperatures ranging between 500 and 1500~K, \nv{caused by radiative heating from the 10--300 $L_\odot$ accretion luminosity liberated from the inner disc near the stellar surface (\citealp{Borchert+2022}, see also \citealp{Zhu+2007})}. This can potentially melt and vaporise the dust content of the disc, depending on pressure and sublimation temperatures. At any rate, the episodic dust thermal processing is expected to alter the mineralogical composition and the subsequent growth of solids \citep{Abraham+2009, Kospal+2020}. If a similar episode occurred in the Solar Nebula after a (hypothetical) flyby, this could account for the presence of Calcium-Aluminium-rich Inclusions (CAIs) in the Solar System \citep{Desch+2002, Audard+2014, Li+2021}.

The fraction of \nv{outbursting} stars with stellar companions remains hard to estimate and it is unlikely that all these extreme accretion events are caused by flybys or encounters \citep{Forgan+2010}. However, \textit{it is surely not a coincidence that the prototype outbursting object is a binary}. In addition, a recent 1.3~mm continuum survey conducted with ALMA shows that, compared to regular Class~II and Class~I objects, the discs of FU-Orionis-like objects are a factor of 2.9--4.4 more massive and a factor of 1.5--4.7 smaller in size \citep{Kospal+2021}. This is suggestive that tidal perturbations could have dramatically shaped these discs. In any case, flybys (especially prograde ones) are able to trigger stellar accretion episodes and spread material around the interacting stars. The former translates into a luminosity outburst and the latter appears as an obscuration event. From the observational perspective, in order to robustly validate the flyby-triggered accretion for FU Orionis-like stars, further monitoring at different wavelengths is required \citep{Borchert+2022, Kospal+2020, Guo+2021}. Yet, given the time and spatial scales involved, this task could easily exceed an astronomer's lifetime (at least for some of us).

Last, some multiple stellar systems exhibit extinction events that may be due to the perturbation of nearby stars. RW~Aurigae \citep{Rodriguez+2013, Dai+2015, Rodriguez+2018} and T~Tauri \citep{Kohler+2016, Kohler+2020} experience recurring occultations, possibly due to the presence of \nv{distributed} material along the line-of-sight. HD~98800 --- a quadruple star --- is a spectacular example with ongoing stellar encounters \citep{Kennedy+2019}. A dimming or occultation event is expected to occur between 2026 and 2031, as the AaAb binary passes behind the BaBa and its disc \citep{Zuniga+2021}. This will be an unprecedented opportunity to characterise the dust and gas properties through photometric and spectroscopic monitoring. The complex structure of the light curve at different wavelengths will provide unique constraints on the 3D geometry and the dust content of the circumbinary disk around BaBb. Discs can also become highly misaligned and warped, shortly after a retrograde flyby. If so, shadows can potentially appear in the disc, due to light obstruction by an inner misaligned disc \citep{Marino+2015, Facchini+2018, Pinilla+2018, Bohn+2022}. On a more speculative note, it has been recently proposed that a close flyby could be responsible for the great dimming of Betelgeuse \citep{Montarges+2021, Aronson+2022}. This indicates that flybys could cause variability and outbursts during the entire life-time of a star. More generally, recent photometric surveys suggest that dynamic interactions --- with either bound or unbound companions --- may control the accretion rates of a substantial fraction of young eruptive stars \citep{Guo+2021}.

\begin{table*}[ht]
    \centering
    \begin{tabular}{|c|c|c|c|c|c|c|}
        \hline
        Name & Distance & Mass ratio: $q=M_2/M_1$ & Projected sep. & $r_{\rm flyby}=r_{\rm peri}/R_{\rm disc}$  & Orbit: $\beta$, $e$ & References \\
        \hline
        SR 24 & $100 \pm 2$ pc & 0.95/1.4=0.7 & 520 au & $\sim1$ & Prograde, $e$? & \cite{Mayama+2010, Mayama+2020} \\
        ISO-Oph 2 & $134\pm 8$ pc & 0.08/0.5=0.16 & 240 au & $\sim2.5$ & Prograde, $e$? & \cite{Gonzalez-Ruilova+2020} \\
        HV \& DO Tau & $138\pm 1$ pc & 0.5/1.35=0.37& 12\,600 au & $285/320\approx0.9$ & $\beta=28^{\rm o}$, $e\sim1$ & \cite{Winter+2018c} \\
        UX Tau & $142 \pm 1$ pc & 0.2/1.0=0.2 & 383 au & $100/90\approx1.1$ & $\beta\approx45^{\rm o}$, $e\sim1$ & \cite{Menard+2020, Zapata+2020} \\
        AS 205 & $142 \pm 3$ pc & 1.28/0.87=1.47 & 168 au & $\sim1$ & Prograde, $e$? & \cite{Kurtovic+2018} \\
        RW Aur & $156\pm 1$ pc & 0.9/1.4=0.64 & 234 au & $70/60\approx1.2$ & $\beta\approx20^{\rm o}$, $e=1$ & \cite{Ghez+1993, Dai+2015, Cabrit+2006} \\
        FU Ori & $408\pm 3$ pc & 1.2/0.6=2.0 & 204 au & $20/50=0.4$ & $\beta\approx45^{\rm o}$, $e\gtrsim1$ & \cite{Perez+2020, Borchert+2022} \\
        Z CMa & $1125\pm30$ pc & 1.8/6.0=0.3 & 4725 au & $3000/840\approx3.6$ & $\beta\approx45^{\rm o}$, $e\sim1$ & \cite{Dong+2022} \\
        Sag. C cloud & 8100 pc & 3.2/31.7=0.1 & $\approx8000$ au & $2000/3000\approx0.7$ & $\beta\approx45^{\rm o}$, $e\sim1$ & \cite{Lu+2022} \\
        \hline
    \end{tabular}
    \caption{Recent observations of discs where a stellar flyby is suspected (see Figure~\ref{fig:flyby-gallery}), ordered by increasing distance from Earth --- using the distances from Gaia DR3 \citep{Gaia+2016} for all the sources but for Z~CMa, for which we use the distance reported by Dong et al. \citep{Dong+2022} instead. We also report the estimated flyby orbit ($q$, $r_{\rm flyby}$, $\beta$, $e$) from literature (last column).}
    \label{tab:flybys}
\end{table*}

\subsection{Application to encounters within multiple stellar systems}
\label{sec:encounters}

Recent dedicated surveys of binary and triple stars \cite{Jensen+1997, Tobin+2016, Cox+2017, Manara+2019, Akeson+2019, Zurlo+2020, Zurlo+2021, Tobin+2022} have revealed that discs in multiple stellar systems are common. However, circumstellar discs in binaries appear to be less massive with respect to disc around single stars \citep{Akeson+2019}. In terms of \nv{disc-to-star mass ratio}, there are no significant differences between circumprimary and circumsecondary discs. Also, the ratio between the (mm-emitting) dust and gas disc radii in binaries are comparable to the same ratio in single stars \citep{Rota+2022}. This is surprising since radial drift is expected to be faster in circumstellar discs in binaries, implying shorter lifetimes for dusty discs \citep{Zagaria+2022}. Also, disc lifetimes increase with increasing binary separation and decreasing disc viscosity (as expected). In this context, stellar encounters within multiple stellar systems are of particular interest since: hierarchical orbital configurations are wide by definition; for large semi-major axis, the number of periodic tidal perturbations experienced by discs ranges between tens and a hundred times at most.

Discs involved in stellar encounters can be considered as discs undergoing several nearly parabolic flybys. Within hierarchical triple systems, two types of discs should be considered: the circumstellar and circumbinary discs around the inner binary, and the circumstellar disc around the third outer body. For the latter, disc truncation is only mildly stronger during encounters when compared to stellar flybys \citep{Umbreit+2011}. This translates into less massive discs; but with almost circular, flatter, and potentially more extended radial profiles \nv{(compared to the original, undisturbed disc)} --- due to the (re-)capture of ejected material during the encounter(s). In contrast, for circumstellar or circumbinary discs perturbed by an external perturber, we expect to retrieve the same kind of disc structures for prograde (spirals, bridges, truncation) and retrograde (warps, misalignment) orbits. The main difference is caused by the periodicity of the perturbation, which has two main consequences: discs are more efficiently truncated, and hence smaller \citep{Artymowicz+1994}; in case of disc misalignment, the perturber's torque can lead to disc precession or libration \citep{Farago+2010, Martin+2014}. For a circumbinary disc in a triple system, the disc is disturbed simultaneously from inside and the outside, but at different frequencies \citep{Martin+2022}.

As for flybys, \textit{accretion from discs in multiple stellar systems may lead to episodic, periodic, and alternated mass accretion \nv{bursts} between the stars} \nv{(alternated meaning alternating between higher $\dot{M}$ on the secondary and higher $\dot{M}$ on the secondary)} according to the orbital parameters \citep[and references therein]{Dunhill+2015, Siwek+2022, Ceppi+2022, Smallwood+2022}. More generally, it has been proposed that stellar multiplicity could account for the correlation between protoplanetary disc masses and accretion rates detected in nearby young star-forming regions \citep[and references therein]{Zagaria+2022}. For instance, a population of binaries could explain the elevated number of high accretors detected in Upper Sco \citep{Zagaria+2022}.  \nv{The effect discussed in \citep{Zagaria+2022} relates specifically to moderate enhancement of the mass accretion rate (to $10^{-8}$--$10^{-7} M_\odot\,{\rm yr}^{-1}$) relative to the millimetre flux, which is a general outcome of binary-disc interaction since disc truncation lowers the mm-flux while simultaneously enhancing $\dot{M}$ due to higher disc surface densities and increased angular momentum transport caused by spiral arms. By contrast, misalignment of captured material in episodic encounters can produce $\dot{M}$ exceeding $10^{-6} M_\odot\,{\rm yr}^{-1}$ \citep{Borchert+2022,Borchert+2022b}. Hence, accretion and luminosity variability constitute distinctive signatures of stellar encounters. Interestingly, these can be then used to infer the presence of unseen perturbers.}

%%%%%%%%%%%%%%%%%%%%%%%%%%%%%%%%%%%%%%%%%%%%%%%%%%%%%%%%%%
\section{Observations of suspected ongoing flybys and encounters}
\label{sec:observations}

\begin{figure*}
\centering
\includegraphics[width=\textwidth]{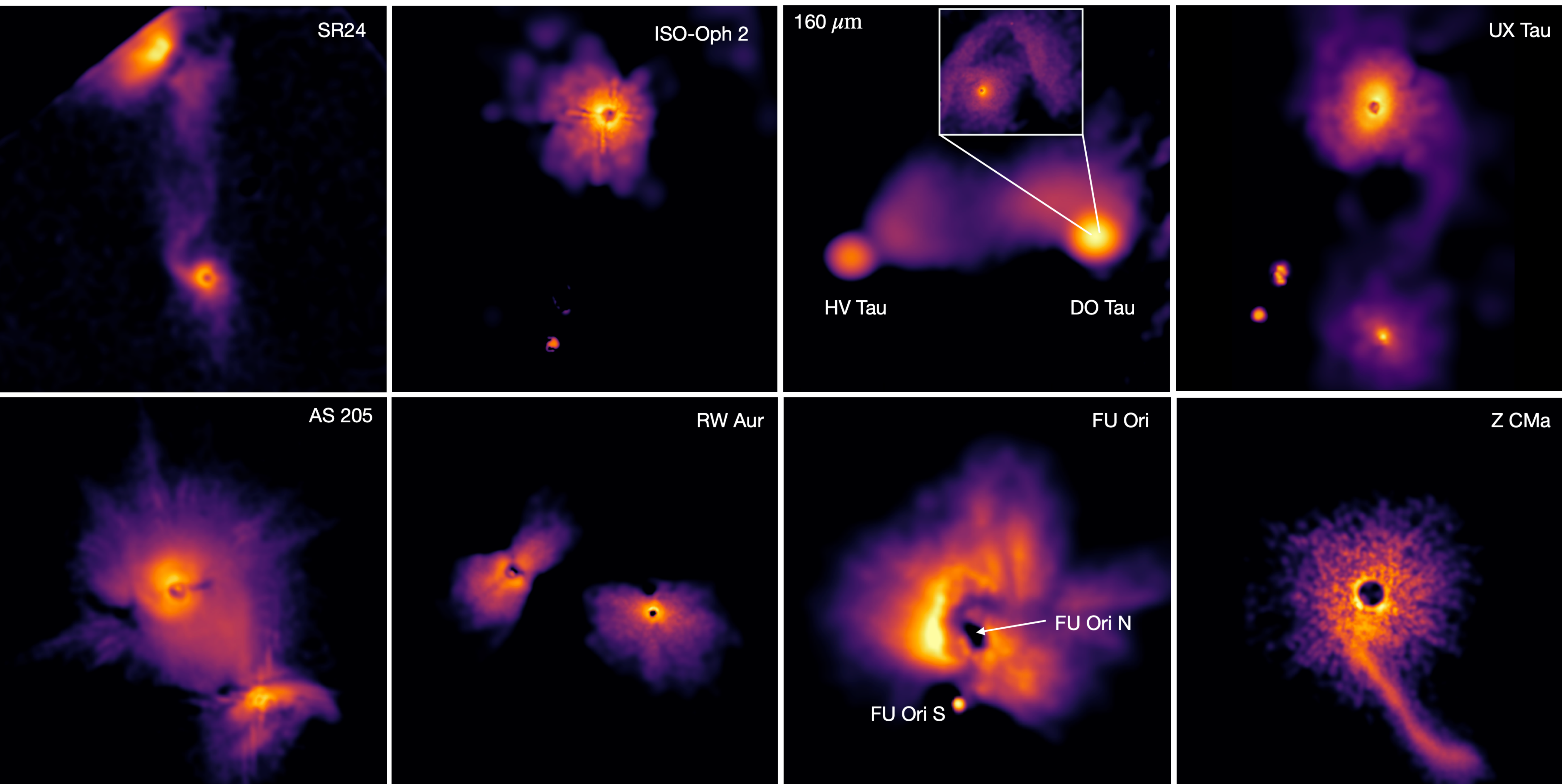}
\caption{Gallery of flyby candidates, shown in scattered light (for HV Tau/DO Tau inset on 160 $\mu {\rm m}$ Herschel image; \cite{Howard+2013}). Z CMa from \citep{Dong+2022}, others via SPHERE archive/F. M\'enard (ISO-Oph 2, DO Tau, RW Aur and FU Ori courtesy of Iain Hammond). Intensity shown with log scaling except for SR24 and AS205 which are shown on a linear scale. Images de-noised following procedure in \citep{Menard+2020}\footnote{https://github.com/danieljprice/denoise}}
\label{fig:flyby-gallery}
\end{figure*}

In Section~\ref{sec:discdynamics}, we reviewed the dynamical signatures of a stellar flyby and the resulting disc morphology and variability. These mainly depend on the flyby orbital orientation $\beta$, the ratio $r_{\rm flyby}$, and the mass ratio $q=M_2/M_1$. Here we focus on the corresponding observational signatures at different wavelengths of discs perturbed by flybys. Based on gas and dust distributions obtained through hydrodynamical simulations, tidally induced spirals should in general be more extended and readily observed in scattered light compared to dust thermal emission \citep{Cuello+2020}. The reason is that scattered light traces micron-sized particles, while the dust thermal emission is sensitive to mm-sized grains --- which \nv{are less perturbed} when the flyby occurs (Figure~\ref{fig:dust}). Last, Doppler motions mapped by molecular line emission (e.g. $^{12}$CO, $^{13}$CO, C$^{18}$O) can reveal disc capture, warping, and misalignment. To effectively map the disc motions, the spatial resolution should be below \nv{0.1} arcsec and the channel maps should have a resolution of at least 1 ${\rm km \, s}^{-1}$ (for typical protoplanetary discs and distances). Below, and in Figure~\ref{fig:flyby-gallery}, we enumerate the young stellar systems with discs in which an ongoing flyby is suspected (see Table~\ref{tab:flybys}) along with estimates of the relevant flyby parameters.

%\begin{figure*}
%\centering
%\includegraphics[width=\textwidth]{flyby-gallery-continuum.pdf}
%\caption{Gallery of flyby suspects, shown in mm-continuum emission on the same scale as Figure~\ref{fig:flyby-gallery}.}
%\label{fig:flyby-gallery-continuum}
%\end{figure*}

\paragraph*{\textbf{SR 24.}} 
Located in the $\rho$-Ophiuchus star forming region, SR~24 (also known as HBC 262) is a hierarchical triple composed of a primary (SR~24S, $M_{\rm S}>1.4\,M_\odot$) and a secondary (SR~24N, $M=$) \citep{Correia+2006} at a separation of 5.''2 \citep{Reipurth&Zinnecker1993}. The latter is itself a binary star: SR~24Nb ($M_{\rm Nb}~=~0.61\,M_\odot$) and SR~24Nc ($M_{\rm Nc}~=~0.34\,M_\odot$) \citep{Correia+2006}, with a projected separation of 0.''197 and an eccentricity of $0.64^{+0.13}_{-0.10}$ \citep{Schaefer+2018}. Dust thermal emission has been recently observed around both SR~24S and SR~24N \citep{Fernandez-Lopez+2017, Pinilla+2017}. ALMA and SMA kinematical observations show three main features: a bridge of gas between SR~24N and SR~24S, an elongated and blue-shifted feature towards the SW of SR~24S, and a gas reservoir towards the NW of SR~24N. This is in agreement with observations in the near-infrared, which show a bridge, a long spiral arm, numerous asymmetries, and shadows \citep{Mayama+2010, Mayama+2020, Weber+2023}. Altogether, the disc morphology strongly suggests that a flyby occurred recently.

\paragraph*{\textbf{ISO-Oph 2.}} Located in the $\rho$-Ophiuchus star forming region, ISO-Oph~2 is a binary system with a separation of 240~au \citep{Cieza+2019}. The primary star harbours a massive ($\sim 40\,M_\oplus$) ring-like disc with a dust cavity of $\sim50$ au in radius. The secondary hosts a lighter ($\sim0.8\,M_\oplus$) disc \citep{Gonzalez-Ruilova+2020}. The combination of 1.3 mm continuum and $^{12}$CO molecular line observations at 0,''02 (3~au) resolution, reveal two non-axisymmetric rings around the primary. Moment 0 and 1 maps obtained from $^{12}$CO data reveal a prominent bridge of gas connecting both discs. A prograde flyby of the southern component has been proposed in order to explain the disc asymmetries observed \citep{Gonzalez-Ruilova+2020}.

\paragraph*{\textbf{HV \& DO Tau.}} Located in the Taurus star-forming region, HV~Tau is a triple system composed of a wide (550~au) binary (HV~Tau~C and HV~Tau~AB) and a tight (10~au) binary (HV~Tau~A and HV~Tau~B). HV~Tau host a disc and it is at 12\,600~au from another disc-hosting star (DO Tau). Winter et al. \citep{Winter+2018c} modelled the interaction between HV~Tau~C and DO~Tau as the decay of a quadruple system. The observations reveal the presence of a bridge between both sources in the 160~$\mu$m emission \citep{Howard+2013}. In this case, a penetrating disc-disc prograde encounter could reproduce most (if not all) the observed features. The mass ratio between the stellar components is poorly constrained, a $q\approx1$ flyby is within errors. Moreover, HV~Tau~C exhibits a high accretion rate \citep{Woitas+1998}, suggestive of a recent prograde flyby.

\begin{figure*}[ht]
    \centering
    \includegraphics[width=\textwidth]{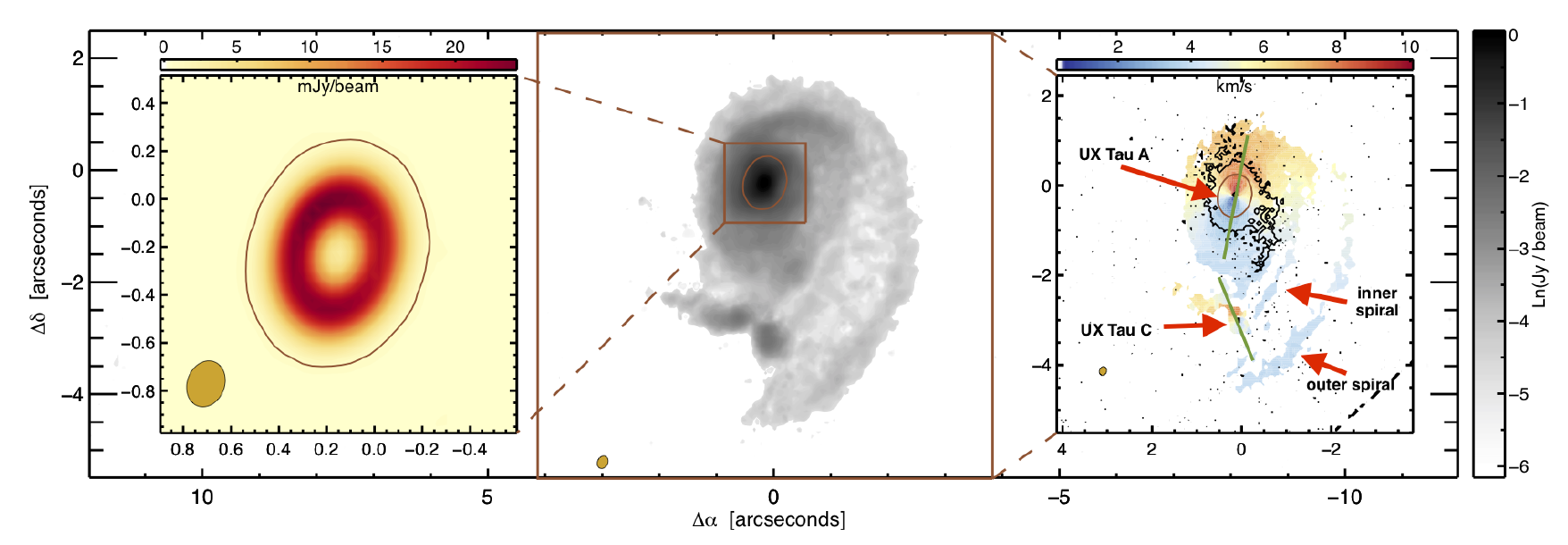}
    \caption{Multi-wavelength observations of the ongoing (grazing) flyby in UX Tauri. The middle panel shows the CO(J=3-2) integrated intensity map (moment 0) of the region spanning UX~Tau~A and C, located at offset [0,0] and [0,-2.7] (respectively) \citep{Zapata+2020}. The left insert shows a zoom of the ringed-disk seen in 0.84~mm continuum emission \citep{Menard+2020}. The right panel shows a map of the peak intensity velocity, where the rotation signatures of the discs around UX Tau A and C are clearly seen \citep{Zapata+2020}. The deep J-band scattered light observation is shown in black contours \citep{Menard+2020}. The beam is shown in the bottom left of each panel. Figure adapted from \citep{Menard+2020}. \nv{Credit: M\'enard et al. 2020, A\&A, 639, L1, reproduced with permission $\copyright$ ESO.}}
    \label{fig:UXTau}
\end{figure*}

\paragraph*{\textbf{UX Tau.}} Located in the Taurus star-forming region, UX~Tau is a young quadruple system where the circumstellar discs around UX~Tau~A and UX~Tau~C exhibit signs of an ongoing dynamical interaction. Near-infrared imaging revealed the presence of large extended spirals in the disc around UX~Tau~A, a prominent bridge between the two stars, and a compact misaligned disc around UX~Tau~C at a separation of 2.''7 \citep[see Fig.~\ref{fig:UXTau}]{Menard+2020}. These features also appear in ALMA CO (J=3-2) observations \citep{Zapata+2020}. Two aspects are striking: i) the disc around UX~Tau~C is misaligned and devoid of mm-sized grains, suggesting that this disc formed during the flyby; ii) the mm-dust disc around UX~Tau~A is more compact that the gas disc and remains nearly unperturbed during the flyby. Hydrodynamical simulations of a recent flyby on an inclined prograde orbit (with respect to the UX~Tau~A disc) reproduce remarkably well all the observed features \citep{Menard+2020}. According to the proposed orbit, the passage at periastron occurred a 1000 years ago approximately.

\paragraph*{\textbf{AS~205.}} Located in the Scorpius star-forming region, AS~205 is a triple where two components have been resolved at 168~au projected separation: AS~205~N ($M=0.87\,M_\odot$) and AS~205~S ($M=1.28\,M_\odot$) \cite[and references therein]{Kurtovic+2018}. The Southern component is itself a binary. Both stars harbour compact discs in the mm emission, which are misaligned to each other \citep{Kurtovic+2018}. The extended emission of gas --- seen in ALMA $^{12}$CO (J=2-1) data --- reveals a bridge of gas between the two sources and two spirals in the dust disc around AS~205~N. Near infrared SPHERE images show similar features (Figure~\ref{fig:flyby-gallery} and \citep{Weber+2023}). Again, the complex disc morphology suggests an inclined prograde flyby where both discs interacted.

\paragraph*{\textbf{RW Aur.}} Located in the Taurus-Auriga Complex star forming region, RW~Aur is a hierarchical young triple system \citep{Ghez+1993}: RW~Aur~A ($M=1.4\,M_\odot$) and a binary RW~Aur~B ($M=0.9\,M_\odot$) (with a and b components). Cabrit et al. \citep{Cabrit+2006} interpreted the prominent tidal arm in the disc around RW~Aur~A (observed in CO) as a consequence of a recent flyby by RW~Aur~B . Hydrodynamical simulations of flybys with $e=1$, $q=0.64,$ and $\beta=20^{\rm o}$ suggest a recent prograde flyby in RW~Aur \cite{Dai+2015}. RW~Aur~A also exhibits a high accretion rate ($10^{-7}\,M_\odot\,{\rm yr}^{-1}$, \citep{Hartigan+1995}), potentially induced by the flyby. More recent observations in the continuum and CO, revealed the presence of addition streamers in the surroundings of the disc of RW~Aur~A \citep{Rodriguez+2018}. This suggests that several periodic perturbations could have occurred, which would favour an encounter scenario in RW~Aur.

\paragraph*{\textbf{\textbf{FU Ori.}}} The case of strong accretors such as FU~Orionis was discussed in Sect.~\ref{sec:variability}. FU~Orionis is a system composed of two stars \citep{Wang+2004}: FU~Ori~N ($M~=~0.3\,M_\odot$ and FU~Ori~S ($M~=~1.2\,M_\odot$) \citep{Beck+2012} with a projected separation of 0.''5 \citep{Perez+2020}. Both stars have gas-rich protoplanetary discs, which are likely interacting with each other based on near infrared observations \citep{Takami+2018, Weber+2023}. In addition, ALMA CO channel maps reveal the presence of a gaseous bridge between both stars and a prominent spiral to the West \citep{Perez+2020}, which is misaligned with respect to the disc mid-plane. Remarkably, the mm emission from these discs is much more compact ($\sim 11$~au) compared to gas and nearly axisymmetric (as in UX~Tau). The small size of these dusty discs is certainly due to a combination of efficient radial drift prior to the encounter and tidal truncation. One of the most striking features of this system is the high accretion rate onto FU~Ori~N ($3.8\times10^{-5}\,M_\odot\,{\rm yr}^{-1}$), which increased in 1936 by a factor of 1000 in $\sim1$ year and has been slowly decreasing ever since. By contrast, FU~Ori~S accretes at an otherwise normal rate of $\approx 10^{-8}\,M_\odot\,{\rm yr}^{-1}$. As noted by Borchert et al. \citep{Borchert+2022}, the more luminous star (FU~Ori~N) is actually the lower mass star. Based on this, it was proposed that the disc around FU~Ori~S could have experienced a penetrating prograde flyby by FU~Ori~S, which triggered a luminosity outburst 86 years ago \citep{Borchert+2022}.

\paragraph*{\textbf{\textbf{Z CMa and Sgr C.}}} Dong et al. \citep{Dong+2022} and Lu et al. \citep{Lu+2022} recently proposed an inclined prograde flyby for two distant discs in very massive systems: Z~CMa  and Sgr C, respectively. The former is a binary star at 1125~pc with a total mass of $6\,M_\odot$, that was possibly perturbed by a external perturber of $1.8\,M_\odot$ which caused the formation of a spiral in the disc. The latter disc has been found around a $32\,M_\odot$ star in the Central Molecular Zone at 8100~pc. The perturber in Sgr~C has a mass of $3.2\,M_\odot$. In both cases, the scenario of a non-penetrating inclined prograde flyby is supported by 3D hydrodynamical simulations. Last, Z~CMa and Sgr~C show signs of outbursts and outflows, which would be in a agreement with the mechanism of flyby-induced accretion (see Sect.~\ref{sec:variability}). These observations suggest that: stellar flybys shape discs within distant star-forming regions (as expected), and massive stars ($M>8\,M_\odot$) with surrounding material are also affected by stellar flybys.

\paragraph*{\textbf{Stellar encounters.}} Amongst all the multiple young stars with highly complex discs, we highlight the following systems: HD~100453 \citep{vanderPlas+2019, Rosotti+2020}, HD~143006 \citep{Benisty+2018}, HD~34700A \citep{Monnier+2019}, HD~98800 \citep{Kennedy+2019}, L1551~IRS~5 \citep{CruzSaenz+2019}, GG~Tau \citep{Keppler+2020}, GW~Ori \citep{Kraus+2020}, IRAS 16293-2422 \citep{Maureira+2020}, HD~139614 \citep{Muro-Arena+2020}, IRAS~04158+2805 \citep{Ragusa+2021} TWA~3 \citep{Czekala+2021}, and V~892 Tau \citep{Long+2021}. This list, which is not meant to be exhaustive, contains systems where dynamical processes --- likely caused by nearby stars --- are currently shaping planet-forming discs \citep[e.g.][]{Doolin+2011, OwenLai2017, MartinLubow2017, Price+2018, CuelloGiuppone2019, Chen+2019, Zhu2019, Poblete+2019, Calcino+2019, Calcino+2020, Gonzalez+2020, Nealon+2020b, Ballabio+2021, Smallwood+2021, Smallwood+2021b}. As opposed to flybys, \textit{discs in multiple stellar systems are periodically perturbed by other stars}, which may eventually lead to misaligned and compact planetary systems.

For bound and eccentric orbits, the perturber spends more time near apoastron than at periastron. However, for reasonable assumptions about the binary eccentricity (e), it turns out that the mean instantaneous projected separation ($\langle d \rangle$) and the semi-major axis ($a$) only differ by approximately 5\% \citep{vanAlbada1968}. For instance, let us consider the following values: \nv{$a=200$ au and $e=0.8$. Despite the fact that apoastron is equal to $a_{\rm apo}=360$ au, the mean projected separation as seen from Earth would be equal to $\langle d \rangle = \exp(-0.056) \, a\approx 95\% \, a$. Hence, even for extreme eccentricities, bound perturbers should be found at distances of the order of $a$ in average. Also, because of tidal truncation, circumstellar disc sizes cannot exceed $0.5\,a$. This kind of information is useful to guide surveys of multiple stellar systems with discs} \citep{Bonavita+2021, Ginski+2021, Bonavita+2022}. 

To sum up, whenever a flyby or an encounter is suspected, obtaining observations at different wavelengths is extremely useful for reconstructing the possible dynamical scenario. For instance, the catalogue of observational signatures in \cite{Cuello+2020} provides a practical guide to distinguish prograde and retrograde encounters. Multi-wavelength observations with \nv{facilities} such as VLT, Subaru, Gemini, ALMA, and VLA are key in order to apply this methodology robustly. Thanks to the advent of new generation high-resolution instruments on large telescopes, we expect more discoveries of discs perturbed by flybys in the near future.

%%%%%%%%%%%%%%%%%%%%%%%%%%%%%%%%%%%%%%%%%%%%%%%%%%%%%%%%%%
\section{Implications for planet formation}
\label{sec:implications}

\subsection{Environmental effects on planet-forming discs}
Given that stellar flybys and radiation can disrupt and remove material from the disc, the environment in which protoplanetary discs form and evolve critically matters \citep{Adams2010}. In this context, the Orion Nebula Cluster (ONC) is a perfect laboratory due to its high stellar density. Using N-body simulations of the ONC, Scally \& Clarke \citep{Scally+2001} first investigated the destruction of protoplanetary discs by close stellar encounters and UV radiation from massive stars. They concluded that flybys do not significantly destroy discs in dense clusters. The fraction of disc destroyed by flybys can reach 10-15\%, mainly caused by high-mass stars \citep{Olczak+2006}. \nv{We note that high-mass stars are also the most affected by encounters (see Eq.~\ref{eq:tau})}. However, even if destruction is \nv{unlikely}, flybys and encounters \nv{can} lead to disc mass loss in crowded regions such as the Trapezium cluster \citep{Olczak+2006}. For parabolic encounters, the relative mean disc mass loss is between 15\% and 80\% for discs in ONC-like environments \citep{Olczak+2006}. Disc sizes are also affected by star-disc interactions in young stellar clusters, especially within regions with elevated stellar density \citep{Rosotti+2014}. More recent studies have shown the dramatic effect of the ultraviolet flux from massive stars on disc size and evolution: in regions with high stellar densities ($>10^4$ stars/pc$^3$), photoevaporation always dominates over tidal truncation \nv{during the phase of gas expulsion and shortly afterwards} \citep{Winter+2018b}; and \nv{simulations predict that} external photoevaporation can destroy between 50 and 80\% \nv{of discs} in regions with densities between $50$ and $100\,M_\odot\,{\rm pc}^{-3}$ \citep{Concha-Ramirez+2019}. Hence, disc survival time-scales determine the time available to form planets, suggesting that planet formation must take place in less than 1~Myr. \nv{This is what is predicted by simulations of typical protoplanetary discs but this might not be necessarily true for all star(s)-disc(s) systems.}

\subsection{Accelerated dust growth and planetesimal formation}
Due to the gravitational perturbation caused by a flyby, the solids within the disc are expected to collide at higher relative speed and follow eccentric orbits (especially for prograde orbits). But, low relative velocities and low eccentricities are more favourable to promote grain growth \citep{Blum2018}. This is why planetesimals are \nv{thought} to be essentially destroyed during or shortly after a flyby. Kobayashi \& Ida \citep{Kobayashi+2001} derived a boundary radius to estimate the distance \nv{beyond} which \textit{in situ} planet formation becomes unlikely after a flyby.  \nv{In order to avoid fragmentation it is necessary that flybys are not too destructive}, i.e. not too disc-penetrating and preferentially distant (especially for prograde orbits).

A way to boost planetesimal formation is to efficiently concentrate solids in dust traps. In particular, the emergence of warps, rings, and clumps in misaligned multiple stellar stellar systems \citep{Cuello+2019b, Poblete+2019, Aly+2020, Aly+2021, Longarini+2021} can stop radial drift leading to a local increase in the dust-to-gas ratio. These kinds of dust traps have been observed in young multiple stellar systems, such as the case of HD~34700 \citep{Benac+2020}. Also, as discussed in Sect.~\ref{sec:variability}, dust can be thermally processed during penetrating stellar flybys \citep{Borchert+2022}. For instance, \nv{an} episode of dust (partial) melting within the disc should alter the crystalline and amorphous structure of dust aggregates and modify their sticking properties \citep{Audard+2014, Kospal+2020, Li+2021}. \nv{Melted, spherical grains are denser and hence decouple from the gas more easily (since S$_{\rm t} \propto$ intrinsic grain density, e.g. \citealp{Weidenschilling1977,Dipierro+2015}) to reach a Stokes number close to unity \nv{for which} concentration in overdensities is most efficient}. Therefore, the interplay between gas and dust in flyby-perturbed discs can lead to the formation of dust traps, which otherwise would not have formed. These locations can potentially become planetesimal factories and accelerate the process of planet formation in multiple stellar systems.

\subsection{Planets on misaligned orbits}
If planets have already formed when the flyby occurs, they \nv{can} also experience a strong gravitational perturbation. Numerical simulations of flyby-perturbed disc with embedded planets show that planet eccentricities increase when the perturber is around the periastron, but are then damped on time-scales of a few $\sim10$~kyr \citep{MarzariPicogna2013}. A similar process is observed for systems of several giant planets as well. In addition, Picogna \& Marzari \citep{PicognaMarzari2014} explored disc-disc interaction during flybys where there is an embedded planet in one of the \nv{discs}. They showed that the discs are warped due to their mutual inclination and the perturber's gravitational pull. The misalignment dissipates as both discs relax to new orbital planes. In addition, planets can i) get closer to their host star \nv{by 15\% to 40\% with respect to their initial semi-major axis}; ii) become eccentric (up to $e\sim0.4$); and iii) end up on misaligned orbits with respect to their initial inclination (up to almost $10^{\rm o}$) \citep{PicognaMarzari2014}. However, as their planet was quickly dragged back within the disc, the interaction with the gaseous disc quickly damped these effects and no apparent trace of the flyby was left in the disc. Yet, for retrograde flybys, the presence of an embedded planet can help to develop a strong misalignment between the inner and outer disc \citep{Nealon+2020b}. Considering the most optimistic parameters for this scenario, the maximum misalignment is of about $45^{\rm o}$ --- but it damps over time due to viscous evolution. At any rate, an embedded planet that \nv{experiences} a flyby is expected to have a modified accretion history (compared to a non-perturbed planet) and a non-zero obliquity. For instance, star-disc misalignment has been recently observed in the $\rho$ Oph and Upper Sco star-forming regions \citep{Davies2019}. Hence, flybys can potentially account for some Hot Jupiters displaying large orbital obliquities \citep{Campante+2016}. 
% Modification of stellar parameters in multiple stellar systems. Hard and soft binaries \citep{Heggie1975, Hills1975}. Possibility to destroy a binary after a flyby. The Heggie-Hills law states that \textit{hard binaries get harder, soft binaries get softer}. This process depends on the stellar cluster density (more details in \citep{Parker+2012}). Potentially, a flyby can destroy a pair of bound stars with circumstellar and/or a circumbinary disc. This would be a rather extreme scenario.

\subsection{Dynamical instability and ejection}
For more evolved stars ($>3$~Myr), once the gaseous protoplanetary disc dissipates, flybys and encounters are expected to \nv{reshape} planetary systems instead of discs \citep{Spurzem+2009}. Simulations of such systems show that: 20\%-40\% of flyby within 100~au lead to the direct ejection of at least one planet from a system containing four giant planets \nv{like} the ones in the Solar System; and flybys lead to the ejection of at least one planet within 100 Myr in 20\%--40\% of cases, due to flyby-induced instabilities \citep{Malberg+2011}. \nv{It is worth noting that all these simulations struggle with low number statistics, so these percentages should be considered with caution. Star-planet and planet-planet scattering can increase the semi-major axis of some planets by a factor ranging between 10 and a few thousand leading to planets on wide and (highly) eccentric orbits. For instance, Solar-like planets initially at a few au can end up with semi-major axis ranging between 100-5000~au \citep{Bailey+2019}). These planets, also called ``Oort'' planets, are generally more prone to develop instabilities over longer time-scales \citep{Brown+2022}.} Remarkably, in some cases, the intruding star is able to \nv{capture} one or several planets \citep{Malberg+2011}. For instance, retrogradely orbiting planets may result from a prograde stellar flyby \citep{Breslau+2019}. More recently, Ndugu et al. \citep{Ndugu+2022} explored the effect of disc truncation and gas removal on the resulting planets, using planet population synthesis techniques. They find that for flyby-perturbed systems, the gas giant formation should be lower compared to unperturbed systems. In their models, the flyby periastron ratio distribution --- which is a good proxy for disc truncation severity --- directly impacts the ratio of hot to cold Jupiters. Overall, flybys reduce the number of cold Jupiters (due to truncation) and the number of hot Jupiters (due to mass removal) \citep{Ndugu+2022}. In addition, Rodet et al. \citep{Rodet+2021}, recently proposed that Hot Jupiters could be formed through the so-called flyby-induced high-e migration mechanism. In particular, this could explain the observed correlation between Hot Jupiters and stellar clustering. Also, it is worth noting that three or four low-mass multiple flybys can produce comparable effects on the planetary orbits as the ones caused by a single flyby with a massive star \citep{Jimenez-Torres2021}. Recently, several dedicated surveys searched for potential stellar perturbers of the Sun with periastron distances of a few pc or less \citep{Mamajek+2015, Bobylev+2020, Dybczynski+2022, Bailer-Jones2022}. Thanks to the {\sc destiny} database \nv{providing results of N-body simulations over wide parameter spaces}, it is also possible to assess the effects of flybys on planetary systems \citep{Bhandare+2019}. Last, the complex gravitational interactions that occur when two planet-harbouring stars perturb each other can result in a broad variety of planetary architectures \citep{Veras+2012, Shara+2016, Bailey+2019,Li+2020}. \nv{Planets can also be stolen or ejected to become free-floating planets \citep{Hills1984,Malmberg+2011,ParkerQuanz2012,Craig+2013,Zheng+2015,Daffern-Powell2022}, the latter of which may also later be recaptured into wide orbits \citep{Perets+2012,Wang+2015}.}

\subsection{Did the Solar System experience a flyby?}

\nv{If the Solar System formed in a leaky cluster, then a stellar flyby could have shaped the Solar Nebula \citep{Ida+2000,Kenyon+2004,Pfalzner+2018b,Pfalzner+2020}}. Such a phenomenon is likely in the supposed Sun natal environment and, more importantly, it would critically determine the disc size and mass \citep{Adams2010, Jilkova+2016, Moore+2020}. After the hypothetical flyby, the probability of a flyby drops drastically \nv{both because of the smaller disc and the rapidly decreasing probability of flybys with age (Figure~\ref{fig:plots-Pfalzner2013})}, which renders a second flyby unlikely beyond 5~Myr. This would explain why the disc and the resulting planetary system could develop and maintain its high circularity in its development \citep{Pfalzner2013} \nv{since the initial flyby would occur at a stage where gas was still present}. On the other hand, in 2015, Mamajek et al. \citep{Mamajek+2015} proposed that the so-called Scholz's star may have penetrated the outer Oort Cloud --- at distances of a few 10\,000 au --- with a 98\% probability. If true, this only had a negligible impact on the flux of long-period comets given the large value of the periastron --- it is in fact beyond what we typically consider a flyby.

\nv{Similarly, the $\approx 35$ au drop in the density of material in the outer Solar System may be a result of an early flyby passing within 100-300 au of the Sun \citep{Ida+2000,Kenyon+2004,Jilkova+2015, Pfalzner+2018b}, though such a scenario is tightly constrained by the observed inclination distribution of trans-Neptunian objects \citep{Batygin+2020, Moore+2020} and stability considerations}. For instance, a flyby perturbation able to modify Neptune's semi-major axis by $\sim0.1\%$ would be strong enough to increase the probability of destabilising the Solar System within 5~Gyrs by one order of magnitude \citep{Brown+2022}.

Last but not least, during close stellar flybys, there is a significant production of \nv{free-floating} objects \nv{crossing our Solar System} \citep{Pfalzner+2021, Breslau+2017}. According to Pfalzner \& Bannister \citep{PfalznerBannister2019}, this mechanism could seed or even accelerate planet formation in discs. The recent discovery of interstellar objects flying-by within the Solar System (e.g. 1I/`Oumuamua \citep{Meech+2017} and 2I/Borisov \citep{Jewitt+2019}) lends support to this idea. \nv{Planet stealing via stellar flybys in the solar birth cluster has also been proposed as the origin of the putative Planet 9 \cite{Mustill+2016}.} On a more speculative note, stellar flybys have also been proposed as an efficient way to travel between stars for technologically advanced civilisations \citep{Hansen+2021}.

%%%%%%%%%%%%%%%%%%%%%%%%%%%%%%%%%%%%%%%%%%%%%%%%%%%%%%%%%%
\section{Conclusions}
\label{sec:conclusions}

The role of stellar flybys and encounters in shaping planet-forming discs can be summarised as follows:
\begin{enumerate}
    \item More than 50\% of stars with protoplanetary discs in a typical star forming environment should experience a stellar encounter or flyby within 1000~au.
    \item Prograde, parabolic, disc-penetrating flybys are the most destructive --- as opposed to retrograde, hyperbolic, non-penetrating flybys. During grazing or penetrating encounters, the perturber may accrete a significant amount of captured material and also form a second-generation disc. \nv{Fresh arrival of material on misaligned orbits compared to the original disc causes extreme accretion and outburst events}. The prototypical star FU~Orionis is thought to be experiencing such a flyby, which would explain its luminosity increase by a factor of 1000 in $\sim1$ year.
    \item Tidal truncation is more effective for prograde flybys and --- depending on the periastron and mass ratio --- can remove up to 80\% of the disc mass. This unavoidably modifies the amount of solids available to form planet(esimal)s. Flybys may also promote the formation of dust traps within the disc, where planetesimals could efficiently grow and eventually turn into planetary cores.
    \item There is a strong link between the flyby orbital parameters and the resulting disc morphology. The comparison between hydrodynamical models and recent multi-wavelength observations has revealed a handful of stars where ongoing or recent flybys are suspected: SR~24, ISO-Oph~2, HV \& DO~Tau, UX~Tau, AS~205, RW~Aur, FU~Ori, Z~CMa, and Sag.~C cloud. More systems should emerge in the near future thanks to new telescopes.
    \item Flybys can enhance planet-planet scattering within (exo)planetary systems,  leading to dynamical instabilities and planetary ejection. The remaining planets could potentially exhibit (highly) eccentric and misaligned orbits. The orbital arrangement of the Solar System (the distribution of Trans-Neptunian Objects in particular) is suggestive that one or several stars may have perturbed the outer Oort cloud. Based on recent surveys, a few stellar candidates have been proposed.
\end{enumerate}

To conclude on a poetical note, as the old man once said in \textit{Close Encounters of the Third Kind} [Spielberg et~al., 1977]: ``¡El sol salió anoche y me cantó!'' (The sun came out last night and sang to me!).

%%%%%%%%%%%%%%%%%%%%%%%%%%%%%%%%%%%%%%%%%%%%%%%%%%%%%%%%%%
\section*{Acknowledgements}
\nv{The authors warmly thank Suzanne Pfalzner, Cathie Clarke, Bo Reipurth, and Richard Larson for useful discussions. We thank Iain Hammond for his expert help in reducing archival data used to make Figure~\ref{fig:flyby-gallery}.} This project has received funding from the European Union’s Horizon 2020 research and innovation programme under the Marie  Sk\l{}odowska-Curie grant agreement No 896319 (SANDS). This research was funded, in part, by ANR (Agence Nationale de la Recherche) of France under contract number ANR-22-ERCS-0002-01. This project has received funding from the European Research Council (ERC) under the European Union Horizon 2020 research and innovation program (grant agreement No. 101042275, project Stellar-MADE). DJP acknowledges funding from the Australian Research Council via DP180104235 and DP220103767.

\section*{Declarations}
The authors have no relevant financial or non-financial interests to disclose. All authors contributed to the study conception and design. The first draft of the manuscript was written by Nicol\'as Cuello and all authors commented on previous versions of the manuscript. All authors contributed figures. All authors read and approved the final manuscript.

\section*{Data availability Statement}
No Data associated in the manuscript.

\bibliography{biblio-flybys}

\begin{thebibliography}{255}%
\makeatletter
\providecommand \@ifxundefined [1]{%
 \@ifx{#1\undefined}
}%
\providecommand \@ifnum [1]{%
 \ifnum #1\expandafter \@firstoftwo
 \else \expandafter \@secondoftwo
 \fi
}%
\providecommand \@ifx [1]{%
 \ifx #1\expandafter \@firstoftwo
 \else \expandafter \@secondoftwo
 \fi
}%
\providecommand \natexlab [1]{#1}%
\providecommand \enquote  [1]{``#1''}%
\providecommand \bibnamefont  [1]{#1}%
\providecommand \bibfnamefont [1]{#1}%
\providecommand \citenamefont [1]{#1}%
\providecommand \href@noop [0]{\@secondoftwo}%
\providecommand \href [0]{\begingroup \@sanitize@url \@href}%
\providecommand \@href[1]{\@@startlink{#1}\@@href}%
\providecommand \@@href[1]{\endgroup#1\@@endlink}%
\providecommand \@sanitize@url [0]{\catcode `\\12\catcode `\$12\catcode
  `\&12\catcode `\#12\catcode `\^12\catcode `\_12\catcode `\%12\relax}%
\providecommand \@@startlink[1]{}%
\providecommand \@@endlink[0]{}%
\providecommand \url  [0]{\begingroup\@sanitize@url \@url }%
\providecommand \@url [1]{\endgroup\@href {#1}{\urlprefix }}%
\providecommand \urlprefix  [0]{URL }%
\providecommand \Eprint [0]{\href }%
\providecommand \doibase [0]{https://doi.org/}%
\providecommand \selectlanguage [0]{\@gobble}%
\providecommand \bibinfo  [0]{\@secondoftwo}%
\providecommand \bibfield  [0]{\@secondoftwo}%
\providecommand \translation [1]{[#1]}%
\providecommand \BibitemOpen [0]{}%
\providecommand \bibitemStop [0]{}%
\providecommand \bibitemNoStop [0]{.\EOS\space}%
\providecommand \EOS [0]{\spacefactor3000\relax}%
\providecommand \BibitemShut  [1]{\csname bibitem#1\endcsname}%
\let\auto@bib@innerbib\@empty
%</preamble>
\bibitem [{\citenamefont {{Lada}}\ and\ \citenamefont
  {{Lada}}(2003)}]{Lada&Lada2003}%
  \BibitemOpen
  \bibfield  {author} {\bibinfo {author} {\bibfnamefont {C.~J.}\ \bibnamefont
  {{Lada}}}\ and\ \bibinfo {author} {\bibfnamefont {E.~A.}\ \bibnamefont
  {{Lada}}},\ }\href {https://doi.org/10.1146/annurev.astro.41.011802.094844}
  {\bibfield  {journal} {\bibinfo  {journal} {\araa}\ }\textbf {\bibinfo
  {volume} {41}},\ \bibinfo {pages} {57} (\bibinfo {year} {2003})},\ \Eprint
  {https://arxiv.org/abs/astro-ph/0301540} {arXiv:astro-ph/0301540 [astro-ph]}
  \BibitemShut {NoStop}%
\bibitem [{\citenamefont {{Tobin}}\ \emph {et~al.}(2016)\citenamefont
  {{Tobin}}, \citenamefont {{Looney}}, \citenamefont {{Li}}, \citenamefont
  {{Chandler}}, \citenamefont {{Dunham}}, \citenamefont {{Segura-Cox}},
  \citenamefont {{Sadavoy}}, \citenamefont {{Melis}}, \citenamefont {{Harris}},
  \citenamefont {{Kratter}},\ and\ \citenamefont {{Perez}}}]{Tobin+2016}%
  \BibitemOpen
  \bibfield  {author} {\bibinfo {author} {\bibfnamefont {J.~J.}\ \bibnamefont
  {{Tobin}}}, \bibinfo {author} {\bibfnamefont {L.~W.}\ \bibnamefont
  {{Looney}}}, \bibinfo {author} {\bibfnamefont {Z.-Y.}\ \bibnamefont {{Li}}},
  \bibinfo {author} {\bibfnamefont {C.~J.}\ \bibnamefont {{Chandler}}},
  \bibinfo {author} {\bibfnamefont {M.~M.}\ \bibnamefont {{Dunham}}}, \bibinfo
  {author} {\bibfnamefont {D.}~\bibnamefont {{Segura-Cox}}}, \bibinfo {author}
  {\bibfnamefont {S.~I.}\ \bibnamefont {{Sadavoy}}}, \bibinfo {author}
  {\bibfnamefont {C.}~\bibnamefont {{Melis}}}, \bibinfo {author} {\bibfnamefont
  {R.~J.}\ \bibnamefont {{Harris}}}, \bibinfo {author} {\bibfnamefont
  {K.}~\bibnamefont {{Kratter}}},\ and\ \bibinfo {author} {\bibfnamefont
  {L.}~\bibnamefont {{Perez}}},\ }\href
  {https://doi.org/10.3847/0004-637X/818/1/73} {\bibfield  {journal} {\bibinfo
  {journal} {\apj}\ }\textbf {\bibinfo {volume} {818}},\ \bibinfo {eid} {73}
  (\bibinfo {year} {2016})},\ \Eprint {https://arxiv.org/abs/1601.00692}
  {arXiv:1601.00692 [astro-ph.SR]} \BibitemShut {NoStop}%
\bibitem [{\citenamefont {{Segura-Cox}}\ \emph {et~al.}(2018)\citenamefont
  {{Segura-Cox}}, \citenamefont {{Looney}}, \citenamefont {{Tobin}},
  \citenamefont {{Li}}, \citenamefont {{Harris}}, \citenamefont {{Sadavoy}},
  \citenamefont {{Dunham}}, \citenamefont {{Chandler}}, \citenamefont
  {{Kratter}}, \citenamefont {{P{\'e}rez}},\ and\ \citenamefont
  {{Melis}}}]{Segura-Cox+2018}%
  \BibitemOpen
  \bibfield  {author} {\bibinfo {author} {\bibfnamefont {D.~M.}\ \bibnamefont
  {{Segura-Cox}}}, \bibinfo {author} {\bibfnamefont {L.~W.}\ \bibnamefont
  {{Looney}}}, \bibinfo {author} {\bibfnamefont {J.~J.}\ \bibnamefont
  {{Tobin}}}, \bibinfo {author} {\bibfnamefont {Z.-Y.}\ \bibnamefont {{Li}}},
  \bibinfo {author} {\bibfnamefont {R.~J.}\ \bibnamefont {{Harris}}}, \bibinfo
  {author} {\bibfnamefont {S.}~\bibnamefont {{Sadavoy}}}, \bibinfo {author}
  {\bibfnamefont {M.~M.}\ \bibnamefont {{Dunham}}}, \bibinfo {author}
  {\bibfnamefont {C.}~\bibnamefont {{Chandler}}}, \bibinfo {author}
  {\bibfnamefont {K.}~\bibnamefont {{Kratter}}}, \bibinfo {author}
  {\bibfnamefont {L.}~\bibnamefont {{P{\'e}rez}}},\ and\ \bibinfo {author}
  {\bibfnamefont {C.}~\bibnamefont {{Melis}}},\ }\href
  {https://doi.org/10.3847/1538-4357/aaddf3} {\bibfield  {journal} {\bibinfo
  {journal} {\apj}\ }\textbf {\bibinfo {volume} {866}},\ \bibinfo {eid} {161}
  (\bibinfo {year} {2018})},\ \Eprint {https://arxiv.org/abs/1808.10438}
  {arXiv:1808.10438 [astro-ph.SR]} \BibitemShut {NoStop}%
\bibitem [{\citenamefont {{Tychoniec}}\ \emph {et~al.}(2018)\citenamefont
  {{Tychoniec}}, \citenamefont {{Tobin}},\ and\ \citenamefont {{Karska et
  al.}}}]{Tychoniec+2018}%
  \BibitemOpen
  \bibfield  {author} {\bibinfo {author} {\bibfnamefont {{\L}.}~\bibnamefont
  {{Tychoniec}}}, \bibinfo {author} {\bibfnamefont {J.~J.}\ \bibnamefont
  {{Tobin}}},\ and\ \bibinfo {author} {\bibfnamefont {A.}~\bibnamefont {{Karska
  et al.}}},\ }\href {https://doi.org/10.3847/1538-4365/aaceae} {\bibfield
  {journal} {\bibinfo  {journal} {\apjs}\ }\textbf {\bibinfo {volume} {238}},\
  \bibinfo {eid} {19} (\bibinfo {year} {2018})},\ \Eprint
  {https://arxiv.org/abs/1806.02434} {arXiv:1806.02434 [astro-ph.SR]}
  \BibitemShut {NoStop}%
\bibitem [{\citenamefont {{Tobin}}\ \emph {et~al.}(2020)\citenamefont
  {{Tobin}}, \citenamefont {{Sheehan}},\ and\ \citenamefont {{Megeath et
  al.}}}]{Tobin+2020}%
  \BibitemOpen
  \bibfield  {author} {\bibinfo {author} {\bibfnamefont {J.~J.}\ \bibnamefont
  {{Tobin}}}, \bibinfo {author} {\bibfnamefont {P.~D.}\ \bibnamefont
  {{Sheehan}}},\ and\ \bibinfo {author} {\bibfnamefont {S.~T.}\ \bibnamefont
  {{Megeath et al.}}},\ }\href {https://doi.org/10.3847/1538-4357/ab6f64}
  {\bibfield  {journal} {\bibinfo  {journal} {\apj}\ }\textbf {\bibinfo
  {volume} {890}},\ \bibinfo {eid} {130} (\bibinfo {year} {2020})},\ \Eprint
  {https://arxiv.org/abs/2001.04468} {arXiv:2001.04468 [astro-ph.GA]}
  \BibitemShut {NoStop}%
\bibitem [{\citenamefont {{Tobin}}\ \emph {et~al.}(2022)\citenamefont
  {{Tobin}}, \citenamefont {{Offner}},\ and\ \citenamefont {{Kratter et
  al.}}}]{Tobin+2022}%
  \BibitemOpen
  \bibfield  {author} {\bibinfo {author} {\bibfnamefont {J.~J.}\ \bibnamefont
  {{Tobin}}}, \bibinfo {author} {\bibfnamefont {S.~S.~R.}\ \bibnamefont
  {{Offner}}},\ and\ \bibinfo {author} {\bibfnamefont {K.~M.}\ \bibnamefont
  {{Kratter et al.}}},\ }\href {https://doi.org/10.3847/1538-4357/ac36d2}
  {\bibfield  {journal} {\bibinfo  {journal} {\apj}\ }\textbf {\bibinfo
  {volume} {925}},\ \bibinfo {eid} {39} (\bibinfo {year} {2022})},\ \Eprint
  {https://arxiv.org/abs/2111.05801} {arXiv:2111.05801 [astro-ph.GA]}
  \BibitemShut {NoStop}%
\bibitem [{\citenamefont {{Manara}}\ \emph {et~al.}(2022)\citenamefont
  {{Manara}}, \citenamefont {{Ansdell}}, \citenamefont {{Rosotti}},
  \citenamefont {{Hughes}}, \citenamefont {{Armitage}}, \citenamefont
  {{Lodato}},\ and\ \citenamefont {{Williams}}}]{Manara+2022}%
  \BibitemOpen
  \bibfield  {author} {\bibinfo {author} {\bibfnamefont {C.~F.}\ \bibnamefont
  {{Manara}}}, \bibinfo {author} {\bibfnamefont {M.}~\bibnamefont {{Ansdell}}},
  \bibinfo {author} {\bibfnamefont {G.~P.}\ \bibnamefont {{Rosotti}}}, \bibinfo
  {author} {\bibfnamefont {A.~M.}\ \bibnamefont {{Hughes}}}, \bibinfo {author}
  {\bibfnamefont {P.~J.}\ \bibnamefont {{Armitage}}}, \bibinfo {author}
  {\bibfnamefont {G.}~\bibnamefont {{Lodato}}},\ and\ \bibinfo {author}
  {\bibfnamefont {J.~P.}\ \bibnamefont {{Williams}}},\ }\href@noop {}
  {\bibfield  {journal} {\bibinfo  {journal} {arXiv e-prints}\ ,\ \bibinfo
  {eid} {arXiv:2203.09930}} (\bibinfo {year} {2022})},\ \Eprint
  {https://arxiv.org/abs/2203.09930} {arXiv:2203.09930 [astro-ph.SR]}
  \BibitemShut {NoStop}%
\bibitem [{\citenamefont {{Bate}}(2018)}]{Bate2018}%
  \BibitemOpen
  \bibfield  {author} {\bibinfo {author} {\bibfnamefont {M.~R.}\ \bibnamefont
  {{Bate}}},\ }\href {https://doi.org/10.1093/mnras/sty169} {\bibfield
  {journal} {\bibinfo  {journal} {\mnras}\ }\textbf {\bibinfo {volume} {475}},\
  \bibinfo {pages} {5618} (\bibinfo {year} {2018})},\ \Eprint
  {https://arxiv.org/abs/1801.07721} {arXiv:1801.07721 [astro-ph.SR]}
  \BibitemShut {NoStop}%
\bibitem [{\citenamefont {{Kuffmeier}}\ \emph {et~al.}(2020)\citenamefont
  {{Kuffmeier}}, \citenamefont {{Goicovic}},\ and\ \citenamefont
  {{Dullemond}}}]{Kuffmeier+2020}%
  \BibitemOpen
  \bibfield  {author} {\bibinfo {author} {\bibfnamefont {M.}~\bibnamefont
  {{Kuffmeier}}}, \bibinfo {author} {\bibfnamefont {F.~G.}\ \bibnamefont
  {{Goicovic}}},\ and\ \bibinfo {author} {\bibfnamefont {C.~P.}\ \bibnamefont
  {{Dullemond}}},\ }\href {https://doi.org/10.1051/0004-6361/201936820}
  {\bibfield  {journal} {\bibinfo  {journal} {\aap}\ }\textbf {\bibinfo
  {volume} {633}},\ \bibinfo {eid} {A3} (\bibinfo {year} {2020})},\ \Eprint
  {https://arxiv.org/abs/1911.04833} {arXiv:1911.04833 [astro-ph.SR]}
  \BibitemShut {NoStop}%
\bibitem [{\citenamefont {{Raghavan}}\ \emph {et~al.}(2010)\citenamefont
  {{Raghavan}}, \citenamefont {{McAlister}}, \citenamefont {{Henry}},
  \citenamefont {{Latham}}, \citenamefont {{Marcy}}, \citenamefont {{Mason}},
  \citenamefont {{Gies}}, \citenamefont {{White}},\ and\ \citenamefont {{ten
  Brummelaar}}}]{Raghavan2010}%
  \BibitemOpen
  \bibfield  {author} {\bibinfo {author} {\bibfnamefont {D.}~\bibnamefont
  {{Raghavan}}}, \bibinfo {author} {\bibfnamefont {H.~A.}\ \bibnamefont
  {{McAlister}}}, \bibinfo {author} {\bibfnamefont {T.~J.}\ \bibnamefont
  {{Henry}}}, \bibinfo {author} {\bibfnamefont {D.~W.}\ \bibnamefont
  {{Latham}}}, \bibinfo {author} {\bibfnamefont {G.~W.}\ \bibnamefont
  {{Marcy}}}, \bibinfo {author} {\bibfnamefont {B.~D.}\ \bibnamefont
  {{Mason}}}, \bibinfo {author} {\bibfnamefont {D.~R.}\ \bibnamefont {{Gies}}},
  \bibinfo {author} {\bibfnamefont {R.~J.}\ \bibnamefont {{White}}},\ and\
  \bibinfo {author} {\bibfnamefont {T.~A.}\ \bibnamefont {{ten Brummelaar}}},\
  }\href {https://doi.org/10.1088/0067-0049/190/1/1} {\bibfield  {journal}
  {\bibinfo  {journal} {\apjs}\ }\textbf {\bibinfo {volume} {190}},\ \bibinfo
  {pages} {1} (\bibinfo {year} {2010})},\ \Eprint
  {https://arxiv.org/abs/1007.0414} {arXiv:1007.0414 [astro-ph.SR]}
  \BibitemShut {NoStop}%
\bibitem [{\citenamefont {{Chen}}\ \emph {et~al.}(2013)\citenamefont {{Chen}},
  \citenamefont {{Arce}}, \citenamefont {{Zhang}}, \citenamefont {{Bourke}},
  \citenamefont {{Launhardt}}, \citenamefont {{J{\o}rgensen}}, \citenamefont
  {{Lee}}, \citenamefont {{Foster}}, \citenamefont {{Dunham}}, \citenamefont
  {{Pineda}},\ and\ \citenamefont {{Henning}}}]{Chen+2013}%
  \BibitemOpen
  \bibfield  {author} {\bibinfo {author} {\bibfnamefont {X.}~\bibnamefont
  {{Chen}}}, \bibinfo {author} {\bibfnamefont {H.~G.}\ \bibnamefont {{Arce}}},
  \bibinfo {author} {\bibfnamefont {Q.}~\bibnamefont {{Zhang}}}, \bibinfo
  {author} {\bibfnamefont {T.~L.}\ \bibnamefont {{Bourke}}}, \bibinfo {author}
  {\bibfnamefont {R.}~\bibnamefont {{Launhardt}}}, \bibinfo {author}
  {\bibfnamefont {J.~K.}\ \bibnamefont {{J{\o}rgensen}}}, \bibinfo {author}
  {\bibfnamefont {C.-F.}\ \bibnamefont {{Lee}}}, \bibinfo {author}
  {\bibfnamefont {J.~B.}\ \bibnamefont {{Foster}}}, \bibinfo {author}
  {\bibfnamefont {M.~M.}\ \bibnamefont {{Dunham}}}, \bibinfo {author}
  {\bibfnamefont {J.~E.}\ \bibnamefont {{Pineda}}},\ and\ \bibinfo {author}
  {\bibfnamefont {T.}~\bibnamefont {{Henning}}},\ }\href
  {https://doi.org/10.1088/0004-637X/768/2/110} {\bibfield  {journal} {\bibinfo
   {journal} {\apj}\ }\textbf {\bibinfo {volume} {768}},\ \bibinfo {eid} {110}
  (\bibinfo {year} {2013})},\ \Eprint {https://arxiv.org/abs/1304.0436}
  {arXiv:1304.0436 [astro-ph.SR]} \BibitemShut {NoStop}%
\bibitem [{\citenamefont {{Duch{\^e}ne}}\ and\ \citenamefont
  {{Kraus}}(2013)}]{Duchene&Kraus2013}%
  \BibitemOpen
  \bibfield  {author} {\bibinfo {author} {\bibfnamefont {G.}~\bibnamefont
  {{Duch{\^e}ne}}}\ and\ \bibinfo {author} {\bibfnamefont {A.}~\bibnamefont
  {{Kraus}}},\ }\href {https://doi.org/10.1146/annurev-astro-081710-102602}
  {\bibfield  {journal} {\bibinfo  {journal} {\araa}\ }\textbf {\bibinfo
  {volume} {51}},\ \bibinfo {pages} {269} (\bibinfo {year} {2013})},\ \Eprint
  {https://arxiv.org/abs/1303.3028} {arXiv:1303.3028 [astro-ph.SR]}
  \BibitemShut {NoStop}%
\bibitem [{\citenamefont {{Offner}}\ \emph {et~al.}(2022)\citenamefont
  {{Offner}}, \citenamefont {{Moe}}, \citenamefont {{Kratter}}, \citenamefont
  {{Sadavoy}}, \citenamefont {{Jensen}},\ and\ \citenamefont
  {{Tobin}}}]{Offner+2022}%
  \BibitemOpen
  \bibfield  {author} {\bibinfo {author} {\bibfnamefont {S.~S.~R.}\
  \bibnamefont {{Offner}}}, \bibinfo {author} {\bibfnamefont {M.}~\bibnamefont
  {{Moe}}}, \bibinfo {author} {\bibfnamefont {K.~M.}\ \bibnamefont
  {{Kratter}}}, \bibinfo {author} {\bibfnamefont {S.~I.}\ \bibnamefont
  {{Sadavoy}}}, \bibinfo {author} {\bibfnamefont {E.~L.~N.}\ \bibnamefont
  {{Jensen}}},\ and\ \bibinfo {author} {\bibfnamefont {J.~J.}\ \bibnamefont
  {{Tobin}}},\ }\href@noop {} {\bibfield  {journal} {\bibinfo  {journal} {arXiv
  e-prints}\ ,\ \bibinfo {eid} {arXiv:2203.10066}} (\bibinfo {year} {2022})},\
  \Eprint {https://arxiv.org/abs/2203.10066} {arXiv:2203.10066 [astro-ph.SR]}
  \BibitemShut {NoStop}%
\bibitem [{\citenamefont {{Gaia Collaboration et
  al.}}(2022)}]{Gaia+2022-multi}%
  \BibitemOpen
  \bibfield  {author} {\bibinfo {author} {\bibnamefont {{Gaia Collaboration et
  al.}}},\ }\href@noop {} {\bibfield  {journal} {\bibinfo  {journal} {arXiv
  e-prints}\ ,\ \bibinfo {eid} {arXiv:2206.05595}} (\bibinfo {year} {2022})},\
  \Eprint {https://arxiv.org/abs/2206.05595} {arXiv:2206.05595 [astro-ph.SR]}
  \BibitemShut {NoStop}%
\bibitem [{\citenamefont {{Duquennoy}}\ and\ \citenamefont
  {{Mayor}}(1991)}]{DuquennoyMayor1991}%
  \BibitemOpen
  \bibfield  {author} {\bibinfo {author} {\bibfnamefont {A.}~\bibnamefont
  {{Duquennoy}}}\ and\ \bibinfo {author} {\bibfnamefont {M.}~\bibnamefont
  {{Mayor}}},\ }\href@noop {} {\bibfield  {journal} {\bibinfo  {journal}
  {\aap}\ }\textbf {\bibinfo {volume} {248}},\ \bibinfo {pages} {485} (\bibinfo
  {year} {1991})}\BibitemShut {NoStop}%
\bibitem [{\citenamefont {{Larson}}(1972)}]{Larson1972}%
  \BibitemOpen
  \bibfield  {author} {\bibinfo {author} {\bibfnamefont {R.~B.}\ \bibnamefont
  {{Larson}}},\ }\href {https://doi.org/10.1093/mnras/156.4.437} {\bibfield
  {journal} {\bibinfo  {journal} {\mnras}\ }\textbf {\bibinfo {volume} {156}},\
  \bibinfo {pages} {437} (\bibinfo {year} {1972})}\BibitemShut {NoStop}%
\bibitem [{\citenamefont {{Reipurth}}\ \emph {et~al.}(2014)\citenamefont
  {{Reipurth}}, \citenamefont {{Clarke}}, \citenamefont {{Boss}}, \citenamefont
  {{Goodwin}}, \citenamefont {{Rodr{\'\i}guez}}, \citenamefont {{Stassun}},
  \citenamefont {{Tokovinin}},\ and\ \citenamefont
  {{Zinnecker}}}]{Reipurth+2014}%
  \BibitemOpen
  \bibfield  {author} {\bibinfo {author} {\bibfnamefont {B.}~\bibnamefont
  {{Reipurth}}}, \bibinfo {author} {\bibfnamefont {C.~J.}\ \bibnamefont
  {{Clarke}}}, \bibinfo {author} {\bibfnamefont {A.~P.}\ \bibnamefont
  {{Boss}}}, \bibinfo {author} {\bibfnamefont {S.~P.}\ \bibnamefont
  {{Goodwin}}}, \bibinfo {author} {\bibfnamefont {L.~F.}\ \bibnamefont
  {{Rodr{\'\i}guez}}}, \bibinfo {author} {\bibfnamefont {K.~G.}\ \bibnamefont
  {{Stassun}}}, \bibinfo {author} {\bibfnamefont {A.}~\bibnamefont
  {{Tokovinin}}},\ and\ \bibinfo {author} {\bibfnamefont {H.}~\bibnamefont
  {{Zinnecker}}},\ }in\ \href
  {https://doi.org/10.2458/azu\_uapress\_9780816531240-ch012} {\emph {\bibinfo
  {booktitle} {Protostars and Planets VI}}},\ \bibinfo {editor} {edited by\
  \bibinfo {editor} {\bibfnamefont {H.}~\bibnamefont {{Beuther}}}, \bibinfo
  {editor} {\bibfnamefont {R.~S.}\ \bibnamefont {{Klessen}}}, \bibinfo {editor}
  {\bibfnamefont {C.~P.}\ \bibnamefont {{Dullemond}}},\ and\ \bibinfo {editor}
  {\bibfnamefont {T.}~\bibnamefont {{Henning}}}}\ (\bibinfo {year} {2014})\ p.\
  \bibinfo {pages} {267},\ \Eprint {https://arxiv.org/abs/1403.1907}
  {arXiv:1403.1907 [astro-ph.SR]} \BibitemShut {NoStop}%
\bibitem [{\citenamefont {{Tokovinin}}(2021)}]{Tokovinin2021}%
  \BibitemOpen
  \bibfield  {author} {\bibinfo {author} {\bibfnamefont {A.}~\bibnamefont
  {{Tokovinin}}},\ }\href {https://doi.org/10.3390/universe7090352} {\bibfield
  {journal} {\bibinfo  {journal} {Universe}\ }\textbf {\bibinfo {volume} {7}},\
  \bibinfo {pages} {352} (\bibinfo {year} {2021})},\ \Eprint
  {https://arxiv.org/abs/2109.09118} {arXiv:2109.09118 [astro-ph.SR]}
  \BibitemShut {NoStop}%
\bibitem [{\citenamefont {{Anosova}}(1986)}]{Anosova1986}%
  \BibitemOpen
  \bibfield  {author} {\bibinfo {author} {\bibfnamefont {J.~P.}\ \bibnamefont
  {{Anosova}}},\ }\href {https://doi.org/10.1007/BF00656037} {\bibfield
  {journal} {\bibinfo  {journal} {\apss}\ }\textbf {\bibinfo {volume} {124}},\
  \bibinfo {pages} {217} (\bibinfo {year} {1986})}\BibitemShut {NoStop}%
\bibitem [{\citenamefont {{Reipurth}}\ \emph {et~al.}(2010)\citenamefont
  {{Reipurth}}, \citenamefont {{Mikkola}}, \citenamefont {{Connelley}},\ and\
  \citenamefont {{Valtonen}}}]{Reipurth+2010}%
  \BibitemOpen
  \bibfield  {author} {\bibinfo {author} {\bibfnamefont {B.}~\bibnamefont
  {{Reipurth}}}, \bibinfo {author} {\bibfnamefont {S.}~\bibnamefont
  {{Mikkola}}}, \bibinfo {author} {\bibfnamefont {M.}~\bibnamefont
  {{Connelley}}},\ and\ \bibinfo {author} {\bibfnamefont {M.}~\bibnamefont
  {{Valtonen}}},\ }\href {https://doi.org/10.1088/2041-8205/725/1/L56}
  {\bibfield  {journal} {\bibinfo  {journal} {\apjl}\ }\textbf {\bibinfo
  {volume} {725}},\ \bibinfo {pages} {L56} (\bibinfo {year} {2010})},\ \Eprint
  {https://arxiv.org/abs/1010.3307} {arXiv:1010.3307 [astro-ph.GA]}
  \BibitemShut {NoStop}%
\bibitem [{\citenamefont {{Schoettler}}\ \emph {et~al.}(2020)\citenamefont
  {{Schoettler}}, \citenamefont {{de Bruijne}}, \citenamefont {{Vaher}},\ and\
  \citenamefont {{Parker}}}]{Schoettler+2020}%
  \BibitemOpen
  \bibfield  {author} {\bibinfo {author} {\bibfnamefont {C.}~\bibnamefont
  {{Schoettler}}}, \bibinfo {author} {\bibfnamefont {J.}~\bibnamefont {{de
  Bruijne}}}, \bibinfo {author} {\bibfnamefont {E.}~\bibnamefont {{Vaher}}},\
  and\ \bibinfo {author} {\bibfnamefont {R.~J.}\ \bibnamefont {{Parker}}},\
  }\href {https://doi.org/10.1093/mnras/staa1228} {\bibfield  {journal}
  {\bibinfo  {journal} {\mnras}\ }\textbf {\bibinfo {volume} {495}},\ \bibinfo
  {pages} {3104} (\bibinfo {year} {2020})},\ \Eprint
  {https://arxiv.org/abs/2004.13730} {arXiv:2004.13730 [astro-ph.SR]}
  \BibitemShut {NoStop}%
\bibitem [{\citenamefont {{Mamajek}}(2009)}]{Mamajek2009}%
  \BibitemOpen
  \bibfield  {author} {\bibinfo {author} {\bibfnamefont {E.~E.}\ \bibnamefont
  {{Mamajek}}},\ }in\ \href {https://doi.org/10.1063/1.3215910} {\emph
  {\bibinfo {booktitle} {Exoplanets and Disks: Their Formation and
  Diversity}}},\ \bibinfo {series} {American Institute of Physics Conference
  Series}, Vol.\ \bibinfo {volume} {1158},\ \bibinfo {editor} {edited by\
  \bibinfo {editor} {\bibfnamefont {T.}~\bibnamefont {{Usuda}}}, \bibinfo
  {editor} {\bibfnamefont {M.}~\bibnamefont {{Tamura}}},\ and\ \bibinfo
  {editor} {\bibfnamefont {M.}~\bibnamefont {{Ishii}}}}\ (\bibinfo {year}
  {2009})\ pp.\ \bibinfo {pages} {3--10},\ \Eprint
  {https://arxiv.org/abs/0906.5011} {arXiv:0906.5011 [astro-ph.EP]}
  \BibitemShut {NoStop}%
\bibitem [{\citenamefont {{Pfalzner}}(2013)}]{Pfalzner2013}%
  \BibitemOpen
  \bibfield  {author} {\bibinfo {author} {\bibfnamefont {S.}~\bibnamefont
  {{Pfalzner}}},\ }\href {https://doi.org/10.1051/0004-6361/201218792}
  {\bibfield  {journal} {\bibinfo  {journal} {\aap}\ }\textbf {\bibinfo
  {volume} {549}},\ \bibinfo {eid} {A82} (\bibinfo {year} {2013})},\ \Eprint
  {https://arxiv.org/abs/1210.8255} {arXiv:1210.8255 [astro-ph.GA]}
  \BibitemShut {NoStop}%
\bibitem [{\citenamefont {{Wurster}}\ \emph {et~al.}(2019)\citenamefont
  {{Wurster}}, \citenamefont {{Bate}},\ and\ \citenamefont
  {{Price}}}]{Wurster+2019}%
  \BibitemOpen
  \bibfield  {author} {\bibinfo {author} {\bibfnamefont {J.}~\bibnamefont
  {{Wurster}}}, \bibinfo {author} {\bibfnamefont {M.~R.}\ \bibnamefont
  {{Bate}}},\ and\ \bibinfo {author} {\bibfnamefont {D.~J.}\ \bibnamefont
  {{Price}}},\ }\href {https://doi.org/10.1093/mnras/stz2215} {\bibfield
  {journal} {\bibinfo  {journal} {\mnras}\ }\textbf {\bibinfo {volume} {489}},\
  \bibinfo {pages} {1719} (\bibinfo {year} {2019})},\ \Eprint
  {https://arxiv.org/abs/1908.03241} {arXiv:1908.03241 [astro-ph.SR]}
  \BibitemShut {NoStop}%
\bibitem [{\citenamefont {{Bate}}(2019)}]{Bate2019}%
  \BibitemOpen
  \bibfield  {author} {\bibinfo {author} {\bibfnamefont {M.~R.}\ \bibnamefont
  {{Bate}}},\ }\href {https://doi.org/10.1093/mnras/stz103} {\bibfield
  {journal} {\bibinfo  {journal} {\mnras}\ }\textbf {\bibinfo {volume} {484}},\
  \bibinfo {pages} {2341} (\bibinfo {year} {2019})},\ \Eprint
  {https://arxiv.org/abs/1901.03713} {arXiv:1901.03713 [astro-ph.SR]}
  \BibitemShut {NoStop}%
\bibitem [{\citenamefont {{Lebreuilly}}\ \emph {et~al.}(2021)\citenamefont
  {{Lebreuilly}}, \citenamefont {{Hennebelle}}, \citenamefont {{Colman}},
  \citenamefont {{Commer{\c{c}}on}}, \citenamefont {{Klessen}}, \citenamefont
  {{Maury}}, \citenamefont {{Molinari}},\ and\ \citenamefont
  {{Testi}}}]{Lebreuilly+2021}%
  \BibitemOpen
  \bibfield  {author} {\bibinfo {author} {\bibfnamefont {U.}~\bibnamefont
  {{Lebreuilly}}}, \bibinfo {author} {\bibfnamefont {P.}~\bibnamefont
  {{Hennebelle}}}, \bibinfo {author} {\bibfnamefont {T.}~\bibnamefont
  {{Colman}}}, \bibinfo {author} {\bibfnamefont {B.}~\bibnamefont
  {{Commer{\c{c}}on}}}, \bibinfo {author} {\bibfnamefont {R.}~\bibnamefont
  {{Klessen}}}, \bibinfo {author} {\bibfnamefont {A.}~\bibnamefont {{Maury}}},
  \bibinfo {author} {\bibfnamefont {S.}~\bibnamefont {{Molinari}}},\ and\
  \bibinfo {author} {\bibfnamefont {L.}~\bibnamefont {{Testi}}},\ }\href
  {https://doi.org/10.3847/2041-8213/ac158c} {\bibfield  {journal} {\bibinfo
  {journal} {\apjl}\ }\textbf {\bibinfo {volume} {917}},\ \bibinfo {eid} {L10}
  (\bibinfo {year} {2021})},\ \Eprint {https://arxiv.org/abs/2107.08491}
  {arXiv:2107.08491 [astro-ph.SR]} \BibitemShut {NoStop}%
\bibitem [{\citenamefont {{Manara}}\ \emph {et~al.}(2019)\citenamefont
  {{Manara}}, \citenamefont {{Tazzari}},\ and\ \citenamefont {{Long et
  al.}}}]{Manara+2019}%
  \BibitemOpen
  \bibfield  {author} {\bibinfo {author} {\bibfnamefont {C.~F.}\ \bibnamefont
  {{Manara}}}, \bibinfo {author} {\bibfnamefont {M.}~\bibnamefont
  {{Tazzari}}},\ and\ \bibinfo {author} {\bibfnamefont {F.}~\bibnamefont {{Long
  et al.}}},\ }\href {https://doi.org/10.1051/0004-6361/201935964} {\bibfield
  {journal} {\bibinfo  {journal} {\aap}\ }\textbf {\bibinfo {volume} {628}},\
  \bibinfo {eid} {A95} (\bibinfo {year} {2019})},\ \Eprint
  {https://arxiv.org/abs/1907.03846} {arXiv:1907.03846 [astro-ph.EP]}
  \BibitemShut {NoStop}%
\bibitem [{\citenamefont {{Akeson}}\ \emph {et~al.}(2019)\citenamefont
  {{Akeson}}, \citenamefont {{Jensen}}, \citenamefont {{Carpenter}},
  \citenamefont {{Ricci}}, \citenamefont {{Laos}}, \citenamefont {{Nogueira}},\
  and\ \citenamefont {{Suen-Lewis}}}]{Akeson+2019}%
  \BibitemOpen
  \bibfield  {author} {\bibinfo {author} {\bibfnamefont {R.~L.}\ \bibnamefont
  {{Akeson}}}, \bibinfo {author} {\bibfnamefont {E.~L.~N.}\ \bibnamefont
  {{Jensen}}}, \bibinfo {author} {\bibfnamefont {J.}~\bibnamefont
  {{Carpenter}}}, \bibinfo {author} {\bibfnamefont {L.}~\bibnamefont
  {{Ricci}}}, \bibinfo {author} {\bibfnamefont {S.}~\bibnamefont {{Laos}}},
  \bibinfo {author} {\bibfnamefont {N.~F.}\ \bibnamefont {{Nogueira}}},\ and\
  \bibinfo {author} {\bibfnamefont {E.~M.}\ \bibnamefont {{Suen-Lewis}}},\
  }\href {https://doi.org/10.3847/1538-4357/aaff6a} {\bibfield  {journal}
  {\bibinfo  {journal} {\apj}\ }\textbf {\bibinfo {volume} {872}},\ \bibinfo
  {eid} {158} (\bibinfo {year} {2019})},\ \Eprint
  {https://arxiv.org/abs/1901.05029} {arXiv:1901.05029 [astro-ph.SR]}
  \BibitemShut {NoStop}%
\bibitem [{\citenamefont {{Czekala}}\ \emph {et~al.}(2019)\citenamefont
  {{Czekala}}, \citenamefont {{Chiang}}, \citenamefont {{Andrews}},
  \citenamefont {{Jensen}}, \citenamefont {{Torres}}, \citenamefont {{Wilner}},
  \citenamefont {{Stassun}},\ and\ \citenamefont {{Macintosh}}}]{Czekala+2019}%
  \BibitemOpen
  \bibfield  {author} {\bibinfo {author} {\bibfnamefont {I.}~\bibnamefont
  {{Czekala}}}, \bibinfo {author} {\bibfnamefont {E.}~\bibnamefont {{Chiang}}},
  \bibinfo {author} {\bibfnamefont {S.~M.}\ \bibnamefont {{Andrews}}}, \bibinfo
  {author} {\bibfnamefont {E.~L.~N.}\ \bibnamefont {{Jensen}}}, \bibinfo
  {author} {\bibfnamefont {G.}~\bibnamefont {{Torres}}}, \bibinfo {author}
  {\bibfnamefont {D.~J.}\ \bibnamefont {{Wilner}}}, \bibinfo {author}
  {\bibfnamefont {K.~G.}\ \bibnamefont {{Stassun}}},\ and\ \bibinfo {author}
  {\bibfnamefont {B.}~\bibnamefont {{Macintosh}}},\ }\href
  {https://doi.org/10.3847/1538-4357/ab287b} {\bibfield  {journal} {\bibinfo
  {journal} {\apj}\ }\textbf {\bibinfo {volume} {883}},\ \bibinfo {eid} {22}
  (\bibinfo {year} {2019})},\ \Eprint {https://arxiv.org/abs/1906.03269}
  {arXiv:1906.03269 [astro-ph.EP]} \BibitemShut {NoStop}%
\bibitem [{\citenamefont {{Zurlo}}\ \emph {et~al.}(2020)\citenamefont
  {{Zurlo}}, \citenamefont {{Cieza}},\ and\ \citenamefont {{P{\'e}rez et
  al.}}}]{Zurlo+2020}%
  \BibitemOpen
  \bibfield  {author} {\bibinfo {author} {\bibfnamefont {A.}~\bibnamefont
  {{Zurlo}}}, \bibinfo {author} {\bibfnamefont {L.~A.}\ \bibnamefont
  {{Cieza}}},\ and\ \bibinfo {author} {\bibfnamefont {S.}~\bibnamefont
  {{P{\'e}rez et al.}}},\ }\href {https://doi.org/10.1093/mnras/staa1886}
  {\bibfield  {journal} {\bibinfo  {journal} {\mnras}\ }\textbf {\bibinfo
  {volume} {496}},\ \bibinfo {pages} {5089} (\bibinfo {year} {2020})},\ \Eprint
  {https://arxiv.org/abs/2006.16259} {arXiv:2006.16259 [astro-ph.SR]}
  \BibitemShut {NoStop}%
\bibitem [{\citenamefont {{Zurlo}}\ \emph {et~al.}(2021)\citenamefont
  {{Zurlo}}, \citenamefont {{Cieza}}, \citenamefont {{Ansdell}}, \citenamefont
  {{Christiaens}}, \citenamefont {{P{\'e}rez}}, \citenamefont {{Lovell}},
  \citenamefont {{Mesa}}, \citenamefont {{Williams}}, \citenamefont
  {{Gonzalez-Ruilova}}, \citenamefont {{Carraro}}, \citenamefont
  {{Ru{\'\i}z-Rodr{\'\i}guez}},\ and\ \citenamefont {{Wyatt}}}]{Zurlo+2021}%
  \BibitemOpen
  \bibfield  {author} {\bibinfo {author} {\bibfnamefont {A.}~\bibnamefont
  {{Zurlo}}}, \bibinfo {author} {\bibfnamefont {L.~A.}\ \bibnamefont
  {{Cieza}}}, \bibinfo {author} {\bibfnamefont {M.}~\bibnamefont {{Ansdell}}},
  \bibinfo {author} {\bibfnamefont {V.}~\bibnamefont {{Christiaens}}}, \bibinfo
  {author} {\bibfnamefont {S.}~\bibnamefont {{P{\'e}rez}}}, \bibinfo {author}
  {\bibfnamefont {J.}~\bibnamefont {{Lovell}}}, \bibinfo {author}
  {\bibfnamefont {D.}~\bibnamefont {{Mesa}}}, \bibinfo {author} {\bibfnamefont
  {J.~P.}\ \bibnamefont {{Williams}}}, \bibinfo {author} {\bibfnamefont
  {C.}~\bibnamefont {{Gonzalez-Ruilova}}}, \bibinfo {author} {\bibfnamefont
  {R.}~\bibnamefont {{Carraro}}}, \bibinfo {author} {\bibfnamefont
  {D.}~\bibnamefont {{Ru{\'\i}z-Rodr{\'\i}guez}}},\ and\ \bibinfo {author}
  {\bibfnamefont {M.}~\bibnamefont {{Wyatt}}},\ }\href
  {https://doi.org/10.1093/mnras/staa3674} {\bibfield  {journal} {\bibinfo
  {journal} {\mnras}\ }\textbf {\bibinfo {volume} {501}},\ \bibinfo {pages}
  {2305} (\bibinfo {year} {2021})},\ \Eprint {https://arxiv.org/abs/2011.12297}
  {arXiv:2011.12297 [astro-ph.EP]} \BibitemShut {NoStop}%
\bibitem [{\citenamefont {{Clarke}}\ and\ \citenamefont
  {{Pringle}}(1993)}]{Clarke&Pringle1993}%
  \BibitemOpen
  \bibfield  {author} {\bibinfo {author} {\bibfnamefont {C.~J.}\ \bibnamefont
  {{Clarke}}}\ and\ \bibinfo {author} {\bibfnamefont {J.~E.}\ \bibnamefont
  {{Pringle}}},\ }\href {https://doi.org/10.1093/mnras/261.1.190} {\bibfield
  {journal} {\bibinfo  {journal} {\mnras}\ }\textbf {\bibinfo {volume} {261}},\
  \bibinfo {pages} {190} (\bibinfo {year} {1993})}\BibitemShut {NoStop}%
\bibitem [{\citenamefont {{Porras}}\ \emph {et~al.}(2003)\citenamefont
  {{Porras}}, \citenamefont {{Christopher}}, \citenamefont {{Allen}},
  \citenamefont {{Di Francesco}}, \citenamefont {{Megeath}},\ and\
  \citenamefont {{Myers}}}]{Porras+2003}%
  \BibitemOpen
  \bibfield  {author} {\bibinfo {author} {\bibfnamefont {A.}~\bibnamefont
  {{Porras}}}, \bibinfo {author} {\bibfnamefont {M.}~\bibnamefont
  {{Christopher}}}, \bibinfo {author} {\bibfnamefont {L.}~\bibnamefont
  {{Allen}}}, \bibinfo {author} {\bibfnamefont {J.}~\bibnamefont {{Di
  Francesco}}}, \bibinfo {author} {\bibfnamefont {S.~T.}\ \bibnamefont
  {{Megeath}}},\ and\ \bibinfo {author} {\bibfnamefont {P.~C.}\ \bibnamefont
  {{Myers}}},\ }\href {https://doi.org/10.1086/377623} {\bibfield  {journal}
  {\bibinfo  {journal} {\aj}\ }\textbf {\bibinfo {volume} {126}},\ \bibinfo
  {pages} {1916} (\bibinfo {year} {2003})},\ \Eprint
  {https://arxiv.org/abs/astro-ph/0307510} {arXiv:astro-ph/0307510 [astro-ph]}
  \BibitemShut {NoStop}%
\bibitem [{\citenamefont {{Winter}}\ \emph
  {et~al.}(2018{\natexlab{a}})\citenamefont {{Winter}}, \citenamefont
  {{Clarke}}, \citenamefont {{Rosotti}},\ and\ \citenamefont
  {{Booth}}}]{Winter+2018a}%
  \BibitemOpen
  \bibfield  {author} {\bibinfo {author} {\bibfnamefont {A.~J.}\ \bibnamefont
  {{Winter}}}, \bibinfo {author} {\bibfnamefont {C.~J.}\ \bibnamefont
  {{Clarke}}}, \bibinfo {author} {\bibfnamefont {G.}~\bibnamefont
  {{Rosotti}}},\ and\ \bibinfo {author} {\bibfnamefont {R.~A.}\ \bibnamefont
  {{Booth}}},\ }\href {https://doi.org/10.1093/mnras/sty012} {\bibfield
  {journal} {\bibinfo  {journal} {\mnras}\ }\textbf {\bibinfo {volume} {475}},\
  \bibinfo {pages} {2314} (\bibinfo {year} {2018}{\natexlab{a}})},\ \Eprint
  {https://arxiv.org/abs/1801.03510} {arXiv:1801.03510 [astro-ph.EP]}
  \BibitemShut {NoStop}%
\bibitem [{\citenamefont {{Galli}}\ \emph {et~al.}(2019)\citenamefont
  {{Galli}}, \citenamefont {{Loinard}}, \citenamefont {{Bouy}}, \citenamefont
  {{Sarro}}, \citenamefont {{Ortiz-Le{\'o}n}}, \citenamefont {{Dzib}},
  \citenamefont {{Olivares}}, \citenamefont {{Heyer}}, \citenamefont
  {{Hernandez}}, \citenamefont {{Rom{\'a}n-Z{\'u}{\~n}iga}}, \citenamefont
  {{Kounkel}},\ and\ \citenamefont {{Covey}}}]{Galli+2019}%
  \BibitemOpen
  \bibfield  {author} {\bibinfo {author} {\bibfnamefont {P.~A.~B.}\
  \bibnamefont {{Galli}}}, \bibinfo {author} {\bibfnamefont {L.}~\bibnamefont
  {{Loinard}}}, \bibinfo {author} {\bibfnamefont {H.}~\bibnamefont {{Bouy}}},
  \bibinfo {author} {\bibfnamefont {L.~M.}\ \bibnamefont {{Sarro}}}, \bibinfo
  {author} {\bibfnamefont {G.~N.}\ \bibnamefont {{Ortiz-Le{\'o}n}}}, \bibinfo
  {author} {\bibfnamefont {S.~A.}\ \bibnamefont {{Dzib}}}, \bibinfo {author}
  {\bibfnamefont {J.}~\bibnamefont {{Olivares}}}, \bibinfo {author}
  {\bibfnamefont {M.}~\bibnamefont {{Heyer}}}, \bibinfo {author} {\bibfnamefont
  {J.}~\bibnamefont {{Hernandez}}}, \bibinfo {author} {\bibfnamefont
  {C.}~\bibnamefont {{Rom{\'a}n-Z{\'u}{\~n}iga}}}, \bibinfo {author}
  {\bibfnamefont {M.}~\bibnamefont {{Kounkel}}},\ and\ \bibinfo {author}
  {\bibfnamefont {K.}~\bibnamefont {{Covey}}},\ }\href
  {https://doi.org/10.1051/0004-6361/201935928} {\bibfield  {journal} {\bibinfo
   {journal} {\aap}\ }\textbf {\bibinfo {volume} {630}},\ \bibinfo {eid} {A137}
  (\bibinfo {year} {2019})},\ \Eprint {https://arxiv.org/abs/1909.01118}
  {arXiv:1909.01118 [astro-ph.SR]} \BibitemShut {NoStop}%
\bibitem [{\citenamefont {{Pfalzner}}(2003)}]{Pfalzner2003}%
  \BibitemOpen
  \bibfield  {author} {\bibinfo {author} {\bibfnamefont {S.}~\bibnamefont
  {{Pfalzner}}},\ }\href {https://doi.org/10.1086/375808} {\bibfield  {journal}
  {\bibinfo  {journal} {\apj}\ }\textbf {\bibinfo {volume} {592}},\ \bibinfo
  {pages} {986} (\bibinfo {year} {2003})}\BibitemShut {NoStop}%
\bibitem [{\citenamefont {{Cuello}}\ \emph
  {et~al.}(2019{\natexlab{a}})\citenamefont {{Cuello}}, \citenamefont
  {{Dipierro}}, \citenamefont {{Mentiplay}}, \citenamefont {{Price}},
  \citenamefont {{Pinte}}, \citenamefont {{Cuadra}}, \citenamefont {{Laibe}},
  \citenamefont {{M{\'e}nard}}, \citenamefont {{Poblete}},\ and\ \citenamefont
  {{Montesinos}}}]{Cuello+2019b}%
  \BibitemOpen
  \bibfield  {author} {\bibinfo {author} {\bibfnamefont {N.}~\bibnamefont
  {{Cuello}}}, \bibinfo {author} {\bibfnamefont {G.}~\bibnamefont
  {{Dipierro}}}, \bibinfo {author} {\bibfnamefont {D.}~\bibnamefont
  {{Mentiplay}}}, \bibinfo {author} {\bibfnamefont {D.~J.}\ \bibnamefont
  {{Price}}}, \bibinfo {author} {\bibfnamefont {C.}~\bibnamefont {{Pinte}}},
  \bibinfo {author} {\bibfnamefont {J.}~\bibnamefont {{Cuadra}}}, \bibinfo
  {author} {\bibfnamefont {G.}~\bibnamefont {{Laibe}}}, \bibinfo {author}
  {\bibfnamefont {F.}~\bibnamefont {{M{\'e}nard}}}, \bibinfo {author}
  {\bibfnamefont {P.~P.}\ \bibnamefont {{Poblete}}},\ and\ \bibinfo {author}
  {\bibfnamefont {M.}~\bibnamefont {{Montesinos}}},\ }\href
  {https://doi.org/10.1093/mnras/sty3325} {\bibfield  {journal} {\bibinfo
  {journal} {\mnras}\ }\textbf {\bibinfo {volume} {483}},\ \bibinfo {pages}
  {4114} (\bibinfo {year} {2019}{\natexlab{a}})},\ \Eprint
  {https://arxiv.org/abs/1812.00961} {arXiv:1812.00961 [astro-ph.EP]}
  \BibitemShut {NoStop}%
\bibitem [{\citenamefont {{Winter}}\ \emph
  {et~al.}(2018{\natexlab{b}})\citenamefont {{Winter}}, \citenamefont
  {{Clarke}}, \citenamefont {{Rosotti}}, \citenamefont {{Ih}}, \citenamefont
  {{Facchini}},\ and\ \citenamefont {{Haworth}}}]{Winter+2018b}%
  \BibitemOpen
  \bibfield  {author} {\bibinfo {author} {\bibfnamefont {A.~J.}\ \bibnamefont
  {{Winter}}}, \bibinfo {author} {\bibfnamefont {C.~J.}\ \bibnamefont
  {{Clarke}}}, \bibinfo {author} {\bibfnamefont {G.}~\bibnamefont {{Rosotti}}},
  \bibinfo {author} {\bibfnamefont {J.}~\bibnamefont {{Ih}}}, \bibinfo {author}
  {\bibfnamefont {S.}~\bibnamefont {{Facchini}}},\ and\ \bibinfo {author}
  {\bibfnamefont {T.~J.}\ \bibnamefont {{Haworth}}},\ }\href
  {https://doi.org/10.1093/mnras/sty984} {\bibfield  {journal} {\bibinfo
  {journal} {\mnras}\ }\textbf {\bibinfo {volume} {478}},\ \bibinfo {pages}
  {2700} (\bibinfo {year} {2018}{\natexlab{b}})},\ \Eprint
  {https://arxiv.org/abs/1804.00013} {arXiv:1804.00013 [astro-ph.SR]}
  \BibitemShut {NoStop}%
\bibitem [{\citenamefont {{Davies}}(2011)}]{Davies2011}%
  \BibitemOpen
  \bibfield  {author} {\bibinfo {author} {\bibfnamefont {M.~B.}\ \bibnamefont
  {{Davies}}},\ }in\ \href {https://doi.org/10.1017/S1743921311020369} {\emph
  {\bibinfo {booktitle} {The Astrophysics of Planetary Systems: Formation,
  Structure, and Dynamical Evolution}}},\ Vol.\ \bibinfo {volume} {276},\
  \bibinfo {editor} {edited by\ \bibinfo {editor} {\bibfnamefont
  {A.}~\bibnamefont {{Sozzetti}}}, \bibinfo {editor} {\bibfnamefont {M.~G.}\
  \bibnamefont {{Lattanzi}}},\ and\ \bibinfo {editor} {\bibfnamefont {A.~P.}\
  \bibnamefont {{Boss}}}}\ (\bibinfo {year} {2011})\ pp.\ \bibinfo {pages}
  {304--307}\BibitemShut {NoStop}%
\bibitem [{\citenamefont {{Moe}}\ and\ \citenamefont {{Di
  Stefano}}(2017)}]{Moe+2017}%
  \BibitemOpen
  \bibfield  {author} {\bibinfo {author} {\bibfnamefont {M.}~\bibnamefont
  {{Moe}}}\ and\ \bibinfo {author} {\bibfnamefont {R.}~\bibnamefont {{Di
  Stefano}}},\ }\href {https://doi.org/10.3847/1538-4365/aa6fb6} {\bibfield
  {journal} {\bibinfo  {journal} {\apjs}\ }\textbf {\bibinfo {volume} {230}},\
  \bibinfo {eid} {15} (\bibinfo {year} {2017})},\ \Eprint
  {https://arxiv.org/abs/1606.05347} {arXiv:1606.05347 [astro-ph.SR]}
  \BibitemShut {NoStop}%
\bibitem [{\citenamefont {{Mu{\~n}oz}}\ \emph {et~al.}(2015)\citenamefont
  {{Mu{\~n}oz}}, \citenamefont {{Kratter}}, \citenamefont {{Vogelsberger}},
  \citenamefont {{Hernquist}},\ and\ \citenamefont {{Springel}}}]{Munoz+2015}%
  \BibitemOpen
  \bibfield  {author} {\bibinfo {author} {\bibfnamefont {D.~J.}\ \bibnamefont
  {{Mu{\~n}oz}}}, \bibinfo {author} {\bibfnamefont {K.}~\bibnamefont
  {{Kratter}}}, \bibinfo {author} {\bibfnamefont {M.}~\bibnamefont
  {{Vogelsberger}}}, \bibinfo {author} {\bibfnamefont {L.}~\bibnamefont
  {{Hernquist}}},\ and\ \bibinfo {author} {\bibfnamefont {V.}~\bibnamefont
  {{Springel}}},\ }\href {https://doi.org/10.1093/mnras/stu2220} {\bibfield
  {journal} {\bibinfo  {journal} {\mnras}\ }\textbf {\bibinfo {volume} {446}},\
  \bibinfo {pages} {2010} (\bibinfo {year} {2015})},\ \Eprint
  {https://arxiv.org/abs/1410.4561} {arXiv:1410.4561 [astro-ph.EP]}
  \BibitemShut {NoStop}%
\bibitem [{\citenamefont {{Pfalzner}}\ and\ \citenamefont
  {{Govind}}(2021)}]{Pfalzner+2021b}%
  \BibitemOpen
  \bibfield  {author} {\bibinfo {author} {\bibfnamefont {S.}~\bibnamefont
  {{Pfalzner}}}\ and\ \bibinfo {author} {\bibfnamefont {A.}~\bibnamefont
  {{Govind}}},\ }\href {https://doi.org/10.3847/1538-4357/ac19aa} {\bibfield
  {journal} {\bibinfo  {journal} {\apj}\ }\textbf {\bibinfo {volume} {921}},\
  \bibinfo {eid} {90} (\bibinfo {year} {2021})},\ \Eprint
  {https://arxiv.org/abs/2108.10296} {arXiv:2108.10296 [astro-ph.SR]}
  \BibitemShut {NoStop}%
\bibitem [{\citenamefont {{Hatchell}}\ \emph {et~al.}(2005)\citenamefont
  {{Hatchell}}, \citenamefont {{Richer}}, \citenamefont {{Fuller}},
  \citenamefont {{Qualtrough}}, \citenamefont {{Ladd}},\ and\ \citenamefont
  {{Chandler}}}]{Hatchel+2005}%
  \BibitemOpen
  \bibfield  {author} {\bibinfo {author} {\bibfnamefont {J.}~\bibnamefont
  {{Hatchell}}}, \bibinfo {author} {\bibfnamefont {J.~S.}\ \bibnamefont
  {{Richer}}}, \bibinfo {author} {\bibfnamefont {G.~A.}\ \bibnamefont
  {{Fuller}}}, \bibinfo {author} {\bibfnamefont {C.~J.}\ \bibnamefont
  {{Qualtrough}}}, \bibinfo {author} {\bibfnamefont {E.~F.}\ \bibnamefont
  {{Ladd}}},\ and\ \bibinfo {author} {\bibfnamefont {C.~J.}\ \bibnamefont
  {{Chandler}}},\ }\href {https://doi.org/10.1051/0004-6361:20041836}
  {\bibfield  {journal} {\bibinfo  {journal} {\aap}\ }\textbf {\bibinfo
  {volume} {440}},\ \bibinfo {pages} {151} (\bibinfo {year}
  {2005})}\BibitemShut {NoStop}%
\bibitem [{\citenamefont {{Joncour}}\ \emph {et~al.}(2017)\citenamefont
  {{Joncour}}, \citenamefont {{Duch{\^e}ne}},\ and\ \citenamefont
  {{Moraux}}}]{Joncour+2017}%
  \BibitemOpen
  \bibfield  {author} {\bibinfo {author} {\bibfnamefont {I.}~\bibnamefont
  {{Joncour}}}, \bibinfo {author} {\bibfnamefont {G.}~\bibnamefont
  {{Duch{\^e}ne}}},\ and\ \bibinfo {author} {\bibfnamefont {E.}~\bibnamefont
  {{Moraux}}},\ }\href {https://doi.org/10.1051/0004-6361/201629398} {\bibfield
   {journal} {\bibinfo  {journal} {\aap}\ }\textbf {\bibinfo {volume} {599}},\
  \bibinfo {eid} {A14} (\bibinfo {year} {2017})},\ \Eprint
  {https://arxiv.org/abs/1612.02098} {arXiv:1612.02098 [astro-ph.SR]}
  \BibitemShut {NoStop}%
\bibitem [{\citenamefont {{Joncour}}\ \emph {et~al.}(2018)\citenamefont
  {{Joncour}}, \citenamefont {{Duch{\^e}ne}}, \citenamefont {{Moraux}},\ and\
  \citenamefont {{Motte}}}]{Joncour+2018}%
  \BibitemOpen
  \bibfield  {author} {\bibinfo {author} {\bibfnamefont {I.}~\bibnamefont
  {{Joncour}}}, \bibinfo {author} {\bibfnamefont {G.}~\bibnamefont
  {{Duch{\^e}ne}}}, \bibinfo {author} {\bibfnamefont {E.}~\bibnamefont
  {{Moraux}}},\ and\ \bibinfo {author} {\bibfnamefont {F.}~\bibnamefont
  {{Motte}}},\ }\href {https://doi.org/10.1051/0004-6361/201833042} {\bibfield
  {journal} {\bibinfo  {journal} {\aap}\ }\textbf {\bibinfo {volume} {620}},\
  \bibinfo {eid} {A27} (\bibinfo {year} {2018})},\ \Eprint
  {https://arxiv.org/abs/1809.02380} {arXiv:1809.02380 [astro-ph.SR]}
  \BibitemShut {NoStop}%
\bibitem [{\citenamefont {{Robitaille}}\ \emph {et~al.}(2020)\citenamefont
  {{Robitaille}}, \citenamefont {{Abdeldayem}}, \citenamefont {{Joncour}},
  \citenamefont {{Moraux}}, \citenamefont {{Motte}}, \citenamefont
  {{Lesaffre}},\ and\ \citenamefont {{Khalil}}}]{Robitaille+2020}%
  \BibitemOpen
  \bibfield  {author} {\bibinfo {author} {\bibfnamefont {J.~F.}\ \bibnamefont
  {{Robitaille}}}, \bibinfo {author} {\bibfnamefont {A.}~\bibnamefont
  {{Abdeldayem}}}, \bibinfo {author} {\bibfnamefont {I.}~\bibnamefont
  {{Joncour}}}, \bibinfo {author} {\bibfnamefont {E.}~\bibnamefont {{Moraux}}},
  \bibinfo {author} {\bibfnamefont {F.}~\bibnamefont {{Motte}}}, \bibinfo
  {author} {\bibfnamefont {P.}~\bibnamefont {{Lesaffre}}},\ and\ \bibinfo
  {author} {\bibfnamefont {A.}~\bibnamefont {{Khalil}}},\ }\href
  {https://doi.org/10.1051/0004-6361/201937085} {\bibfield  {journal} {\bibinfo
   {journal} {\aap}\ }\textbf {\bibinfo {volume} {641}},\ \bibinfo {eid} {A138}
  (\bibinfo {year} {2020})},\ \Eprint {https://arxiv.org/abs/2007.08206}
  {arXiv:2007.08206 [astro-ph.GA]} \BibitemShut {NoStop}%
\bibitem [{\citenamefont {{Ardila}}\ \emph {et~al.}(2005)\citenamefont
  {{Ardila}}, \citenamefont {{Lubow}}, \citenamefont {{Golimowski}},
  \citenamefont {{Krist}}, \citenamefont {{Clampin}}, \citenamefont {{Ford}},
  \citenamefont {{Hartig}},\ and\ \citenamefont {{Illingworth et
  al.}}}]{Ardila+2005}%
  \BibitemOpen
  \bibfield  {author} {\bibinfo {author} {\bibfnamefont {D.~R.}\ \bibnamefont
  {{Ardila}}}, \bibinfo {author} {\bibfnamefont {S.~H.}\ \bibnamefont
  {{Lubow}}}, \bibinfo {author} {\bibfnamefont {D.~A.}\ \bibnamefont
  {{Golimowski}}}, \bibinfo {author} {\bibfnamefont {J.~E.}\ \bibnamefont
  {{Krist}}}, \bibinfo {author} {\bibfnamefont {M.}~\bibnamefont {{Clampin}}},
  \bibinfo {author} {\bibfnamefont {H.~C.}\ \bibnamefont {{Ford}}}, \bibinfo
  {author} {\bibfnamefont {G.~F.}\ \bibnamefont {{Hartig}}},\ and\ \bibinfo
  {author} {\bibnamefont {{Illingworth et al.}}},\ }\href
  {https://doi.org/10.1086/430395} {\bibfield  {journal} {\bibinfo  {journal}
  {\apj}\ }\textbf {\bibinfo {volume} {627}},\ \bibinfo {pages} {986} (\bibinfo
  {year} {2005})},\ \Eprint {https://arxiv.org/abs/astro-ph/0503445}
  {arXiv:astro-ph/0503445 [astro-ph]} \BibitemShut {NoStop}%
\bibitem [{\citenamefont {{Reche}}\ \emph {et~al.}(2009)\citenamefont
  {{Reche}}, \citenamefont {{Beust}},\ and\ \citenamefont
  {{Augereau}}}]{Reche+2009}%
  \BibitemOpen
  \bibfield  {author} {\bibinfo {author} {\bibfnamefont {R.}~\bibnamefont
  {{Reche}}}, \bibinfo {author} {\bibfnamefont {H.}~\bibnamefont {{Beust}}},\
  and\ \bibinfo {author} {\bibfnamefont {J.~C.}\ \bibnamefont {{Augereau}}},\
  }\href {https://doi.org/10.1051/0004-6361:200810419} {\bibfield  {journal}
  {\bibinfo  {journal} {\aap}\ }\textbf {\bibinfo {volume} {493}},\ \bibinfo
  {pages} {661} (\bibinfo {year} {2009})},\ \Eprint
  {https://arxiv.org/abs/0809.4421} {arXiv:0809.4421 [astro-ph]} \BibitemShut
  {NoStop}%
\bibitem [{\citenamefont {{White}}\ \emph {et~al.}(2018)\citenamefont
  {{White}}, \citenamefont {{Boley}}, \citenamefont {{MacGregor}},
  \citenamefont {{Hughes}},\ and\ \citenamefont {{Wilner}}}]{White+2018}%
  \BibitemOpen
  \bibfield  {author} {\bibinfo {author} {\bibfnamefont {J.~A.}\ \bibnamefont
  {{White}}}, \bibinfo {author} {\bibfnamefont {A.~C.}\ \bibnamefont
  {{Boley}}}, \bibinfo {author} {\bibfnamefont {M.~A.}\ \bibnamefont
  {{MacGregor}}}, \bibinfo {author} {\bibfnamefont {A.~M.}\ \bibnamefont
  {{Hughes}}},\ and\ \bibinfo {author} {\bibfnamefont {D.~J.}\ \bibnamefont
  {{Wilner}}},\ }\href {https://doi.org/10.1093/mnras/stx3098} {\bibfield
  {journal} {\bibinfo  {journal} {\mnras}\ }\textbf {\bibinfo {volume} {474}},\
  \bibinfo {pages} {4500} (\bibinfo {year} {2018})},\ \Eprint
  {https://arxiv.org/abs/1711.07489} {arXiv:1711.07489 [astro-ph.EP]}
  \BibitemShut {NoStop}%
\bibitem [{\citenamefont {{Mamajek}}\ \emph {et~al.}(2015)\citenamefont
  {{Mamajek}}, \citenamefont {{Barenfeld}}, \citenamefont {{Ivanov}},
  \citenamefont {{Kniazev}}, \citenamefont {{V{\"a}is{\"a}nen}}, \citenamefont
  {{Beletsky}},\ and\ \citenamefont {{Boffin}}}]{Mamajek+2015}%
  \BibitemOpen
  \bibfield  {author} {\bibinfo {author} {\bibfnamefont {E.~E.}\ \bibnamefont
  {{Mamajek}}}, \bibinfo {author} {\bibfnamefont {S.~A.}\ \bibnamefont
  {{Barenfeld}}}, \bibinfo {author} {\bibfnamefont {V.~D.}\ \bibnamefont
  {{Ivanov}}}, \bibinfo {author} {\bibfnamefont {A.~Y.}\ \bibnamefont
  {{Kniazev}}}, \bibinfo {author} {\bibfnamefont {P.}~\bibnamefont
  {{V{\"a}is{\"a}nen}}}, \bibinfo {author} {\bibfnamefont {Y.}~\bibnamefont
  {{Beletsky}}},\ and\ \bibinfo {author} {\bibfnamefont {H.~M.~J.}\
  \bibnamefont {{Boffin}}},\ }\href
  {https://doi.org/10.1088/2041-8205/800/1/L17} {\bibfield  {journal} {\bibinfo
   {journal} {\apjl}\ }\textbf {\bibinfo {volume} {800}},\ \bibinfo {eid} {L17}
  (\bibinfo {year} {2015})},\ \Eprint {https://arxiv.org/abs/1502.04655}
  {arXiv:1502.04655 [astro-ph.SR]} \BibitemShut {NoStop}%
\bibitem [{\citenamefont {{Pfalzner}}\ and\ \citenamefont
  {{Vincke}}(2020)}]{Pfalzner+2020}%
  \BibitemOpen
  \bibfield  {author} {\bibinfo {author} {\bibfnamefont {S.}~\bibnamefont
  {{Pfalzner}}}\ and\ \bibinfo {author} {\bibfnamefont {K.}~\bibnamefont
  {{Vincke}}},\ }\href {https://doi.org/10.3847/1538-4357/ab9533} {\bibfield
  {journal} {\bibinfo  {journal} {\apj}\ }\textbf {\bibinfo {volume} {897}},\
  \bibinfo {eid} {60} (\bibinfo {year} {2020})},\ \Eprint
  {https://arxiv.org/abs/2005.11260} {arXiv:2005.11260 [astro-ph.EP]}
  \BibitemShut {NoStop}%
\bibitem [{\citenamefont {{Heller}}(1993)}]{Heller1993}%
  \BibitemOpen
  \bibfield  {author} {\bibinfo {author} {\bibfnamefont {C.~H.}\ \bibnamefont
  {{Heller}}},\ }\href {https://doi.org/10.1086/172591} {\bibfield  {journal}
  {\bibinfo  {journal} {\apj}\ }\textbf {\bibinfo {volume} {408}},\ \bibinfo
  {pages} {337} (\bibinfo {year} {1993})}\BibitemShut {NoStop}%
\bibitem [{\citenamefont {{Ida}}\ \emph {et~al.}(2000)\citenamefont {{Ida}},
  \citenamefont {{Larwood}},\ and\ \citenamefont {{Burkert}}}]{Ida+2000}%
  \BibitemOpen
  \bibfield  {author} {\bibinfo {author} {\bibfnamefont {S.}~\bibnamefont
  {{Ida}}}, \bibinfo {author} {\bibfnamefont {J.}~\bibnamefont {{Larwood}}},\
  and\ \bibinfo {author} {\bibfnamefont {A.}~\bibnamefont {{Burkert}}},\ }\href
  {https://doi.org/10.1086/308179} {\bibfield  {journal} {\bibinfo  {journal}
  {\apj}\ }\textbf {\bibinfo {volume} {528}},\ \bibinfo {pages} {351} (\bibinfo
  {year} {2000})},\ \Eprint {https://arxiv.org/abs/astro-ph/9907217}
  {arXiv:astro-ph/9907217 [astro-ph]} \BibitemShut {NoStop}%
\bibitem [{\citenamefont {{Kenyon}}\ and\ \citenamefont
  {{Bromley}}(2004)}]{Kenyon+2004}%
  \BibitemOpen
  \bibfield  {author} {\bibinfo {author} {\bibfnamefont {S.~J.}\ \bibnamefont
  {{Kenyon}}}\ and\ \bibinfo {author} {\bibfnamefont {B.~C.}\ \bibnamefont
  {{Bromley}}},\ }\href {https://doi.org/10.1038/nature03136} {\bibfield
  {journal} {\bibinfo  {journal} {\nat}\ }\textbf {\bibinfo {volume} {432}},\
  \bibinfo {pages} {598} (\bibinfo {year} {2004})},\ \Eprint
  {https://arxiv.org/abs/astro-ph/0412030} {arXiv:astro-ph/0412030 [astro-ph]}
  \BibitemShut {NoStop}%
\bibitem [{\citenamefont {{Kobayashi}}\ \emph {et~al.}(2005)\citenamefont
  {{Kobayashi}}, \citenamefont {{Ida}},\ and\ \citenamefont
  {{Tanaka}}}]{Kobayashi+2005}%
  \BibitemOpen
  \bibfield  {author} {\bibinfo {author} {\bibfnamefont {H.}~\bibnamefont
  {{Kobayashi}}}, \bibinfo {author} {\bibfnamefont {S.}~\bibnamefont {{Ida}}},\
  and\ \bibinfo {author} {\bibfnamefont {H.}~\bibnamefont {{Tanaka}}},\ }\href
  {https://doi.org/10.1016/j.icarus.2005.02.017} {\bibfield  {journal}
  {\bibinfo  {journal} {\icarus}\ }\textbf {\bibinfo {volume} {177}},\ \bibinfo
  {pages} {246} (\bibinfo {year} {2005})}\BibitemShut {NoStop}%
\bibitem [{\citenamefont {{Gaia Collaboration}}\ \emph
  {et~al.}(2016)\citenamefont {{Gaia Collaboration}}, \citenamefont {{Prusti}},
  \citenamefont {{de Bruijne}}, \citenamefont {{Brown}},\ and\ \citenamefont
  {{Vallenari et al.}}}]{Gaia+2016}%
  \BibitemOpen
  \bibfield  {author} {\bibinfo {author} {\bibnamefont {{Gaia Collaboration}}},
  \bibinfo {author} {\bibfnamefont {T.}~\bibnamefont {{Prusti}}}, \bibinfo
  {author} {\bibfnamefont {J.~H.~J.}\ \bibnamefont {{de Bruijne}}}, \bibinfo
  {author} {\bibfnamefont {A.~G.~A.}\ \bibnamefont {{Brown}}},\ and\ \bibinfo
  {author} {\bibfnamefont {A.}~\bibnamefont {{Vallenari et al.}}},\ }\href
  {https://doi.org/10.1051/0004-6361/201629272} {\bibfield  {journal} {\bibinfo
   {journal} {\aap}\ }\textbf {\bibinfo {volume} {595}},\ \bibinfo {eid} {A1}
  (\bibinfo {year} {2016})},\ \Eprint {https://arxiv.org/abs/1609.04153}
  {arXiv:1609.04153 [astro-ph.IM]} \BibitemShut {NoStop}%
\bibitem [{\citenamefont {{Bailer-Jones}}(2015)}]{Bailer-Jones2015}%
  \BibitemOpen
  \bibfield  {author} {\bibinfo {author} {\bibfnamefont {C.~A.~L.}\
  \bibnamefont {{Bailer-Jones}}},\ }\href
  {https://doi.org/10.1051/0004-6361/201425221} {\bibfield  {journal} {\bibinfo
   {journal} {\aap}\ }\textbf {\bibinfo {volume} {575}},\ \bibinfo {eid} {A35}
  (\bibinfo {year} {2015})},\ \Eprint {https://arxiv.org/abs/1412.3648}
  {arXiv:1412.3648 [astro-ph.SR]} \BibitemShut {NoStop}%
\bibitem [{\citenamefont {{Bailer-Jones}}(2018)}]{Bailer-Jones2018}%
  \BibitemOpen
  \bibfield  {author} {\bibinfo {author} {\bibfnamefont {C.~A.~L.}\
  \bibnamefont {{Bailer-Jones}}},\ }\href
  {https://doi.org/10.1051/0004-6361/201731453} {\bibfield  {journal} {\bibinfo
   {journal} {\aap}\ }\textbf {\bibinfo {volume} {609}},\ \bibinfo {eid} {A8}
  (\bibinfo {year} {2018})},\ \Eprint {https://arxiv.org/abs/1708.08595}
  {arXiv:1708.08595 [astro-ph.SR]} \BibitemShut {NoStop}%
\bibitem [{\citenamefont {{Bailer-Jones}}\ \emph {et~al.}(2018)\citenamefont
  {{Bailer-Jones}}, \citenamefont {{Rybizki}}, \citenamefont {{Andrae}},\ and\
  \citenamefont {{Fouesneau}}}]{Bailer-Jones+2018}%
  \BibitemOpen
  \bibfield  {author} {\bibinfo {author} {\bibfnamefont {C.~A.~L.}\
  \bibnamefont {{Bailer-Jones}}}, \bibinfo {author} {\bibfnamefont
  {J.}~\bibnamefont {{Rybizki}}}, \bibinfo {author} {\bibfnamefont
  {R.}~\bibnamefont {{Andrae}}},\ and\ \bibinfo {author} {\bibfnamefont
  {M.}~\bibnamefont {{Fouesneau}}},\ }\href
  {https://doi.org/10.1051/0004-6361/201833456} {\bibfield  {journal} {\bibinfo
   {journal} {\aap}\ }\textbf {\bibinfo {volume} {616}},\ \bibinfo {eid} {A37}
  (\bibinfo {year} {2018})},\ \Eprint {https://arxiv.org/abs/1805.07581}
  {arXiv:1805.07581 [astro-ph.SR]} \BibitemShut {NoStop}%
\bibitem [{\citenamefont {{Ma}}\ \emph {et~al.}(2022)\citenamefont {{Ma}},
  \citenamefont {{De Rosa}},\ and\ \citenamefont {{Kalas}}}]{Ma+2022}%
  \BibitemOpen
  \bibfield  {author} {\bibinfo {author} {\bibfnamefont {Y.}~\bibnamefont
  {{Ma}}}, \bibinfo {author} {\bibfnamefont {R.~J.}\ \bibnamefont {{De
  Rosa}}},\ and\ \bibinfo {author} {\bibfnamefont {P.}~\bibnamefont
  {{Kalas}}},\ }\href@noop {} {\bibfield  {journal} {\bibinfo  {journal} {arXiv
  e-prints}\ ,\ \bibinfo {eid} {arXiv:2202.00922}} (\bibinfo {year} {2022})},\
  \Eprint {https://arxiv.org/abs/2202.00922} {arXiv:2202.00922 [astro-ph.EP]}
  \BibitemShut {NoStop}%
\bibitem [{\citenamefont {{Hansen}}(2022)}]{Hansen2022}%
  \BibitemOpen
  \bibfield  {author} {\bibinfo {author} {\bibfnamefont {B.~M.~S.}\
  \bibnamefont {{Hansen}}},\ }\href {https://doi.org/10.3847/1538-3881/ac3a8b}
  {\bibfield  {journal} {\bibinfo  {journal} {\aj}\ }\textbf {\bibinfo {volume}
  {163}},\ \bibinfo {eid} {44} (\bibinfo {year} {2022})},\ \Eprint
  {https://arxiv.org/abs/2112.00852} {arXiv:2112.00852 [astro-ph.SR]}
  \BibitemShut {NoStop}%
\bibitem [{\citenamefont {{Shuai}}\ \emph {et~al.}(2022)\citenamefont
  {{Shuai}}, \citenamefont {{Ren}}, \citenamefont {{Dong}}, \citenamefont
  {{Zhou}}, \citenamefont {{Pueyo}}, \citenamefont {{De Rosa}}, \citenamefont
  {{Fang}},\ and\ \citenamefont {{Mawet}}}]{Shuai+2022}%
  \BibitemOpen
  \bibfield  {author} {\bibinfo {author} {\bibfnamefont {L.}~\bibnamefont
  {{Shuai}}}, \bibinfo {author} {\bibfnamefont {B.~B.}\ \bibnamefont {{Ren}}},
  \bibinfo {author} {\bibfnamefont {R.}~\bibnamefont {{Dong}}}, \bibinfo
  {author} {\bibfnamefont {X.}~\bibnamefont {{Zhou}}}, \bibinfo {author}
  {\bibfnamefont {L.}~\bibnamefont {{Pueyo}}}, \bibinfo {author} {\bibfnamefont
  {R.~J.}\ \bibnamefont {{De Rosa}}}, \bibinfo {author} {\bibfnamefont
  {T.}~\bibnamefont {{Fang}}},\ and\ \bibinfo {author} {\bibfnamefont
  {D.}~\bibnamefont {{Mawet}}},\ }\href@noop {} {\bibfield  {journal} {\bibinfo
   {journal} {arXiv e-prints}\ ,\ \bibinfo {eid} {arXiv:2210.03725}} (\bibinfo
  {year} {2022})},\ \Eprint {https://arxiv.org/abs/2210.03725}
  {arXiv:2210.03725 [astro-ph.EP]} \BibitemShut {NoStop}%
\bibitem [{\citenamefont {{Reipurth}}\ and\ \citenamefont
  {{Mikkola}}(2012)}]{Reipurth&Mikkola2012}%
  \BibitemOpen
  \bibfield  {author} {\bibinfo {author} {\bibfnamefont {B.}~\bibnamefont
  {{Reipurth}}}\ and\ \bibinfo {author} {\bibfnamefont {S.}~\bibnamefont
  {{Mikkola}}},\ }\href {https://doi.org/10.1038/nature11662} {\bibfield
  {journal} {\bibinfo  {journal} {\nat}\ }\textbf {\bibinfo {volume} {492}},\
  \bibinfo {pages} {221} (\bibinfo {year} {2012})},\ \Eprint
  {https://arxiv.org/abs/1212.1246} {arXiv:1212.1246 [astro-ph.GA]}
  \BibitemShut {NoStop}%
\bibitem [{\citenamefont {{Tokovinin}}(2014)}]{Tokovinin2014}%
  \BibitemOpen
  \bibfield  {author} {\bibinfo {author} {\bibfnamefont {A.}~\bibnamefont
  {{Tokovinin}}},\ }\href {https://doi.org/10.1088/0004-6256/147/4/87}
  {\bibfield  {journal} {\bibinfo  {journal} {\aj}\ }\textbf {\bibinfo {volume}
  {147}},\ \bibinfo {eid} {87} (\bibinfo {year} {2014})},\ \Eprint
  {https://arxiv.org/abs/1401.6827} {arXiv:1401.6827 [astro-ph.SR]}
  \BibitemShut {NoStop}%
\bibitem [{\citenamefont {{Halbwachs}}\ \emph {et~al.}(2017)\citenamefont
  {{Halbwachs}}, \citenamefont {{Mayor}},\ and\ \citenamefont
  {{Udry}}}]{Halbwachs2017}%
  \BibitemOpen
  \bibfield  {author} {\bibinfo {author} {\bibfnamefont {J.~L.}\ \bibnamefont
  {{Halbwachs}}}, \bibinfo {author} {\bibfnamefont {M.}~\bibnamefont
  {{Mayor}}},\ and\ \bibinfo {author} {\bibfnamefont {S.}~\bibnamefont
  {{Udry}}},\ }\href {https://doi.org/10.1093/mnras/stw2683} {\bibfield
  {journal} {\bibinfo  {journal} {\mnras}\ }\textbf {\bibinfo {volume} {464}},\
  \bibinfo {pages} {4966} (\bibinfo {year} {2017})},\ \Eprint
  {https://arxiv.org/abs/1610.04423} {arXiv:1610.04423 [astro-ph.SR]}
  \BibitemShut {NoStop}%
\bibitem [{\citenamefont {{Riddle}}\ \emph {et~al.}(2015)\citenamefont
  {{Riddle}}, \citenamefont {{Tokovinin}}, \citenamefont {{Mason}},
  \citenamefont {{Hartkopf}}, \citenamefont {{Roberts}}, \citenamefont
  {{Baranec}}, \citenamefont {{Law}}, \citenamefont {{Bui}}, \citenamefont
  {{Burse}}, \citenamefont {{Das}}, \citenamefont {{Dekany}}, \citenamefont
  {{Kulkarni}}, \citenamefont {{Punnadi}}, \citenamefont {{Ramaprakash}},\ and\
  \citenamefont {{Tendulkar}}}]{Riddle2015}%
  \BibitemOpen
  \bibfield  {author} {\bibinfo {author} {\bibfnamefont {R.~L.}\ \bibnamefont
  {{Riddle}}}, \bibinfo {author} {\bibfnamefont {A.}~\bibnamefont
  {{Tokovinin}}}, \bibinfo {author} {\bibfnamefont {B.~D.}\ \bibnamefont
  {{Mason}}}, \bibinfo {author} {\bibfnamefont {W.~I.}\ \bibnamefont
  {{Hartkopf}}}, \bibinfo {author} {\bibfnamefont {J.}~\bibnamefont
  {{Roberts}}, \bibfnamefont {Lewis~C.}}, \bibinfo {author} {\bibfnamefont
  {C.}~\bibnamefont {{Baranec}}}, \bibinfo {author} {\bibfnamefont {N.~M.}\
  \bibnamefont {{Law}}}, \bibinfo {author} {\bibfnamefont {K.}~\bibnamefont
  {{Bui}}}, \bibinfo {author} {\bibfnamefont {M.~P.}\ \bibnamefont {{Burse}}},
  \bibinfo {author} {\bibfnamefont {H.~K.}\ \bibnamefont {{Das}}}, \bibinfo
  {author} {\bibfnamefont {R.~G.}\ \bibnamefont {{Dekany}}}, \bibinfo {author}
  {\bibfnamefont {S.}~\bibnamefont {{Kulkarni}}}, \bibinfo {author}
  {\bibfnamefont {S.}~\bibnamefont {{Punnadi}}}, \bibinfo {author}
  {\bibfnamefont {A.~N.}\ \bibnamefont {{Ramaprakash}}},\ and\ \bibinfo
  {author} {\bibfnamefont {S.~P.}\ \bibnamefont {{Tendulkar}}},\ }\href
  {https://doi.org/10.1088/0004-637X/799/1/4} {\bibfield  {journal} {\bibinfo
  {journal} {\apj}\ }\textbf {\bibinfo {volume} {799}},\ \bibinfo {eid} {4}
  (\bibinfo {year} {2015})},\ \Eprint {https://arxiv.org/abs/1411.0682}
  {arXiv:1411.0682 [astro-ph.SR]} \BibitemShut {NoStop}%
\bibitem [{\citenamefont {{Clarke}}(2020)}]{Clarke2020}%
  \BibitemOpen
  \bibfield  {author} {\bibinfo {author} {\bibfnamefont {C.~J.}\ \bibnamefont
  {{Clarke}}},\ }\href {https://doi.org/10.1093/mnrasl/slz161} {\bibfield
  {journal} {\bibinfo  {journal} {\mnras}\ }\textbf {\bibinfo {volume} {491}},\
  \bibinfo {pages} {L72} (\bibinfo {year} {2020})},\ \Eprint
  {https://arxiv.org/abs/1910.10256} {arXiv:1910.10256 [astro-ph.SR]}
  \BibitemShut {NoStop}%
\bibitem [{\citenamefont {{McBride}}\ and\ \citenamefont
  {{Kounkel}}(2019)}]{McBride+2019}%
  \BibitemOpen
  \bibfield  {author} {\bibinfo {author} {\bibfnamefont {A.}~\bibnamefont
  {{McBride}}}\ and\ \bibinfo {author} {\bibfnamefont {M.}~\bibnamefont
  {{Kounkel}}},\ }\href {https://doi.org/10.3847/1538-4357/ab3df9} {\bibfield
  {journal} {\bibinfo  {journal} {\apj}\ }\textbf {\bibinfo {volume} {884}},\
  \bibinfo {eid} {6} (\bibinfo {year} {2019})},\ \Eprint
  {https://arxiv.org/abs/1908.07550} {arXiv:1908.07550 [astro-ph.SR]}
  \BibitemShut {NoStop}%
\bibitem [{\citenamefont {{Ostriker}}(1994)}]{Ostriker1994}%
  \BibitemOpen
  \bibfield  {author} {\bibinfo {author} {\bibfnamefont {E.~C.}\ \bibnamefont
  {{Ostriker}}},\ }\href {https://doi.org/10.1086/173890} {\bibfield  {journal}
  {\bibinfo  {journal} {\apj}\ }\textbf {\bibinfo {volume} {424}},\ \bibinfo
  {pages} {292} (\bibinfo {year} {1994})}\BibitemShut {NoStop}%
\bibitem [{\citenamefont {{Toomre}}\ and\ \citenamefont
  {{Toomre}}(1972)}]{Toomre1972}%
  \BibitemOpen
  \bibfield  {author} {\bibinfo {author} {\bibfnamefont {A.}~\bibnamefont
  {{Toomre}}}\ and\ \bibinfo {author} {\bibfnamefont {J.}~\bibnamefont
  {{Toomre}}},\ }\href {https://doi.org/10.1086/151823} {\bibfield  {journal}
  {\bibinfo  {journal} {\apj}\ }\textbf {\bibinfo {volume} {178}},\ \bibinfo
  {pages} {623} (\bibinfo {year} {1972})}\BibitemShut {NoStop}%
\bibitem [{\citenamefont {{D'Onghia}}\ \emph {et~al.}(2010)\citenamefont
  {{D'Onghia}}, \citenamefont {{Vogelsberger}}, \citenamefont
  {{Faucher-Giguere}},\ and\ \citenamefont {{Hernquist}}}]{DOnghia+2010}%
  \BibitemOpen
  \bibfield  {author} {\bibinfo {author} {\bibfnamefont {E.}~\bibnamefont
  {{D'Onghia}}}, \bibinfo {author} {\bibfnamefont {M.}~\bibnamefont
  {{Vogelsberger}}}, \bibinfo {author} {\bibfnamefont {C.-A.}\ \bibnamefont
  {{Faucher-Giguere}}},\ and\ \bibinfo {author} {\bibfnamefont
  {L.}~\bibnamefont {{Hernquist}}},\ }\href
  {https://doi.org/10.1088/0004-637X/725/1/353} {\bibfield  {journal} {\bibinfo
   {journal} {\apj}\ }\textbf {\bibinfo {volume} {725}},\ \bibinfo {pages}
  {353} (\bibinfo {year} {2010})},\ \Eprint {https://arxiv.org/abs/1009.3927}
  {arXiv:1009.3927 [astro-ph.CO]} \BibitemShut {NoStop}%
\bibitem [{\citenamefont {{Cuello}}\ \emph {et~al.}(2020)\citenamefont
  {{Cuello}}, \citenamefont {{Louvet}}, \citenamefont {{Mentiplay}},
  \citenamefont {{Pinte}}, \citenamefont {{Price}}, \citenamefont {{Winter}},
  \citenamefont {{Nealon}}, \citenamefont {{M{\'e}nard}}, \citenamefont
  {{Lodato}}, \citenamefont {{Dipierro}}, \citenamefont {{Christiaens}},
  \citenamefont {{Montesinos}}, \citenamefont {{Cuadra}}, \citenamefont
  {{Laibe}}, \citenamefont {{Cieza}}, \citenamefont {{Dong}},\ and\
  \citenamefont {{Alexander}}}]{Cuello+2020}%
  \BibitemOpen
  \bibfield  {author} {\bibinfo {author} {\bibfnamefont {N.}~\bibnamefont
  {{Cuello}}}, \bibinfo {author} {\bibfnamefont {F.}~\bibnamefont {{Louvet}}},
  \bibinfo {author} {\bibfnamefont {D.}~\bibnamefont {{Mentiplay}}}, \bibinfo
  {author} {\bibfnamefont {C.}~\bibnamefont {{Pinte}}}, \bibinfo {author}
  {\bibfnamefont {D.~J.}\ \bibnamefont {{Price}}}, \bibinfo {author}
  {\bibfnamefont {A.~J.}\ \bibnamefont {{Winter}}}, \bibinfo {author}
  {\bibfnamefont {R.}~\bibnamefont {{Nealon}}}, \bibinfo {author}
  {\bibfnamefont {F.}~\bibnamefont {{M{\'e}nard}}}, \bibinfo {author}
  {\bibfnamefont {G.}~\bibnamefont {{Lodato}}}, \bibinfo {author}
  {\bibfnamefont {G.}~\bibnamefont {{Dipierro}}}, \bibinfo {author}
  {\bibfnamefont {V.}~\bibnamefont {{Christiaens}}}, \bibinfo {author}
  {\bibfnamefont {M.}~\bibnamefont {{Montesinos}}}, \bibinfo {author}
  {\bibfnamefont {J.}~\bibnamefont {{Cuadra}}}, \bibinfo {author}
  {\bibfnamefont {G.}~\bibnamefont {{Laibe}}}, \bibinfo {author} {\bibfnamefont
  {L.}~\bibnamefont {{Cieza}}}, \bibinfo {author} {\bibfnamefont
  {R.}~\bibnamefont {{Dong}}},\ and\ \bibinfo {author} {\bibfnamefont
  {R.}~\bibnamefont {{Alexander}}},\ }\href
  {https://doi.org/10.1093/mnras/stz2938} {\bibfield  {journal} {\bibinfo
  {journal} {\mnras}\ }\textbf {\bibinfo {volume} {491}},\ \bibinfo {pages}
  {504} (\bibinfo {year} {2020})},\ \Eprint {https://arxiv.org/abs/1910.06822}
  {arXiv:1910.06822 [astro-ph.EP]} \BibitemShut {NoStop}%
\bibitem [{\citenamefont {{Binney}}\ and\ \citenamefont
  {{Tremaine}}(2008)}]{BinneyTremaine2008}%
  \BibitemOpen
  \bibfield  {author} {\bibinfo {author} {\bibfnamefont {J.}~\bibnamefont
  {{Binney}}}\ and\ \bibinfo {author} {\bibfnamefont {S.}~\bibnamefont
  {{Tremaine}}},\ }\href@noop {} {\emph {\bibinfo {title} {{Galactic Dynamics:
  Second Edition}}}}\ (\bibinfo  {publisher} {{Princeton University Press}},\
  \bibinfo {year} {2008})\BibitemShut {NoStop}%
\bibitem [{\citenamefont {{Dong}}\ \emph {et~al.}(2016)\citenamefont {{Dong}},
  \citenamefont {{Fung}},\ and\ \citenamefont {{Chiang}}}]{Dong+2016}%
  \BibitemOpen
  \bibfield  {author} {\bibinfo {author} {\bibfnamefont {R.}~\bibnamefont
  {{Dong}}}, \bibinfo {author} {\bibfnamefont {J.}~\bibnamefont {{Fung}}},\
  and\ \bibinfo {author} {\bibfnamefont {E.}~\bibnamefont {{Chiang}}},\ }\href
  {https://doi.org/10.3847/0004-637X/826/1/75} {\bibfield  {journal} {\bibinfo
  {journal} {\apj}\ }\textbf {\bibinfo {volume} {826}},\ \bibinfo {eid} {75}
  (\bibinfo {year} {2016})},\ \Eprint {https://arxiv.org/abs/1602.04814}
  {arXiv:1602.04814 [astro-ph.EP]} \BibitemShut {NoStop}%
\bibitem [{\citenamefont {{Zhu}}\ and\ \citenamefont
  {{Zhang}}(2022)}]{Zhu+2022}%
  \BibitemOpen
  \bibfield  {author} {\bibinfo {author} {\bibfnamefont {Z.}~\bibnamefont
  {{Zhu}}}\ and\ \bibinfo {author} {\bibfnamefont {R.~M.}\ \bibnamefont
  {{Zhang}}},\ }\href {https://doi.org/10.1093/mnras/stab3641} {\bibfield
  {journal} {\bibinfo  {journal} {\mnras}\ }\textbf {\bibinfo {volume} {510}},\
  \bibinfo {pages} {3986} (\bibinfo {year} {2022})},\ \Eprint
  {https://arxiv.org/abs/2112.03311} {arXiv:2112.03311 [astro-ph.EP]}
  \BibitemShut {NoStop}%
\bibitem [{\citenamefont {{Breslau}}\ \emph {et~al.}(2014)\citenamefont
  {{Breslau}}, \citenamefont {{Steinhausen}}, \citenamefont {{Vincke}},\ and\
  \citenamefont {{Pfalzner}}}]{Breslau+2014}%
  \BibitemOpen
  \bibfield  {author} {\bibinfo {author} {\bibfnamefont {A.}~\bibnamefont
  {{Breslau}}}, \bibinfo {author} {\bibfnamefont {M.}~\bibnamefont
  {{Steinhausen}}}, \bibinfo {author} {\bibfnamefont {K.}~\bibnamefont
  {{Vincke}}},\ and\ \bibinfo {author} {\bibfnamefont {S.}~\bibnamefont
  {{Pfalzner}}},\ }\href {https://doi.org/10.1051/0004-6361/201323043}
  {\bibfield  {journal} {\bibinfo  {journal} {\aap}\ }\textbf {\bibinfo
  {volume} {565}},\ \bibinfo {eid} {A130} (\bibinfo {year} {2014})},\ \Eprint
  {https://arxiv.org/abs/1403.8099} {arXiv:1403.8099 [astro-ph.GA]}
  \BibitemShut {NoStop}%
\bibitem [{\citenamefont {{Vincke}}\ \emph {et~al.}(2015)\citenamefont
  {{Vincke}}, \citenamefont {{Breslau}},\ and\ \citenamefont
  {{Pfalzner}}}]{Vincke+2015}%
  \BibitemOpen
  \bibfield  {author} {\bibinfo {author} {\bibfnamefont {K.}~\bibnamefont
  {{Vincke}}}, \bibinfo {author} {\bibfnamefont {A.}~\bibnamefont
  {{Breslau}}},\ and\ \bibinfo {author} {\bibfnamefont {S.}~\bibnamefont
  {{Pfalzner}}},\ }\href {https://doi.org/10.1051/0004-6361/201425552}
  {\bibfield  {journal} {\bibinfo  {journal} {\aap}\ }\textbf {\bibinfo
  {volume} {577}},\ \bibinfo {eid} {A115} (\bibinfo {year} {2015})},\ \Eprint
  {https://arxiv.org/abs/1504.06092} {arXiv:1504.06092 [astro-ph.SR]}
  \BibitemShut {NoStop}%
\bibitem [{\citenamefont {{Vincke}}\ and\ \citenamefont
  {{Pfalzner}}(2016)}]{Vincke+2016}%
  \BibitemOpen
  \bibfield  {author} {\bibinfo {author} {\bibfnamefont {K.}~\bibnamefont
  {{Vincke}}}\ and\ \bibinfo {author} {\bibfnamefont {S.}~\bibnamefont
  {{Pfalzner}}},\ }\href {https://doi.org/10.3847/0004-637X/828/1/48}
  {\bibfield  {journal} {\bibinfo  {journal} {\apj}\ }\textbf {\bibinfo
  {volume} {828}},\ \bibinfo {eid} {48} (\bibinfo {year} {2016})},\ \Eprint
  {https://arxiv.org/abs/1606.07431} {arXiv:1606.07431 [astro-ph.GA]}
  \BibitemShut {NoStop}%
\bibitem [{\citenamefont {{Bhandare}}\ \emph {et~al.}(2016)\citenamefont
  {{Bhandare}}, \citenamefont {{Breslau}},\ and\ \citenamefont
  {{Pfalzner}}}]{Bhandare+2016}%
  \BibitemOpen
  \bibfield  {author} {\bibinfo {author} {\bibfnamefont {A.}~\bibnamefont
  {{Bhandare}}}, \bibinfo {author} {\bibfnamefont {A.}~\bibnamefont
  {{Breslau}}},\ and\ \bibinfo {author} {\bibfnamefont {S.}~\bibnamefont
  {{Pfalzner}}},\ }\href {https://doi.org/10.1051/0004-6361/201628086}
  {\bibfield  {journal} {\bibinfo  {journal} {\aap}\ }\textbf {\bibinfo
  {volume} {594}},\ \bibinfo {eid} {A53} (\bibinfo {year} {2016})},\ \Eprint
  {https://arxiv.org/abs/1608.03239} {arXiv:1608.03239 [astro-ph.EP]}
  \BibitemShut {NoStop}%
\bibitem [{\citenamefont {{Breslau}}\ \emph {et~al.}(2017)\citenamefont
  {{Breslau}}, \citenamefont {{Vincke}},\ and\ \citenamefont
  {{Pfalzner}}}]{Breslau+2017}%
  \BibitemOpen
  \bibfield  {author} {\bibinfo {author} {\bibfnamefont {A.}~\bibnamefont
  {{Breslau}}}, \bibinfo {author} {\bibfnamefont {K.}~\bibnamefont
  {{Vincke}}},\ and\ \bibinfo {author} {\bibfnamefont {S.}~\bibnamefont
  {{Pfalzner}}},\ }\href {https://doi.org/10.1051/0004-6361/201526068}
  {\bibfield  {journal} {\bibinfo  {journal} {\aap}\ }\textbf {\bibinfo
  {volume} {599}},\ \bibinfo {eid} {A91} (\bibinfo {year} {2017})},\ \Eprint
  {https://arxiv.org/abs/1701.00855} {arXiv:1701.00855 [astro-ph.EP]}
  \BibitemShut {NoStop}%
\bibitem [{\citenamefont {{J{\'\i}lkov{\'a}}}\ \emph
  {et~al.}(2016)\citenamefont {{J{\'\i}lkov{\'a}}}, \citenamefont {{Hamers}},
  \citenamefont {{Hammer}},\ and\ \citenamefont {{Portegies
  Zwart}}}]{Jilkova+2016}%
  \BibitemOpen
  \bibfield  {author} {\bibinfo {author} {\bibfnamefont {L.}~\bibnamefont
  {{J{\'\i}lkov{\'a}}}}, \bibinfo {author} {\bibfnamefont {A.~S.}\ \bibnamefont
  {{Hamers}}}, \bibinfo {author} {\bibfnamefont {M.}~\bibnamefont {{Hammer}}},\
  and\ \bibinfo {author} {\bibfnamefont {S.}~\bibnamefont {{Portegies
  Zwart}}},\ }\href {https://doi.org/10.1093/mnras/stw264} {\bibfield
  {journal} {\bibinfo  {journal} {\mnras}\ }\textbf {\bibinfo {volume} {457}},\
  \bibinfo {pages} {4218} (\bibinfo {year} {2016})},\ \Eprint
  {https://arxiv.org/abs/1601.08171} {arXiv:1601.08171 [astro-ph.EP]}
  \BibitemShut {NoStop}%
\bibitem [{\citenamefont {{Picogna}}\ and\ \citenamefont
  {{Marzari}}(2014)}]{PicognaMarzari2014}%
  \BibitemOpen
  \bibfield  {author} {\bibinfo {author} {\bibfnamefont {G.}~\bibnamefont
  {{Picogna}}}\ and\ \bibinfo {author} {\bibfnamefont {F.}~\bibnamefont
  {{Marzari}}},\ }\href {https://doi.org/10.1051/0004-6361/201322816}
  {\bibfield  {journal} {\bibinfo  {journal} {\aap}\ }\textbf {\bibinfo
  {volume} {564}},\ \bibinfo {eid} {A28} (\bibinfo {year} {2014})},\ \Eprint
  {https://arxiv.org/abs/1402.0077} {arXiv:1402.0077 [astro-ph.EP]}
  \BibitemShut {NoStop}%
\bibitem [{\citenamefont {{Dullemond}}\ \emph {et~al.}(2019)\citenamefont
  {{Dullemond}}, \citenamefont {{K{\"u}ffmeier}}, \citenamefont {{Goicovic}},
  \citenamefont {{Fukagawa}}, \citenamefont {{Oehl}},\ and\ \citenamefont
  {{Kramer}}}]{Dullemond+2019}%
  \BibitemOpen
  \bibfield  {author} {\bibinfo {author} {\bibfnamefont {C.~P.}\ \bibnamefont
  {{Dullemond}}}, \bibinfo {author} {\bibfnamefont {M.}~\bibnamefont
  {{K{\"u}ffmeier}}}, \bibinfo {author} {\bibfnamefont {F.}~\bibnamefont
  {{Goicovic}}}, \bibinfo {author} {\bibfnamefont {M.}~\bibnamefont
  {{Fukagawa}}}, \bibinfo {author} {\bibfnamefont {V.}~\bibnamefont {{Oehl}}},\
  and\ \bibinfo {author} {\bibfnamefont {M.}~\bibnamefont {{Kramer}}},\ }\href
  {https://doi.org/10.1051/0004-6361/201832632} {\bibfield  {journal} {\bibinfo
   {journal} {\aap}\ }\textbf {\bibinfo {volume} {628}},\ \bibinfo {eid} {A20}
  (\bibinfo {year} {2019})},\ \Eprint {https://arxiv.org/abs/1911.05158}
  {arXiv:1911.05158 [astro-ph.EP]} \BibitemShut {NoStop}%
\bibitem [{\citenamefont {{Vorobyov}}\ \emph {et~al.}(2020)\citenamefont
  {{Vorobyov}}, \citenamefont {{Skliarevskii}}, \citenamefont {{Elbakyan}},
  \citenamefont {{Takami}}, \citenamefont {{Liu}}, \citenamefont {{Liu}},\ and\
  \citenamefont {{Akiyama}}}]{Vorobyov+2020}%
  \BibitemOpen
  \bibfield  {author} {\bibinfo {author} {\bibfnamefont {E.~I.}\ \bibnamefont
  {{Vorobyov}}}, \bibinfo {author} {\bibfnamefont {A.~M.}\ \bibnamefont
  {{Skliarevskii}}}, \bibinfo {author} {\bibfnamefont {V.~G.}\ \bibnamefont
  {{Elbakyan}}}, \bibinfo {author} {\bibfnamefont {M.}~\bibnamefont
  {{Takami}}}, \bibinfo {author} {\bibfnamefont {H.~B.}\ \bibnamefont {{Liu}}},
  \bibinfo {author} {\bibfnamefont {S.-Y.}\ \bibnamefont {{Liu}}},\ and\
  \bibinfo {author} {\bibfnamefont {E.}~\bibnamefont {{Akiyama}}},\ }\href
  {https://doi.org/10.1051/0004-6361/201936990} {\bibfield  {journal} {\bibinfo
   {journal} {\aap}\ }\textbf {\bibinfo {volume} {635}},\ \bibinfo {eid} {A196}
  (\bibinfo {year} {2020})},\ \Eprint {https://arxiv.org/abs/2002.05764}
  {arXiv:2002.05764 [astro-ph.EP]} \BibitemShut {NoStop}%
\bibitem [{\citenamefont {{Pinte}}\ \emph {et~al.}(2022)\citenamefont
  {{Pinte}}, \citenamefont {{Teague}}, \citenamefont {{Flaherty}},
  \citenamefont {{Hall}}, \citenamefont {{Facchini}},\ and\ \citenamefont
  {{Casassus}}}]{Pinte+2022}%
  \BibitemOpen
  \bibfield  {author} {\bibinfo {author} {\bibfnamefont {C.}~\bibnamefont
  {{Pinte}}}, \bibinfo {author} {\bibfnamefont {R.}~\bibnamefont {{Teague}}},
  \bibinfo {author} {\bibfnamefont {K.}~\bibnamefont {{Flaherty}}}, \bibinfo
  {author} {\bibfnamefont {C.}~\bibnamefont {{Hall}}}, \bibinfo {author}
  {\bibfnamefont {S.}~\bibnamefont {{Facchini}}},\ and\ \bibinfo {author}
  {\bibfnamefont {S.}~\bibnamefont {{Casassus}}},\ }\href@noop {} {\bibfield
  {journal} {\bibinfo  {journal} {arXiv e-prints}\ ,\ \bibinfo {eid}
  {arXiv:2203.09528}} (\bibinfo {year} {2022})},\ \Eprint
  {https://arxiv.org/abs/2203.09528} {arXiv:2203.09528 [astro-ph.EP]}
  \BibitemShut {NoStop}%
\bibitem [{\citenamefont {{Terquem}}\ and\ \citenamefont
  {{Bertout}}(1993)}]{Terquem+1993}%
  \BibitemOpen
  \bibfield  {author} {\bibinfo {author} {\bibfnamefont {C.}~\bibnamefont
  {{Terquem}}}\ and\ \bibinfo {author} {\bibfnamefont {C.}~\bibnamefont
  {{Bertout}}},\ }\href@noop {} {\bibfield  {journal} {\bibinfo  {journal}
  {\aap}\ }\textbf {\bibinfo {volume} {274}},\ \bibinfo {pages} {291} (\bibinfo
  {year} {1993})}\BibitemShut {NoStop}%
\bibitem [{\citenamefont {{Xiang-Gruess}}(2016)}]{Xiang-Gruess2016}%
  \BibitemOpen
  \bibfield  {author} {\bibinfo {author} {\bibfnamefont {M.}~\bibnamefont
  {{Xiang-Gruess}}},\ }\href {https://doi.org/10.1093/mnras/stv2514} {\bibfield
   {journal} {\bibinfo  {journal} {\mnras}\ }\textbf {\bibinfo {volume}
  {455}},\ \bibinfo {pages} {3086} (\bibinfo {year} {2016})},\ \Eprint
  {https://arxiv.org/abs/1510.07458} {arXiv:1510.07458 [astro-ph.EP]}
  \BibitemShut {NoStop}%
\bibitem [{\citenamefont {{Nixon}}\ \emph {et~al.}(2013)\citenamefont
  {{Nixon}}, \citenamefont {{King}},\ and\ \citenamefont
  {{Price}}}]{Nixon+2013}%
  \BibitemOpen
  \bibfield  {author} {\bibinfo {author} {\bibfnamefont {C.}~\bibnamefont
  {{Nixon}}}, \bibinfo {author} {\bibfnamefont {A.}~\bibnamefont {{King}}},\
  and\ \bibinfo {author} {\bibfnamefont {D.}~\bibnamefont {{Price}}},\ }\href
  {https://doi.org/10.1093/mnras/stt1136} {\bibfield  {journal} {\bibinfo
  {journal} {\mnras}\ }\textbf {\bibinfo {volume} {434}},\ \bibinfo {pages}
  {1946} (\bibinfo {year} {2013})},\ \Eprint {https://arxiv.org/abs/1307.0010}
  {arXiv:1307.0010 [astro-ph.HE]} \BibitemShut {NoStop}%
\bibitem [{\citenamefont {{Marino}}\ \emph {et~al.}(2015)\citenamefont
  {{Marino}}, \citenamefont {{Perez}},\ and\ \citenamefont
  {{Casassus}}}]{Marino+2015}%
  \BibitemOpen
  \bibfield  {author} {\bibinfo {author} {\bibfnamefont {S.}~\bibnamefont
  {{Marino}}}, \bibinfo {author} {\bibfnamefont {S.}~\bibnamefont {{Perez}}},\
  and\ \bibinfo {author} {\bibfnamefont {S.}~\bibnamefont {{Casassus}}},\
  }\href {https://doi.org/10.1088/2041-8205/798/2/L44} {\bibfield  {journal}
  {\bibinfo  {journal} {\apjl}\ }\textbf {\bibinfo {volume} {798}},\ \bibinfo
  {eid} {L44} (\bibinfo {year} {2015})},\ \Eprint
  {https://arxiv.org/abs/1412.4632} {arXiv:1412.4632 [astro-ph.EP]}
  \BibitemShut {NoStop}%
\bibitem [{\citenamefont {{Mayama}}\ \emph {et~al.}(2018)\citenamefont
  {{Mayama}}, \citenamefont {{Akiyama}},\ and\ \citenamefont {{Pani{\'c} et
  al.}}}]{Mayama+2018}%
  \BibitemOpen
  \bibfield  {author} {\bibinfo {author} {\bibfnamefont {S.}~\bibnamefont
  {{Mayama}}}, \bibinfo {author} {\bibfnamefont {E.}~\bibnamefont
  {{Akiyama}}},\ and\ \bibinfo {author} {\bibfnamefont {O.}~\bibnamefont
  {{Pani{\'c} et al.}}},\ }\href {https://doi.org/10.3847/2041-8213/aae88b}
  {\bibfield  {journal} {\bibinfo  {journal} {\apjl}\ }\textbf {\bibinfo
  {volume} {868}},\ \bibinfo {eid} {L3} (\bibinfo {year} {2018})},\ \Eprint
  {https://arxiv.org/abs/1810.06941} {arXiv:1810.06941 [astro-ph.EP]}
  \BibitemShut {NoStop}%
\bibitem [{\citenamefont {{Nealon}}\ \emph {et~al.}(2019)\citenamefont
  {{Nealon}}, \citenamefont {{Pinte}}, \citenamefont {{Alexander}},
  \citenamefont {{Mentiplay}},\ and\ \citenamefont {{Dipierro}}}]{Nealon+2019}%
  \BibitemOpen
  \bibfield  {author} {\bibinfo {author} {\bibfnamefont {R.}~\bibnamefont
  {{Nealon}}}, \bibinfo {author} {\bibfnamefont {C.}~\bibnamefont {{Pinte}}},
  \bibinfo {author} {\bibfnamefont {R.}~\bibnamefont {{Alexander}}}, \bibinfo
  {author} {\bibfnamefont {D.}~\bibnamefont {{Mentiplay}}},\ and\ \bibinfo
  {author} {\bibfnamefont {G.}~\bibnamefont {{Dipierro}}},\ }\href
  {https://doi.org/10.1093/mnras/stz346} {\bibfield  {journal} {\bibinfo
  {journal} {\mnras}\ }\textbf {\bibinfo {volume} {484}},\ \bibinfo {pages}
  {4951} (\bibinfo {year} {2019})},\ \Eprint {https://arxiv.org/abs/1902.00036}
  {arXiv:1902.00036 [astro-ph.EP]} \BibitemShut {NoStop}%
\bibitem [{\citenamefont {{Nealon}}\ \emph
  {et~al.}(2020{\natexlab{a}})\citenamefont {{Nealon}}, \citenamefont
  {{Price}},\ and\ \citenamefont {{Pinte}}}]{Nealon+2020c}%
  \BibitemOpen
  \bibfield  {author} {\bibinfo {author} {\bibfnamefont {R.}~\bibnamefont
  {{Nealon}}}, \bibinfo {author} {\bibfnamefont {D.~J.}\ \bibnamefont
  {{Price}}},\ and\ \bibinfo {author} {\bibfnamefont {C.}~\bibnamefont
  {{Pinte}}},\ }\href {https://doi.org/10.1093/mnrasl/slaa026} {\bibfield
  {journal} {\bibinfo  {journal} {\mnras}\ }\textbf {\bibinfo {volume} {493}},\
  \bibinfo {pages} {L143} (\bibinfo {year} {2020}{\natexlab{a}})},\ \Eprint
  {https://arxiv.org/abs/2002.02983} {arXiv:2002.02983 [astro-ph.EP]}
  \BibitemShut {NoStop}%
\bibitem [{\citenamefont {{Facchini}}\ \emph {et~al.}(2018)\citenamefont
  {{Facchini}}, \citenamefont {{Juh{\'a}sz}},\ and\ \citenamefont
  {{Lodato}}}]{Facchini+2018}%
  \BibitemOpen
  \bibfield  {author} {\bibinfo {author} {\bibfnamefont {S.}~\bibnamefont
  {{Facchini}}}, \bibinfo {author} {\bibfnamefont {A.}~\bibnamefont
  {{Juh{\'a}sz}}},\ and\ \bibinfo {author} {\bibfnamefont {G.}~\bibnamefont
  {{Lodato}}},\ }\href {https://doi.org/10.1093/mnras/stx2523} {\bibfield
  {journal} {\bibinfo  {journal} {\mnras}\ }\textbf {\bibinfo {volume} {473}},\
  \bibinfo {pages} {4459} (\bibinfo {year} {2018})},\ \Eprint
  {https://arxiv.org/abs/1709.08369} {arXiv:1709.08369 [astro-ph.SR]}
  \BibitemShut {NoStop}%
\bibitem [{\citenamefont {{Montesinos}}\ and\ \citenamefont
  {{Cuello}}(2018)}]{MontesinosCuello2018}%
  \BibitemOpen
  \bibfield  {author} {\bibinfo {author} {\bibfnamefont {M.}~\bibnamefont
  {{Montesinos}}}\ and\ \bibinfo {author} {\bibfnamefont {N.}~\bibnamefont
  {{Cuello}}},\ }\href {https://doi.org/10.1093/mnrasl/sly001} {\bibfield
  {journal} {\bibinfo  {journal} {\mnras}\ }\textbf {\bibinfo {volume} {475}},\
  \bibinfo {pages} {L35} (\bibinfo {year} {2018})},\ \Eprint
  {https://arxiv.org/abs/1712.09157} {arXiv:1712.09157 [astro-ph.EP]}
  \BibitemShut {NoStop}%
\bibitem [{\citenamefont {{Cuello}}\ \emph
  {et~al.}(2019{\natexlab{b}})\citenamefont {{Cuello}}, \citenamefont
  {{Montesinos}}, \citenamefont {{Stammler}}, \citenamefont {{Louvet}},\ and\
  \citenamefont {{Cuadra}}}]{Cuello+2019a}%
  \BibitemOpen
  \bibfield  {author} {\bibinfo {author} {\bibfnamefont {N.}~\bibnamefont
  {{Cuello}}}, \bibinfo {author} {\bibfnamefont {M.}~\bibnamefont
  {{Montesinos}}}, \bibinfo {author} {\bibfnamefont {S.~M.}\ \bibnamefont
  {{Stammler}}}, \bibinfo {author} {\bibfnamefont {F.}~\bibnamefont
  {{Louvet}}},\ and\ \bibinfo {author} {\bibfnamefont {J.}~\bibnamefont
  {{Cuadra}}},\ }\href {https://doi.org/10.1051/0004-6361/201731732} {\bibfield
   {journal} {\bibinfo  {journal} {\aap}\ }\textbf {\bibinfo {volume} {622}},\
  \bibinfo {eid} {A43} (\bibinfo {year} {2019}{\natexlab{b}})},\ \Eprint
  {https://arxiv.org/abs/1811.08441} {arXiv:1811.08441 [astro-ph.EP]}
  \BibitemShut {NoStop}%
\bibitem [{\citenamefont {{Bohn}}\ \emph {et~al.}(2022)\citenamefont {{Bohn}},
  \citenamefont {{Benisty}},\ and\ \citenamefont {{Perraut et
  al.}}}]{Bohn+2022}%
  \BibitemOpen
  \bibfield  {author} {\bibinfo {author} {\bibfnamefont {A.~J.}\ \bibnamefont
  {{Bohn}}}, \bibinfo {author} {\bibfnamefont {M.}~\bibnamefont {{Benisty}}},\
  and\ \bibinfo {author} {\bibfnamefont {K.}~\bibnamefont {{Perraut et al.}}},\
  }\href {https://doi.org/10.1051/0004-6361/202142070} {\bibfield  {journal}
  {\bibinfo  {journal} {\aap}\ }\textbf {\bibinfo {volume} {658}},\ \bibinfo
  {eid} {A183} (\bibinfo {year} {2022})},\ \Eprint
  {https://arxiv.org/abs/2112.00123} {arXiv:2112.00123 [astro-ph.EP]}
  \BibitemShut {NoStop}%
\bibitem [{\citenamefont {{Young}}\ \emph {et~al.}(2021)\citenamefont
  {{Young}}, \citenamefont {{Alexander}}, \citenamefont {{Walsh}},
  \citenamefont {{Nealon}}, \citenamefont {{Booth}},\ and\ \citenamefont
  {{Pinte}}}]{Young+2021}%
  \BibitemOpen
  \bibfield  {author} {\bibinfo {author} {\bibfnamefont {A.~K.}\ \bibnamefont
  {{Young}}}, \bibinfo {author} {\bibfnamefont {R.}~\bibnamefont
  {{Alexander}}}, \bibinfo {author} {\bibfnamefont {C.}~\bibnamefont
  {{Walsh}}}, \bibinfo {author} {\bibfnamefont {R.}~\bibnamefont {{Nealon}}},
  \bibinfo {author} {\bibfnamefont {A.}~\bibnamefont {{Booth}}},\ and\ \bibinfo
  {author} {\bibfnamefont {C.}~\bibnamefont {{Pinte}}},\ }\href
  {https://doi.org/10.1093/mnras/stab1675} {\bibfield  {journal} {\bibinfo
  {journal} {\mnras}\ }\textbf {\bibinfo {volume} {505}},\ \bibinfo {pages}
  {4821} (\bibinfo {year} {2021})},\ \Eprint {https://arxiv.org/abs/2106.02660}
  {arXiv:2106.02660 [astro-ph.EP]} \BibitemShut {NoStop}%
\bibitem [{\citenamefont {{Young}}\ \emph {et~al.}(2022)\citenamefont
  {{Young}}, \citenamefont {{Alexander}}, \citenamefont {{Rosotti}},\ and\
  \citenamefont {{Pinte}}}]{Young+2022}%
  \BibitemOpen
  \bibfield  {author} {\bibinfo {author} {\bibfnamefont {A.~K.}\ \bibnamefont
  {{Young}}}, \bibinfo {author} {\bibfnamefont {R.}~\bibnamefont
  {{Alexander}}}, \bibinfo {author} {\bibfnamefont {G.}~\bibnamefont
  {{Rosotti}}},\ and\ \bibinfo {author} {\bibfnamefont {C.}~\bibnamefont
  {{Pinte}}},\ }\href {https://doi.org/10.1093/mnras/stac840} {\bibfield
  {journal} {\bibinfo  {journal} {\mnras}\ }\textbf {\bibinfo {volume} {513}},\
  \bibinfo {pages} {487} (\bibinfo {year} {2022})},\ \Eprint
  {https://arxiv.org/abs/2203.12348} {arXiv:2203.12348 [astro-ph.EP]}
  \BibitemShut {NoStop}%
\bibitem [{\citenamefont {{Dong}}\ \emph {et~al.}(2022)\citenamefont {{Dong}},
  \citenamefont {{Liu}}, \citenamefont {{Cuello}}, \citenamefont {{Pinte}},
  \citenamefont {{{\'A}brah{\'a}m}},\ and\ \citenamefont {{Vorobyov et
  al.}}}]{Dong+2022}%
  \BibitemOpen
  \bibfield  {author} {\bibinfo {author} {\bibfnamefont {R.}~\bibnamefont
  {{Dong}}}, \bibinfo {author} {\bibfnamefont {H.~B.}\ \bibnamefont {{Liu}}},
  \bibinfo {author} {\bibfnamefont {N.}~\bibnamefont {{Cuello}}}, \bibinfo
  {author} {\bibfnamefont {C.}~\bibnamefont {{Pinte}}}, \bibinfo {author}
  {\bibfnamefont {P.}~\bibnamefont {{{\'A}brah{\'a}m}}},\ and\ \bibinfo
  {author} {\bibfnamefont {E.}~\bibnamefont {{Vorobyov et al.}}},\ }\bibfield
  {journal} {\bibinfo  {journal} {Nature Astronomy}\ }\href
  {https://doi.org/10.1038/s41550-021-01558-y} {10.1038/s41550-021-01558-y}
  (\bibinfo {year} {2022}),\ \Eprint {https://arxiv.org/abs/2201.05617}
  {arXiv:2201.05617 [astro-ph.SR]} \BibitemShut {NoStop}%
\bibitem [{\citenamefont {{M{\'e}nard}}\ \emph {et~al.}(2020)\citenamefont
  {{M{\'e}nard}}, \citenamefont {{Cuello}},\ and\ \citenamefont {{Ginski et
  al.}}}]{Menard+2020}%
  \BibitemOpen
  \bibfield  {author} {\bibinfo {author} {\bibfnamefont {F.}~\bibnamefont
  {{M{\'e}nard}}}, \bibinfo {author} {\bibfnamefont {N.}~\bibnamefont
  {{Cuello}}},\ and\ \bibinfo {author} {\bibfnamefont {C.}~\bibnamefont
  {{Ginski et al.}}},\ }\href {https://doi.org/10.1051/0004-6361/202038356}
  {\bibfield  {journal} {\bibinfo  {journal} {\aap}\ }\textbf {\bibinfo
  {volume} {639}},\ \bibinfo {eid} {L1} (\bibinfo {year} {2020})},\ \Eprint
  {https://arxiv.org/abs/2006.02439} {arXiv:2006.02439 [astro-ph.SR]}
  \BibitemShut {NoStop}%
\bibitem [{\citenamefont {{Zapata}}\ \emph {et~al.}(2020)\citenamefont
  {{Zapata}}, \citenamefont {{Rodr{\'\i}guez}}, \citenamefont
  {{Fern{\'a}ndez-L{\'o}pez}}, \citenamefont {{Palau}}, \citenamefont
  {{Estalella}}, \citenamefont {{Osorio}}, \citenamefont {{Anglada}},\ and\
  \citenamefont {{Huelamo}}}]{Zapata+2020}%
  \BibitemOpen
  \bibfield  {author} {\bibinfo {author} {\bibfnamefont {L.~A.}\ \bibnamefont
  {{Zapata}}}, \bibinfo {author} {\bibfnamefont {L.~F.}\ \bibnamefont
  {{Rodr{\'\i}guez}}}, \bibinfo {author} {\bibfnamefont {M.}~\bibnamefont
  {{Fern{\'a}ndez-L{\'o}pez}}}, \bibinfo {author} {\bibfnamefont
  {A.}~\bibnamefont {{Palau}}}, \bibinfo {author} {\bibfnamefont
  {R.}~\bibnamefont {{Estalella}}}, \bibinfo {author} {\bibfnamefont
  {M.}~\bibnamefont {{Osorio}}}, \bibinfo {author} {\bibfnamefont
  {G.}~\bibnamefont {{Anglada}}},\ and\ \bibinfo {author} {\bibfnamefont
  {N.}~\bibnamefont {{Huelamo}}},\ }\href
  {https://doi.org/10.3847/1538-4357/ab8fac} {\bibfield  {journal} {\bibinfo
  {journal} {\apj}\ }\textbf {\bibinfo {volume} {896}},\ \bibinfo {eid} {132}
  (\bibinfo {year} {2020})}\BibitemShut {NoStop}%
\bibitem [{\citenamefont {{Bonnell}}\ and\ \citenamefont
  {{Bastien}}(1992)}]{Bonnell+1992}%
  \BibitemOpen
  \bibfield  {author} {\bibinfo {author} {\bibfnamefont {I.}~\bibnamefont
  {{Bonnell}}}\ and\ \bibinfo {author} {\bibfnamefont {P.}~\bibnamefont
  {{Bastien}}},\ }\href {https://doi.org/10.1086/186663} {\bibfield  {journal}
  {\bibinfo  {journal} {\apjl}\ }\textbf {\bibinfo {volume} {401}},\ \bibinfo
  {pages} {L31} (\bibinfo {year} {1992})}\BibitemShut {NoStop}%
\bibitem [{\citenamefont {{Aspin}}\ and\ \citenamefont
  {{Reipurth}}(2003)}]{Aspin+2003}%
  \BibitemOpen
  \bibfield  {author} {\bibinfo {author} {\bibfnamefont {C.}~\bibnamefont
  {{Aspin}}}\ and\ \bibinfo {author} {\bibfnamefont {B.}~\bibnamefont
  {{Reipurth}}},\ }\href {https://doi.org/10.1086/379172} {\bibfield  {journal}
  {\bibinfo  {journal} {\aj}\ }\textbf {\bibinfo {volume} {126}},\ \bibinfo
  {pages} {2936} (\bibinfo {year} {2003})}\BibitemShut {NoStop}%
\bibitem [{\citenamefont {{Pfalzner}}(2008)}]{Pfalzner2008}%
  \BibitemOpen
  \bibfield  {author} {\bibinfo {author} {\bibfnamefont {S.}~\bibnamefont
  {{Pfalzner}}},\ }\href {https://doi.org/10.1051/0004-6361:200810879}
  {\bibfield  {journal} {\bibinfo  {journal} {\aap}\ }\textbf {\bibinfo
  {volume} {492}},\ \bibinfo {pages} {735} (\bibinfo {year} {2008})},\ \Eprint
  {https://arxiv.org/abs/0810.2854} {arXiv:0810.2854 [astro-ph]} \BibitemShut
  {NoStop}%
\bibitem [{\citenamefont {{Borchert}}\ \emph
  {et~al.}(2022{\natexlab{a}})\citenamefont {{Borchert}}, \citenamefont
  {{Price}}, \citenamefont {{Pinte}},\ and\ \citenamefont
  {{Cuello}}}]{Borchert+2022}%
  \BibitemOpen
  \bibfield  {author} {\bibinfo {author} {\bibfnamefont {E.~M.~A.}\
  \bibnamefont {{Borchert}}}, \bibinfo {author} {\bibfnamefont {D.~J.}\
  \bibnamefont {{Price}}}, \bibinfo {author} {\bibfnamefont {C.}~\bibnamefont
  {{Pinte}}},\ and\ \bibinfo {author} {\bibfnamefont {N.}~\bibnamefont
  {{Cuello}}},\ }\href {https://doi.org/10.1093/mnrasl/slab123} {\bibfield
  {journal} {\bibinfo  {journal} {\mnras}\ }\textbf {\bibinfo {volume} {510}},\
  \bibinfo {pages} {L37} (\bibinfo {year} {2022}{\natexlab{a}})},\ \Eprint
  {https://arxiv.org/abs/2111.12723} {arXiv:2111.12723 [astro-ph.GA]}
  \BibitemShut {NoStop}%
\bibitem [{\citenamefont {{Lin}}\ \emph {et~al.}(1985)\citenamefont {{Lin}},
  \citenamefont {{Papaloizou}},\ and\ \citenamefont {{Faulkner}}}]{Lin+1985}%
  \BibitemOpen
  \bibfield  {author} {\bibinfo {author} {\bibfnamefont {D.~N.~C.}\
  \bibnamefont {{Lin}}}, \bibinfo {author} {\bibfnamefont {J.}~\bibnamefont
  {{Papaloizou}}},\ and\ \bibinfo {author} {\bibfnamefont {J.}~\bibnamefont
  {{Faulkner}}},\ }\href {https://doi.org/10.1093/mnras/212.1.105} {\bibfield
  {journal} {\bibinfo  {journal} {\mnras}\ }\textbf {\bibinfo {volume} {212}},\
  \bibinfo {pages} {105} (\bibinfo {year} {1985})}\BibitemShut {NoStop}%
\bibitem [{\citenamefont {{Clarke}}\ \emph {et~al.}(1990)\citenamefont
  {{Clarke}}, \citenamefont {{Lin}},\ and\ \citenamefont
  {{Pringle}}}]{Clarke+1990}%
  \BibitemOpen
  \bibfield  {author} {\bibinfo {author} {\bibfnamefont {C.~J.}\ \bibnamefont
  {{Clarke}}}, \bibinfo {author} {\bibfnamefont {D.~N.~C.}\ \bibnamefont
  {{Lin}}},\ and\ \bibinfo {author} {\bibfnamefont {J.~E.}\ \bibnamefont
  {{Pringle}}},\ }\href {https://doi.org/10.1093/mnras/242.3.439} {\bibfield
  {journal} {\bibinfo  {journal} {\mnras}\ }\textbf {\bibinfo {volume} {242}},\
  \bibinfo {pages} {439} (\bibinfo {year} {1990})}\BibitemShut {NoStop}%
\bibitem [{\citenamefont {{Borchert}}\ \emph
  {et~al.}(2022{\natexlab{b}})\citenamefont {{Borchert}}, \citenamefont
  {{Price}}, \citenamefont {{Pinte}},\ and\ \citenamefont
  {{Cuello}}}]{Borchert+2022b}%
  \BibitemOpen
  \bibfield  {author} {\bibinfo {author} {\bibfnamefont {E.~M.~A.}\
  \bibnamefont {{Borchert}}}, \bibinfo {author} {\bibfnamefont {D.~J.}\
  \bibnamefont {{Price}}}, \bibinfo {author} {\bibfnamefont {C.}~\bibnamefont
  {{Pinte}}},\ and\ \bibinfo {author} {\bibfnamefont {N.}~\bibnamefont
  {{Cuello}}},\ }\href {https://doi.org/10.1093/mnras/stac2872} {\bibfield
  {journal} {\bibinfo  {journal} {\mnras}\ }\textbf {\bibinfo {volume} {517}},\
  \bibinfo {pages} {4436} (\bibinfo {year} {2022}{\natexlab{b}})},\ \Eprint
  {https://arxiv.org/abs/2210.01143} {arXiv:2210.01143 [astro-ph.SR]}
  \BibitemShut {NoStop}%
\bibitem [{\citenamefont {{Hartmann}}\ and\ \citenamefont
  {{Kenyon}}(1996)}]{Hartmann+1996}%
  \BibitemOpen
  \bibfield  {author} {\bibinfo {author} {\bibfnamefont {L.}~\bibnamefont
  {{Hartmann}}}\ and\ \bibinfo {author} {\bibfnamefont {S.~J.}\ \bibnamefont
  {{Kenyon}}},\ }\href {https://doi.org/10.1146/annurev.astro.34.1.207}
  {\bibfield  {journal} {\bibinfo  {journal} {\araa}\ }\textbf {\bibinfo
  {volume} {34}},\ \bibinfo {pages} {207} (\bibinfo {year} {1996})}\BibitemShut
  {NoStop}%
\bibitem [{\citenamefont {{Audard}}\ \emph {et~al.}(2014)\citenamefont
  {{Audard}}, \citenamefont {{{\'A}brah{\'a}m}}, \citenamefont {{Dunham}},
  \citenamefont {{Green}}, \citenamefont {{Grosso}}, \citenamefont
  {{Hamaguchi}}, \citenamefont {{Kastner}}, \citenamefont {{K{\'o}sp{\'a}l}},
  \citenamefont {{Lodato}}, \citenamefont {{Romanova}}, \citenamefont
  {{Skinner}}, \citenamefont {{Vorobyov}},\ and\ \citenamefont
  {{Zhu}}}]{Audard+2014}%
  \BibitemOpen
  \bibfield  {author} {\bibinfo {author} {\bibfnamefont {M.}~\bibnamefont
  {{Audard}}}, \bibinfo {author} {\bibfnamefont {P.}~\bibnamefont
  {{{\'A}brah{\'a}m}}}, \bibinfo {author} {\bibfnamefont {M.~M.}\ \bibnamefont
  {{Dunham}}}, \bibinfo {author} {\bibfnamefont {J.~D.}\ \bibnamefont
  {{Green}}}, \bibinfo {author} {\bibfnamefont {N.}~\bibnamefont {{Grosso}}},
  \bibinfo {author} {\bibfnamefont {K.}~\bibnamefont {{Hamaguchi}}}, \bibinfo
  {author} {\bibfnamefont {J.~H.}\ \bibnamefont {{Kastner}}}, \bibinfo {author}
  {\bibfnamefont {{\'A}.}~\bibnamefont {{K{\'o}sp{\'a}l}}}, \bibinfo {author}
  {\bibfnamefont {G.}~\bibnamefont {{Lodato}}}, \bibinfo {author}
  {\bibfnamefont {M.~M.}\ \bibnamefont {{Romanova}}}, \bibinfo {author}
  {\bibfnamefont {S.~L.}\ \bibnamefont {{Skinner}}}, \bibinfo {author}
  {\bibfnamefont {E.~I.}\ \bibnamefont {{Vorobyov}}},\ and\ \bibinfo {author}
  {\bibfnamefont {Z.}~\bibnamefont {{Zhu}}},\ }in\ \href
  {https://doi.org/10.2458/azu\_uapress\_9780816531240-ch017} {\emph {\bibinfo
  {booktitle} {Protostars and Planets VI}}},\ \bibinfo {editor} {edited by\
  \bibinfo {editor} {\bibfnamefont {H.}~\bibnamefont {{Beuther}}}, \bibinfo
  {editor} {\bibfnamefont {R.~S.}\ \bibnamefont {{Klessen}}}, \bibinfo {editor}
  {\bibfnamefont {C.~P.}\ \bibnamefont {{Dullemond}}},\ and\ \bibinfo {editor}
  {\bibfnamefont {T.}~\bibnamefont {{Henning}}}}\ (\bibinfo {year} {2014})\ p.\
  \bibinfo {pages} {387},\ \Eprint {https://arxiv.org/abs/1401.3368}
  {arXiv:1401.3368 [astro-ph.SR]} \BibitemShut {NoStop}%
\bibitem [{\citenamefont {{Wang}}\ \emph {et~al.}(2004)\citenamefont {{Wang}},
  \citenamefont {{Apai}}, \citenamefont {{Henning}},\ and\ \citenamefont
  {{Pascucci}}}]{Wang+2004}%
  \BibitemOpen
  \bibfield  {author} {\bibinfo {author} {\bibfnamefont {H.}~\bibnamefont
  {{Wang}}}, \bibinfo {author} {\bibfnamefont {D.}~\bibnamefont {{Apai}}},
  \bibinfo {author} {\bibfnamefont {T.}~\bibnamefont {{Henning}}},\ and\
  \bibinfo {author} {\bibfnamefont {I.}~\bibnamefont {{Pascucci}}},\ }\href
  {https://doi.org/10.1086/381705} {\bibfield  {journal} {\bibinfo  {journal}
  {\apjl}\ }\textbf {\bibinfo {volume} {601}},\ \bibinfo {pages} {L83}
  (\bibinfo {year} {2004})},\ \Eprint {https://arxiv.org/abs/astro-ph/0311606}
  {arXiv:astro-ph/0311606 [astro-ph]} \BibitemShut {NoStop}%
\bibitem [{\citenamefont {{Pignatale}}\ \emph {et~al.}(2017)\citenamefont
  {{Pignatale}}, \citenamefont {{Gonzalez}}, \citenamefont {{Cuello}},
  \citenamefont {{Bourdon}},\ and\ \citenamefont
  {{Fitoussi}}}]{Pignatale+2017}%
  \BibitemOpen
  \bibfield  {author} {\bibinfo {author} {\bibfnamefont {F.~C.}\ \bibnamefont
  {{Pignatale}}}, \bibinfo {author} {\bibfnamefont {J.~F.}\ \bibnamefont
  {{Gonzalez}}}, \bibinfo {author} {\bibfnamefont {N.}~\bibnamefont
  {{Cuello}}}, \bibinfo {author} {\bibfnamefont {B.}~\bibnamefont
  {{Bourdon}}},\ and\ \bibinfo {author} {\bibfnamefont {C.}~\bibnamefont
  {{Fitoussi}}},\ }\href {https://doi.org/10.1093/mnras/stx801} {\bibfield
  {journal} {\bibinfo  {journal} {\mnras}\ }\textbf {\bibinfo {volume} {469}},\
  \bibinfo {pages} {237} (\bibinfo {year} {2017})},\ \Eprint
  {https://arxiv.org/abs/1703.10385} {arXiv:1703.10385 [astro-ph.EP]}
  \BibitemShut {NoStop}%
\bibitem [{\citenamefont {{Weidenschilling}}(1977)}]{Weidenschilling1977}%
  \BibitemOpen
  \bibfield  {author} {\bibinfo {author} {\bibfnamefont {S.~J.}\ \bibnamefont
  {{Weidenschilling}}},\ }\href {https://doi.org/10.1093/mnras/180.2.57}
  {\bibfield  {journal} {\bibinfo  {journal} {\mnras}\ }\textbf {\bibinfo
  {volume} {180}},\ \bibinfo {pages} {57} (\bibinfo {year} {1977})}\BibitemShut
  {NoStop}%
\bibitem [{\citenamefont {{Barri{\`e}re-Fouchet}}\ \emph
  {et~al.}(2005)\citenamefont {{Barri{\`e}re-Fouchet}}, \citenamefont
  {{Gonzalez}}, \citenamefont {{Murray}}, \citenamefont {{Humble}},\ and\
  \citenamefont {{Maddison}}}]{Fouchet+2005}%
  \BibitemOpen
  \bibfield  {author} {\bibinfo {author} {\bibfnamefont {L.}~\bibnamefont
  {{Barri{\`e}re-Fouchet}}}, \bibinfo {author} {\bibfnamefont {J.~F.}\
  \bibnamefont {{Gonzalez}}}, \bibinfo {author} {\bibfnamefont {J.~R.}\
  \bibnamefont {{Murray}}}, \bibinfo {author} {\bibfnamefont {R.~J.}\
  \bibnamefont {{Humble}}},\ and\ \bibinfo {author} {\bibfnamefont {S.~T.}\
  \bibnamefont {{Maddison}}},\ }\href
  {https://doi.org/10.1051/0004-6361:20042249} {\bibfield  {journal} {\bibinfo
  {journal} {\aap}\ }\textbf {\bibinfo {volume} {443}},\ \bibinfo {pages} {185}
  (\bibinfo {year} {2005})},\ \Eprint {https://arxiv.org/abs/astro-ph/0508452}
  {arXiv:astro-ph/0508452 [astro-ph]} \BibitemShut {NoStop}%
\bibitem [{\citenamefont {{Laibe}}\ \emph {et~al.}(2012)\citenamefont
  {{Laibe}}, \citenamefont {{Gonzalez}},\ and\ \citenamefont
  {{Maddison}}}]{Laibe+2012}%
  \BibitemOpen
  \bibfield  {author} {\bibinfo {author} {\bibfnamefont {G.}~\bibnamefont
  {{Laibe}}}, \bibinfo {author} {\bibfnamefont {J.~F.}\ \bibnamefont
  {{Gonzalez}}},\ and\ \bibinfo {author} {\bibfnamefont {S.~T.}\ \bibnamefont
  {{Maddison}}},\ }\href {https://doi.org/10.1051/0004-6361/201015349}
  {\bibfield  {journal} {\bibinfo  {journal} {\aap}\ }\textbf {\bibinfo
  {volume} {537}},\ \bibinfo {eid} {A61} (\bibinfo {year} {2012})},\ \Eprint
  {https://arxiv.org/abs/1111.3083} {arXiv:1111.3083 [astro-ph.EP]}
  \BibitemShut {NoStop}%
\bibitem [{\citenamefont {{Pignatale}}\ \emph {et~al.}(2019)\citenamefont
  {{Pignatale}}, \citenamefont {{Gonzalez}}, \citenamefont {{Bourdon}},\ and\
  \citenamefont {{Fitoussi}}}]{Pignatale+2019}%
  \BibitemOpen
  \bibfield  {author} {\bibinfo {author} {\bibfnamefont {F.~C.}\ \bibnamefont
  {{Pignatale}}}, \bibinfo {author} {\bibfnamefont {J.~F.}\ \bibnamefont
  {{Gonzalez}}}, \bibinfo {author} {\bibfnamefont {B.}~\bibnamefont
  {{Bourdon}}},\ and\ \bibinfo {author} {\bibfnamefont {C.}~\bibnamefont
  {{Fitoussi}}},\ }\href {https://doi.org/10.1093/mnras/stz2883} {\bibfield
  {journal} {\bibinfo  {journal} {\mnras}\ }\textbf {\bibinfo {volume} {490}},\
  \bibinfo {pages} {4428} (\bibinfo {year} {2019})},\ \Eprint
  {https://arxiv.org/abs/1910.04717} {arXiv:1910.04717 [astro-ph.EP]}
  \BibitemShut {NoStop}%
\bibitem [{\citenamefont {{Silsbee}}\ and\ \citenamefont
  {{Rafikov}}(2021)}]{Silsbee+2021}%
  \BibitemOpen
  \bibfield  {author} {\bibinfo {author} {\bibfnamefont {K.}~\bibnamefont
  {{Silsbee}}}\ and\ \bibinfo {author} {\bibfnamefont {R.~R.}\ \bibnamefont
  {{Rafikov}}},\ }\href {https://doi.org/10.1051/0004-6361/202141139}
  {\bibfield  {journal} {\bibinfo  {journal} {\aap}\ }\textbf {\bibinfo
  {volume} {652}},\ \bibinfo {eid} {A104} (\bibinfo {year} {2021})},\ \Eprint
  {https://arxiv.org/abs/2107.11389} {arXiv:2107.11389 [astro-ph.EP]}
  \BibitemShut {NoStop}%
\bibitem [{\citenamefont {{Gonzalez}}\ \emph {et~al.}(2017)\citenamefont
  {{Gonzalez}}, \citenamefont {{Laibe}},\ and\ \citenamefont
  {{Maddison}}}]{Gonzalez+2017}%
  \BibitemOpen
  \bibfield  {author} {\bibinfo {author} {\bibfnamefont {J.~F.}\ \bibnamefont
  {{Gonzalez}}}, \bibinfo {author} {\bibfnamefont {G.}~\bibnamefont
  {{Laibe}}},\ and\ \bibinfo {author} {\bibfnamefont {S.~T.}\ \bibnamefont
  {{Maddison}}},\ }\href {https://doi.org/10.1093/mnras/stx016} {\bibfield
  {journal} {\bibinfo  {journal} {\mnras}\ }\textbf {\bibinfo {volume} {467}},\
  \bibinfo {pages} {1984} (\bibinfo {year} {2017})},\ \Eprint
  {https://arxiv.org/abs/1701.01115} {arXiv:1701.01115 [astro-ph.EP]}
  \BibitemShut {NoStop}%
\bibitem [{\citenamefont {{Vericel}}\ and\ \citenamefont
  {{Gonzalez}}(2020)}]{Vericel+2020}%
  \BibitemOpen
  \bibfield  {author} {\bibinfo {author} {\bibfnamefont {A.}~\bibnamefont
  {{Vericel}}}\ and\ \bibinfo {author} {\bibfnamefont {J.-F.}\ \bibnamefont
  {{Gonzalez}}},\ }\href {https://doi.org/10.1093/mnras/stz3444} {\bibfield
  {journal} {\bibinfo  {journal} {\mnras}\ }\textbf {\bibinfo {volume} {492}},\
  \bibinfo {pages} {210} (\bibinfo {year} {2020})},\ \Eprint
  {https://arxiv.org/abs/1912.02464} {arXiv:1912.02464 [astro-ph.EP]}
  \BibitemShut {NoStop}%
\bibitem [{\citenamefont {{Vittone}}\ and\ \citenamefont
  {{Errico}}(2005)}]{Vitton&Errico2005}%
  \BibitemOpen
  \bibfield  {author} {\bibinfo {author} {\bibfnamefont {A.~A.}\ \bibnamefont
  {{Vittone}}}\ and\ \bibinfo {author} {\bibfnamefont {L.}~\bibnamefont
  {{Errico}}},\ }\href@noop {} {\bibfield  {journal} {\bibinfo  {journal}
  {\memsai}\ }\textbf {\bibinfo {volume} {76}},\ \bibinfo {pages} {320}
  (\bibinfo {year} {2005})}\BibitemShut {NoStop}%
\bibitem [{\citenamefont {{Reipurth}}\ and\ \citenamefont
  {{Aspin}}(2004)}]{Reipurth&Aspin2004}%
  \BibitemOpen
  \bibfield  {author} {\bibinfo {author} {\bibfnamefont {B.}~\bibnamefont
  {{Reipurth}}}\ and\ \bibinfo {author} {\bibfnamefont {C.}~\bibnamefont
  {{Aspin}}},\ }\href {https://doi.org/10.1086/422250} {\bibfield  {journal}
  {\bibinfo  {journal} {\apjl}\ }\textbf {\bibinfo {volume} {608}},\ \bibinfo
  {pages} {L65} (\bibinfo {year} {2004})}\BibitemShut {NoStop}%
\bibitem [{\citenamefont {{Forgan}}\ and\ \citenamefont
  {{Rice}}(2010)}]{Forgan+2010}%
  \BibitemOpen
  \bibfield  {author} {\bibinfo {author} {\bibfnamefont {D.}~\bibnamefont
  {{Forgan}}}\ and\ \bibinfo {author} {\bibfnamefont {K.}~\bibnamefont
  {{Rice}}},\ }\href {https://doi.org/10.1111/j.1365-2966.2009.15974.x}
  {\bibfield  {journal} {\bibinfo  {journal} {\mnras}\ }\textbf {\bibinfo
  {volume} {402}},\ \bibinfo {pages} {1349} (\bibinfo {year} {2010})},\ \Eprint
  {https://arxiv.org/abs/0911.0531} {arXiv:0911.0531 [astro-ph.SR]}
  \BibitemShut {NoStop}%
\bibitem [{\citenamefont {{Vorobyov}}\ \emph {et~al.}(2021)\citenamefont
  {{Vorobyov}}, \citenamefont {{Elbakyan}}, \citenamefont {{Liu}},\ and\
  \citenamefont {{Takami}}}]{Vorobyov+2021}%
  \BibitemOpen
  \bibfield  {author} {\bibinfo {author} {\bibfnamefont {E.~I.}\ \bibnamefont
  {{Vorobyov}}}, \bibinfo {author} {\bibfnamefont {V.~G.}\ \bibnamefont
  {{Elbakyan}}}, \bibinfo {author} {\bibfnamefont {H.~B.}\ \bibnamefont
  {{Liu}}},\ and\ \bibinfo {author} {\bibfnamefont {M.}~\bibnamefont
  {{Takami}}},\ }\href {https://doi.org/10.1051/0004-6361/202039391} {\bibfield
   {journal} {\bibinfo  {journal} {\aap}\ }\textbf {\bibinfo {volume} {647}},\
  \bibinfo {eid} {A44} (\bibinfo {year} {2021})},\ \Eprint
  {https://arxiv.org/abs/2101.01596} {arXiv:2101.01596 [astro-ph.SR]}
  \BibitemShut {NoStop}%
\bibitem [{\citenamefont {{Shakura}}\ and\ \citenamefont
  {{Sunyaev}}(1973)}]{Shakura+1973}%
  \BibitemOpen
  \bibfield  {author} {\bibinfo {author} {\bibfnamefont {N.~I.}\ \bibnamefont
  {{Shakura}}}\ and\ \bibinfo {author} {\bibfnamefont {R.~A.}\ \bibnamefont
  {{Sunyaev}}},\ }\href@noop {} {\bibfield  {journal} {\bibinfo  {journal}
  {\aap}\ }\textbf {\bibinfo {volume} {24}},\ \bibinfo {pages} {337} (\bibinfo
  {year} {1973})}\BibitemShut {NoStop}%
\bibitem [{\citenamefont {{Clarke}}\ and\ \citenamefont
  {{Pringle}}(2006)}]{Clarke&Pringle2006}%
  \BibitemOpen
  \bibfield  {author} {\bibinfo {author} {\bibfnamefont {C.~J.}\ \bibnamefont
  {{Clarke}}}\ and\ \bibinfo {author} {\bibfnamefont {J.~E.}\ \bibnamefont
  {{Pringle}}},\ }\href {https://doi.org/10.1111/j.1745-3933.2006.00177.x}
  {\bibfield  {journal} {\bibinfo  {journal} {\mnras}\ }\textbf {\bibinfo
  {volume} {370}},\ \bibinfo {pages} {L10} (\bibinfo {year} {2006})},\ \Eprint
  {https://arxiv.org/abs/astro-ph/0604196} {arXiv:astro-ph/0604196 [astro-ph]}
  \BibitemShut {NoStop}%
\bibitem [{\citenamefont {{Zhu}}\ \emph {et~al.}(2007)\citenamefont {{Zhu}},
  \citenamefont {{Hartmann}}, \citenamefont {{Calvet}}, \citenamefont
  {{Hernandez}}, \citenamefont {{Muzerolle}},\ and\ \citenamefont
  {{Tannirkulam}}}]{Zhu+2007}%
  \BibitemOpen
  \bibfield  {author} {\bibinfo {author} {\bibfnamefont {Z.}~\bibnamefont
  {{Zhu}}}, \bibinfo {author} {\bibfnamefont {L.}~\bibnamefont {{Hartmann}}},
  \bibinfo {author} {\bibfnamefont {N.}~\bibnamefont {{Calvet}}}, \bibinfo
  {author} {\bibfnamefont {J.}~\bibnamefont {{Hernandez}}}, \bibinfo {author}
  {\bibfnamefont {J.}~\bibnamefont {{Muzerolle}}},\ and\ \bibinfo {author}
  {\bibfnamefont {A.-K.}\ \bibnamefont {{Tannirkulam}}},\ }\href
  {https://doi.org/10.1086/521345} {\bibfield  {journal} {\bibinfo  {journal}
  {\apj}\ }\textbf {\bibinfo {volume} {669}},\ \bibinfo {pages} {483} (\bibinfo
  {year} {2007})},\ \Eprint {https://arxiv.org/abs/0707.3429} {arXiv:0707.3429
  [astro-ph]} \BibitemShut {NoStop}%
\bibitem [{\citenamefont {{{\'A}brah{\'a}m}}\ \emph {et~al.}(2009)\citenamefont
  {{{\'A}brah{\'a}m}}, \citenamefont {{Juh{\'a}sz}}, \citenamefont
  {{Dullemond}}, \citenamefont {{K{\'o}sp{\'a}l}}, \citenamefont {{van
  Boekel}}, \citenamefont {{Bouwman}}, \citenamefont {{Henning}}, \citenamefont
  {{Mo{\'o}r}}, \citenamefont {{Mosoni}}, \citenamefont {{Sicilia-Aguilar}},\
  and\ \citenamefont {{Sipos}}}]{Abraham+2009}%
  \BibitemOpen
  \bibfield  {author} {\bibinfo {author} {\bibfnamefont {P.}~\bibnamefont
  {{{\'A}brah{\'a}m}}}, \bibinfo {author} {\bibfnamefont {A.}~\bibnamefont
  {{Juh{\'a}sz}}}, \bibinfo {author} {\bibfnamefont {C.~P.}\ \bibnamefont
  {{Dullemond}}}, \bibinfo {author} {\bibfnamefont {{\'A}.}~\bibnamefont
  {{K{\'o}sp{\'a}l}}}, \bibinfo {author} {\bibfnamefont {R.}~\bibnamefont {{van
  Boekel}}}, \bibinfo {author} {\bibfnamefont {J.}~\bibnamefont {{Bouwman}}},
  \bibinfo {author} {\bibfnamefont {T.}~\bibnamefont {{Henning}}}, \bibinfo
  {author} {\bibfnamefont {A.}~\bibnamefont {{Mo{\'o}r}}}, \bibinfo {author}
  {\bibfnamefont {L.}~\bibnamefont {{Mosoni}}}, \bibinfo {author}
  {\bibfnamefont {A.}~\bibnamefont {{Sicilia-Aguilar}}},\ and\ \bibinfo
  {author} {\bibfnamefont {N.}~\bibnamefont {{Sipos}}},\ }\href
  {https://doi.org/10.1038/nature08004} {\bibfield  {journal} {\bibinfo
  {journal} {\nat}\ }\textbf {\bibinfo {volume} {459}},\ \bibinfo {pages} {224}
  (\bibinfo {year} {2009})},\ \Eprint {https://arxiv.org/abs/0906.3161}
  {arXiv:0906.3161 [astro-ph.SR]} \BibitemShut {NoStop}%
\bibitem [{\citenamefont {{K{\'o}sp{\'a}l}}\ \emph {et~al.}(2020)\citenamefont
  {{K{\'o}sp{\'a}l}}, \citenamefont {{{\'A}brah{\'a}m}}, \citenamefont
  {{Carmona}}, \citenamefont {{Chen}}, \citenamefont {{Green}}, \citenamefont
  {{van Boekel}},\ and\ \citenamefont {{White}}}]{Kospal+2020}%
  \BibitemOpen
  \bibfield  {author} {\bibinfo {author} {\bibfnamefont {{\'A}.}~\bibnamefont
  {{K{\'o}sp{\'a}l}}}, \bibinfo {author} {\bibfnamefont {P.}~\bibnamefont
  {{{\'A}brah{\'a}m}}}, \bibinfo {author} {\bibfnamefont {A.}~\bibnamefont
  {{Carmona}}}, \bibinfo {author} {\bibfnamefont {L.}~\bibnamefont {{Chen}}},
  \bibinfo {author} {\bibfnamefont {J.~D.}\ \bibnamefont {{Green}}}, \bibinfo
  {author} {\bibfnamefont {R.}~\bibnamefont {{van Boekel}}},\ and\ \bibinfo
  {author} {\bibfnamefont {J.~A.}\ \bibnamefont {{White}}},\ }\href
  {https://doi.org/10.3847/2041-8213/ab93d4} {\bibfield  {journal} {\bibinfo
  {journal} {\apjl}\ }\textbf {\bibinfo {volume} {895}},\ \bibinfo {eid} {L48}
  (\bibinfo {year} {2020})},\ \Eprint {https://arxiv.org/abs/2005.09364}
  {arXiv:2005.09364 [astro-ph.SR]} \BibitemShut {NoStop}%
\bibitem [{\citenamefont {{Desch}}\ and\ \citenamefont
  {{Connolly}}(2002)}]{Desch+2002}%
  \BibitemOpen
  \bibfield  {author} {\bibinfo {author} {\bibfnamefont {S.~J.}\ \bibnamefont
  {{Desch}}}\ and\ \bibinfo {author} {\bibfnamefont {J.}~\bibnamefont
  {{Connolly}}, \bibfnamefont {H.~C.}},\ }\href
  {https://doi.org/10.1111/j.1945-5100.2002.tb01104.x} {\bibfield  {journal}
  {\bibinfo  {journal} {Meteoritics \& Planetary Science}\ }\textbf {\bibinfo
  {volume} {37}},\ \bibinfo {pages} {183} (\bibinfo {year} {2002})}\BibitemShut
  {NoStop}%
\bibitem [{\citenamefont {{Li}}\ \emph {et~al.}(2021)\citenamefont {{Li}},
  \citenamefont {{Zhu}}, \citenamefont {{Huang}}, \citenamefont {{Sui}},
  \citenamefont {{Petaev}},\ and\ \citenamefont {{Steffen}}}]{Li+2021}%
  \BibitemOpen
  \bibfield  {author} {\bibinfo {author} {\bibfnamefont {M.}~\bibnamefont
  {{Li}}}, \bibinfo {author} {\bibfnamefont {Z.}~\bibnamefont {{Zhu}}},
  \bibinfo {author} {\bibfnamefont {S.}~\bibnamefont {{Huang}}}, \bibinfo
  {author} {\bibfnamefont {N.}~\bibnamefont {{Sui}}}, \bibinfo {author}
  {\bibfnamefont {M.~I.}\ \bibnamefont {{Petaev}}},\ and\ \bibinfo {author}
  {\bibfnamefont {J.~H.}\ \bibnamefont {{Steffen}}},\ }\href@noop {} {\bibfield
   {journal} {\bibinfo  {journal} {arXiv e-prints}\ ,\ \bibinfo {eid}
  {arXiv:2111.03798}} (\bibinfo {year} {2021})},\ \Eprint
  {https://arxiv.org/abs/2111.03798} {arXiv:2111.03798 [astro-ph.EP]}
  \BibitemShut {NoStop}%
\bibitem [{\citenamefont {{K{\'o}sp{\'a}l}}\ \emph {et~al.}(2021)\citenamefont
  {{K{\'o}sp{\'a}l}}, \citenamefont {{Cruz-S{\'a}enz de Miera}},\ and\
  \citenamefont {{White et al.}}}]{Kospal+2021}%
  \BibitemOpen
  \bibfield  {author} {\bibinfo {author} {\bibfnamefont {{\'A}.}~\bibnamefont
  {{K{\'o}sp{\'a}l}}}, \bibinfo {author} {\bibfnamefont {F.}~\bibnamefont
  {{Cruz-S{\'a}enz de Miera}}},\ and\ \bibinfo {author} {\bibfnamefont {J.~A.}\
  \bibnamefont {{White et al.}}},\ }\href
  {https://doi.org/10.3847/1538-4365/ac0f09} {\bibfield  {journal} {\bibinfo
  {journal} {\apjs}\ }\textbf {\bibinfo {volume} {256}},\ \bibinfo {eid} {30}
  (\bibinfo {year} {2021})},\ \Eprint {https://arxiv.org/abs/2106.14409}
  {arXiv:2106.14409 [astro-ph.SR]} \BibitemShut {NoStop}%
\bibitem [{\citenamefont {{Guo}}\ \emph {et~al.}(2021)\citenamefont {{Guo}},
  \citenamefont {{Lucas}}, \citenamefont {{Contreras Pe{\~n}a}}, \citenamefont
  {{Smith}}, \citenamefont {{Morris}}, \citenamefont {{Kurtev}}, \citenamefont
  {{Borissova}}, \citenamefont {{Alonso-Garc{\'\i}a}}, \citenamefont
  {{Minniti}}, \citenamefont {{Chen{\'e}}}, \citenamefont {{Kumar}},
  \citenamefont {{Caratti o Garatti}}, \citenamefont {{Froebrich}},\ and\
  \citenamefont {{Stimson}}}]{Guo+2021}%
  \BibitemOpen
  \bibfield  {author} {\bibinfo {author} {\bibfnamefont {Z.}~\bibnamefont
  {{Guo}}}, \bibinfo {author} {\bibfnamefont {P.~W.}\ \bibnamefont {{Lucas}}},
  \bibinfo {author} {\bibfnamefont {C.}~\bibnamefont {{Contreras Pe{\~n}a}}},
  \bibinfo {author} {\bibfnamefont {L.~C.}\ \bibnamefont {{Smith}}}, \bibinfo
  {author} {\bibfnamefont {C.}~\bibnamefont {{Morris}}}, \bibinfo {author}
  {\bibfnamefont {R.~G.}\ \bibnamefont {{Kurtev}}}, \bibinfo {author}
  {\bibfnamefont {J.}~\bibnamefont {{Borissova}}}, \bibinfo {author}
  {\bibfnamefont {J.}~\bibnamefont {{Alonso-Garc{\'\i}a}}}, \bibinfo {author}
  {\bibfnamefont {D.}~\bibnamefont {{Minniti}}}, \bibinfo {author}
  {\bibfnamefont {A.~N.}\ \bibnamefont {{Chen{\'e}}}}, \bibinfo {author}
  {\bibfnamefont {M.~S.~N.}\ \bibnamefont {{Kumar}}}, \bibinfo {author}
  {\bibfnamefont {A.}~\bibnamefont {{Caratti o Garatti}}}, \bibinfo {author}
  {\bibfnamefont {D.}~\bibnamefont {{Froebrich}}},\ and\ \bibinfo {author}
  {\bibfnamefont {W.~H.}\ \bibnamefont {{Stimson}}},\ }\href
  {https://doi.org/10.1093/mnras/stab882} {\bibfield  {journal} {\bibinfo
  {journal} {\mnras}\ }\textbf {\bibinfo {volume} {504}},\ \bibinfo {pages}
  {830} (\bibinfo {year} {2021})},\ \Eprint {https://arxiv.org/abs/2103.13335}
  {arXiv:2103.13335 [astro-ph.SR]} \BibitemShut {NoStop}%
\bibitem [{\citenamefont {{Rodriguez}}\ \emph {et~al.}(2013)\citenamefont
  {{Rodriguez}}, \citenamefont {{Pepper}}, \citenamefont {{Stassun}},
  \citenamefont {{Siverd}}, \citenamefont {{Cargile}}, \citenamefont
  {{Beatty}},\ and\ \citenamefont {{Gaudi}}}]{Rodriguez+2013}%
  \BibitemOpen
  \bibfield  {author} {\bibinfo {author} {\bibfnamefont {J.~E.}\ \bibnamefont
  {{Rodriguez}}}, \bibinfo {author} {\bibfnamefont {J.}~\bibnamefont
  {{Pepper}}}, \bibinfo {author} {\bibfnamefont {K.~G.}\ \bibnamefont
  {{Stassun}}}, \bibinfo {author} {\bibfnamefont {R.~J.}\ \bibnamefont
  {{Siverd}}}, \bibinfo {author} {\bibfnamefont {P.}~\bibnamefont {{Cargile}}},
  \bibinfo {author} {\bibfnamefont {T.~G.}\ \bibnamefont {{Beatty}}},\ and\
  \bibinfo {author} {\bibfnamefont {B.~S.}\ \bibnamefont {{Gaudi}}},\ }\href
  {https://doi.org/10.1088/0004-6256/146/5/112} {\bibfield  {journal} {\bibinfo
   {journal} {\aj}\ }\textbf {\bibinfo {volume} {146}},\ \bibinfo {eid} {112}
  (\bibinfo {year} {2013})},\ \Eprint {https://arxiv.org/abs/1308.2017}
  {arXiv:1308.2017 [astro-ph.SR]} \BibitemShut {NoStop}%
\bibitem [{\citenamefont {{Dai}}\ \emph {et~al.}(2015)\citenamefont {{Dai}},
  \citenamefont {{Facchini}}, \citenamefont {{Clarke}},\ and\ \citenamefont
  {{Haworth}}}]{Dai+2015}%
  \BibitemOpen
  \bibfield  {author} {\bibinfo {author} {\bibfnamefont {F.}~\bibnamefont
  {{Dai}}}, \bibinfo {author} {\bibfnamefont {S.}~\bibnamefont {{Facchini}}},
  \bibinfo {author} {\bibfnamefont {C.~J.}\ \bibnamefont {{Clarke}}},\ and\
  \bibinfo {author} {\bibfnamefont {T.~J.}\ \bibnamefont {{Haworth}}},\ }\href
  {https://doi.org/10.1093/mnras/stv403} {\bibfield  {journal} {\bibinfo
  {journal} {\mnras}\ }\textbf {\bibinfo {volume} {449}},\ \bibinfo {pages}
  {1996} (\bibinfo {year} {2015})},\ \Eprint {https://arxiv.org/abs/1502.06649}
  {arXiv:1502.06649 [astro-ph.SR]} \BibitemShut {NoStop}%
\bibitem [{\citenamefont {{Rodriguez}}\ \emph {et~al.}(2018)\citenamefont
  {{Rodriguez}}, \citenamefont {{Loomis}}, \citenamefont {{Cabrit}},
  \citenamefont {{Haworth}}, \citenamefont {{Facchini}}, \citenamefont
  {{Dougados}}, \citenamefont {{Booth}}, \citenamefont {{Jensen}},
  \citenamefont {{Clarke}}, \citenamefont {{Stassun}}, \citenamefont {{Dent}},\
  and\ \citenamefont {{Pety}}}]{Rodriguez+2018}%
  \BibitemOpen
  \bibfield  {author} {\bibinfo {author} {\bibfnamefont {J.~E.}\ \bibnamefont
  {{Rodriguez}}}, \bibinfo {author} {\bibfnamefont {R.}~\bibnamefont
  {{Loomis}}}, \bibinfo {author} {\bibfnamefont {S.}~\bibnamefont {{Cabrit}}},
  \bibinfo {author} {\bibfnamefont {T.~J.}\ \bibnamefont {{Haworth}}}, \bibinfo
  {author} {\bibfnamefont {S.}~\bibnamefont {{Facchini}}}, \bibinfo {author}
  {\bibfnamefont {C.}~\bibnamefont {{Dougados}}}, \bibinfo {author}
  {\bibfnamefont {R.~A.}\ \bibnamefont {{Booth}}}, \bibinfo {author}
  {\bibfnamefont {E.~L.~N.}\ \bibnamefont {{Jensen}}}, \bibinfo {author}
  {\bibfnamefont {C.~J.}\ \bibnamefont {{Clarke}}}, \bibinfo {author}
  {\bibfnamefont {K.~G.}\ \bibnamefont {{Stassun}}}, \bibinfo {author}
  {\bibfnamefont {W.~R.~F.}\ \bibnamefont {{Dent}}},\ and\ \bibinfo {author}
  {\bibfnamefont {J.}~\bibnamefont {{Pety}}},\ }\href
  {https://doi.org/10.3847/1538-4357/aac08f} {\bibfield  {journal} {\bibinfo
  {journal} {\apj}\ }\textbf {\bibinfo {volume} {859}},\ \bibinfo {eid} {150}
  (\bibinfo {year} {2018})},\ \Eprint {https://arxiv.org/abs/1804.09190}
  {arXiv:1804.09190 [astro-ph.SR]} \BibitemShut {NoStop}%
\bibitem [{\citenamefont {{K{\"o}hler}}\ \emph {et~al.}(2016)\citenamefont
  {{K{\"o}hler}}, \citenamefont {{Kasper}}, \citenamefont {{Herbst}},
  \citenamefont {{Ratzka}},\ and\ \citenamefont {{Bertrang}}}]{Kohler+2016}%
  \BibitemOpen
  \bibfield  {author} {\bibinfo {author} {\bibfnamefont {R.}~\bibnamefont
  {{K{\"o}hler}}}, \bibinfo {author} {\bibfnamefont {M.}~\bibnamefont
  {{Kasper}}}, \bibinfo {author} {\bibfnamefont {T.~M.}\ \bibnamefont
  {{Herbst}}}, \bibinfo {author} {\bibfnamefont {T.}~\bibnamefont {{Ratzka}}},\
  and\ \bibinfo {author} {\bibfnamefont {G.~H.~M.}\ \bibnamefont
  {{Bertrang}}},\ }\href {https://doi.org/10.1051/0004-6361/201527125}
  {\bibfield  {journal} {\bibinfo  {journal} {\aap}\ }\textbf {\bibinfo
  {volume} {587}},\ \bibinfo {eid} {A35} (\bibinfo {year} {2016})},\ \Eprint
  {https://arxiv.org/abs/1512.05736} {arXiv:1512.05736 [astro-ph.SR]}
  \BibitemShut {NoStop}%
\bibitem [{\citenamefont {{K{\"o}hler}}\ and\ \citenamefont
  {{Kubiak}}(2020)}]{Kohler+2020}%
  \BibitemOpen
  \bibfield  {author} {\bibinfo {author} {\bibfnamefont {R.}~\bibnamefont
  {{K{\"o}hler}}}\ and\ \bibinfo {author} {\bibfnamefont {K.}~\bibnamefont
  {{Kubiak}}},\ }\href {https://doi.org/10.3847/2515-5172/ab963b} {\bibfield
  {journal} {\bibinfo  {journal} {Research Notes of the American Astronomical
  Society}\ }\textbf {\bibinfo {volume} {4}},\ \bibinfo {eid} {73} (\bibinfo
  {year} {2020})}\BibitemShut {NoStop}%
\bibitem [{\citenamefont {{Kennedy}}\ \emph {et~al.}(2019)\citenamefont
  {{Kennedy}}, \citenamefont {{Matr{\`a}}}, \citenamefont {{Facchini}},
  \citenamefont {{Milli}}, \citenamefont {{Pani{\'c}}}, \citenamefont
  {{Price}}, \citenamefont {{Wilner}}, \citenamefont {{Wyatt}},\ and\
  \citenamefont {{Yelverton}}}]{Kennedy+2019}%
  \BibitemOpen
  \bibfield  {author} {\bibinfo {author} {\bibfnamefont {G.~M.}\ \bibnamefont
  {{Kennedy}}}, \bibinfo {author} {\bibfnamefont {L.}~\bibnamefont
  {{Matr{\`a}}}}, \bibinfo {author} {\bibfnamefont {S.}~\bibnamefont
  {{Facchini}}}, \bibinfo {author} {\bibfnamefont {J.}~\bibnamefont {{Milli}}},
  \bibinfo {author} {\bibfnamefont {O.}~\bibnamefont {{Pani{\'c}}}}, \bibinfo
  {author} {\bibfnamefont {D.}~\bibnamefont {{Price}}}, \bibinfo {author}
  {\bibfnamefont {D.~J.}\ \bibnamefont {{Wilner}}}, \bibinfo {author}
  {\bibfnamefont {M.~C.}\ \bibnamefont {{Wyatt}}},\ and\ \bibinfo {author}
  {\bibfnamefont {B.~M.}\ \bibnamefont {{Yelverton}}},\ }\href
  {https://doi.org/10.1038/s41550-018-0667-x} {\bibfield  {journal} {\bibinfo
  {journal} {Nature Astronomy}\ }\textbf {\bibinfo {volume} {3}},\ \bibinfo
  {pages} {230} (\bibinfo {year} {2019})}\BibitemShut {NoStop}%
\bibitem [{\citenamefont {{Z{\'u}{\~n}iga-Fern{\'a}ndez}}\ \emph
  {et~al.}(2021)\citenamefont {{Z{\'u}{\~n}iga-Fern{\'a}ndez}}, \citenamefont
  {{Olofsson}}, \citenamefont {{Bayo}}, \citenamefont {{Haubois}},
  \citenamefont {{Corral-Santana}}, \citenamefont {{Lopera-Mej{\'\i}a}},
  \citenamefont {{Ronco}}, \citenamefont {{Tokovinin}}, \citenamefont
  {{Gallenne}}, \citenamefont {{Kennedy}},\ and\ \citenamefont
  {{Berger}}}]{Zuniga+2021}%
  \BibitemOpen
  \bibfield  {author} {\bibinfo {author} {\bibfnamefont {S.}~\bibnamefont
  {{Z{\'u}{\~n}iga-Fern{\'a}ndez}}}, \bibinfo {author} {\bibfnamefont
  {J.}~\bibnamefont {{Olofsson}}}, \bibinfo {author} {\bibfnamefont
  {A.}~\bibnamefont {{Bayo}}}, \bibinfo {author} {\bibfnamefont
  {X.}~\bibnamefont {{Haubois}}}, \bibinfo {author} {\bibfnamefont {J.~M.}\
  \bibnamefont {{Corral-Santana}}}, \bibinfo {author} {\bibfnamefont
  {A.}~\bibnamefont {{Lopera-Mej{\'\i}a}}}, \bibinfo {author} {\bibfnamefont
  {M.~P.}\ \bibnamefont {{Ronco}}}, \bibinfo {author} {\bibfnamefont
  {A.}~\bibnamefont {{Tokovinin}}}, \bibinfo {author} {\bibfnamefont
  {A.}~\bibnamefont {{Gallenne}}}, \bibinfo {author} {\bibfnamefont {G.~M.}\
  \bibnamefont {{Kennedy}}},\ and\ \bibinfo {author} {\bibfnamefont {J.~P.}\
  \bibnamefont {{Berger}}},\ }\href
  {https://doi.org/10.1051/0004-6361/202141985} {\bibfield  {journal} {\bibinfo
   {journal} {\aap}\ }\textbf {\bibinfo {volume} {655}},\ \bibinfo {eid} {A15}
  (\bibinfo {year} {2021})},\ \Eprint {https://arxiv.org/abs/2109.02841}
  {arXiv:2109.02841 [astro-ph.SR]} \BibitemShut {NoStop}%
\bibitem [{\citenamefont {{Pinilla}}\ \emph {et~al.}(2018)\citenamefont
  {{Pinilla}}, \citenamefont {{Benisty}}, \citenamefont {{de Boer}},
  \citenamefont {{Manara}}, \citenamefont {{Bouvier}}, \citenamefont
  {{Dominik}}, \citenamefont {{Ginski}}, \citenamefont {{Loomis}},\ and\
  \citenamefont {{Sicilia Aguilar}}}]{Pinilla+2018}%
  \BibitemOpen
  \bibfield  {author} {\bibinfo {author} {\bibfnamefont {P.}~\bibnamefont
  {{Pinilla}}}, \bibinfo {author} {\bibfnamefont {M.}~\bibnamefont
  {{Benisty}}}, \bibinfo {author} {\bibfnamefont {J.}~\bibnamefont {{de
  Boer}}}, \bibinfo {author} {\bibfnamefont {C.~F.}\ \bibnamefont {{Manara}}},
  \bibinfo {author} {\bibfnamefont {J.}~\bibnamefont {{Bouvier}}}, \bibinfo
  {author} {\bibfnamefont {C.}~\bibnamefont {{Dominik}}}, \bibinfo {author}
  {\bibfnamefont {C.}~\bibnamefont {{Ginski}}}, \bibinfo {author}
  {\bibfnamefont {R.~A.}\ \bibnamefont {{Loomis}}},\ and\ \bibinfo {author}
  {\bibfnamefont {A.}~\bibnamefont {{Sicilia Aguilar}}},\ }\href
  {https://doi.org/10.3847/1538-4357/aae824} {\bibfield  {journal} {\bibinfo
  {journal} {\apj}\ }\textbf {\bibinfo {volume} {868}},\ \bibinfo {eid} {85}
  (\bibinfo {year} {2018})},\ \Eprint {https://arxiv.org/abs/1810.05172}
  {arXiv:1810.05172 [astro-ph.EP]} \BibitemShut {NoStop}%
\bibitem [{\citenamefont {{Montarg{\`e}s}}\ \emph {et~al.}(2021)\citenamefont
  {{Montarg{\`e}s}}, \citenamefont {{Cannon}}, \citenamefont {{Lagadec}},
  \citenamefont {{de Koter}}, \citenamefont {{Kervella}}, \citenamefont
  {{Sanchez-Bermudez}}, \citenamefont {{Paladini}},\ and\ \citenamefont
  {{Cantalloube et al}}}]{Montarges+2021}%
  \BibitemOpen
  \bibfield  {author} {\bibinfo {author} {\bibfnamefont {M.}~\bibnamefont
  {{Montarg{\`e}s}}}, \bibinfo {author} {\bibfnamefont {E.}~\bibnamefont
  {{Cannon}}}, \bibinfo {author} {\bibfnamefont {E.}~\bibnamefont {{Lagadec}}},
  \bibinfo {author} {\bibfnamefont {A.}~\bibnamefont {{de Koter}}}, \bibinfo
  {author} {\bibfnamefont {P.}~\bibnamefont {{Kervella}}}, \bibinfo {author}
  {\bibfnamefont {J.}~\bibnamefont {{Sanchez-Bermudez}}}, \bibinfo {author}
  {\bibfnamefont {C.}~\bibnamefont {{Paladini}}},\ and\ \bibinfo {author}
  {\bibfnamefont {F.}~\bibnamefont {{Cantalloube et al}}},\ }\href
  {https://doi.org/10.1038/s41586-021-03546-8} {\bibfield  {journal} {\bibinfo
  {journal} {\nat}\ }\textbf {\bibinfo {volume} {594}},\ \bibinfo {pages} {365}
  (\bibinfo {year} {2021})},\ \Eprint {https://arxiv.org/abs/2201.10551}
  {arXiv:2201.10551 [astro-ph.SR]} \BibitemShut {NoStop}%
\bibitem [{\citenamefont {{Aronson}}\ \emph {et~al.}(2022)\citenamefont
  {{Aronson}}, \citenamefont {{Baumgarte}},\ and\ \citenamefont
  {{Shapiro}}}]{Aronson+2022}%
  \BibitemOpen
  \bibfield  {author} {\bibinfo {author} {\bibfnamefont {H.}~\bibnamefont
  {{Aronson}}}, \bibinfo {author} {\bibfnamefont {T.~W.}\ \bibnamefont
  {{Baumgarte}}},\ and\ \bibinfo {author} {\bibfnamefont {S.~L.}\ \bibnamefont
  {{Shapiro}}},\ }\href@noop {} {\bibfield  {journal} {\bibinfo  {journal}
  {arXiv e-prints}\ ,\ \bibinfo {eid} {arXiv:2201.08438}} (\bibinfo {year}
  {2022})},\ \Eprint {https://arxiv.org/abs/2201.08438} {arXiv:2201.08438
  [astro-ph.SR]} \BibitemShut {NoStop}%
\bibitem [{\citenamefont {{Mayama}}\ \emph {et~al.}(2010)\citenamefont
  {{Mayama}}, \citenamefont {{Tamura}}, \citenamefont {{Hanawa}}, \citenamefont
  {{Matsumoto}}, \citenamefont {{Ishii}}, \citenamefont {{Pyo}}, \citenamefont
  {{Suto}}, \citenamefont {{Naoi}}, \citenamefont {{Kudo}}, \citenamefont
  {{Hashimoto}}, \citenamefont {{Nishiyama}}, \citenamefont {{Kuzuhara}},\ and\
  \citenamefont {{Hayashi}}}]{Mayama+2010}%
  \BibitemOpen
  \bibfield  {author} {\bibinfo {author} {\bibfnamefont {S.}~\bibnamefont
  {{Mayama}}}, \bibinfo {author} {\bibfnamefont {M.}~\bibnamefont {{Tamura}}},
  \bibinfo {author} {\bibfnamefont {T.}~\bibnamefont {{Hanawa}}}, \bibinfo
  {author} {\bibfnamefont {T.}~\bibnamefont {{Matsumoto}}}, \bibinfo {author}
  {\bibfnamefont {M.}~\bibnamefont {{Ishii}}}, \bibinfo {author} {\bibfnamefont
  {T.-S.}\ \bibnamefont {{Pyo}}}, \bibinfo {author} {\bibfnamefont
  {H.}~\bibnamefont {{Suto}}}, \bibinfo {author} {\bibfnamefont
  {T.}~\bibnamefont {{Naoi}}}, \bibinfo {author} {\bibfnamefont
  {T.}~\bibnamefont {{Kudo}}}, \bibinfo {author} {\bibfnamefont
  {J.}~\bibnamefont {{Hashimoto}}}, \bibinfo {author} {\bibfnamefont
  {S.}~\bibnamefont {{Nishiyama}}}, \bibinfo {author} {\bibfnamefont
  {M.}~\bibnamefont {{Kuzuhara}}},\ and\ \bibinfo {author} {\bibfnamefont
  {M.}~\bibnamefont {{Hayashi}}},\ }\href
  {https://doi.org/10.1126/science.1179679} {\bibfield  {journal} {\bibinfo
  {journal} {Science}\ }\textbf {\bibinfo {volume} {327}},\ \bibinfo {pages}
  {306} (\bibinfo {year} {2010})}\BibitemShut {NoStop}%
\bibitem [{\citenamefont {{Mayama}}\ \emph {et~al.}(2020)\citenamefont
  {{Mayama}}, \citenamefont {{P{\'e}rez}}, \citenamefont {{Kusakabe}},
  \citenamefont {{Muto}}, \citenamefont {{Tsukagoshi}}, \citenamefont
  {{Sitko}}, \citenamefont {{Takami}}, \citenamefont {{Hashimoto}},
  \citenamefont {{Dong}}, \citenamefont {{Kwon}}, \citenamefont {{Hayashi}},
  \citenamefont {{Kudo}}, \citenamefont {{Kuzuhara}}, \citenamefont
  {{Follette}}, \citenamefont {{Fukagawa}}, \citenamefont {{Momose}},
  \citenamefont {{Oh}}, \citenamefont {{de Leon}}, \citenamefont {{Akiyama}},
  \citenamefont {{Wisniewski}},\ and\ \citenamefont {{Yang et
  al.}}}]{Mayama+2020}%
  \BibitemOpen
  \bibfield  {author} {\bibinfo {author} {\bibfnamefont {S.}~\bibnamefont
  {{Mayama}}}, \bibinfo {author} {\bibfnamefont {S.}~\bibnamefont
  {{P{\'e}rez}}}, \bibinfo {author} {\bibfnamefont {N.}~\bibnamefont
  {{Kusakabe}}}, \bibinfo {author} {\bibfnamefont {T.}~\bibnamefont {{Muto}}},
  \bibinfo {author} {\bibfnamefont {T.}~\bibnamefont {{Tsukagoshi}}}, \bibinfo
  {author} {\bibfnamefont {M.~L.}\ \bibnamefont {{Sitko}}}, \bibinfo {author}
  {\bibfnamefont {M.}~\bibnamefont {{Takami}}}, \bibinfo {author}
  {\bibfnamefont {J.}~\bibnamefont {{Hashimoto}}}, \bibinfo {author}
  {\bibfnamefont {R.}~\bibnamefont {{Dong}}}, \bibinfo {author} {\bibfnamefont
  {J.}~\bibnamefont {{Kwon}}}, \bibinfo {author} {\bibfnamefont {S.~S.}\
  \bibnamefont {{Hayashi}}}, \bibinfo {author} {\bibfnamefont {T.}~\bibnamefont
  {{Kudo}}}, \bibinfo {author} {\bibfnamefont {M.}~\bibnamefont {{Kuzuhara}}},
  \bibinfo {author} {\bibfnamefont {K.}~\bibnamefont {{Follette}}}, \bibinfo
  {author} {\bibfnamefont {M.}~\bibnamefont {{Fukagawa}}}, \bibinfo {author}
  {\bibfnamefont {M.}~\bibnamefont {{Momose}}}, \bibinfo {author}
  {\bibfnamefont {D.}~\bibnamefont {{Oh}}}, \bibinfo {author} {\bibfnamefont
  {J.}~\bibnamefont {{de Leon}}}, \bibinfo {author} {\bibfnamefont
  {E.}~\bibnamefont {{Akiyama}}}, \bibinfo {author} {\bibfnamefont {J.~P.}\
  \bibnamefont {{Wisniewski}}},\ and\ \bibinfo {author} {\bibnamefont {{Yang et
  al.}}},\ }\href {https://doi.org/10.3847/1538-3881/ab5850} {\bibfield
  {journal} {\bibinfo  {journal} {\aj}\ }\textbf {\bibinfo {volume} {159}},\
  \bibinfo {eid} {12} (\bibinfo {year} {2020})},\ \Eprint
  {https://arxiv.org/abs/1911.10941} {arXiv:1911.10941 [astro-ph.SR]}
  \BibitemShut {NoStop}%
\bibitem [{\citenamefont {{Gonz{\'a}lez-Ruilova}}\ \emph
  {et~al.}(2020)\citenamefont {{Gonz{\'a}lez-Ruilova}}, \citenamefont
  {{Cieza}}, \citenamefont {{Hales}}, \citenamefont {{P{\'e}rez}},
  \citenamefont {{Zurlo}}, \citenamefont {{Arce-Tord}}, \citenamefont
  {{Casassus}}, \citenamefont {{C{\'a}novas}}, \citenamefont {{Flock}},
  \citenamefont {{Herczeg}}, \citenamefont {{Pinilla}}, \citenamefont
  {{Price}}, \citenamefont {{Principe}}, \citenamefont
  {{Ru{\'\i}z-Rodr{\'\i}guez}},\ and\ \citenamefont
  {{Williams}}}]{Gonzalez-Ruilova+2020}%
  \BibitemOpen
  \bibfield  {author} {\bibinfo {author} {\bibfnamefont {C.}~\bibnamefont
  {{Gonz{\'a}lez-Ruilova}}}, \bibinfo {author} {\bibfnamefont {L.~A.}\
  \bibnamefont {{Cieza}}}, \bibinfo {author} {\bibfnamefont {A.~S.}\
  \bibnamefont {{Hales}}}, \bibinfo {author} {\bibfnamefont {S.}~\bibnamefont
  {{P{\'e}rez}}}, \bibinfo {author} {\bibfnamefont {A.}~\bibnamefont
  {{Zurlo}}}, \bibinfo {author} {\bibfnamefont {C.}~\bibnamefont
  {{Arce-Tord}}}, \bibinfo {author} {\bibfnamefont {S.}~\bibnamefont
  {{Casassus}}}, \bibinfo {author} {\bibfnamefont {H.}~\bibnamefont
  {{C{\'a}novas}}}, \bibinfo {author} {\bibfnamefont {M.}~\bibnamefont
  {{Flock}}}, \bibinfo {author} {\bibfnamefont {G.~J.}\ \bibnamefont
  {{Herczeg}}}, \bibinfo {author} {\bibfnamefont {P.}~\bibnamefont
  {{Pinilla}}}, \bibinfo {author} {\bibfnamefont {D.~J.}\ \bibnamefont
  {{Price}}}, \bibinfo {author} {\bibfnamefont {D.~A.}\ \bibnamefont
  {{Principe}}}, \bibinfo {author} {\bibfnamefont {D.}~\bibnamefont
  {{Ru{\'\i}z-Rodr{\'\i}guez}}},\ and\ \bibinfo {author} {\bibfnamefont
  {J.~P.}\ \bibnamefont {{Williams}}},\ }\href
  {https://doi.org/10.3847/2041-8213/abbcce} {\bibfield  {journal} {\bibinfo
  {journal} {\apjl}\ }\textbf {\bibinfo {volume} {902}},\ \bibinfo {eid} {L33}
  (\bibinfo {year} {2020})},\ \Eprint {https://arxiv.org/abs/2010.03650}
  {arXiv:2010.03650 [astro-ph.EP]} \BibitemShut {NoStop}%
\bibitem [{\citenamefont {{Winter}}\ \emph
  {et~al.}(2018{\natexlab{c}})\citenamefont {{Winter}}, \citenamefont
  {{Booth}},\ and\ \citenamefont {{Clarke}}}]{Winter+2018c}%
  \BibitemOpen
  \bibfield  {author} {\bibinfo {author} {\bibfnamefont {A.~J.}\ \bibnamefont
  {{Winter}}}, \bibinfo {author} {\bibfnamefont {R.~A.}\ \bibnamefont
  {{Booth}}},\ and\ \bibinfo {author} {\bibfnamefont {C.~J.}\ \bibnamefont
  {{Clarke}}},\ }\href {https://doi.org/10.1093/mnras/sty1866} {\bibfield
  {journal} {\bibinfo  {journal} {\mnras}\ }\textbf {\bibinfo {volume} {479}},\
  \bibinfo {pages} {5522} (\bibinfo {year} {2018}{\natexlab{c}})},\ \Eprint
  {https://arxiv.org/abs/1807.04295} {arXiv:1807.04295 [astro-ph.SR]}
  \BibitemShut {NoStop}%
\bibitem [{\citenamefont {{Kurtovic}}\ \emph {et~al.}(2018)\citenamefont
  {{Kurtovic}}, \citenamefont {{P{\'e}rez}}, \citenamefont {{Benisty}},
  \citenamefont {{Zhu}}, \citenamefont {{Zhang}}, \citenamefont {{Huang}},
  \citenamefont {{Andrews}}, \citenamefont {{Dullemond}}, \citenamefont
  {{Isella}}, \citenamefont {{Bai}}, \citenamefont {{Carpenter}}, \citenamefont
  {{Guzm{\'a}n}}, \citenamefont {{Ricci}},\ and\ \citenamefont
  {{Wilner}}}]{Kurtovic+2018}%
  \BibitemOpen
  \bibfield  {author} {\bibinfo {author} {\bibfnamefont {N.~T.}\ \bibnamefont
  {{Kurtovic}}}, \bibinfo {author} {\bibfnamefont {L.~M.}\ \bibnamefont
  {{P{\'e}rez}}}, \bibinfo {author} {\bibfnamefont {M.}~\bibnamefont
  {{Benisty}}}, \bibinfo {author} {\bibfnamefont {Z.}~\bibnamefont {{Zhu}}},
  \bibinfo {author} {\bibfnamefont {S.}~\bibnamefont {{Zhang}}}, \bibinfo
  {author} {\bibfnamefont {J.}~\bibnamefont {{Huang}}}, \bibinfo {author}
  {\bibfnamefont {S.~M.}\ \bibnamefont {{Andrews}}}, \bibinfo {author}
  {\bibfnamefont {C.~P.}\ \bibnamefont {{Dullemond}}}, \bibinfo {author}
  {\bibfnamefont {A.}~\bibnamefont {{Isella}}}, \bibinfo {author}
  {\bibfnamefont {X.-N.}\ \bibnamefont {{Bai}}}, \bibinfo {author}
  {\bibfnamefont {J.~M.}\ \bibnamefont {{Carpenter}}}, \bibinfo {author}
  {\bibfnamefont {V.~V.}\ \bibnamefont {{Guzm{\'a}n}}}, \bibinfo {author}
  {\bibfnamefont {L.}~\bibnamefont {{Ricci}}},\ and\ \bibinfo {author}
  {\bibfnamefont {D.~J.}\ \bibnamefont {{Wilner}}},\ }\href
  {https://doi.org/10.3847/2041-8213/aaf746} {\bibfield  {journal} {\bibinfo
  {journal} {\apjl}\ }\textbf {\bibinfo {volume} {869}},\ \bibinfo {eid} {L44}
  (\bibinfo {year} {2018})},\ \Eprint {https://arxiv.org/abs/1812.04536}
  {arXiv:1812.04536 [astro-ph.SR]} \BibitemShut {NoStop}%
\bibitem [{\citenamefont {{Ghez}}\ \emph {et~al.}(1993)\citenamefont {{Ghez}},
  \citenamefont {{Neugebauer}},\ and\ \citenamefont {{Matthews}}}]{Ghez+1993}%
  \BibitemOpen
  \bibfield  {author} {\bibinfo {author} {\bibfnamefont {A.~M.}\ \bibnamefont
  {{Ghez}}}, \bibinfo {author} {\bibfnamefont {G.}~\bibnamefont
  {{Neugebauer}}},\ and\ \bibinfo {author} {\bibfnamefont {K.}~\bibnamefont
  {{Matthews}}},\ }\href {https://doi.org/10.1086/116782} {\bibfield  {journal}
  {\bibinfo  {journal} {\aj}\ }\textbf {\bibinfo {volume} {106}},\ \bibinfo
  {pages} {2005} (\bibinfo {year} {1993})}\BibitemShut {NoStop}%
\bibitem [{\citenamefont {{Cabrit}}\ \emph {et~al.}(2006)\citenamefont
  {{Cabrit}}, \citenamefont {{Pety}}, \citenamefont {{Pesenti}},\ and\
  \citenamefont {{Dougados}}}]{Cabrit+2006}%
  \BibitemOpen
  \bibfield  {author} {\bibinfo {author} {\bibfnamefont {S.}~\bibnamefont
  {{Cabrit}}}, \bibinfo {author} {\bibfnamefont {J.}~\bibnamefont {{Pety}}},
  \bibinfo {author} {\bibfnamefont {N.}~\bibnamefont {{Pesenti}}},\ and\
  \bibinfo {author} {\bibfnamefont {C.}~\bibnamefont {{Dougados}}},\ }\href
  {https://doi.org/10.1051/0004-6361:20054047} {\bibfield  {journal} {\bibinfo
  {journal} {\aap}\ }\textbf {\bibinfo {volume} {452}},\ \bibinfo {pages} {897}
  (\bibinfo {year} {2006})}\BibitemShut {NoStop}%
\bibitem [{\citenamefont {{P{\'e}rez}}\ \emph {et~al.}(2020)\citenamefont
  {{P{\'e}rez}}, \citenamefont {{Hales}}, \citenamefont {{Liu}}, \citenamefont
  {{Zhu}}, \citenamefont {{Casassus}}, \citenamefont {{Williams}},
  \citenamefont {{Zurlo}}, \citenamefont {{Cuello}}, \citenamefont {{Cieza}},\
  and\ \citenamefont {{Principe}}}]{Perez+2020}%
  \BibitemOpen
  \bibfield  {author} {\bibinfo {author} {\bibfnamefont {S.}~\bibnamefont
  {{P{\'e}rez}}}, \bibinfo {author} {\bibfnamefont {A.}~\bibnamefont
  {{Hales}}}, \bibinfo {author} {\bibfnamefont {H.~B.}\ \bibnamefont {{Liu}}},
  \bibinfo {author} {\bibfnamefont {Z.}~\bibnamefont {{Zhu}}}, \bibinfo
  {author} {\bibfnamefont {S.}~\bibnamefont {{Casassus}}}, \bibinfo {author}
  {\bibfnamefont {J.}~\bibnamefont {{Williams}}}, \bibinfo {author}
  {\bibfnamefont {A.}~\bibnamefont {{Zurlo}}}, \bibinfo {author} {\bibfnamefont
  {N.}~\bibnamefont {{Cuello}}}, \bibinfo {author} {\bibfnamefont
  {L.}~\bibnamefont {{Cieza}}},\ and\ \bibinfo {author} {\bibfnamefont
  {D.}~\bibnamefont {{Principe}}},\ }\href
  {https://doi.org/10.3847/1538-4357/ab5c1b} {\bibfield  {journal} {\bibinfo
  {journal} {\apj}\ }\textbf {\bibinfo {volume} {889}},\ \bibinfo {eid} {59}
  (\bibinfo {year} {2020})},\ \Eprint {https://arxiv.org/abs/1911.11282}
  {arXiv:1911.11282 [astro-ph.EP]} \BibitemShut {NoStop}%
\bibitem [{\citenamefont {{Lu}}\ \emph {et~al.}(2022)\citenamefont {{Lu}},
  \citenamefont {{Li}}, \citenamefont {{Zhang}},\ and\ \citenamefont
  {{Lin}}}]{Lu+2022}%
  \BibitemOpen
  \bibfield  {author} {\bibinfo {author} {\bibfnamefont {X.}~\bibnamefont
  {{Lu}}}, \bibinfo {author} {\bibfnamefont {G.-X.}\ \bibnamefont {{Li}}},
  \bibinfo {author} {\bibfnamefont {Q.}~\bibnamefont {{Zhang}}},\ and\ \bibinfo
  {author} {\bibfnamefont {Y.}~\bibnamefont {{Lin}}},\ }\bibfield  {journal}
  {\bibinfo  {journal} {Nature Astronomy}\ }\href
  {https://doi.org/10.1038/s41550-022-01681-4} {10.1038/s41550-022-01681-4}
  (\bibinfo {year} {2022}),\ \Eprint {https://arxiv.org/abs/2206.00202}
  {arXiv:2206.00202 [astro-ph.GA]} \BibitemShut {NoStop}%
\bibitem [{\citenamefont {{Jensen}}\ and\ \citenamefont
  {{Mathieu}}(1997)}]{Jensen+1997}%
  \BibitemOpen
  \bibfield  {author} {\bibinfo {author} {\bibfnamefont {E.~L.~N.}\
  \bibnamefont {{Jensen}}}\ and\ \bibinfo {author} {\bibfnamefont {R.~D.}\
  \bibnamefont {{Mathieu}}},\ }\href {https://doi.org/10.1086/118475}
  {\bibfield  {journal} {\bibinfo  {journal} {\aj}\ }\textbf {\bibinfo {volume}
  {114}},\ \bibinfo {pages} {301} (\bibinfo {year} {1997})}\BibitemShut
  {NoStop}%
\bibitem [{\citenamefont {{Cox}}\ \emph {et~al.}(2017)\citenamefont {{Cox}},
  \citenamefont {{Harris}}, \citenamefont {{Looney}}, \citenamefont {{Chiang}},
  \citenamefont {{Chandler}}, \citenamefont {{Kratter}}, \citenamefont {{Li}},
  \citenamefont {{Perez}},\ and\ \citenamefont {{Tobin}}}]{Cox+2017}%
  \BibitemOpen
  \bibfield  {author} {\bibinfo {author} {\bibfnamefont {E.~G.}\ \bibnamefont
  {{Cox}}}, \bibinfo {author} {\bibfnamefont {R.~J.}\ \bibnamefont {{Harris}}},
  \bibinfo {author} {\bibfnamefont {L.~W.}\ \bibnamefont {{Looney}}}, \bibinfo
  {author} {\bibfnamefont {H.-F.}\ \bibnamefont {{Chiang}}}, \bibinfo {author}
  {\bibfnamefont {C.}~\bibnamefont {{Chandler}}}, \bibinfo {author}
  {\bibfnamefont {K.}~\bibnamefont {{Kratter}}}, \bibinfo {author}
  {\bibfnamefont {Z.-Y.}\ \bibnamefont {{Li}}}, \bibinfo {author}
  {\bibfnamefont {L.}~\bibnamefont {{Perez}}},\ and\ \bibinfo {author}
  {\bibfnamefont {J.~J.}\ \bibnamefont {{Tobin}}},\ }\href
  {https://doi.org/10.3847/1538-4357/aa97e2} {\bibfield  {journal} {\bibinfo
  {journal} {\apj}\ }\textbf {\bibinfo {volume} {851}},\ \bibinfo {eid} {83}
  (\bibinfo {year} {2017})},\ \Eprint {https://arxiv.org/abs/1711.03974}
  {arXiv:1711.03974 [astro-ph.SR]} \BibitemShut {NoStop}%
\bibitem [{\citenamefont {{Rota}}\ \emph {et~al.}(2022)\citenamefont {{Rota}},
  \citenamefont {{Manara}},\ and\ \citenamefont {{Miotello et
  al.}}}]{Rota+2022}%
  \BibitemOpen
  \bibfield  {author} {\bibinfo {author} {\bibfnamefont {A.~A.}\ \bibnamefont
  {{Rota}}}, \bibinfo {author} {\bibfnamefont {C.~F.}\ \bibnamefont
  {{Manara}}},\ and\ \bibinfo {author} {\bibfnamefont {A.}~\bibnamefont
  {{Miotello et al.}}},\ }\href@noop {} {\bibfield  {journal} {\bibinfo
  {journal} {arXiv e-prints}\ ,\ \bibinfo {eid} {arXiv:2201.03588}} (\bibinfo
  {year} {2022})},\ \Eprint {https://arxiv.org/abs/2201.03588}
  {arXiv:2201.03588 [astro-ph.EP]} \BibitemShut {NoStop}%
\bibitem [{\citenamefont {{Zagaria}}\ \emph {et~al.}(2022)\citenamefont
  {{Zagaria}}, \citenamefont {{Clarke}}, \citenamefont {{Rosotti}},\ and\
  \citenamefont {{Manara}}}]{Zagaria+2022}%
  \BibitemOpen
  \bibfield  {author} {\bibinfo {author} {\bibfnamefont {F.}~\bibnamefont
  {{Zagaria}}}, \bibinfo {author} {\bibfnamefont {C.~J.}\ \bibnamefont
  {{Clarke}}}, \bibinfo {author} {\bibfnamefont {G.~P.}\ \bibnamefont
  {{Rosotti}}},\ and\ \bibinfo {author} {\bibfnamefont {C.~F.}\ \bibnamefont
  {{Manara}}},\ }\href {https://doi.org/10.1093/mnras/stac621} {\bibfield
  {journal} {\bibinfo  {journal} {\mnras}\ }\textbf {\bibinfo {volume} {512}},\
  \bibinfo {pages} {3538} (\bibinfo {year} {2022})},\ \Eprint
  {https://arxiv.org/abs/2203.01986} {arXiv:2203.01986 [astro-ph.SR]}
  \BibitemShut {NoStop}%
\bibitem [{\citenamefont {{Umbreit}}\ \emph {et~al.}(2011)\citenamefont
  {{Umbreit}}, \citenamefont {{Spurzem}}, \citenamefont {{Henning}},
  \citenamefont {{Klahr}},\ and\ \citenamefont {{Mikkola}}}]{Umbreit+2011}%
  \BibitemOpen
  \bibfield  {author} {\bibinfo {author} {\bibfnamefont {S.}~\bibnamefont
  {{Umbreit}}}, \bibinfo {author} {\bibfnamefont {R.}~\bibnamefont
  {{Spurzem}}}, \bibinfo {author} {\bibfnamefont {T.}~\bibnamefont
  {{Henning}}}, \bibinfo {author} {\bibfnamefont {H.}~\bibnamefont {{Klahr}}},\
  and\ \bibinfo {author} {\bibfnamefont {S.}~\bibnamefont {{Mikkola}}},\ }\href
  {https://doi.org/10.1088/0004-637X/743/2/106} {\bibfield  {journal} {\bibinfo
   {journal} {\apj}\ }\textbf {\bibinfo {volume} {743}},\ \bibinfo {eid} {106}
  (\bibinfo {year} {2011})}\BibitemShut {NoStop}%
\bibitem [{\citenamefont {{Artymowicz}}\ and\ \citenamefont
  {{Lubow}}(1994)}]{Artymowicz+1994}%
  \BibitemOpen
  \bibfield  {author} {\bibinfo {author} {\bibfnamefont {P.}~\bibnamefont
  {{Artymowicz}}}\ and\ \bibinfo {author} {\bibfnamefont {S.~H.}\ \bibnamefont
  {{Lubow}}},\ }\href {https://doi.org/10.1086/173679} {\bibfield  {journal}
  {\bibinfo  {journal} {\apj}\ }\textbf {\bibinfo {volume} {421}},\ \bibinfo
  {pages} {651} (\bibinfo {year} {1994})}\BibitemShut {NoStop}%
\bibitem [{\citenamefont {{Farago}}\ and\ \citenamefont
  {{Laskar}}(2010)}]{Farago+2010}%
  \BibitemOpen
  \bibfield  {author} {\bibinfo {author} {\bibfnamefont {F.}~\bibnamefont
  {{Farago}}}\ and\ \bibinfo {author} {\bibfnamefont {J.}~\bibnamefont
  {{Laskar}}},\ }\href {https://doi.org/10.1111/j.1365-2966.2009.15711.x}
  {\bibfield  {journal} {\bibinfo  {journal} {\mnras}\ }\textbf {\bibinfo
  {volume} {401}},\ \bibinfo {pages} {1189} (\bibinfo {year} {2010})},\ \Eprint
  {https://arxiv.org/abs/0909.2287} {arXiv:0909.2287 [astro-ph.EP]}
  \BibitemShut {NoStop}%
\bibitem [{\citenamefont {{Martin}}\ \emph {et~al.}(2014)\citenamefont
  {{Martin}}, \citenamefont {{Nixon}}, \citenamefont {{Lubow}}, \citenamefont
  {{Armitage}}, \citenamefont {{Price}}, \citenamefont {{Do{\u{g}}an}},\ and\
  \citenamefont {{King}}}]{Martin+2014}%
  \BibitemOpen
  \bibfield  {author} {\bibinfo {author} {\bibfnamefont {R.~G.}\ \bibnamefont
  {{Martin}}}, \bibinfo {author} {\bibfnamefont {C.}~\bibnamefont {{Nixon}}},
  \bibinfo {author} {\bibfnamefont {S.~H.}\ \bibnamefont {{Lubow}}}, \bibinfo
  {author} {\bibfnamefont {P.~J.}\ \bibnamefont {{Armitage}}}, \bibinfo
  {author} {\bibfnamefont {D.~J.}\ \bibnamefont {{Price}}}, \bibinfo {author}
  {\bibfnamefont {S.}~\bibnamefont {{Do{\u{g}}an}}},\ and\ \bibinfo {author}
  {\bibfnamefont {A.}~\bibnamefont {{King}}},\ }\href
  {https://doi.org/10.1088/2041-8205/792/2/L33} {\bibfield  {journal} {\bibinfo
   {journal} {\apjl}\ }\textbf {\bibinfo {volume} {792}},\ \bibinfo {eid} {L33}
  (\bibinfo {year} {2014})},\ \Eprint {https://arxiv.org/abs/1409.1226}
  {arXiv:1409.1226 [astro-ph.EP]} \BibitemShut {NoStop}%
\bibitem [{\citenamefont {{Martin}}\ \emph {et~al.}(2022)\citenamefont
  {{Martin}}, \citenamefont {{Lepp}}, \citenamefont {{Lubow}}, \citenamefont
  {{Kenworthy}}, \citenamefont {{Kennedy}},\ and\ \citenamefont
  {{Vallet}}}]{Martin+2022}%
  \BibitemOpen
  \bibfield  {author} {\bibinfo {author} {\bibfnamefont {R.~G.}\ \bibnamefont
  {{Martin}}}, \bibinfo {author} {\bibfnamefont {S.}~\bibnamefont {{Lepp}}},
  \bibinfo {author} {\bibfnamefont {S.~H.}\ \bibnamefont {{Lubow}}}, \bibinfo
  {author} {\bibfnamefont {M.~A.}\ \bibnamefont {{Kenworthy}}}, \bibinfo
  {author} {\bibfnamefont {G.~M.}\ \bibnamefont {{Kennedy}}},\ and\ \bibinfo
  {author} {\bibfnamefont {D.}~\bibnamefont {{Vallet}}},\ }\href
  {https://doi.org/10.3847/2041-8213/ac54b4} {\bibfield  {journal} {\bibinfo
  {journal} {\apjl}\ }\textbf {\bibinfo {volume} {927}},\ \bibinfo {eid} {L26}
  (\bibinfo {year} {2022})},\ \Eprint {https://arxiv.org/abs/2202.06878}
  {arXiv:2202.06878 [astro-ph.EP]} \BibitemShut {NoStop}%
\bibitem [{\citenamefont {{Dunhill}}\ \emph {et~al.}(2015)\citenamefont
  {{Dunhill}}, \citenamefont {{Cuadra}},\ and\ \citenamefont
  {{Dougados}}}]{Dunhill+2015}%
  \BibitemOpen
  \bibfield  {author} {\bibinfo {author} {\bibfnamefont {A.~C.}\ \bibnamefont
  {{Dunhill}}}, \bibinfo {author} {\bibfnamefont {J.}~\bibnamefont
  {{Cuadra}}},\ and\ \bibinfo {author} {\bibfnamefont {C.}~\bibnamefont
  {{Dougados}}},\ }\href {https://doi.org/10.1093/mnras/stv284} {\bibfield
  {journal} {\bibinfo  {journal} {\mnras}\ }\textbf {\bibinfo {volume} {448}},\
  \bibinfo {pages} {3545} (\bibinfo {year} {2015})},\ \Eprint
  {https://arxiv.org/abs/1411.0687} {arXiv:1411.0687 [astro-ph.SR]}
  \BibitemShut {NoStop}%
\bibitem [{\citenamefont {{Siwek}}\ \emph {et~al.}(2022)\citenamefont
  {{Siwek}}, \citenamefont {{Weinberger}}, \citenamefont {{Munoz}},\ and\
  \citenamefont {{Hernquist}}}]{Siwek+2022}%
  \BibitemOpen
  \bibfield  {author} {\bibinfo {author} {\bibfnamefont {M.}~\bibnamefont
  {{Siwek}}}, \bibinfo {author} {\bibfnamefont {R.}~\bibnamefont
  {{Weinberger}}}, \bibinfo {author} {\bibfnamefont {D.}~\bibnamefont
  {{Munoz}}},\ and\ \bibinfo {author} {\bibfnamefont {L.}~\bibnamefont
  {{Hernquist}}},\ }\href@noop {} {\bibfield  {journal} {\bibinfo  {journal}
  {arXiv e-prints}\ ,\ \bibinfo {eid} {arXiv:2203.02514}} (\bibinfo {year}
  {2022})},\ \Eprint {https://arxiv.org/abs/2203.02514} {arXiv:2203.02514
  [astro-ph.HE]} \BibitemShut {NoStop}%
\bibitem [{\citenamefont {{Ceppi}}\ \emph {et~al.}(2022)\citenamefont
  {{Ceppi}}, \citenamefont {{Cuello}}, \citenamefont {{Lodato}}, \citenamefont
  {{Clarke}}, \citenamefont {{Toci}},\ and\ \citenamefont
  {{Price}}}]{Ceppi+2022}%
  \BibitemOpen
  \bibfield  {author} {\bibinfo {author} {\bibfnamefont {S.}~\bibnamefont
  {{Ceppi}}}, \bibinfo {author} {\bibfnamefont {N.}~\bibnamefont {{Cuello}}},
  \bibinfo {author} {\bibfnamefont {G.}~\bibnamefont {{Lodato}}}, \bibinfo
  {author} {\bibfnamefont {C.}~\bibnamefont {{Clarke}}}, \bibinfo {author}
  {\bibfnamefont {C.}~\bibnamefont {{Toci}}},\ and\ \bibinfo {author}
  {\bibfnamefont {D.~J.}\ \bibnamefont {{Price}}},\ }\href
  {https://doi.org/10.1093/mnras/stac1390} {\bibfield  {journal} {\bibinfo
  {journal} {\mnras}\ }\textbf {\bibinfo {volume} {514}},\ \bibinfo {pages}
  {906} (\bibinfo {year} {2022})},\ \Eprint {https://arxiv.org/abs/2205.08784}
  {arXiv:2205.08784 [astro-ph.SR]} \BibitemShut {NoStop}%
\bibitem [{\citenamefont {{Smallwood}}\ \emph {et~al.}(2022)\citenamefont
  {{Smallwood}}, \citenamefont {{Lubow}},\ and\ \citenamefont
  {{Martin}}}]{Smallwood+2022}%
  \BibitemOpen
  \bibfield  {author} {\bibinfo {author} {\bibfnamefont {J.~L.}\ \bibnamefont
  {{Smallwood}}}, \bibinfo {author} {\bibfnamefont {S.~H.}\ \bibnamefont
  {{Lubow}}},\ and\ \bibinfo {author} {\bibfnamefont {R.~G.}\ \bibnamefont
  {{Martin}}},\ }\href {https://doi.org/10.1093/mnras/stac1416} {\bibfield
  {journal} {\bibinfo  {journal} {\mnras}\ }\textbf {\bibinfo {volume} {514}},\
  \bibinfo {pages} {1249} (\bibinfo {year} {2022})},\ \Eprint
  {https://arxiv.org/abs/2205.09183} {arXiv:2205.09183 [astro-ph.EP]}
  \BibitemShut {NoStop}%
\bibitem [{\citenamefont {{Howard}}\ \emph {et~al.}(2013)\citenamefont
  {{Howard}}, \citenamefont {{Sandell}}, \citenamefont {{Vacca}}, \citenamefont
  {{Duch{\^e}ne}}, \citenamefont {{Mathews}}, \citenamefont {{Augereau}},
  \citenamefont {{Barrado}}, \citenamefont {{Dent}}, \citenamefont {{Eiroa}},
  \citenamefont {{Grady}}, \citenamefont {{Kamp}}, \citenamefont {{Meeus}},
  \citenamefont {{M{\'e}nard}}, \citenamefont {{Pinte}}, \citenamefont
  {{Podio}}, \citenamefont {{Riviere-Marichalar}}, \citenamefont {{Roberge}},
  \citenamefont {{Thi}}, \citenamefont {{Vicente}},\ and\ \citenamefont
  {{Williams}}}]{Howard+2013}%
  \BibitemOpen
  \bibfield  {author} {\bibinfo {author} {\bibfnamefont {C.~D.}\ \bibnamefont
  {{Howard}}}, \bibinfo {author} {\bibfnamefont {G.}~\bibnamefont {{Sandell}}},
  \bibinfo {author} {\bibfnamefont {W.~D.}\ \bibnamefont {{Vacca}}}, \bibinfo
  {author} {\bibfnamefont {G.}~\bibnamefont {{Duch{\^e}ne}}}, \bibinfo {author}
  {\bibfnamefont {G.}~\bibnamefont {{Mathews}}}, \bibinfo {author}
  {\bibfnamefont {J.-C.}\ \bibnamefont {{Augereau}}}, \bibinfo {author}
  {\bibfnamefont {D.}~\bibnamefont {{Barrado}}}, \bibinfo {author}
  {\bibfnamefont {W.~R.~F.}\ \bibnamefont {{Dent}}}, \bibinfo {author}
  {\bibfnamefont {C.}~\bibnamefont {{Eiroa}}}, \bibinfo {author} {\bibfnamefont
  {C.}~\bibnamefont {{Grady}}}, \bibinfo {author} {\bibfnamefont
  {I.}~\bibnamefont {{Kamp}}}, \bibinfo {author} {\bibfnamefont
  {G.}~\bibnamefont {{Meeus}}}, \bibinfo {author} {\bibfnamefont
  {F.}~\bibnamefont {{M{\'e}nard}}}, \bibinfo {author} {\bibfnamefont
  {C.}~\bibnamefont {{Pinte}}}, \bibinfo {author} {\bibfnamefont
  {L.}~\bibnamefont {{Podio}}}, \bibinfo {author} {\bibfnamefont
  {P.}~\bibnamefont {{Riviere-Marichalar}}}, \bibinfo {author} {\bibfnamefont
  {A.}~\bibnamefont {{Roberge}}}, \bibinfo {author} {\bibfnamefont {W.-F.}\
  \bibnamefont {{Thi}}}, \bibinfo {author} {\bibfnamefont {S.}~\bibnamefont
  {{Vicente}}},\ and\ \bibinfo {author} {\bibfnamefont {J.~P.}\ \bibnamefont
  {{Williams}}},\ }\href {https://doi.org/10.1088/0004-637X/776/1/21}
  {\bibfield  {journal} {\bibinfo  {journal} {\apj}\ }\textbf {\bibinfo
  {volume} {776}},\ \bibinfo {eid} {21} (\bibinfo {year} {2013})},\ \Eprint
  {https://arxiv.org/abs/1308.6019} {arXiv:1308.6019 [astro-ph.GA]}
  \BibitemShut {NoStop}%
\bibitem [{\citenamefont {{Correia}}\ \emph {et~al.}(2006)\citenamefont
  {{Correia}}, \citenamefont {{Zinnecker}}, \citenamefont {{Ratzka}},\ and\
  \citenamefont {{Sterzik}}}]{Correia+2006}%
  \BibitemOpen
  \bibfield  {author} {\bibinfo {author} {\bibfnamefont {S.}~\bibnamefont
  {{Correia}}}, \bibinfo {author} {\bibfnamefont {H.}~\bibnamefont
  {{Zinnecker}}}, \bibinfo {author} {\bibfnamefont {T.}~\bibnamefont
  {{Ratzka}}},\ and\ \bibinfo {author} {\bibfnamefont {M.~F.}\ \bibnamefont
  {{Sterzik}}},\ }\href {https://doi.org/10.1051/0004-6361:20065545} {\bibfield
   {journal} {\bibinfo  {journal} {\aap}\ }\textbf {\bibinfo {volume} {459}},\
  \bibinfo {pages} {909} (\bibinfo {year} {2006})},\ \Eprint
  {https://arxiv.org/abs/astro-ph/0608674} {arXiv:astro-ph/0608674 [astro-ph]}
  \BibitemShut {NoStop}%
\bibitem [{\citenamefont {{Reipurth}}\ and\ \citenamefont
  {{Zinnecker}}(1993)}]{Reipurth&Zinnecker1993}%
  \BibitemOpen
  \bibfield  {author} {\bibinfo {author} {\bibfnamefont {B.}~\bibnamefont
  {{Reipurth}}}\ and\ \bibinfo {author} {\bibfnamefont {H.}~\bibnamefont
  {{Zinnecker}}},\ }\href@noop {} {\bibfield  {journal} {\bibinfo  {journal}
  {\aap}\ }\textbf {\bibinfo {volume} {278}},\ \bibinfo {pages} {81} (\bibinfo
  {year} {1993})}\BibitemShut {NoStop}%
\bibitem [{\citenamefont {{Schaefer}}\ \emph {et~al.}(2018)\citenamefont
  {{Schaefer}}, \citenamefont {{Prato}},\ and\ \citenamefont
  {{Simon}}}]{Schaefer+2018}%
  \BibitemOpen
  \bibfield  {author} {\bibinfo {author} {\bibfnamefont {G.~H.}\ \bibnamefont
  {{Schaefer}}}, \bibinfo {author} {\bibfnamefont {L.}~\bibnamefont
  {{Prato}}},\ and\ \bibinfo {author} {\bibfnamefont {M.}~\bibnamefont
  {{Simon}}},\ }\href {https://doi.org/10.3847/1538-3881/aaa59a} {\bibfield
  {journal} {\bibinfo  {journal} {\aj}\ }\textbf {\bibinfo {volume} {155}},\
  \bibinfo {eid} {109} (\bibinfo {year} {2018})},\ \Eprint
  {https://arxiv.org/abs/1802.02747} {arXiv:1802.02747 [astro-ph.SR]}
  \BibitemShut {NoStop}%
\bibitem [{\citenamefont {{Fern{\'a}ndez-L{\'o}pez}}\ \emph
  {et~al.}(2017)\citenamefont {{Fern{\'a}ndez-L{\'o}pez}}, \citenamefont
  {{Zapata}},\ and\ \citenamefont {{Gabbasov}}}]{Fernandez-Lopez+2017}%
  \BibitemOpen
  \bibfield  {author} {\bibinfo {author} {\bibfnamefont {M.}~\bibnamefont
  {{Fern{\'a}ndez-L{\'o}pez}}}, \bibinfo {author} {\bibfnamefont {L.~A.}\
  \bibnamefont {{Zapata}}},\ and\ \bibinfo {author} {\bibfnamefont
  {R.}~\bibnamefont {{Gabbasov}}},\ }\href
  {https://doi.org/10.3847/1538-4357/aa7d51} {\bibfield  {journal} {\bibinfo
  {journal} {\apj}\ }\textbf {\bibinfo {volume} {845}},\ \bibinfo {eid} {10}
  (\bibinfo {year} {2017})},\ \Eprint {https://arxiv.org/abs/1707.01128}
  {arXiv:1707.01128 [astro-ph.SR]} \BibitemShut {NoStop}%
\bibitem [{\citenamefont {{Pinilla}}\ \emph {et~al.}(2017)\citenamefont
  {{Pinilla}}, \citenamefont {{P{\'e}rez}}, \citenamefont {{Andrews}},
  \citenamefont {{van der Marel}}, \citenamefont {{van Dishoeck}},
  \citenamefont {{Ataiee}}, \citenamefont {{Benisty}}, \citenamefont
  {{Birnstiel}}, \citenamefont {{Juh{\'a}sz}}, \citenamefont {{Natta}},
  \citenamefont {{Ricci}},\ and\ \citenamefont {{Testi}}}]{Pinilla+2017}%
  \BibitemOpen
  \bibfield  {author} {\bibinfo {author} {\bibfnamefont {P.}~\bibnamefont
  {{Pinilla}}}, \bibinfo {author} {\bibfnamefont {L.~M.}\ \bibnamefont
  {{P{\'e}rez}}}, \bibinfo {author} {\bibfnamefont {S.}~\bibnamefont
  {{Andrews}}}, \bibinfo {author} {\bibfnamefont {N.}~\bibnamefont {{van der
  Marel}}}, \bibinfo {author} {\bibfnamefont {E.~F.}\ \bibnamefont {{van
  Dishoeck}}}, \bibinfo {author} {\bibfnamefont {S.}~\bibnamefont {{Ataiee}}},
  \bibinfo {author} {\bibfnamefont {M.}~\bibnamefont {{Benisty}}}, \bibinfo
  {author} {\bibfnamefont {T.}~\bibnamefont {{Birnstiel}}}, \bibinfo {author}
  {\bibfnamefont {A.}~\bibnamefont {{Juh{\'a}sz}}}, \bibinfo {author}
  {\bibfnamefont {A.}~\bibnamefont {{Natta}}}, \bibinfo {author} {\bibfnamefont
  {L.}~\bibnamefont {{Ricci}}},\ and\ \bibinfo {author} {\bibfnamefont
  {L.}~\bibnamefont {{Testi}}},\ }\href
  {https://doi.org/10.3847/1538-4357/aa6973} {\bibfield  {journal} {\bibinfo
  {journal} {\apj}\ }\textbf {\bibinfo {volume} {839}},\ \bibinfo {eid} {99}
  (\bibinfo {year} {2017})},\ \Eprint {https://arxiv.org/abs/1703.09227}
  {arXiv:1703.09227 [astro-ph.EP]} \BibitemShut {NoStop}%
\bibitem [{\citenamefont {{Weber}}\ \emph {et~al.}(2023)\citenamefont
  {{Weber}}, \citenamefont {{P{\'e}rez}}, \citenamefont {{Guidi}},
  \citenamefont {{Kurtovic}}, \citenamefont {{Zurlo}}, \citenamefont
  {{Garufi}}, \citenamefont {{Pinilla}}, \citenamefont {{Mayama}},
  \citenamefont {{van Holstein}}, \citenamefont {{Dullemond}}, \citenamefont
  {{Cuello}}, \citenamefont {{Principe}}, \citenamefont {{Cieza}},
  \citenamefont {{Gonz{\'a}lez-Ruilova}},\ and\ \citenamefont
  {{Girard}}}]{Weber+2023}%
  \BibitemOpen
  \bibfield  {author} {\bibinfo {author} {\bibfnamefont {P.}~\bibnamefont
  {{Weber}}}, \bibinfo {author} {\bibfnamefont {S.}~\bibnamefont
  {{P{\'e}rez}}}, \bibinfo {author} {\bibfnamefont {G.}~\bibnamefont
  {{Guidi}}}, \bibinfo {author} {\bibfnamefont {N.~T.}\ \bibnamefont
  {{Kurtovic}}}, \bibinfo {author} {\bibfnamefont {A.}~\bibnamefont {{Zurlo}}},
  \bibinfo {author} {\bibfnamefont {A.}~\bibnamefont {{Garufi}}}, \bibinfo
  {author} {\bibfnamefont {P.}~\bibnamefont {{Pinilla}}}, \bibinfo {author}
  {\bibfnamefont {S.}~\bibnamefont {{Mayama}}}, \bibinfo {author}
  {\bibfnamefont {R.~G.}\ \bibnamefont {{van Holstein}}}, \bibinfo {author}
  {\bibfnamefont {C.~P.}\ \bibnamefont {{Dullemond}}}, \bibinfo {author}
  {\bibfnamefont {N.}~\bibnamefont {{Cuello}}}, \bibinfo {author}
  {\bibfnamefont {D.}~\bibnamefont {{Principe}}}, \bibinfo {author}
  {\bibfnamefont {L.}~\bibnamefont {{Cieza}}}, \bibinfo {author} {\bibfnamefont
  {C.}~\bibnamefont {{Gonz{\'a}lez-Ruilova}}},\ and\ \bibinfo {author}
  {\bibfnamefont {J.}~\bibnamefont {{Girard}}},\ }\href
  {https://doi.org/10.1093/mnras/stac3478} {\bibfield  {journal} {\bibinfo
  {journal} {\mnras}\ }\textbf {\bibinfo {volume} {518}},\ \bibinfo {pages}
  {5620} (\bibinfo {year} {2023})},\ \Eprint {https://arxiv.org/abs/2211.14322}
  {arXiv:2211.14322 [astro-ph.SR]} \BibitemShut {NoStop}%
\bibitem [{\citenamefont {{Cieza}}\ \emph {et~al.}(2019)\citenamefont
  {{Cieza}}, \citenamefont {{Ru{\'\i}z-Rodr{\'\i}guez}}, \citenamefont
  {{Hales}}, \citenamefont {{Casassus}}, \citenamefont {{P{\'e}rez}},
  \citenamefont {{Gonzalez-Ruilova}}, \citenamefont {{C{\'a}novas}},
  \citenamefont {{Williams}},\ and\ \citenamefont {{Zurlo et
  al.}}}]{Cieza+2019}%
  \BibitemOpen
  \bibfield  {author} {\bibinfo {author} {\bibfnamefont {L.~A.}\ \bibnamefont
  {{Cieza}}}, \bibinfo {author} {\bibfnamefont {D.}~\bibnamefont
  {{Ru{\'\i}z-Rodr{\'\i}guez}}}, \bibinfo {author} {\bibfnamefont
  {A.}~\bibnamefont {{Hales}}}, \bibinfo {author} {\bibfnamefont
  {S.}~\bibnamefont {{Casassus}}}, \bibinfo {author} {\bibfnamefont
  {S.}~\bibnamefont {{P{\'e}rez}}}, \bibinfo {author} {\bibfnamefont
  {C.}~\bibnamefont {{Gonzalez-Ruilova}}}, \bibinfo {author} {\bibfnamefont
  {H.}~\bibnamefont {{C{\'a}novas}}}, \bibinfo {author} {\bibfnamefont {J.~P.}\
  \bibnamefont {{Williams}}},\ and\ \bibinfo {author} {\bibfnamefont
  {A.}~\bibnamefont {{Zurlo et al.}}},\ }\href
  {https://doi.org/10.1093/mnras/sty2653} {\bibfield  {journal} {\bibinfo
  {journal} {\mnras}\ }\textbf {\bibinfo {volume} {482}},\ \bibinfo {pages}
  {698} (\bibinfo {year} {2019})},\ \Eprint {https://arxiv.org/abs/1809.08844}
  {arXiv:1809.08844 [astro-ph.EP]} \BibitemShut {NoStop}%
\bibitem [{\citenamefont {{Woitas}}\ and\ \citenamefont
  {{Leinert}}(1998)}]{Woitas+1998}%
  \BibitemOpen
  \bibfield  {author} {\bibinfo {author} {\bibfnamefont {J.}~\bibnamefont
  {{Woitas}}}\ and\ \bibinfo {author} {\bibfnamefont {C.}~\bibnamefont
  {{Leinert}}},\ }\href@noop {} {\bibfield  {journal} {\bibinfo  {journal}
  {\aap}\ }\textbf {\bibinfo {volume} {338}},\ \bibinfo {pages} {122} (\bibinfo
  {year} {1998})}\BibitemShut {NoStop}%
\bibitem [{\citenamefont {{Hartigan}}\ \emph {et~al.}(1995)\citenamefont
  {{Hartigan}}, \citenamefont {{Edwards}},\ and\ \citenamefont
  {{Ghandour}}}]{Hartigan+1995}%
  \BibitemOpen
  \bibfield  {author} {\bibinfo {author} {\bibfnamefont {P.}~\bibnamefont
  {{Hartigan}}}, \bibinfo {author} {\bibfnamefont {S.}~\bibnamefont
  {{Edwards}}},\ and\ \bibinfo {author} {\bibfnamefont {L.}~\bibnamefont
  {{Ghandour}}},\ }\href {https://doi.org/10.1086/176344} {\bibfield  {journal}
  {\bibinfo  {journal} {\apj}\ }\textbf {\bibinfo {volume} {452}},\ \bibinfo
  {pages} {736} (\bibinfo {year} {1995})}\BibitemShut {NoStop}%
\bibitem [{\citenamefont {{Beck}}\ and\ \citenamefont
  {{Aspin}}(2012)}]{Beck+2012}%
  \BibitemOpen
  \bibfield  {author} {\bibinfo {author} {\bibfnamefont {T.~L.}\ \bibnamefont
  {{Beck}}}\ and\ \bibinfo {author} {\bibfnamefont {C.}~\bibnamefont
  {{Aspin}}},\ }\href {https://doi.org/10.1088/0004-6256/143/3/55} {\bibfield
  {journal} {\bibinfo  {journal} {\aj}\ }\textbf {\bibinfo {volume} {143}},\
  \bibinfo {eid} {55} (\bibinfo {year} {2012})}\BibitemShut {NoStop}%
\bibitem [{\citenamefont {{Takami}}\ \emph {et~al.}(2018)\citenamefont
  {{Takami}}, \citenamefont {{Fu}},\ and\ \citenamefont {{Liu et
  al.}}}]{Takami+2018}%
  \BibitemOpen
  \bibfield  {author} {\bibinfo {author} {\bibfnamefont {M.}~\bibnamefont
  {{Takami}}}, \bibinfo {author} {\bibfnamefont {G.}~\bibnamefont {{Fu}}},\
  and\ \bibinfo {author} {\bibfnamefont {H.~B.}\ \bibnamefont {{Liu et al.}}},\
  }\href {https://doi.org/10.3847/1538-4357/aad2e1} {\bibfield  {journal}
  {\bibinfo  {journal} {\apj}\ }\textbf {\bibinfo {volume} {864}},\ \bibinfo
  {eid} {20} (\bibinfo {year} {2018})},\ \Eprint
  {https://arxiv.org/abs/1807.03499} {arXiv:1807.03499 [astro-ph.SR]}
  \BibitemShut {NoStop}%
\bibitem [{\citenamefont {{van der Plas}}\ \emph {et~al.}(2019)\citenamefont
  {{van der Plas}}, \citenamefont {{M{\'e}nard}}, \citenamefont {{Gonzalez}},
  \citenamefont {{Perez}}, \citenamefont {{Rodet}}, \citenamefont {{Pinte}},
  \citenamefont {{Cieza}}, \citenamefont {{Casassus}},\ and\ \citenamefont
  {{Benisty}}}]{vanderPlas+2019}%
  \BibitemOpen
  \bibfield  {author} {\bibinfo {author} {\bibfnamefont {G.}~\bibnamefont {{van
  der Plas}}}, \bibinfo {author} {\bibfnamefont {F.}~\bibnamefont
  {{M{\'e}nard}}}, \bibinfo {author} {\bibfnamefont {J.~F.}\ \bibnamefont
  {{Gonzalez}}}, \bibinfo {author} {\bibfnamefont {S.}~\bibnamefont {{Perez}}},
  \bibinfo {author} {\bibfnamefont {L.}~\bibnamefont {{Rodet}}}, \bibinfo
  {author} {\bibfnamefont {C.}~\bibnamefont {{Pinte}}}, \bibinfo {author}
  {\bibfnamefont {L.}~\bibnamefont {{Cieza}}}, \bibinfo {author} {\bibfnamefont
  {S.}~\bibnamefont {{Casassus}}},\ and\ \bibinfo {author} {\bibfnamefont
  {M.}~\bibnamefont {{Benisty}}},\ }\href
  {https://doi.org/10.1051/0004-6361/201834134} {\bibfield  {journal} {\bibinfo
   {journal} {\aap}\ }\textbf {\bibinfo {volume} {624}},\ \bibinfo {eid} {A33}
  (\bibinfo {year} {2019})},\ \Eprint {https://arxiv.org/abs/1902.00720}
  {arXiv:1902.00720 [astro-ph.SR]} \BibitemShut {NoStop}%
\bibitem [{\citenamefont {{Rosotti}}\ \emph {et~al.}(2020)\citenamefont
  {{Rosotti}}, \citenamefont {{Benisty}}, \citenamefont {{Juh{\'a}sz}},
  \citenamefont {{Teague}}, \citenamefont {{Clarke}}, \citenamefont
  {{Dominik}}, \citenamefont {{Dullemond}}, \citenamefont {{Klaassen}},
  \citenamefont {{Matr{\`a}}},\ and\ \citenamefont {{Stolker}}}]{Rosotti+2020}%
  \BibitemOpen
  \bibfield  {author} {\bibinfo {author} {\bibfnamefont {G.~P.}\ \bibnamefont
  {{Rosotti}}}, \bibinfo {author} {\bibfnamefont {M.}~\bibnamefont
  {{Benisty}}}, \bibinfo {author} {\bibfnamefont {A.}~\bibnamefont
  {{Juh{\'a}sz}}}, \bibinfo {author} {\bibfnamefont {R.}~\bibnamefont
  {{Teague}}}, \bibinfo {author} {\bibfnamefont {C.}~\bibnamefont {{Clarke}}},
  \bibinfo {author} {\bibfnamefont {C.}~\bibnamefont {{Dominik}}}, \bibinfo
  {author} {\bibfnamefont {C.~P.}\ \bibnamefont {{Dullemond}}}, \bibinfo
  {author} {\bibfnamefont {P.~D.}\ \bibnamefont {{Klaassen}}}, \bibinfo
  {author} {\bibfnamefont {L.}~\bibnamefont {{Matr{\`a}}}},\ and\ \bibinfo
  {author} {\bibfnamefont {T.}~\bibnamefont {{Stolker}}},\ }\href
  {https://doi.org/10.1093/mnras/stz3090} {\bibfield  {journal} {\bibinfo
  {journal} {\mnras}\ }\textbf {\bibinfo {volume} {491}},\ \bibinfo {pages}
  {1335} (\bibinfo {year} {2020})},\ \Eprint {https://arxiv.org/abs/1911.00518}
  {arXiv:1911.00518 [astro-ph.EP]} \BibitemShut {NoStop}%
\bibitem [{\citenamefont {{Benisty}}\ \emph {et~al.}(2018)\citenamefont
  {{Benisty}}, \citenamefont {{Juh{\'a}sz}},\ and\ \citenamefont {{Facchini et
  al.}}}]{Benisty+2018}%
  \BibitemOpen
  \bibfield  {author} {\bibinfo {author} {\bibfnamefont {M.}~\bibnamefont
  {{Benisty}}}, \bibinfo {author} {\bibfnamefont {A.}~\bibnamefont
  {{Juh{\'a}sz}}},\ and\ \bibinfo {author} {\bibfnamefont {S.}~\bibnamefont
  {{Facchini et al.}}},\ }\href {https://doi.org/10.1051/0004-6361/201833913}
  {\bibfield  {journal} {\bibinfo  {journal} {\aap}\ }\textbf {\bibinfo
  {volume} {619}},\ \bibinfo {eid} {A171} (\bibinfo {year} {2018})},\ \Eprint
  {https://arxiv.org/abs/1809.01082} {arXiv:1809.01082 [astro-ph.EP]}
  \BibitemShut {NoStop}%
\bibitem [{\citenamefont {{Monnier}}\ \emph {et~al.}(2019)\citenamefont
  {{Monnier}}, \citenamefont {{Harries}},\ and\ \citenamefont {{Bae et
  al.}}}]{Monnier+2019}%
  \BibitemOpen
  \bibfield  {author} {\bibinfo {author} {\bibfnamefont {J.~D.}\ \bibnamefont
  {{Monnier}}}, \bibinfo {author} {\bibfnamefont {T.~J.}\ \bibnamefont
  {{Harries}}},\ and\ \bibinfo {author} {\bibfnamefont {J.}~\bibnamefont {{Bae
  et al.}}},\ }\href {https://doi.org/10.3847/1538-4357/aafe87} {\bibfield
  {journal} {\bibinfo  {journal} {\apj}\ }\textbf {\bibinfo {volume} {872}},\
  \bibinfo {eid} {122} (\bibinfo {year} {2019})},\ \Eprint
  {https://arxiv.org/abs/1901.02467} {arXiv:1901.02467 [astro-ph.EP]}
  \BibitemShut {NoStop}%
\bibitem [{\citenamefont {{Cruz-S{\'a}enz de Miera}}\ \emph
  {et~al.}(2019)\citenamefont {{Cruz-S{\'a}enz de Miera}}, \citenamefont
  {{K{\'o}sp{\'a}l}}, \citenamefont {{{\'A}brah{\'a}m}}, \citenamefont
  {{Liu}},\ and\ \citenamefont {{Takami}}}]{CruzSaenz+2019}%
  \BibitemOpen
  \bibfield  {author} {\bibinfo {author} {\bibfnamefont {F.}~\bibnamefont
  {{Cruz-S{\'a}enz de Miera}}}, \bibinfo {author} {\bibfnamefont
  {{\'A}.}~\bibnamefont {{K{\'o}sp{\'a}l}}}, \bibinfo {author} {\bibfnamefont
  {P.}~\bibnamefont {{{\'A}brah{\'a}m}}}, \bibinfo {author} {\bibfnamefont
  {H.~B.}\ \bibnamefont {{Liu}}},\ and\ \bibinfo {author} {\bibfnamefont
  {M.}~\bibnamefont {{Takami}}},\ }\href
  {https://doi.org/10.3847/2041-8213/ab39ea} {\bibfield  {journal} {\bibinfo
  {journal} {\apjl}\ }\textbf {\bibinfo {volume} {882}},\ \bibinfo {eid} {L4}
  (\bibinfo {year} {2019})},\ \Eprint {https://arxiv.org/abs/1908.04649}
  {arXiv:1908.04649 [astro-ph.SR]} \BibitemShut {NoStop}%
\bibitem [{\citenamefont {{Keppler}}\ \emph {et~al.}(2020)\citenamefont
  {{Keppler}}, \citenamefont {{Penzlin}}, \citenamefont {{Benisty}},
  \citenamefont {{van Boekel}}, \citenamefont {{Henning}}, \citenamefont {{van
  Holstein}}, \citenamefont {{Kley}}, \citenamefont {{Garufi}},\ and\
  \citenamefont {{Ginski et al.}}}]{Keppler+2020}%
  \BibitemOpen
  \bibfield  {author} {\bibinfo {author} {\bibfnamefont {M.}~\bibnamefont
  {{Keppler}}}, \bibinfo {author} {\bibfnamefont {A.}~\bibnamefont
  {{Penzlin}}}, \bibinfo {author} {\bibfnamefont {M.}~\bibnamefont
  {{Benisty}}}, \bibinfo {author} {\bibfnamefont {R.}~\bibnamefont {{van
  Boekel}}}, \bibinfo {author} {\bibfnamefont {T.}~\bibnamefont {{Henning}}},
  \bibinfo {author} {\bibfnamefont {R.~G.}\ \bibnamefont {{van Holstein}}},
  \bibinfo {author} {\bibfnamefont {W.}~\bibnamefont {{Kley}}}, \bibinfo
  {author} {\bibfnamefont {A.}~\bibnamefont {{Garufi}}},\ and\ \bibinfo
  {author} {\bibnamefont {{Ginski et al.}}},\ }\href
  {https://doi.org/10.1051/0004-6361/202038032} {\bibfield  {journal} {\bibinfo
   {journal} {\aap}\ }\textbf {\bibinfo {volume} {639}},\ \bibinfo {eid} {A62}
  (\bibinfo {year} {2020})},\ \Eprint {https://arxiv.org/abs/2005.09037}
  {arXiv:2005.09037 [astro-ph.SR]} \BibitemShut {NoStop}%
\bibitem [{\citenamefont {{Kraus}}\ \emph {et~al.}(2020)\citenamefont
  {{Kraus}}, \citenamefont {{Kreplin}}, \citenamefont {{Young}}, \citenamefont
  {{Bate}}, \citenamefont {{Monnier}}, \citenamefont {{Harries}}, \citenamefont
  {{Avenhaus}}, \citenamefont {{Kluska}}, \citenamefont {{Laws}}, \citenamefont
  {{Rich}},\ and\ \citenamefont {{Willson et al.}}}]{Kraus+2020}%
  \BibitemOpen
  \bibfield  {author} {\bibinfo {author} {\bibfnamefont {S.}~\bibnamefont
  {{Kraus}}}, \bibinfo {author} {\bibfnamefont {A.}~\bibnamefont {{Kreplin}}},
  \bibinfo {author} {\bibfnamefont {A.~K.}\ \bibnamefont {{Young}}}, \bibinfo
  {author} {\bibfnamefont {M.~R.}\ \bibnamefont {{Bate}}}, \bibinfo {author}
  {\bibfnamefont {J.~D.}\ \bibnamefont {{Monnier}}}, \bibinfo {author}
  {\bibfnamefont {T.~J.}\ \bibnamefont {{Harries}}}, \bibinfo {author}
  {\bibfnamefont {H.}~\bibnamefont {{Avenhaus}}}, \bibinfo {author}
  {\bibfnamefont {J.}~\bibnamefont {{Kluska}}}, \bibinfo {author}
  {\bibfnamefont {A.~S.~E.}\ \bibnamefont {{Laws}}}, \bibinfo {author}
  {\bibfnamefont {E.~A.}\ \bibnamefont {{Rich}}},\ and\ \bibinfo {author}
  {\bibfnamefont {M.}~\bibnamefont {{Willson et al.}}},\ }\href
  {https://doi.org/10.1126/science.aba4633} {\bibfield  {journal} {\bibinfo
  {journal} {Science}\ }\textbf {\bibinfo {volume} {369}},\ \bibinfo {pages}
  {1233} (\bibinfo {year} {2020})},\ \Eprint {https://arxiv.org/abs/2004.01204}
  {arXiv:2004.01204 [astro-ph.SR]} \BibitemShut {NoStop}%
\bibitem [{\citenamefont {{Maureira}}\ \emph {et~al.}(2020)\citenamefont
  {{Maureira}}, \citenamefont {{Pineda}}, \citenamefont {{Segura-Cox}},
  \citenamefont {{Caselli}}, \citenamefont {{Testi}}, \citenamefont {{Lodato}},
  \citenamefont {{Loinard}},\ and\ \citenamefont
  {{Hern{\'a}ndez-G{\'o}mez}}}]{Maureira+2020}%
  \BibitemOpen
  \bibfield  {author} {\bibinfo {author} {\bibfnamefont {M.~J.}\ \bibnamefont
  {{Maureira}}}, \bibinfo {author} {\bibfnamefont {J.~E.}\ \bibnamefont
  {{Pineda}}}, \bibinfo {author} {\bibfnamefont {D.~M.}\ \bibnamefont
  {{Segura-Cox}}}, \bibinfo {author} {\bibfnamefont {P.}~\bibnamefont
  {{Caselli}}}, \bibinfo {author} {\bibfnamefont {L.}~\bibnamefont {{Testi}}},
  \bibinfo {author} {\bibfnamefont {G.}~\bibnamefont {{Lodato}}}, \bibinfo
  {author} {\bibfnamefont {L.}~\bibnamefont {{Loinard}}},\ and\ \bibinfo
  {author} {\bibfnamefont {A.}~\bibnamefont {{Hern{\'a}ndez-G{\'o}mez}}},\
  }\href {https://doi.org/10.3847/1538-4357/ab960b} {\bibfield  {journal}
  {\bibinfo  {journal} {\apj}\ }\textbf {\bibinfo {volume} {897}},\ \bibinfo
  {eid} {59} (\bibinfo {year} {2020})},\ \Eprint
  {https://arxiv.org/abs/2005.11954} {arXiv:2005.11954 [astro-ph.SR]}
  \BibitemShut {NoStop}%
\bibitem [{\citenamefont {{Muro-Arena}}\ \emph {et~al.}(2020)\citenamefont
  {{Muro-Arena}}, \citenamefont {{Benisty}}, \citenamefont {{Ginski}},
  \citenamefont {{Dominik}}, \citenamefont {{Facchini}}, \citenamefont
  {{Villenave}},\ and\ \citenamefont {{van Boekel et al.}}}]{Muro-Arena+2020}%
  \BibitemOpen
  \bibfield  {author} {\bibinfo {author} {\bibfnamefont {G.~A.}\ \bibnamefont
  {{Muro-Arena}}}, \bibinfo {author} {\bibfnamefont {M.}~\bibnamefont
  {{Benisty}}}, \bibinfo {author} {\bibfnamefont {C.}~\bibnamefont {{Ginski}}},
  \bibinfo {author} {\bibfnamefont {C.}~\bibnamefont {{Dominik}}}, \bibinfo
  {author} {\bibfnamefont {S.}~\bibnamefont {{Facchini}}}, \bibinfo {author}
  {\bibfnamefont {M.}~\bibnamefont {{Villenave}}},\ and\ \bibinfo {author}
  {\bibfnamefont {R.}~\bibnamefont {{van Boekel et al.}}},\ }\href
  {https://doi.org/10.1051/0004-6361/201936509} {\bibfield  {journal} {\bibinfo
   {journal} {\aap}\ }\textbf {\bibinfo {volume} {635}},\ \bibinfo {eid} {A121}
  (\bibinfo {year} {2020})},\ \Eprint {https://arxiv.org/abs/1911.09612}
  {arXiv:1911.09612 [astro-ph.EP]} \BibitemShut {NoStop}%
\bibitem [{\citenamefont {{Ragusa}}\ \emph {et~al.}(2021)\citenamefont
  {{Ragusa}}, \citenamefont {{Fasano}}, \citenamefont {{Toci}}, \citenamefont
  {{Duch{\^e}ne}}, \citenamefont {{Cuello}}, \citenamefont {{Villenave}},
  \citenamefont {{van der Plas}}, \citenamefont {{Lodato}}, \citenamefont
  {{M{\'e}nard}}, \citenamefont {{Price}}, \citenamefont {{Pinte}},
  \citenamefont {{Stapelfeldt}},\ and\ \citenamefont {{Wolff}}}]{Ragusa+2021}%
  \BibitemOpen
  \bibfield  {author} {\bibinfo {author} {\bibfnamefont {E.}~\bibnamefont
  {{Ragusa}}}, \bibinfo {author} {\bibfnamefont {D.}~\bibnamefont {{Fasano}}},
  \bibinfo {author} {\bibfnamefont {C.}~\bibnamefont {{Toci}}}, \bibinfo
  {author} {\bibfnamefont {G.}~\bibnamefont {{Duch{\^e}ne}}}, \bibinfo {author}
  {\bibfnamefont {N.}~\bibnamefont {{Cuello}}}, \bibinfo {author}
  {\bibfnamefont {M.}~\bibnamefont {{Villenave}}}, \bibinfo {author}
  {\bibfnamefont {G.}~\bibnamefont {{van der Plas}}}, \bibinfo {author}
  {\bibfnamefont {G.}~\bibnamefont {{Lodato}}}, \bibinfo {author}
  {\bibfnamefont {F.}~\bibnamefont {{M{\'e}nard}}}, \bibinfo {author}
  {\bibfnamefont {D.~J.}\ \bibnamefont {{Price}}}, \bibinfo {author}
  {\bibfnamefont {C.}~\bibnamefont {{Pinte}}}, \bibinfo {author} {\bibfnamefont
  {K.}~\bibnamefont {{Stapelfeldt}}},\ and\ \bibinfo {author} {\bibfnamefont
  {S.}~\bibnamefont {{Wolff}}},\ }\href
  {https://doi.org/10.1093/mnras/stab2179} {\bibfield  {journal} {\bibinfo
  {journal} {\mnras}\ }\textbf {\bibinfo {volume} {507}},\ \bibinfo {pages}
  {1157} (\bibinfo {year} {2021})},\ \Eprint {https://arxiv.org/abs/2107.13566}
  {arXiv:2107.13566 [astro-ph.SR]} \BibitemShut {NoStop}%
\bibitem [{\citenamefont {{Czekala}}\ \emph {et~al.}(2021)\citenamefont
  {{Czekala}}, \citenamefont {{Ribas}}, \citenamefont {{Cuello}}, \citenamefont
  {{Chiang}}, \citenamefont {{Mac{\'\i}as}}, \citenamefont {{Duch{\^e}ne}},
  \citenamefont {{Andrews}},\ and\ \citenamefont {{Espaillat}}}]{Czekala+2021}%
  \BibitemOpen
  \bibfield  {author} {\bibinfo {author} {\bibfnamefont {I.}~\bibnamefont
  {{Czekala}}}, \bibinfo {author} {\bibfnamefont {{\'A}.}~\bibnamefont
  {{Ribas}}}, \bibinfo {author} {\bibfnamefont {N.}~\bibnamefont {{Cuello}}},
  \bibinfo {author} {\bibfnamefont {E.}~\bibnamefont {{Chiang}}}, \bibinfo
  {author} {\bibfnamefont {E.}~\bibnamefont {{Mac{\'\i}as}}}, \bibinfo {author}
  {\bibfnamefont {G.}~\bibnamefont {{Duch{\^e}ne}}}, \bibinfo {author}
  {\bibfnamefont {S.~M.}\ \bibnamefont {{Andrews}}},\ and\ \bibinfo {author}
  {\bibfnamefont {C.~C.}\ \bibnamefont {{Espaillat}}},\ }\href
  {https://doi.org/10.3847/1538-4357/abebe3} {\bibfield  {journal} {\bibinfo
  {journal} {\apj}\ }\textbf {\bibinfo {volume} {912}},\ \bibinfo {eid} {6}
  (\bibinfo {year} {2021})},\ \Eprint {https://arxiv.org/abs/2102.11875}
  {arXiv:2102.11875 [astro-ph.EP]} \BibitemShut {NoStop}%
\bibitem [{\citenamefont {{Long}}\ \emph {et~al.}(2021)\citenamefont {{Long}},
  \citenamefont {{Andrews}},\ and\ \citenamefont {{Vega et al.}}}]{Long+2021}%
  \BibitemOpen
  \bibfield  {author} {\bibinfo {author} {\bibfnamefont {F.}~\bibnamefont
  {{Long}}}, \bibinfo {author} {\bibfnamefont {S.~M.}\ \bibnamefont
  {{Andrews}}},\ and\ \bibinfo {author} {\bibfnamefont {J.}~\bibnamefont {{Vega
  et al.}}},\ }\href {https://doi.org/10.3847/1538-4357/abff53} {\bibfield
  {journal} {\bibinfo  {journal} {\apj}\ }\textbf {\bibinfo {volume} {915}},\
  \bibinfo {eid} {131} (\bibinfo {year} {2021})},\ \Eprint
  {https://arxiv.org/abs/2105.02918} {arXiv:2105.02918 [astro-ph.EP]}
  \BibitemShut {NoStop}%
\bibitem [{\citenamefont {{Doolin}}\ and\ \citenamefont
  {{Blundell}}(2011)}]{Doolin+2011}%
  \BibitemOpen
  \bibfield  {author} {\bibinfo {author} {\bibfnamefont {S.}~\bibnamefont
  {{Doolin}}}\ and\ \bibinfo {author} {\bibfnamefont {K.~M.}\ \bibnamefont
  {{Blundell}}},\ }\href {https://doi.org/10.1111/j.1365-2966.2011.19657.x}
  {\bibfield  {journal} {\bibinfo  {journal} {\mnras}\ }\textbf {\bibinfo
  {volume} {418}},\ \bibinfo {pages} {2656} (\bibinfo {year} {2011})},\ \Eprint
  {https://arxiv.org/abs/1108.4144} {arXiv:1108.4144 [astro-ph.SR]}
  \BibitemShut {NoStop}%
\bibitem [{\citenamefont {{Owen}}\ and\ \citenamefont
  {{Lai}}(2017)}]{OwenLai2017}%
  \BibitemOpen
  \bibfield  {author} {\bibinfo {author} {\bibfnamefont {J.~E.}\ \bibnamefont
  {{Owen}}}\ and\ \bibinfo {author} {\bibfnamefont {D.}~\bibnamefont {{Lai}}},\
  }\href {https://doi.org/10.1093/mnras/stx1033} {\bibfield  {journal}
  {\bibinfo  {journal} {\mnras}\ }\textbf {\bibinfo {volume} {469}},\ \bibinfo
  {pages} {2834} (\bibinfo {year} {2017})},\ \Eprint
  {https://arxiv.org/abs/1703.09250} {arXiv:1703.09250 [astro-ph.SR]}
  \BibitemShut {NoStop}%
\bibitem [{\citenamefont {{Martin}}\ and\ \citenamefont
  {{Lubow}}(2017)}]{MartinLubow2017}%
  \BibitemOpen
  \bibfield  {author} {\bibinfo {author} {\bibfnamefont {R.~G.}\ \bibnamefont
  {{Martin}}}\ and\ \bibinfo {author} {\bibfnamefont {S.~H.}\ \bibnamefont
  {{Lubow}}},\ }\href {https://doi.org/10.3847/2041-8213/835/2/L28} {\bibfield
  {journal} {\bibinfo  {journal} {\apjl}\ }\textbf {\bibinfo {volume} {835}},\
  \bibinfo {eid} {L28} (\bibinfo {year} {2017})},\ \Eprint
  {https://arxiv.org/abs/1702.00545} {arXiv:1702.00545 [astro-ph.EP]}
  \BibitemShut {NoStop}%
\bibitem [{\citenamefont {{Price}}\ \emph {et~al.}(2018)\citenamefont
  {{Price}}, \citenamefont {{Cuello}},\ and\ \citenamefont {{Pinte et
  al.}}}]{Price+2018}%
  \BibitemOpen
  \bibfield  {author} {\bibinfo {author} {\bibfnamefont {D.~J.}\ \bibnamefont
  {{Price}}}, \bibinfo {author} {\bibfnamefont {N.}~\bibnamefont {{Cuello}}},\
  and\ \bibinfo {author} {\bibfnamefont {C.}~\bibnamefont {{Pinte et al.}}},\
  }\href {https://doi.org/10.1093/mnras/sty647} {\bibfield  {journal} {\bibinfo
   {journal} {\mnras}\ }\textbf {\bibinfo {volume} {477}},\ \bibinfo {pages}
  {1270} (\bibinfo {year} {2018})},\ \Eprint {https://arxiv.org/abs/1803.02484}
  {arXiv:1803.02484 [astro-ph.SR]} \BibitemShut {NoStop}%
\bibitem [{\citenamefont {{Cuello}}\ and\ \citenamefont
  {{Giuppone}}(2019)}]{CuelloGiuppone2019}%
  \BibitemOpen
  \bibfield  {author} {\bibinfo {author} {\bibfnamefont {N.}~\bibnamefont
  {{Cuello}}}\ and\ \bibinfo {author} {\bibfnamefont {C.~A.}\ \bibnamefont
  {{Giuppone}}},\ }\href {https://doi.org/10.1051/0004-6361/201833976}
  {\bibfield  {journal} {\bibinfo  {journal} {\aap}\ }\textbf {\bibinfo
  {volume} {628}},\ \bibinfo {eid} {A119} (\bibinfo {year} {2019})},\ \Eprint
  {https://arxiv.org/abs/1906.10579} {arXiv:1906.10579 [astro-ph.EP]}
  \BibitemShut {NoStop}%
\bibitem [{\citenamefont {{Chen}}\ \emph {et~al.}(2019)\citenamefont {{Chen}},
  \citenamefont {{Franchini}}, \citenamefont {{Lubow}},\ and\ \citenamefont
  {{Martin}}}]{Chen+2019}%
  \BibitemOpen
  \bibfield  {author} {\bibinfo {author} {\bibfnamefont {C.}~\bibnamefont
  {{Chen}}}, \bibinfo {author} {\bibfnamefont {A.}~\bibnamefont {{Franchini}}},
  \bibinfo {author} {\bibfnamefont {S.~H.}\ \bibnamefont {{Lubow}}},\ and\
  \bibinfo {author} {\bibfnamefont {R.~G.}\ \bibnamefont {{Martin}}},\ }\href
  {https://doi.org/10.1093/mnras/stz2948} {\bibfield  {journal} {\bibinfo
  {journal} {\mnras}\ }\textbf {\bibinfo {volume} {490}},\ \bibinfo {pages}
  {5634} (\bibinfo {year} {2019})},\ \Eprint {https://arxiv.org/abs/1908.06331}
  {arXiv:1908.06331 [astro-ph.EP]} \BibitemShut {NoStop}%
\bibitem [{\citenamefont {{Zhu}}(2019)}]{Zhu2019}%
  \BibitemOpen
  \bibfield  {author} {\bibinfo {author} {\bibfnamefont {Z.}~\bibnamefont
  {{Zhu}}},\ }\href {https://doi.org/10.1093/mnras/sty3358} {\bibfield
  {journal} {\bibinfo  {journal} {\mnras}\ }\textbf {\bibinfo {volume} {483}},\
  \bibinfo {pages} {4221} (\bibinfo {year} {2019})},\ \Eprint
  {https://arxiv.org/abs/1812.01262} {arXiv:1812.01262 [astro-ph.EP]}
  \BibitemShut {NoStop}%
\bibitem [{\citenamefont {{Poblete}}\ \emph {et~al.}(2019)\citenamefont
  {{Poblete}}, \citenamefont {{Cuello}},\ and\ \citenamefont
  {{Cuadra}}}]{Poblete+2019}%
  \BibitemOpen
  \bibfield  {author} {\bibinfo {author} {\bibfnamefont {P.~P.}\ \bibnamefont
  {{Poblete}}}, \bibinfo {author} {\bibfnamefont {N.}~\bibnamefont
  {{Cuello}}},\ and\ \bibinfo {author} {\bibfnamefont {J.}~\bibnamefont
  {{Cuadra}}},\ }\href {https://doi.org/10.1093/mnras/stz2297} {\bibfield
  {journal} {\bibinfo  {journal} {\mnras}\ }\textbf {\bibinfo {volume} {489}},\
  \bibinfo {pages} {2204} (\bibinfo {year} {2019})},\ \Eprint
  {https://arxiv.org/abs/1908.05784} {arXiv:1908.05784 [astro-ph.EP]}
  \BibitemShut {NoStop}%
\bibitem [{\citenamefont {{Calcino}}\ \emph {et~al.}(2019)\citenamefont
  {{Calcino}}, \citenamefont {{Price}}, \citenamefont {{Pinte}}, \citenamefont
  {{van der Marel}}, \citenamefont {{Ragusa}}, \citenamefont {{Dipierro}},
  \citenamefont {{Cuello}},\ and\ \citenamefont
  {{Christiaens}}}]{Calcino+2019}%
  \BibitemOpen
  \bibfield  {author} {\bibinfo {author} {\bibfnamefont {J.}~\bibnamefont
  {{Calcino}}}, \bibinfo {author} {\bibfnamefont {D.~J.}\ \bibnamefont
  {{Price}}}, \bibinfo {author} {\bibfnamefont {C.}~\bibnamefont {{Pinte}}},
  \bibinfo {author} {\bibfnamefont {N.}~\bibnamefont {{van der Marel}}},
  \bibinfo {author} {\bibfnamefont {E.}~\bibnamefont {{Ragusa}}}, \bibinfo
  {author} {\bibfnamefont {G.}~\bibnamefont {{Dipierro}}}, \bibinfo {author}
  {\bibfnamefont {N.}~\bibnamefont {{Cuello}}},\ and\ \bibinfo {author}
  {\bibfnamefont {V.}~\bibnamefont {{Christiaens}}},\ }\href
  {https://doi.org/10.1093/mnras/stz2770} {\bibfield  {journal} {\bibinfo
  {journal} {\mnras}\ }\textbf {\bibinfo {volume} {490}},\ \bibinfo {pages}
  {2579} (\bibinfo {year} {2019})},\ \Eprint {https://arxiv.org/abs/1910.00161}
  {arXiv:1910.00161 [astro-ph.EP]} \BibitemShut {NoStop}%
\bibitem [{\citenamefont {{Calcino}}\ \emph {et~al.}(2020)\citenamefont
  {{Calcino}}, \citenamefont {{Christiaens}}, \citenamefont {{Price}},
  \citenamefont {{Pinte}}, \citenamefont {{Davis}}, \citenamefont {{van der
  Marel}},\ and\ \citenamefont {{Cuello}}}]{Calcino+2020}%
  \BibitemOpen
  \bibfield  {author} {\bibinfo {author} {\bibfnamefont {J.}~\bibnamefont
  {{Calcino}}}, \bibinfo {author} {\bibfnamefont {V.}~\bibnamefont
  {{Christiaens}}}, \bibinfo {author} {\bibfnamefont {D.~J.}\ \bibnamefont
  {{Price}}}, \bibinfo {author} {\bibfnamefont {C.}~\bibnamefont {{Pinte}}},
  \bibinfo {author} {\bibfnamefont {T.~M.}\ \bibnamefont {{Davis}}}, \bibinfo
  {author} {\bibfnamefont {N.}~\bibnamefont {{van der Marel}}},\ and\ \bibinfo
  {author} {\bibfnamefont {N.}~\bibnamefont {{Cuello}}},\ }\href
  {https://doi.org/10.1093/mnras/staa2468} {\bibfield  {journal} {\bibinfo
  {journal} {\mnras}\ }\textbf {\bibinfo {volume} {498}},\ \bibinfo {pages}
  {639} (\bibinfo {year} {2020})},\ \Eprint {https://arxiv.org/abs/2007.06155}
  {arXiv:2007.06155 [astro-ph.EP]} \BibitemShut {NoStop}%
\bibitem [{\citenamefont {{Gonzalez}}\ \emph {et~al.}(2020)\citenamefont
  {{Gonzalez}}, \citenamefont {{van der Plas}}, \citenamefont {{Pinte}},
  \citenamefont {{Cuello}}, \citenamefont {{Nealon}}, \citenamefont
  {{M{\'e}nard}}, \citenamefont {{Revol}}, \citenamefont {{Rodet}},
  \citenamefont {{Langlois}},\ and\ \citenamefont {{Maire}}}]{Gonzalez+2020}%
  \BibitemOpen
  \bibfield  {author} {\bibinfo {author} {\bibfnamefont {J.-F.}\ \bibnamefont
  {{Gonzalez}}}, \bibinfo {author} {\bibfnamefont {G.}~\bibnamefont {{van der
  Plas}}}, \bibinfo {author} {\bibfnamefont {C.}~\bibnamefont {{Pinte}}},
  \bibinfo {author} {\bibfnamefont {N.}~\bibnamefont {{Cuello}}}, \bibinfo
  {author} {\bibfnamefont {R.}~\bibnamefont {{Nealon}}}, \bibinfo {author}
  {\bibfnamefont {F.}~\bibnamefont {{M{\'e}nard}}}, \bibinfo {author}
  {\bibfnamefont {A.}~\bibnamefont {{Revol}}}, \bibinfo {author} {\bibfnamefont
  {L.}~\bibnamefont {{Rodet}}}, \bibinfo {author} {\bibfnamefont
  {M.}~\bibnamefont {{Langlois}}},\ and\ \bibinfo {author} {\bibfnamefont
  {A.-L.}\ \bibnamefont {{Maire}}},\ }\href
  {https://doi.org/10.1093/mnras/staa2938} {\bibfield  {journal} {\bibinfo
  {journal} {\mnras}\ }\textbf {\bibinfo {volume} {499}},\ \bibinfo {pages}
  {3837} (\bibinfo {year} {2020})},\ \Eprint {https://arxiv.org/abs/2009.10504}
  {arXiv:2009.10504 [astro-ph.EP]} \BibitemShut {NoStop}%
\bibitem [{\citenamefont {{Nealon}}\ \emph
  {et~al.}(2020{\natexlab{b}})\citenamefont {{Nealon}}, \citenamefont
  {{Cuello}}, \citenamefont {{Gonzalez}}, \citenamefont {{van der Plas}},
  \citenamefont {{Pinte}}, \citenamefont {{Alexander}}, \citenamefont
  {{M{\'e}nard}},\ and\ \citenamefont {{Price}}}]{Nealon+2020b}%
  \BibitemOpen
  \bibfield  {author} {\bibinfo {author} {\bibfnamefont {R.}~\bibnamefont
  {{Nealon}}}, \bibinfo {author} {\bibfnamefont {N.}~\bibnamefont {{Cuello}}},
  \bibinfo {author} {\bibfnamefont {J.-F.}\ \bibnamefont {{Gonzalez}}},
  \bibinfo {author} {\bibfnamefont {G.}~\bibnamefont {{van der Plas}}},
  \bibinfo {author} {\bibfnamefont {C.}~\bibnamefont {{Pinte}}}, \bibinfo
  {author} {\bibfnamefont {R.}~\bibnamefont {{Alexander}}}, \bibinfo {author}
  {\bibfnamefont {F.}~\bibnamefont {{M{\'e}nard}}},\ and\ \bibinfo {author}
  {\bibfnamefont {D.~J.}\ \bibnamefont {{Price}}},\ }\href
  {https://doi.org/10.1093/mnras/staa2721} {\bibfield  {journal} {\bibinfo
  {journal} {\mnras}\ }\textbf {\bibinfo {volume} {499}},\ \bibinfo {pages}
  {3857} (\bibinfo {year} {2020}{\natexlab{b}})},\ \Eprint
  {https://arxiv.org/abs/2009.10505} {arXiv:2009.10505 [astro-ph.EP]}
  \BibitemShut {NoStop}%
\bibitem [{\citenamefont {{Ballabio}}\ \emph {et~al.}(2021)\citenamefont
  {{Ballabio}}, \citenamefont {{Nealon}}, \citenamefont {{Alexander}},
  \citenamefont {{Cuello}}, \citenamefont {{Pinte}},\ and\ \citenamefont
  {{Price}}}]{Ballabio+2021}%
  \BibitemOpen
  \bibfield  {author} {\bibinfo {author} {\bibfnamefont {G.}~\bibnamefont
  {{Ballabio}}}, \bibinfo {author} {\bibfnamefont {R.}~\bibnamefont
  {{Nealon}}}, \bibinfo {author} {\bibfnamefont {R.~D.}\ \bibnamefont
  {{Alexander}}}, \bibinfo {author} {\bibfnamefont {N.}~\bibnamefont
  {{Cuello}}}, \bibinfo {author} {\bibfnamefont {C.}~\bibnamefont {{Pinte}}},\
  and\ \bibinfo {author} {\bibfnamefont {D.~J.}\ \bibnamefont {{Price}}},\
  }\href {https://doi.org/10.1093/mnras/stab922} {\bibfield  {journal}
  {\bibinfo  {journal} {\mnras}\ }\textbf {\bibinfo {volume} {504}},\ \bibinfo
  {pages} {888} (\bibinfo {year} {2021})},\ \Eprint
  {https://arxiv.org/abs/2103.16213} {arXiv:2103.16213 [astro-ph.EP]}
  \BibitemShut {NoStop}%
\bibitem [{\citenamefont {{Smallwood}}\ \emph
  {et~al.}(2021{\natexlab{a}})\citenamefont {{Smallwood}}, \citenamefont
  {{Martin}},\ and\ \citenamefont {{Lubow}}}]{Smallwood+2021}%
  \BibitemOpen
  \bibfield  {author} {\bibinfo {author} {\bibfnamefont {J.~L.}\ \bibnamefont
  {{Smallwood}}}, \bibinfo {author} {\bibfnamefont {R.~G.}\ \bibnamefont
  {{Martin}}},\ and\ \bibinfo {author} {\bibfnamefont {S.~H.}\ \bibnamefont
  {{Lubow}}},\ }\href {https://doi.org/10.3847/2041-8213/abd4d6} {\bibfield
  {journal} {\bibinfo  {journal} {\apjl}\ }\textbf {\bibinfo {volume} {907}},\
  \bibinfo {eid} {L14} (\bibinfo {year} {2021}{\natexlab{a}})},\ \Eprint
  {https://arxiv.org/abs/2012.11068} {arXiv:2012.11068 [astro-ph.EP]}
  \BibitemShut {NoStop}%
\bibitem [{\citenamefont {{Smallwood}}\ \emph
  {et~al.}(2021{\natexlab{b}})\citenamefont {{Smallwood}}, \citenamefont
  {{Nealon}}, \citenamefont {{Chen}}, \citenamefont {{Martin}}, \citenamefont
  {{Bi}}, \citenamefont {{Dong}},\ and\ \citenamefont
  {{Pinte}}}]{Smallwood+2021b}%
  \BibitemOpen
  \bibfield  {author} {\bibinfo {author} {\bibfnamefont {J.~L.}\ \bibnamefont
  {{Smallwood}}}, \bibinfo {author} {\bibfnamefont {R.}~\bibnamefont
  {{Nealon}}}, \bibinfo {author} {\bibfnamefont {C.}~\bibnamefont {{Chen}}},
  \bibinfo {author} {\bibfnamefont {R.~G.}\ \bibnamefont {{Martin}}}, \bibinfo
  {author} {\bibfnamefont {J.}~\bibnamefont {{Bi}}}, \bibinfo {author}
  {\bibfnamefont {R.}~\bibnamefont {{Dong}}},\ and\ \bibinfo {author}
  {\bibfnamefont {C.}~\bibnamefont {{Pinte}}},\ }\href
  {https://doi.org/10.1093/mnras/stab2624} {\bibfield  {journal} {\bibinfo
  {journal} {\mnras}\ }\textbf {\bibinfo {volume} {508}},\ \bibinfo {pages}
  {392} (\bibinfo {year} {2021}{\natexlab{b}})},\ \Eprint
  {https://arxiv.org/abs/2109.09776} {arXiv:2109.09776 [astro-ph.EP]}
  \BibitemShut {NoStop}%
\bibitem [{\citenamefont {{van Albada}}(1968)}]{vanAlbada1968}%
  \BibitemOpen
  \bibfield  {author} {\bibinfo {author} {\bibfnamefont {T.~S.}\ \bibnamefont
  {{van Albada}}},\ }\href@noop {} {\bibfield  {journal} {\bibinfo  {journal}
  {\bain}\ }\textbf {\bibinfo {volume} {20}},\ \bibinfo {pages} {47} (\bibinfo
  {year} {1968})}\BibitemShut {NoStop}%
\bibitem [{\citenamefont {{Bonavita}}\ \emph {et~al.}(2021)\citenamefont
  {{Bonavita}}, \citenamefont {{Gratton}},\ and\ \citenamefont {{Desidera et
  al.}}}]{Bonavita+2021}%
  \BibitemOpen
  \bibfield  {author} {\bibinfo {author} {\bibfnamefont {M.}~\bibnamefont
  {{Bonavita}}}, \bibinfo {author} {\bibfnamefont {R.}~\bibnamefont
  {{Gratton}}},\ and\ \bibinfo {author} {\bibfnamefont {S.}~\bibnamefont
  {{Desidera et al.}}},\ }\href@noop {} {\bibfield  {journal} {\bibinfo
  {journal} {arXiv e-prints}\ ,\ \bibinfo {eid} {arXiv:2103.13706}} (\bibinfo
  {year} {2021})},\ \Eprint {https://arxiv.org/abs/2103.13706}
  {arXiv:2103.13706 [astro-ph.SR]} \BibitemShut {NoStop}%
\bibitem [{\citenamefont {{Ginski}}\ \emph {et~al.}(2021)\citenamefont
  {{Ginski}}, \citenamefont {{Mugrauer}}, \citenamefont {{Adam}}, \citenamefont
  {{Vogt}},\ and\ \citenamefont {{van Holstein}}}]{Ginski+2021}%
  \BibitemOpen
  \bibfield  {author} {\bibinfo {author} {\bibfnamefont {C.}~\bibnamefont
  {{Ginski}}}, \bibinfo {author} {\bibfnamefont {M.}~\bibnamefont
  {{Mugrauer}}}, \bibinfo {author} {\bibfnamefont {C.}~\bibnamefont {{Adam}}},
  \bibinfo {author} {\bibfnamefont {N.}~\bibnamefont {{Vogt}}},\ and\ \bibinfo
  {author} {\bibfnamefont {R.~G.}\ \bibnamefont {{van Holstein}}},\ }\href
  {https://doi.org/10.1051/0004-6361/202038964} {\bibfield  {journal} {\bibinfo
   {journal} {\aap}\ }\textbf {\bibinfo {volume} {649}},\ \bibinfo {eid} {A156}
  (\bibinfo {year} {2021})},\ \Eprint {https://arxiv.org/abs/2009.10363}
  {arXiv:2009.10363 [astro-ph.EP]} \BibitemShut {NoStop}%
\bibitem [{\citenamefont {{Bonavita}}\ \emph {et~al.}(2022)\citenamefont
  {{Bonavita}}, \citenamefont {{Fontanive}},\ and\ \citenamefont {{Gratton et
  al.}}}]{Bonavita+2022}%
  \BibitemOpen
  \bibfield  {author} {\bibinfo {author} {\bibfnamefont {M.}~\bibnamefont
  {{Bonavita}}}, \bibinfo {author} {\bibfnamefont {C.}~\bibnamefont
  {{Fontanive}}},\ and\ \bibinfo {author} {\bibfnamefont {R.}~\bibnamefont
  {{Gratton et al.}}},\ }\href {https://doi.org/10.1093/mnras/stac1250}
  {\bibfield  {journal} {\bibinfo  {journal} {\mnras}\ }\textbf {\bibinfo
  {volume} {513}},\ \bibinfo {pages} {5588} (\bibinfo {year} {2022})},\ \Eprint
  {https://arxiv.org/abs/2205.02213} {arXiv:2205.02213 [astro-ph.SR]}
  \BibitemShut {NoStop}%
\bibitem [{\citenamefont {{Adams}}(2010)}]{Adams2010}%
  \BibitemOpen
  \bibfield  {author} {\bibinfo {author} {\bibfnamefont {F.~C.}\ \bibnamefont
  {{Adams}}},\ }\href {https://doi.org/10.1146/annurev-astro-081309-130830}
  {\bibfield  {journal} {\bibinfo  {journal} {\araa}\ }\textbf {\bibinfo
  {volume} {48}},\ \bibinfo {pages} {47} (\bibinfo {year} {2010})},\ \Eprint
  {https://arxiv.org/abs/1001.5444} {arXiv:1001.5444 [astro-ph.SR]}
  \BibitemShut {NoStop}%
\bibitem [{\citenamefont {{Scally}}\ and\ \citenamefont
  {{Clarke}}(2001)}]{Scally+2001}%
  \BibitemOpen
  \bibfield  {author} {\bibinfo {author} {\bibfnamefont {A.}~\bibnamefont
  {{Scally}}}\ and\ \bibinfo {author} {\bibfnamefont {C.}~\bibnamefont
  {{Clarke}}},\ }\href {https://doi.org/10.1046/j.1365-8711.2001.04274.x}
  {\bibfield  {journal} {\bibinfo  {journal} {\mnras}\ }\textbf {\bibinfo
  {volume} {325}},\ \bibinfo {pages} {449} (\bibinfo {year} {2001})},\ \Eprint
  {https://arxiv.org/abs/astro-ph/0012098} {arXiv:astro-ph/0012098 [astro-ph]}
  \BibitemShut {NoStop}%
\bibitem [{\citenamefont {{Olczak}}\ \emph {et~al.}(2006)\citenamefont
  {{Olczak}}, \citenamefont {{Pfalzner}},\ and\ \citenamefont
  {{Spurzem}}}]{Olczak+2006}%
  \BibitemOpen
  \bibfield  {author} {\bibinfo {author} {\bibfnamefont {C.}~\bibnamefont
  {{Olczak}}}, \bibinfo {author} {\bibfnamefont {S.}~\bibnamefont
  {{Pfalzner}}},\ and\ \bibinfo {author} {\bibfnamefont {R.}~\bibnamefont
  {{Spurzem}}},\ }\href {https://doi.org/10.1086/501044} {\bibfield  {journal}
  {\bibinfo  {journal} {\apj}\ }\textbf {\bibinfo {volume} {642}},\ \bibinfo
  {pages} {1140} (\bibinfo {year} {2006})},\ \Eprint
  {https://arxiv.org/abs/astro-ph/0601166} {arXiv:astro-ph/0601166 [astro-ph]}
  \BibitemShut {NoStop}%
\bibitem [{\citenamefont {{Rosotti}}\ \emph {et~al.}(2014)\citenamefont
  {{Rosotti}}, \citenamefont {{Dale}}, \citenamefont {{de Juan Ovelar}},
  \citenamefont {{Hubber}}, \citenamefont {{Kruijssen}}, \citenamefont
  {{Ercolano}},\ and\ \citenamefont {{Walch}}}]{Rosotti+2014}%
  \BibitemOpen
  \bibfield  {author} {\bibinfo {author} {\bibfnamefont {G.~P.}\ \bibnamefont
  {{Rosotti}}}, \bibinfo {author} {\bibfnamefont {J.~E.}\ \bibnamefont
  {{Dale}}}, \bibinfo {author} {\bibfnamefont {M.}~\bibnamefont {{de Juan
  Ovelar}}}, \bibinfo {author} {\bibfnamefont {D.~A.}\ \bibnamefont
  {{Hubber}}}, \bibinfo {author} {\bibfnamefont {J.~M.~D.}\ \bibnamefont
  {{Kruijssen}}}, \bibinfo {author} {\bibfnamefont {B.}~\bibnamefont
  {{Ercolano}}},\ and\ \bibinfo {author} {\bibfnamefont {S.}~\bibnamefont
  {{Walch}}},\ }\href {https://doi.org/10.1093/mnras/stu679} {\bibfield
  {journal} {\bibinfo  {journal} {\mnras}\ }\textbf {\bibinfo {volume} {441}},\
  \bibinfo {pages} {2094} (\bibinfo {year} {2014})},\ \Eprint
  {https://arxiv.org/abs/1404.1931} {arXiv:1404.1931 [astro-ph.EP]}
  \BibitemShut {NoStop}%
\bibitem [{\citenamefont {{Concha-Ram{\'\i}rez}}\ \emph
  {et~al.}(2019)\citenamefont {{Concha-Ram{\'\i}rez}}, \citenamefont
  {{Wilhelm}}, \citenamefont {{Portegies Zwart}},\ and\ \citenamefont
  {{Haworth}}}]{Concha-Ramirez+2019}%
  \BibitemOpen
  \bibfield  {author} {\bibinfo {author} {\bibfnamefont {F.}~\bibnamefont
  {{Concha-Ram{\'\i}rez}}}, \bibinfo {author} {\bibfnamefont {M.~J.~C.}\
  \bibnamefont {{Wilhelm}}}, \bibinfo {author} {\bibfnamefont {S.}~\bibnamefont
  {{Portegies Zwart}}},\ and\ \bibinfo {author} {\bibfnamefont {T.~J.}\
  \bibnamefont {{Haworth}}},\ }\href {https://doi.org/10.1093/mnras/stz2973}
  {\bibfield  {journal} {\bibinfo  {journal} {\mnras}\ }\textbf {\bibinfo
  {volume} {490}},\ \bibinfo {pages} {5678} (\bibinfo {year} {2019})},\ \Eprint
  {https://arxiv.org/abs/1907.03760} {arXiv:1907.03760 [astro-ph.EP]}
  \BibitemShut {NoStop}%
\bibitem [{\citenamefont {{Blum}}(2018)}]{Blum2018}%
  \BibitemOpen
  \bibfield  {author} {\bibinfo {author} {\bibfnamefont {J.}~\bibnamefont
  {{Blum}}},\ }\href {https://doi.org/10.1007/s11214-018-0486-5} {\bibfield
  {journal} {\bibinfo  {journal} {\ssr}\ }\textbf {\bibinfo {volume} {214}},\
  \bibinfo {eid} {52} (\bibinfo {year} {2018})},\ \Eprint
  {https://arxiv.org/abs/1802.00221} {arXiv:1802.00221 [astro-ph.EP]}
  \BibitemShut {NoStop}%
\bibitem [{\citenamefont {{Kobayashi}}\ and\ \citenamefont
  {{Ida}}(2001)}]{Kobayashi+2001}%
  \BibitemOpen
  \bibfield  {author} {\bibinfo {author} {\bibfnamefont {H.}~\bibnamefont
  {{Kobayashi}}}\ and\ \bibinfo {author} {\bibfnamefont {S.}~\bibnamefont
  {{Ida}}},\ }\href {https://doi.org/10.1006/icar.2001.6700} {\bibfield
  {journal} {\bibinfo  {journal} {\icarus}\ }\textbf {\bibinfo {volume}
  {153}},\ \bibinfo {pages} {416} (\bibinfo {year} {2001})},\ \Eprint
  {https://arxiv.org/abs/astro-ph/0107086} {arXiv:astro-ph/0107086 [astro-ph]}
  \BibitemShut {NoStop}%
\bibitem [{\citenamefont {{Aly}}\ and\ \citenamefont
  {{Lodato}}(2020)}]{Aly+2020}%
  \BibitemOpen
  \bibfield  {author} {\bibinfo {author} {\bibfnamefont {H.}~\bibnamefont
  {{Aly}}}\ and\ \bibinfo {author} {\bibfnamefont {G.}~\bibnamefont
  {{Lodato}}},\ }\href {https://doi.org/10.1093/mnras/stz3633} {\bibfield
  {journal} {\bibinfo  {journal} {\mnras}\ }\textbf {\bibinfo {volume} {492}},\
  \bibinfo {pages} {3306} (\bibinfo {year} {2020})},\ \Eprint
  {https://arxiv.org/abs/2001.03066} {arXiv:2001.03066 [astro-ph.EP]}
  \BibitemShut {NoStop}%
\bibitem [{\citenamefont {{Aly}}\ \emph {et~al.}(2021)\citenamefont {{Aly}},
  \citenamefont {{Gonzalez}}, \citenamefont {{Nealon}}, \citenamefont
  {{Longarini}}, \citenamefont {{Lodato}},\ and\ \citenamefont
  {{Price}}}]{Aly+2021}%
  \BibitemOpen
  \bibfield  {author} {\bibinfo {author} {\bibfnamefont {H.}~\bibnamefont
  {{Aly}}}, \bibinfo {author} {\bibfnamefont {J.-F.}\ \bibnamefont
  {{Gonzalez}}}, \bibinfo {author} {\bibfnamefont {R.}~\bibnamefont
  {{Nealon}}}, \bibinfo {author} {\bibfnamefont {C.}~\bibnamefont
  {{Longarini}}}, \bibinfo {author} {\bibfnamefont {G.}~\bibnamefont
  {{Lodato}}},\ and\ \bibinfo {author} {\bibfnamefont {D.~J.}\ \bibnamefont
  {{Price}}},\ }\href {https://doi.org/10.1093/mnras/stab2794} {\bibfield
  {journal} {\bibinfo  {journal} {\mnras}\ }\textbf {\bibinfo {volume} {508}},\
  \bibinfo {pages} {2743} (\bibinfo {year} {2021})},\ \Eprint
  {https://arxiv.org/abs/2109.13256} {arXiv:2109.13256 [astro-ph.EP]}
  \BibitemShut {NoStop}%
\bibitem [{\citenamefont {{Longarini}}\ \emph {et~al.}(2021)\citenamefont
  {{Longarini}}, \citenamefont {{Lodato}}, \citenamefont {{Toci}},\ and\
  \citenamefont {{Aly}}}]{Longarini+2021}%
  \BibitemOpen
  \bibfield  {author} {\bibinfo {author} {\bibfnamefont {C.}~\bibnamefont
  {{Longarini}}}, \bibinfo {author} {\bibfnamefont {G.}~\bibnamefont
  {{Lodato}}}, \bibinfo {author} {\bibfnamefont {C.}~\bibnamefont {{Toci}}},\
  and\ \bibinfo {author} {\bibfnamefont {H.}~\bibnamefont {{Aly}}},\ }\href
  {https://doi.org/10.1093/mnras/stab843} {\bibfield  {journal} {\bibinfo
  {journal} {\mnras}\ }\textbf {\bibinfo {volume} {503}},\ \bibinfo {pages}
  {4930} (\bibinfo {year} {2021})},\ \Eprint {https://arxiv.org/abs/2103.12084}
  {arXiv:2103.12084 [astro-ph.EP]} \BibitemShut {NoStop}%
\bibitem [{\citenamefont {{Benac}}\ \emph {et~al.}(2020)\citenamefont
  {{Benac}}, \citenamefont {{Matr{\`a}}}, \citenamefont {{Wilner}},
  \citenamefont {{Jim{\`e}nez-Donaire}}, \citenamefont {{Monnier}},
  \citenamefont {{Harries}}, \citenamefont {{Laws}}, \citenamefont {{Rich}},\
  and\ \citenamefont {{Zhang}}}]{Benac+2020}%
  \BibitemOpen
  \bibfield  {author} {\bibinfo {author} {\bibfnamefont {P.}~\bibnamefont
  {{Benac}}}, \bibinfo {author} {\bibfnamefont {L.}~\bibnamefont
  {{Matr{\`a}}}}, \bibinfo {author} {\bibfnamefont {D.~J.}\ \bibnamefont
  {{Wilner}}}, \bibinfo {author} {\bibfnamefont {M.~J.}\ \bibnamefont
  {{Jim{\`e}nez-Donaire}}}, \bibinfo {author} {\bibfnamefont {J.~D.}\
  \bibnamefont {{Monnier}}}, \bibinfo {author} {\bibfnamefont {T.~J.}\
  \bibnamefont {{Harries}}}, \bibinfo {author} {\bibfnamefont {A.}~\bibnamefont
  {{Laws}}}, \bibinfo {author} {\bibfnamefont {E.~A.}\ \bibnamefont {{Rich}}},\
  and\ \bibinfo {author} {\bibfnamefont {Q.}~\bibnamefont {{Zhang}}},\ }\href
  {https://doi.org/10.3847/1538-4357/abc74b} {\bibfield  {journal} {\bibinfo
  {journal} {\apj}\ }\textbf {\bibinfo {volume} {905}},\ \bibinfo {eid} {120}
  (\bibinfo {year} {2020})},\ \Eprint {https://arxiv.org/abs/2011.03489}
  {arXiv:2011.03489 [astro-ph.EP]} \BibitemShut {NoStop}%
\bibitem [{\citenamefont {{Dipierro}}\ \emph {et~al.}(2015)\citenamefont
  {{Dipierro}}, \citenamefont {{Price}}, \citenamefont {{Laibe}}, \citenamefont
  {{Hirsh}}, \citenamefont {{Cerioli}},\ and\ \citenamefont
  {{Lodato}}}]{Dipierro+2015}%
  \BibitemOpen
  \bibfield  {author} {\bibinfo {author} {\bibfnamefont {G.}~\bibnamefont
  {{Dipierro}}}, \bibinfo {author} {\bibfnamefont {D.}~\bibnamefont {{Price}}},
  \bibinfo {author} {\bibfnamefont {G.}~\bibnamefont {{Laibe}}}, \bibinfo
  {author} {\bibfnamefont {K.}~\bibnamefont {{Hirsh}}}, \bibinfo {author}
  {\bibfnamefont {A.}~\bibnamefont {{Cerioli}}},\ and\ \bibinfo {author}
  {\bibfnamefont {G.}~\bibnamefont {{Lodato}}},\ }\href
  {https://doi.org/10.1093/mnrasl/slv105} {\bibfield  {journal} {\bibinfo
  {journal} {\mnras}\ }\textbf {\bibinfo {volume} {453}},\ \bibinfo {pages}
  {L73} (\bibinfo {year} {2015})},\ \Eprint {https://arxiv.org/abs/1507.06719}
  {arXiv:1507.06719 [astro-ph.EP]} \BibitemShut {NoStop}%
\bibitem [{\citenamefont {{Marzari}}\ and\ \citenamefont
  {{Picogna}}(2013)}]{MarzariPicogna2013}%
  \BibitemOpen
  \bibfield  {author} {\bibinfo {author} {\bibfnamefont {F.}~\bibnamefont
  {{Marzari}}}\ and\ \bibinfo {author} {\bibfnamefont {G.}~\bibnamefont
  {{Picogna}}},\ }\href {https://doi.org/10.1051/0004-6361/201220436}
  {\bibfield  {journal} {\bibinfo  {journal} {\aap}\ }\textbf {\bibinfo
  {volume} {550}},\ \bibinfo {eid} {A64} (\bibinfo {year} {2013})},\ \Eprint
  {https://arxiv.org/abs/1212.1561} {arXiv:1212.1561 [astro-ph.EP]}
  \BibitemShut {NoStop}%
\bibitem [{\citenamefont {{Davies}}(2019)}]{Davies2019}%
  \BibitemOpen
  \bibfield  {author} {\bibinfo {author} {\bibfnamefont {C.~L.}\ \bibnamefont
  {{Davies}}},\ }\href {https://doi.org/10.1093/mnras/stz086} {\bibfield
  {journal} {\bibinfo  {journal} {\mnras}\ }\textbf {\bibinfo {volume} {484}},\
  \bibinfo {pages} {1926} (\bibinfo {year} {2019})},\ \Eprint
  {https://arxiv.org/abs/1901.01929} {arXiv:1901.01929 [astro-ph.SR]}
  \BibitemShut {NoStop}%
\bibitem [{\citenamefont {{Campante}}\ \emph {et~al.}(2016)\citenamefont
  {{Campante}}, \citenamefont {{Lund}}, \citenamefont {{Kuszlewicz}},
  \citenamefont {{Davies}}, \citenamefont {{Chaplin}}, \citenamefont
  {{Albrecht}},\ and\ \citenamefont {{Winn et al.}}}]{Campante+2016}%
  \BibitemOpen
  \bibfield  {author} {\bibinfo {author} {\bibfnamefont {T.~L.}\ \bibnamefont
  {{Campante}}}, \bibinfo {author} {\bibfnamefont {M.~N.}\ \bibnamefont
  {{Lund}}}, \bibinfo {author} {\bibfnamefont {J.~S.}\ \bibnamefont
  {{Kuszlewicz}}}, \bibinfo {author} {\bibfnamefont {G.~R.}\ \bibnamefont
  {{Davies}}}, \bibinfo {author} {\bibfnamefont {W.~J.}\ \bibnamefont
  {{Chaplin}}}, \bibinfo {author} {\bibfnamefont {S.}~\bibnamefont
  {{Albrecht}}},\ and\ \bibinfo {author} {\bibfnamefont {J.~N.}\ \bibnamefont
  {{Winn et al.}}},\ }\href {https://doi.org/10.3847/0004-637X/819/1/85}
  {\bibfield  {journal} {\bibinfo  {journal} {\apj}\ }\textbf {\bibinfo
  {volume} {819}},\ \bibinfo {eid} {85} (\bibinfo {year} {2016})},\ \Eprint
  {https://arxiv.org/abs/1601.06052} {arXiv:1601.06052 [astro-ph.EP]}
  \BibitemShut {NoStop}%
\bibitem [{\citenamefont {{Spurzem}}\ \emph {et~al.}(2009)\citenamefont
  {{Spurzem}}, \citenamefont {{Giersz}}, \citenamefont {{Heggie}},\ and\
  \citenamefont {{Lin}}}]{Spurzem+2009}%
  \BibitemOpen
  \bibfield  {author} {\bibinfo {author} {\bibfnamefont {R.}~\bibnamefont
  {{Spurzem}}}, \bibinfo {author} {\bibfnamefont {M.}~\bibnamefont {{Giersz}}},
  \bibinfo {author} {\bibfnamefont {D.~C.}\ \bibnamefont {{Heggie}}},\ and\
  \bibinfo {author} {\bibfnamefont {D.~N.~C.}\ \bibnamefont {{Lin}}},\ }\href
  {https://doi.org/10.1088/0004-637X/697/1/458} {\bibfield  {journal} {\bibinfo
   {journal} {\apj}\ }\textbf {\bibinfo {volume} {697}},\ \bibinfo {pages}
  {458} (\bibinfo {year} {2009})},\ \Eprint
  {https://arxiv.org/abs/astro-ph/0612757} {arXiv:astro-ph/0612757 [astro-ph]}
  \BibitemShut {NoStop}%
\bibitem [{\citenamefont {{Malmberg}}\ \emph
  {et~al.}(2011{\natexlab{a}})\citenamefont {{Malmberg}}, \citenamefont
  {{Davies}},\ and\ \citenamefont {{Heggie}}}]{Malberg+2011}%
  \BibitemOpen
  \bibfield  {author} {\bibinfo {author} {\bibfnamefont {D.}~\bibnamefont
  {{Malmberg}}}, \bibinfo {author} {\bibfnamefont {M.~B.}\ \bibnamefont
  {{Davies}}},\ and\ \bibinfo {author} {\bibfnamefont {D.~C.}\ \bibnamefont
  {{Heggie}}},\ }\href {https://doi.org/10.1111/j.1365-2966.2010.17730.x}
  {\bibfield  {journal} {\bibinfo  {journal} {\mnras}\ }\textbf {\bibinfo
  {volume} {411}},\ \bibinfo {pages} {859} (\bibinfo {year}
  {2011}{\natexlab{a}})},\ \Eprint {https://arxiv.org/abs/1009.4196}
  {arXiv:1009.4196 [astro-ph.EP]} \BibitemShut {NoStop}%
\bibitem [{\citenamefont {{Bailey}}\ and\ \citenamefont
  {{Fabrycky}}(2019)}]{Bailey+2019}%
  \BibitemOpen
  \bibfield  {author} {\bibinfo {author} {\bibfnamefont {N.}~\bibnamefont
  {{Bailey}}}\ and\ \bibinfo {author} {\bibfnamefont {D.}~\bibnamefont
  {{Fabrycky}}},\ }\href {https://doi.org/10.3847/1538-3881/ab2d2a} {\bibfield
  {journal} {\bibinfo  {journal} {\aj}\ }\textbf {\bibinfo {volume} {158}},\
  \bibinfo {eid} {94} (\bibinfo {year} {2019})},\ \Eprint
  {https://arxiv.org/abs/1905.07044} {arXiv:1905.07044 [astro-ph.EP]}
  \BibitemShut {NoStop}%
\bibitem [{\citenamefont {{Brown}}\ and\ \citenamefont
  {{Rein}}(2022)}]{Brown+2022}%
  \BibitemOpen
  \bibfield  {author} {\bibinfo {author} {\bibfnamefont {G.}~\bibnamefont
  {{Brown}}}\ and\ \bibinfo {author} {\bibfnamefont {H.}~\bibnamefont
  {{Rein}}},\ }\bibfield  {journal} {\bibinfo  {journal} {\mnras}\ }\href
  {https://doi.org/10.1093/mnras/stac1763} {10.1093/mnras/stac1763} (\bibinfo
  {year} {2022}),\ \Eprint {https://arxiv.org/abs/2206.14240} {arXiv:2206.14240
  [astro-ph.EP]} \BibitemShut {NoStop}%
\bibitem [{\citenamefont {{Breslau}}\ and\ \citenamefont
  {{Pfalzner}}(2019)}]{Breslau+2019}%
  \BibitemOpen
  \bibfield  {author} {\bibinfo {author} {\bibfnamefont {A.}~\bibnamefont
  {{Breslau}}}\ and\ \bibinfo {author} {\bibfnamefont {S.}~\bibnamefont
  {{Pfalzner}}},\ }\href {https://doi.org/10.1051/0004-6361/201833729}
  {\bibfield  {journal} {\bibinfo  {journal} {\aap}\ }\textbf {\bibinfo
  {volume} {621}},\ \bibinfo {eid} {A101} (\bibinfo {year} {2019})},\ \Eprint
  {https://arxiv.org/abs/1812.04104} {arXiv:1812.04104 [astro-ph.EP]}
  \BibitemShut {NoStop}%
\bibitem [{\citenamefont {{Ndugu}}\ \emph {et~al.}(2022)\citenamefont
  {{Ndugu}}, \citenamefont {{Abedigamba}},\ and\ \citenamefont
  {{Andama}}}]{Ndugu+2022}%
  \BibitemOpen
  \bibfield  {author} {\bibinfo {author} {\bibfnamefont {N.}~\bibnamefont
  {{Ndugu}}}, \bibinfo {author} {\bibfnamefont {O.~P.}\ \bibnamefont
  {{Abedigamba}}},\ and\ \bibinfo {author} {\bibfnamefont {G.}~\bibnamefont
  {{Andama}}},\ }\href {https://doi.org/10.1093/mnras/stac569} {\bibfield
  {journal} {\bibinfo  {journal} {\mnras}\ }\textbf {\bibinfo {volume} {512}},\
  \bibinfo {pages} {861} (\bibinfo {year} {2022})},\ \Eprint
  {https://arxiv.org/abs/2202.11935} {arXiv:2202.11935 [astro-ph.EP]}
  \BibitemShut {NoStop}%
\bibitem [{\citenamefont {{Rodet}}\ \emph {et~al.}(2021)\citenamefont
  {{Rodet}}, \citenamefont {{Su}},\ and\ \citenamefont {{Lai}}}]{Rodet+2021}%
  \BibitemOpen
  \bibfield  {author} {\bibinfo {author} {\bibfnamefont {L.}~\bibnamefont
  {{Rodet}}}, \bibinfo {author} {\bibfnamefont {Y.}~\bibnamefont {{Su}}},\ and\
  \bibinfo {author} {\bibfnamefont {D.}~\bibnamefont {{Lai}}},\ }\href
  {https://doi.org/10.3847/1538-4357/abf8a7} {\bibfield  {journal} {\bibinfo
  {journal} {\apj}\ }\textbf {\bibinfo {volume} {913}},\ \bibinfo {eid} {104}
  (\bibinfo {year} {2021})},\ \Eprint {https://arxiv.org/abs/2102.07898}
  {arXiv:2102.07898 [astro-ph.EP]} \BibitemShut {NoStop}%
\bibitem [{\citenamefont {{Jim{\'e}nez-Torres}}(2021)}]{Jimenez-Torres2021}%
  \BibitemOpen
  \bibfield  {author} {\bibinfo {author} {\bibfnamefont {J.~J.}\ \bibnamefont
  {{Jim{\'e}nez-Torres}}},\ }\href
  {https://doi.org/10.1051/0004-6361/202141323} {\bibfield  {journal} {\bibinfo
   {journal} {\aap}\ }\textbf {\bibinfo {volume} {653}},\ \bibinfo {eid} {A168}
  (\bibinfo {year} {2021})}\BibitemShut {NoStop}%
\bibitem [{\citenamefont {{Bobylev}}\ and\ \citenamefont
  {{Bajkova}}(2020)}]{Bobylev+2020}%
  \BibitemOpen
  \bibfield  {author} {\bibinfo {author} {\bibfnamefont {V.~V.}\ \bibnamefont
  {{Bobylev}}}\ and\ \bibinfo {author} {\bibfnamefont {A.~T.}\ \bibnamefont
  {{Bajkova}}},\ }\href {https://doi.org/10.1134/S1063773720040039} {\bibfield
  {journal} {\bibinfo  {journal} {Astronomy Letters}\ }\textbf {\bibinfo
  {volume} {46}},\ \bibinfo {pages} {245} (\bibinfo {year} {2020})},\ \Eprint
  {https://arxiv.org/abs/2006.16555} {arXiv:2006.16555 [astro-ph.GA]}
  \BibitemShut {NoStop}%
\bibitem [{\citenamefont {{Dybczy{\'n}ski}}\ \emph {et~al.}(2022)\citenamefont
  {{Dybczy{\'n}ski}}, \citenamefont {{Berski}}, \citenamefont {{Tokarek}},
  \citenamefont {{Podlewska-Gaca}}, \citenamefont {{Langner}},\ and\
  \citenamefont {{Bartczak}}}]{Dybczynski+2022}%
  \BibitemOpen
  \bibfield  {author} {\bibinfo {author} {\bibfnamefont {P.~A.}\ \bibnamefont
  {{Dybczy{\'n}ski}}}, \bibinfo {author} {\bibfnamefont {F.}~\bibnamefont
  {{Berski}}}, \bibinfo {author} {\bibfnamefont {J.}~\bibnamefont {{Tokarek}}},
  \bibinfo {author} {\bibfnamefont {E.}~\bibnamefont {{Podlewska-Gaca}}},
  \bibinfo {author} {\bibfnamefont {K.}~\bibnamefont {{Langner}}},\ and\
  \bibinfo {author} {\bibfnamefont {P.}~\bibnamefont {{Bartczak}}},\
  }\href@noop {} {\bibfield  {journal} {\bibinfo  {journal} {arXiv e-prints}\
  ,\ \bibinfo {eid} {arXiv:2206.11047}} (\bibinfo {year} {2022})},\ \Eprint
  {https://arxiv.org/abs/2206.11047} {arXiv:2206.11047 [astro-ph.SR]}
  \BibitemShut {NoStop}%
\bibitem [{\citenamefont {{Bailer-Jones}}(2022)}]{Bailer-Jones2022}%
  \BibitemOpen
  \bibfield  {author} {\bibinfo {author} {\bibfnamefont {C.~A.~L.}\
  \bibnamefont {{Bailer-Jones}}},\ }\href@noop {} {\bibfield  {journal}
  {\bibinfo  {journal} {arXiv e-prints}\ ,\ \bibinfo {eid} {arXiv:2207.06258}}
  (\bibinfo {year} {2022})},\ \Eprint {https://arxiv.org/abs/2207.06258}
  {arXiv:2207.06258 [astro-ph.SR]} \BibitemShut {NoStop}%
\bibitem [{\citenamefont {{Bhandare}}\ and\ \citenamefont
  {{Pfalzner}}(2019)}]{Bhandare+2019}%
  \BibitemOpen
  \bibfield  {author} {\bibinfo {author} {\bibfnamefont {A.}~\bibnamefont
  {{Bhandare}}}\ and\ \bibinfo {author} {\bibfnamefont {S.}~\bibnamefont
  {{Pfalzner}}},\ }\href {https://doi.org/10.1186/s40668-019-0030-3} {\bibfield
   {journal} {\bibinfo  {journal} {Computational Astrophysics and Cosmology}\
  }\textbf {\bibinfo {volume} {6}},\ \bibinfo {eid} {3} (\bibinfo {year}
  {2019})},\ \Eprint {https://arxiv.org/abs/1909.08404} {arXiv:1909.08404
  [astro-ph.EP]} \BibitemShut {NoStop}%
\bibitem [{\citenamefont {{Veras}}\ and\ \citenamefont
  {{Moeckel}}(2012)}]{Veras+2012}%
  \BibitemOpen
  \bibfield  {author} {\bibinfo {author} {\bibfnamefont {D.}~\bibnamefont
  {{Veras}}}\ and\ \bibinfo {author} {\bibfnamefont {N.}~\bibnamefont
  {{Moeckel}}},\ }\href {https://doi.org/10.1111/j.1365-2966.2012.21552.x}
  {\bibfield  {journal} {\bibinfo  {journal} {\mnras}\ }\textbf {\bibinfo
  {volume} {425}},\ \bibinfo {pages} {680} (\bibinfo {year} {2012})},\ \Eprint
  {https://arxiv.org/abs/1206.4694} {arXiv:1206.4694 [astro-ph.EP]}
  \BibitemShut {NoStop}%
\bibitem [{\citenamefont {{Shara}}\ \emph {et~al.}(2016)\citenamefont
  {{Shara}}, \citenamefont {{Hurley}},\ and\ \citenamefont
  {{Mardling}}}]{Shara+2016}%
  \BibitemOpen
  \bibfield  {author} {\bibinfo {author} {\bibfnamefont {M.~M.}\ \bibnamefont
  {{Shara}}}, \bibinfo {author} {\bibfnamefont {J.~R.}\ \bibnamefont
  {{Hurley}}},\ and\ \bibinfo {author} {\bibfnamefont {R.~A.}\ \bibnamefont
  {{Mardling}}},\ }\href {https://doi.org/10.3847/0004-637X/816/2/59}
  {\bibfield  {journal} {\bibinfo  {journal} {\apj}\ }\textbf {\bibinfo
  {volume} {816}},\ \bibinfo {eid} {59} (\bibinfo {year} {2016})},\ \Eprint
  {https://arxiv.org/abs/1411.7061} {arXiv:1411.7061 [astro-ph.EP]}
  \BibitemShut {NoStop}%
\bibitem [{\citenamefont {{Li}}\ \emph {et~al.}(2020)\citenamefont {{Li}},
  \citenamefont {{Mustill}},\ and\ \citenamefont {{Davies}}}]{Li+2020}%
  \BibitemOpen
  \bibfield  {author} {\bibinfo {author} {\bibfnamefont {D.}~\bibnamefont
  {{Li}}}, \bibinfo {author} {\bibfnamefont {A.~J.}\ \bibnamefont
  {{Mustill}}},\ and\ \bibinfo {author} {\bibfnamefont {M.~B.}\ \bibnamefont
  {{Davies}}},\ }\href {https://doi.org/10.1093/mnras/staa1622} {\bibfield
  {journal} {\bibinfo  {journal} {\mnras}\ }\textbf {\bibinfo {volume} {496}},\
  \bibinfo {pages} {1149} (\bibinfo {year} {2020})},\ \Eprint
  {https://arxiv.org/abs/2002.09271} {arXiv:2002.09271 [astro-ph.EP]}
  \BibitemShut {NoStop}%
\bibitem [{\citenamefont {{Hills}}(1984)}]{Hills1984}%
  \BibitemOpen
  \bibfield  {author} {\bibinfo {author} {\bibfnamefont {J.~G.}\ \bibnamefont
  {{Hills}}},\ }\href {https://doi.org/10.1086/113659} {\bibfield  {journal}
  {\bibinfo  {journal} {\aj}\ }\textbf {\bibinfo {volume} {89}},\ \bibinfo
  {pages} {1559} (\bibinfo {year} {1984})}\BibitemShut {NoStop}%
\bibitem [{\citenamefont {{Malmberg}}\ \emph
  {et~al.}(2011{\natexlab{b}})\citenamefont {{Malmberg}}, \citenamefont
  {{Davies}},\ and\ \citenamefont {{Heggie}}}]{Malmberg+2011}%
  \BibitemOpen
  \bibfield  {author} {\bibinfo {author} {\bibfnamefont {D.}~\bibnamefont
  {{Malmberg}}}, \bibinfo {author} {\bibfnamefont {M.~B.}\ \bibnamefont
  {{Davies}}},\ and\ \bibinfo {author} {\bibfnamefont {D.~C.}\ \bibnamefont
  {{Heggie}}},\ }\href {https://doi.org/10.1111/j.1365-2966.2010.17730.x}
  {\bibfield  {journal} {\bibinfo  {journal} {\mnras}\ }\textbf {\bibinfo
  {volume} {411}},\ \bibinfo {pages} {859} (\bibinfo {year}
  {2011}{\natexlab{b}})},\ \Eprint {https://arxiv.org/abs/1009.4196}
  {arXiv:1009.4196 [astro-ph.EP]} \BibitemShut {NoStop}%
\bibitem [{\citenamefont {{Parker}}\ and\ \citenamefont
  {{Quanz}}(2012)}]{ParkerQuanz2012}%
  \BibitemOpen
  \bibfield  {author} {\bibinfo {author} {\bibfnamefont {R.~J.}\ \bibnamefont
  {{Parker}}}\ and\ \bibinfo {author} {\bibfnamefont {S.~P.}\ \bibnamefont
  {{Quanz}}},\ }\href {https://doi.org/10.1111/j.1365-2966.2011.19911.x}
  {\bibfield  {journal} {\bibinfo  {journal} {\mnras}\ }\textbf {\bibinfo
  {volume} {419}},\ \bibinfo {pages} {2448} (\bibinfo {year} {2012})},\ \Eprint
  {https://arxiv.org/abs/1109.6007} {arXiv:1109.6007 [astro-ph.EP]}
  \BibitemShut {NoStop}%
\bibitem [{\citenamefont {{Craig}}\ and\ \citenamefont
  {{Krumholz}}(2013)}]{Craig+2013}%
  \BibitemOpen
  \bibfield  {author} {\bibinfo {author} {\bibfnamefont {J.}~\bibnamefont
  {{Craig}}}\ and\ \bibinfo {author} {\bibfnamefont {M.~R.}\ \bibnamefont
  {{Krumholz}}},\ }\href {https://doi.org/10.1088/0004-637X/769/2/150}
  {\bibfield  {journal} {\bibinfo  {journal} {\apj}\ }\textbf {\bibinfo
  {volume} {769}},\ \bibinfo {eid} {150} (\bibinfo {year} {2013})},\ \Eprint
  {https://arxiv.org/abs/1304.7683} {arXiv:1304.7683 [astro-ph.EP]}
  \BibitemShut {NoStop}%
\bibitem [{\citenamefont {{Zheng}}\ \emph {et~al.}(2015)\citenamefont
  {{Zheng}}, \citenamefont {{Kouwenhoven}},\ and\ \citenamefont
  {{Wang}}}]{Zheng+2015}%
  \BibitemOpen
  \bibfield  {author} {\bibinfo {author} {\bibfnamefont {X.}~\bibnamefont
  {{Zheng}}}, \bibinfo {author} {\bibfnamefont {M.~B.~N.}\ \bibnamefont
  {{Kouwenhoven}}},\ and\ \bibinfo {author} {\bibfnamefont {L.}~\bibnamefont
  {{Wang}}},\ }\href {https://doi.org/10.1093/mnras/stv1832} {\bibfield
  {journal} {\bibinfo  {journal} {\mnras}\ }\textbf {\bibinfo {volume} {453}},\
  \bibinfo {pages} {2759} (\bibinfo {year} {2015})},\ \Eprint
  {https://arxiv.org/abs/1508.01593} {arXiv:1508.01593 [astro-ph.EP]}
  \BibitemShut {NoStop}%
\bibitem [{\citenamefont {{Daffern-Powell}}\ \emph {et~al.}(2022)\citenamefont
  {{Daffern-Powell}}, \citenamefont {{Parker}},\ and\ \citenamefont
  {{Quanz}}}]{Daffern-Powell2022}%
  \BibitemOpen
  \bibfield  {author} {\bibinfo {author} {\bibfnamefont {E.~C.}\ \bibnamefont
  {{Daffern-Powell}}}, \bibinfo {author} {\bibfnamefont {R.~J.}\ \bibnamefont
  {{Parker}}},\ and\ \bibinfo {author} {\bibfnamefont {S.~P.}\ \bibnamefont
  {{Quanz}}},\ }\href {https://doi.org/10.1093/mnras/stac1392} {\bibfield
  {journal} {\bibinfo  {journal} {\mnras}\ }\textbf {\bibinfo {volume} {514}},\
  \bibinfo {pages} {920} (\bibinfo {year} {2022})},\ \Eprint
  {https://arxiv.org/abs/2205.07895} {arXiv:2205.07895 [astro-ph.EP]}
  \BibitemShut {NoStop}%
\bibitem [{\citenamefont {{Perets}}\ and\ \citenamefont
  {{Kouwenhoven}}(2012)}]{Perets+2012}%
  \BibitemOpen
  \bibfield  {author} {\bibinfo {author} {\bibfnamefont {H.~B.}\ \bibnamefont
  {{Perets}}}\ and\ \bibinfo {author} {\bibfnamefont {M.~B.~N.}\ \bibnamefont
  {{Kouwenhoven}}},\ }\href {https://doi.org/10.1088/0004-637X/750/1/83}
  {\bibfield  {journal} {\bibinfo  {journal} {\apj}\ }\textbf {\bibinfo
  {volume} {750}},\ \bibinfo {eid} {83} (\bibinfo {year} {2012})},\ \Eprint
  {https://arxiv.org/abs/1202.2362} {arXiv:1202.2362 [astro-ph.EP]}
  \BibitemShut {NoStop}%
\bibitem [{\citenamefont {{Wang}}\ \emph {et~al.}(2015)\citenamefont {{Wang}},
  \citenamefont {{Kouwenhoven}}, \citenamefont {{Zheng}}, \citenamefont
  {{Church}},\ and\ \citenamefont {{Davies}}}]{Wang+2015}%
  \BibitemOpen
  \bibfield  {author} {\bibinfo {author} {\bibfnamefont {L.}~\bibnamefont
  {{Wang}}}, \bibinfo {author} {\bibfnamefont {M.~B.~N.}\ \bibnamefont
  {{Kouwenhoven}}}, \bibinfo {author} {\bibfnamefont {X.}~\bibnamefont
  {{Zheng}}}, \bibinfo {author} {\bibfnamefont {R.~P.}\ \bibnamefont
  {{Church}}},\ and\ \bibinfo {author} {\bibfnamefont {M.~B.}\ \bibnamefont
  {{Davies}}},\ }\href {https://doi.org/10.1093/mnras/stv542} {\bibfield
  {journal} {\bibinfo  {journal} {\mnras}\ }\textbf {\bibinfo {volume} {449}},\
  \bibinfo {pages} {3543} (\bibinfo {year} {2015})},\ \Eprint
  {https://arxiv.org/abs/1503.03077} {arXiv:1503.03077 [astro-ph.EP]}
  \BibitemShut {NoStop}%
\bibitem [{\citenamefont {{Pfalzner}}\ \emph {et~al.}(2018)\citenamefont
  {{Pfalzner}}, \citenamefont {{Bhandare}}, \citenamefont {{Vincke}},\ and\
  \citenamefont {{Lacerda}}}]{Pfalzner+2018b}%
  \BibitemOpen
  \bibfield  {author} {\bibinfo {author} {\bibfnamefont {S.}~\bibnamefont
  {{Pfalzner}}}, \bibinfo {author} {\bibfnamefont {A.}~\bibnamefont
  {{Bhandare}}}, \bibinfo {author} {\bibfnamefont {K.}~\bibnamefont
  {{Vincke}}},\ and\ \bibinfo {author} {\bibfnamefont {P.}~\bibnamefont
  {{Lacerda}}},\ }\href {https://doi.org/10.3847/1538-4357/aad23c} {\bibfield
  {journal} {\bibinfo  {journal} {\apj}\ }\textbf {\bibinfo {volume} {863}},\
  \bibinfo {eid} {45} (\bibinfo {year} {2018})},\ \Eprint
  {https://arxiv.org/abs/1807.02960} {arXiv:1807.02960 [astro-ph.GA]}
  \BibitemShut {NoStop}%
\bibitem [{\citenamefont {{Moore}}\ \emph {et~al.}(2020)\citenamefont
  {{Moore}}, \citenamefont {{Li}},\ and\ \citenamefont {{Adams}}}]{Moore+2020}%
  \BibitemOpen
  \bibfield  {author} {\bibinfo {author} {\bibfnamefont {N.~W.~H.}\
  \bibnamefont {{Moore}}}, \bibinfo {author} {\bibfnamefont {G.}~\bibnamefont
  {{Li}}},\ and\ \bibinfo {author} {\bibfnamefont {F.~C.}\ \bibnamefont
  {{Adams}}},\ }\href {https://doi.org/10.3847/1538-4357/abb08f} {\bibfield
  {journal} {\bibinfo  {journal} {\apj}\ }\textbf {\bibinfo {volume} {901}},\
  \bibinfo {eid} {92} (\bibinfo {year} {2020})},\ \Eprint
  {https://arxiv.org/abs/2007.15666} {arXiv:2007.15666 [astro-ph.EP]}
  \BibitemShut {NoStop}%
\bibitem [{\citenamefont {{J{\'\i}lkov{\'a}}}\ \emph
  {et~al.}(2015)\citenamefont {{J{\'\i}lkov{\'a}}}, \citenamefont {{Portegies
  Zwart}}, \citenamefont {{Pijloo}},\ and\ \citenamefont
  {{Hammer}}}]{Jilkova+2015}%
  \BibitemOpen
  \bibfield  {author} {\bibinfo {author} {\bibfnamefont {L.}~\bibnamefont
  {{J{\'\i}lkov{\'a}}}}, \bibinfo {author} {\bibfnamefont {S.}~\bibnamefont
  {{Portegies Zwart}}}, \bibinfo {author} {\bibfnamefont {T.}~\bibnamefont
  {{Pijloo}}},\ and\ \bibinfo {author} {\bibfnamefont {M.}~\bibnamefont
  {{Hammer}}},\ }\href {https://doi.org/10.1093/mnras/stv1803} {\bibfield
  {journal} {\bibinfo  {journal} {\mnras}\ }\textbf {\bibinfo {volume} {453}},\
  \bibinfo {pages} {3157} (\bibinfo {year} {2015})},\ \Eprint
  {https://arxiv.org/abs/1506.03105} {arXiv:1506.03105 [astro-ph.EP]}
  \BibitemShut {NoStop}%
\bibitem [{\citenamefont {{Batygin}}\ \emph {et~al.}(2020)\citenamefont
  {{Batygin}}, \citenamefont {{Adams}}, \citenamefont {{Batygin}},\ and\
  \citenamefont {{Petigura}}}]{Batygin+2020}%
  \BibitemOpen
  \bibfield  {author} {\bibinfo {author} {\bibfnamefont {K.}~\bibnamefont
  {{Batygin}}}, \bibinfo {author} {\bibfnamefont {F.~C.}\ \bibnamefont
  {{Adams}}}, \bibinfo {author} {\bibfnamefont {Y.~K.}\ \bibnamefont
  {{Batygin}}},\ and\ \bibinfo {author} {\bibfnamefont {E.~A.}\ \bibnamefont
  {{Petigura}}},\ }\href {https://doi.org/10.3847/1538-3881/ab665d} {\bibfield
  {journal} {\bibinfo  {journal} {\aj}\ }\textbf {\bibinfo {volume} {159}},\
  \bibinfo {eid} {101} (\bibinfo {year} {2020})},\ \Eprint
  {https://arxiv.org/abs/2002.05656} {arXiv:2002.05656 [astro-ph.EP]}
  \BibitemShut {NoStop}%
\bibitem [{\citenamefont {{Pfalzner}}\ \emph {et~al.}(2021)\citenamefont
  {{Pfalzner}}, \citenamefont {{Aizpuru Vargas}}, \citenamefont {{Bhandare}},\
  and\ \citenamefont {{Veras}}}]{Pfalzner+2021}%
  \BibitemOpen
  \bibfield  {author} {\bibinfo {author} {\bibfnamefont {S.}~\bibnamefont
  {{Pfalzner}}}, \bibinfo {author} {\bibfnamefont {L.~L.}\ \bibnamefont
  {{Aizpuru Vargas}}}, \bibinfo {author} {\bibfnamefont {A.}~\bibnamefont
  {{Bhandare}}},\ and\ \bibinfo {author} {\bibfnamefont {D.}~\bibnamefont
  {{Veras}}},\ }\href {https://doi.org/10.1051/0004-6361/202140587} {\bibfield
  {journal} {\bibinfo  {journal} {\aap}\ }\textbf {\bibinfo {volume} {651}},\
  \bibinfo {eid} {A38} (\bibinfo {year} {2021})},\ \Eprint
  {https://arxiv.org/abs/2104.06845} {arXiv:2104.06845 [astro-ph.SR]}
  \BibitemShut {NoStop}%
\bibitem [{\citenamefont {{Pfalzner}}\ and\ \citenamefont
  {{Bannister}}(2019)}]{PfalznerBannister2019}%
  \BibitemOpen
  \bibfield  {author} {\bibinfo {author} {\bibfnamefont {S.}~\bibnamefont
  {{Pfalzner}}}\ and\ \bibinfo {author} {\bibfnamefont {M.~T.}\ \bibnamefont
  {{Bannister}}},\ }\href {https://doi.org/10.3847/2041-8213/ab0fa0} {\bibfield
   {journal} {\bibinfo  {journal} {\apjl}\ }\textbf {\bibinfo {volume} {874}},\
  \bibinfo {eid} {L34} (\bibinfo {year} {2019})},\ \Eprint
  {https://arxiv.org/abs/1903.04451} {arXiv:1903.04451 [astro-ph.EP]}
  \BibitemShut {NoStop}%
\bibitem [{\citenamefont {{Meech}}\ \emph {et~al.}(2017)\citenamefont
  {{Meech}}, \citenamefont {{Weryk}},\ and\ \citenamefont {{Micheli et
  al.}}}]{Meech+2017}%
  \BibitemOpen
  \bibfield  {author} {\bibinfo {author} {\bibfnamefont {K.~J.}\ \bibnamefont
  {{Meech}}}, \bibinfo {author} {\bibfnamefont {R.}~\bibnamefont {{Weryk}}},\
  and\ \bibinfo {author} {\bibfnamefont {M.}~\bibnamefont {{Micheli et al.}}},\
  }\href {https://doi.org/10.1038/nature25020} {\bibfield  {journal} {\bibinfo
  {journal} {\nat}\ }\textbf {\bibinfo {volume} {552}},\ \bibinfo {pages} {378}
  (\bibinfo {year} {2017})}\BibitemShut {NoStop}%
\bibitem [{\citenamefont {{Jewitt}}\ and\ \citenamefont
  {{Luu}}(2019)}]{Jewitt+2019}%
  \BibitemOpen
  \bibfield  {author} {\bibinfo {author} {\bibfnamefont {D.}~\bibnamefont
  {{Jewitt}}}\ and\ \bibinfo {author} {\bibfnamefont {J.}~\bibnamefont
  {{Luu}}},\ }\href {https://doi.org/10.3847/2041-8213/ab530b} {\bibfield
  {journal} {\bibinfo  {journal} {\apjl}\ }\textbf {\bibinfo {volume} {886}},\
  \bibinfo {eid} {L29} (\bibinfo {year} {2019})},\ \Eprint
  {https://arxiv.org/abs/1910.02547} {arXiv:1910.02547 [astro-ph.EP]}
  \BibitemShut {NoStop}%
\bibitem [{\citenamefont {{Mustill}}\ \emph {et~al.}(2016)\citenamefont
  {{Mustill}}, \citenamefont {{Raymond}},\ and\ \citenamefont
  {{Davies}}}]{Mustill+2016}%
  \BibitemOpen
  \bibfield  {author} {\bibinfo {author} {\bibfnamefont {A.~J.}\ \bibnamefont
  {{Mustill}}}, \bibinfo {author} {\bibfnamefont {S.~N.}\ \bibnamefont
  {{Raymond}}},\ and\ \bibinfo {author} {\bibfnamefont {M.~B.}\ \bibnamefont
  {{Davies}}},\ }\href {https://doi.org/10.1093/mnrasl/slw075} {\bibfield
  {journal} {\bibinfo  {journal} {\mnras}\ }\textbf {\bibinfo {volume} {460}},\
  \bibinfo {pages} {L109} (\bibinfo {year} {2016})},\ \Eprint
  {https://arxiv.org/abs/1603.07247} {arXiv:1603.07247 [astro-ph.EP]}
  \BibitemShut {NoStop}%
\bibitem [{\citenamefont {{Hansen}}\ and\ \citenamefont
  {{Zuckerman}}(2021)}]{Hansen+2021}%
  \BibitemOpen
  \bibfield  {author} {\bibinfo {author} {\bibfnamefont {B.~M.~S.}\
  \bibnamefont {{Hansen}}}\ and\ \bibinfo {author} {\bibfnamefont
  {B.}~\bibnamefont {{Zuckerman}}},\ }\href
  {https://doi.org/10.3847/1538-3881/abd547} {\bibfield  {journal} {\bibinfo
  {journal} {\aj}\ }\textbf {\bibinfo {volume} {161}},\ \bibinfo {eid} {145}
  (\bibinfo {year} {2021})},\ \Eprint {https://arxiv.org/abs/2102.05703}
  {arXiv:2102.05703 [astro-ph.SR]} \BibitemShut {NoStop}%
\end{thebibliography}%

\end{document}